\pdfoutput=1
\documentclass[12pt]{article}
\usepackage{amsmath}
\usepackage{amssymb}
\usepackage[dvips]{graphics}
\usepackage{epsfig}
\usepackage{calc}
\usepackage{amsfonts}
\usepackage{graphicx}
\usepackage[ansinew]{inputenc}
\usepackage{color}

\newcommand{\appendixA}{\setcounter{equation}{0}
\def\theequation{\rm{A}.\arabic{equation}}\section*}

\catcode`\@=11
\def\marginnote#1{}
\def\draftlabel#1{{\@bsphack\if@filesw {\let\thepage\relax
   \xdef\@gtempa{\write\@auxout{\string
      \newlabel{#1}{{\@currentlabel}{\thepage}}}}}\@gtempa
   \if@nobreak \ifvmode\nobreak\fi\fi\fi\@esphack}
        \gdef\@eqnlabel{#1}}
\def\@eqnlabel{}
\def\@vacuum{}
\def\draftmarginnote#1{\marginpar{\raggedright\scriptsize\tt#1}}
\def\draft{\oddsidemargin -.5truein
        \def\@oddfoot{\sl preliminary draft \hfil
        \rm\thepage\hfil\sl\today
        }
        \let\@evenfoot\@oddfoot \overfullrule 3pt
        \let\label=\draftlabel
        \let\marginnote=\draftmarginnote
   \def\@eqnnum{(\theequation)\rlap{\kern\marginparsep\tt\@eqnlabel}
\global\let\@eqnlabel\@vacuum}  }

\def\preprint{\twocolumn\sloppy\flushbottom\parindent 1em
        \leftmargini 2em\leftmarginv .5em\leftmarginvi .5em
        \oddsidemargin -.5in    \evensidemargin -.5in
        \columnsep 15mm \footheight 0pt
        \textwidth 250mmin      \topmargin  -.4in
        \headheight 12pt \topskip .4in
        \textheight 175mm
        \footskip 0pt
        \def\@oddhead{\thepage\hfil\addtocounter{page}{1}\thepage}
        \let\@evenhead\@oddhead \def\@oddfoot{} \def\@evenfoot{} }

\def\titlepage{\@restonecolfalse\if@twocolumn\@restonecoltrue\onecolumn
     \else \newpage \fi \thispagestyle{empty}\c@page\z@
        \def\thefootnote{\fnsymbol{footnote}} }

\def\endtitlepage{\if@restonecol\twocolumn \else  \fi
        \def\thefootnote{\arabic{footnote}} \setcounter{footnote}{0}}

\catcode`@=12 \relax
\def\bea{\begin{array}}
\def\eea{\end{array}}
\def\half{\frac{1}{2}}
\def\bra#1{\left\langle #1\right|}
\def\ket#1{\left| #1\right\rangle}

\def\ov{\overline}

\relax
\def\be{\begin{equation}}
\def\ee{\end{equation}}
\def\ba{\begin{eqnarray}}
\def\ea{\end{eqnarray}}
\def\del{\partial}
\def\d{{\rm d}}
\def\tr{\,{\rm tr}\,}
\def\Tr{\,{\rm Tr}\,}

\def\k{\kappa}
\def\r{\rho}
\def\a{\alpha}
\def\b{\beta}
\def\g{\gamma}

\def\G{\Gamma}
\def\dd{\delta}
\def\D{\Delta}
\def\e{\epsilon}
\def\f{\phi}

\def\p{\psi}
\def\pb{\bar\p}

\def\m{\mu}
\def\n{\nu}
\def\o{\omega}

\def\l{\lambda}
\def\L{\Lambda}
\def\s{\sigma}
\def\S{\Sigma}

\def\cN{{\cal N}}
\def\cM{{\cal M}}
\def\cL{{\cal L}}

\def\cO{{\cal O}}
\def\cD{{\cal D}}
\def\cR{{\cal R}}

\def\t{\theta}

\def\wt{\widetilde}
\def\wh{\widehat}
\def\w{\wedge}
\def\st{{}^*}
\def\cT{{\cal T}}
\def\vf{\varphi}

\def\qsl{{q\hskip-2.mm /}}
\def\ksl{{k\hskip-2.mm /}}
\def\psl{{p\hskip-1.8mm /}}
\def\Asl{{A\hskip-2.2mm /}}
\def\dsl{{\del\hskip-2.2mm /}}

\def\Dsl{{D\hskip-2.7mm /}\,}
\def\Dslf{{D\hskip-2.2mm /}\,} 

\def\trR{{\rm tr}_\cR\,}
\def\trD{{\rm tr}_D\,}
\def\ind{{\rm ind}\,}
\def\Rb{{\overline R}}
\def\Det{{\rm Det}\,}
\def\cU{{\cal U}}
\def\Ub{{\overline U}}
\def\cUb{{\overline\cU}}

\def\ca{{\cal A}}

\def\cF{{\textsf F}}
\def\sa{{\textsf A}}
\def\cA{{\large\textmd a}}
\def\cW{{\cal W}}
\def\cG{{\cal G}}
\def\cC{{\cal C}}
\def\oR{{\overline \cR}}
\def\zb{{\bar z}}
\def\vh{{\widehat v}}

\relax

\renewcommand{\theequation}{\thesection.\arabic{equation}}
\begin{document}
\topmargin-2.4cm
%
%
%
%
\begin{titlepage}
\begin{flushright}
LPTENS-08/05\\
January 2008\\
\end{flushright}
\vskip 2.5cm

\begin{center}{\huge\bf Lectures on Anomalies\footnote{
Based on lectures given at the joint Amsterdam-Brussels-Paris graduate school in theoretical high-energy physics
}}
\\
\vskip 1.5cm 
{\bf Adel Bilal}\\
\vskip.3cm 
Laboratoire de Physique Th\'eorique,
\'Ecole Normale Sup\'erieure - CNRS UMR8549\footnote{
Unit\'e mixte du CNRS et de l'Ecole Normale Sup\'erieure associ\'ee \`a l'UPMC Univ Paris 06 (Pierre et Marie Curie)
}\\
24 rue Lhomond, 75231 Paris Cedex 05, France

\end{center}
\vskip .5cm

\begin{center}
{\bf Abstract}
\end{center}
\begin{quote}
These  lectures on anomalies are relatively self-contained and intended for graduate students in theoretical high-energy physics who are familiar with the basics of 
quantum field theory. More elaborate concepts are introduced when needed.\hfill\break
We begin with several derivations of the abelian anomaly: anomalous transformation of the measure, explicit computation of the triangle Feynman diagram, relation to the index of the Euclidean Dirac operator. The chiral (non-abelian) gauge anomaly is derived by evaluating the anomalous triangle diagram with three non-abelian gauge fields coupled to a chiral fermion. We discuss in detail the relation between anomaly, current non-conservation and non-invariance of the effective action, with special emphasis on the derivation of the anomalous Slavnov-Taylor/Ward identities. We show why anomalies always are finite and local. A general characterization is given of gauge groups and fermion representations which may lead to anomalies in four dimensions, and the issue of anomaly cancellation is discussed, in particular the classical example of the standard model.\hfill\break 
Then, in a second part, we move to more formal developments and arbitrary even dimensions. After introducing a few basic notions of differential geometry, in particular the gauge bundle and characteristic classes, we derive the descent equations. We prove the Wess-Zumino consistency condition and show that relevant anomalies correspond to BRST cohomologies at ghost number one. We discuss why and how anomalies are related via the descent equations to characteristic classes in two more dimensions. The computation of the anomalies in terms of the index of an appropriate Dirac operator in these higher dimensions is outlined. Finally we derive the gauge and gravitational anomalies in arbitrary even dimensions from the appropriate index and explain the anomaly cancellations in ten-dimensional IIB supergravity and in the field theory limits of type I and heterotic superstrings.
\end{quote}

\end{titlepage}
\setcounter{footnote}{0} 
\setlength{\baselineskip}{.6cm}
\newpage
\setcounter{page}{0}
\begin{small}
\tableofcontents
\end{small}

\setcounter{page}{0}
\newpage
%

\setcounter{section}{0}

\section{Introduction\label{intro}}
\setcounter{equation}{0}


Symmetries play an important role in physics in general and in quantum field theory in particular. A symmetry of the classical action is a transformation of the fields that leaves the action invariant. Standard examples are Lorentz, or more generally Poincar\'e transformations, and gauge transformations in gauge theories. One must then ask whether these symmetries are still valid in the quantum theory. 

In the functional integral formulation of quantum field theory, symmetries of the classical action are easily seen to translate into the  Ward identities for the correlation functions or the Slavnov-Taylor identities for the quantum effective action. An important assumption in the proof is that the functional integral measure also is invariant under the symmetry. If this is not true, these Slavnov-Taylor or Ward identities are violated by a so-called anomaly. 

Alternatively, if one is computing (diverging) Feynman diagrams, one has to introduce some regularization and it may happen that no regularization preserves all of the symmetries. Then there is no guarantee that the renormalized Green's functions still display the analogue of the classical symmetry, i.e. satisfy the Ward identities. If they don't, there is an anomaly. Equivalently, one often checks whether a classically conserved current is still conserved at the quantum level. A non-conserved current signals a non-invariance of the quantum effective action, i.e. an anomaly. 

If a global symmetry is anomalous it only implies that  classical selection rules are not obeyed in the quantum theory and classically forbidden processes may actually occur. 

On the other hand, in a (non-abelian) gauge theory, the gauge symmetry is crucial in demonstrating unitarity and renormalizability, and an anomaly in the gauge symmetry would be a disaster. Hence,  in a consistent gauge theory, if present, such anomalies must cancel when adding the contributions due to the various chiral fermions. 

\vskip3.mm

\centerline{* \ * \ *}

\vskip2.mm
These lectures are divided into two parts. The first part (sections \ref{Notation} to \ref{frecan}) is very detailed and mainly concerned with four-dimensional gauge theories, while the second part (starting with section \ref{formaldev}) deals with more formal developments in arbitrary (even) dimensions.

We begin (section \ref{Notation}) by quickly reviewing a few facts about non-abelian gauge symmetries, mainly to fix our notation. In section \ref{fmeasure}, we discuss the possible non-invariance of the functional integral measure for fermions under (global) chiral transformations and how this is related to the abelian anomaly. We explicitly obtain the anomaly from an appropriate regularization of the functional Jacobian determinant under these transformations. Then we derive in detail how this anomaly is linked to the non-conservation of the axial current and to the non-invariance of the effective action. We show the relation with instantons and the index of the (Euclidean) Dirac operator. 

In section \ref{noneffa}, we consider anomalies in general, with emphasis on anomalies under non-abelian gauge transformations. We derive the (anomalous) Slavnov-Taylor identities for gauge theories and, again, relate the anomalous parts to the non-invariance of the fermion measures. Furthermore, the non-invariance of an appropriate effective action and the covariant divergence of the gauge current are shown to be directly given by the anomaly. We spell out in detail the anomalous Ward identities for the correlation functions and how to extract the anomaly from a calculation of  one-loop diagrams (including all the signs and $i$'s). 

In section \ref{tria}, we explicitly evaluate the relevant triangle diagrams in some detail. The Feynman diagram computation for the abelian anomaly exactly reproduces the result already obtained in section \ref{fmeasure}. We then explain why an anomaly under (non-abelian) gauge transformations is expected if the theory contains chiral fermions, and similarly compute the corresponding triangle diagram. To evaluate these Feynman diagrams we use Pauli-Villars regularization which very clearly shows where the anomaly arises in the computation. We exhibit the anomalous and non-anomalous parts of these triangle diagrams and show that they are indeed in agreement with the general structure predicted by the anomalous Ward, resp. Slavnov-Taylor identities derived in section \ref{noneffa}.

In section \ref{lfa}, we show in general why anomalies are necessarily local and finite, and discuss the notion of relevant anomalies. Locality means that the {\it variation} of the effective action is a local functional of the gauge fields.  Finiteness means in particular that the anomalous parts of the one-loop triangle diagrams have a regulator independent limit without the need to add any (non-invariant) counterterms. 

In section \ref{frecan}, we study which gauge groups and which fermion representations lead to anomalies and how to correctly add up the contributions of the different fermions and anti-fermions. We consider the example of the standard model and show that all anomalies cancel within each generation of fermions. This concludes the first part.

The second part of these lectures discusses more formal developments related to gauge and gravitational anomalies in arbitrary even dimensions. In section \ref{formaldev}, we introduce various notions from differential geometry with emphasis on characteristic classes and Chern-Simons forms, necessary to derive the descent equations. 

In section \ref{secWZBRST}, we derive the Wess-Zumino consistency conditions which we reformulate in a BRST language showing that relevant anomalies can be identified with BRST cohomology classes at ghost number one. Then we show how the consistency conditions can be solved in terms of the descent equations, thus characterizing the anomalies by invariant polynomials in two more dimensions, leaving only the overall coefficient undetermined.  

In section \ref{indexth}, we outline how the anomalies in $d$ dimensions are related to the index of  appropriate (Euclidean) Dirac operators in $d+2$ dimensions, which in turn is related to the invariant polynomials. This relation naturally involves the descent equations and fixes the so-far undetermined overall coefficient. We carefully discuss the continuation between Euclidean and Minkowski signature (which is quite subtle for the topological terms), and indeed find perfect agreement with the result of the explicit triangle computation of section \ref{tria} for $d=4$.

In section \ref{gravmixed}, we show how to understand gravitational anomalies as anomalies under local Lorentz transformations, allowing us to treat them in (almost) complete analogy with anomalies under gauge-transformations. We discuss how all the gauge, gravitational and mixed gauge-gravitational anomalies are related to appropriate indices for which we give explicit formulae. 

Finally, in section \ref{anomcanc}, we specialize to ten dimensional gauge, gravitational and mixed gauge-gravitational anomalies. We discuss their cancellation  in ten-dimensional IIB supergravity and in the (field theory limits of) type I and heterotic superstrings. This includes a discussion of anomaly cancellation by Green-Schwarz-type mechanisms and by anomaly inflow.

\vskip3.mm

\centerline{* \ * \ *}

\vskip2.mm
These  notes are based on lectures on anomalies that were part of an advanced quantum field theory course for graduate students in theoretical high-energy physics who were already familiar with the basics of quantum field theory. Staying close to the spirit of lectures, we have made no effort to provide any historical introduction or to give appropriate references. On the other hand, we have made a reasonable effort to make these lectures as self-contained as possible in the sense that we have tried to prove - or at least motivate - most of the claims and statements that are made, rather than refer to the literature.

Of course, the literature on anomalies is abundant: many textbooks on quantum field theory contain at least a chapter on anomalies. Our presentation of anomalies in four dimensions in the first part of these lectures has been much inspired by the treatment in the textbook \cite{Weinbook} whose conventions and notations we mostly adopted. Useful references for the second part of these lectures are e.g. \cite{AGG, AGW, Nak}. Finally, there exist quite a few other lectures, textbooks or reprint volumes on anomalies emphasizing different aspects. A very partial list is \cite{TJZW}-\cite{Harv}.

\newpage
{\huge\centerline{\bf Part I :}
\vskip3.mm
\centerline{\bf Anomalies in non-abelian gauge theories}}

\vskip15.mm

\section{Notations and conventions for gauge theories\label{Notation}}

We begin by introducing some notation and summarizing our conventions  for (non-abelian) gauge theories. 

\subsection{Lie algebra and representations\label{Lialg}}

We take the generators $t_\a$ of the Lie algebra to be hermitian, $t_\a^\dag = t_\a$, and let
\be\label{liealg}
[t_\a,t_\b]=i C^\g_{\ \a\b} t_\g \ ,
\ee
with real structure constants $C^\g_{\ \a\b}$ which satisfy the Jacobi identity $ C^\dd_{\ [\a\b} C^\e_{\ \g]\dd}=0$. If we use a specific representation $\cR$  we write $t_\a^\cR$ or $(t^\cR_\a)^k_{\ l}$ for the
$\dim \cR\times \dim \cR$ matrices of the representation. For compact Lie algebras (i.e.~if $\tr t_\a t_\b$ is positive-definite), all finite dimensional representations are hermitian. This is the case of most interest in gauge theories and, hence, $(t^\cR_\a)^\dag = t^\cR_\a$, but we will not need to assume this in general.\footnote{
Of course, when discussing local Lorentz transformations in sect.~\ref{gravmixed}, the relevant algebra $SO(3,1)$ is not compact and its generators are not all hermitian. Alternatively though, one can work in the Euclidean where the relevant algebra is $SO(4) \simeq SU(2)\times SU(2)$ which is compact. 
} 

The matrices of the adjoint representation are given by
\be\label{adjoint}
(t^{\rm adj}_\a)^\b_{\ \g} =i C^\b_{\ \a\g} \ .
\ee
They satisfy the algebra (\ref{liealg}) thanks to the Jacobi identity. 
One often says that the adjoint representation acts by commutation. This means the following: if some field transforms in the adjoint representation one has e.g. $\dd \f^\g = i \wt\f^\g$ with
\be\label{adj}
\wt\f^\g= (\e^\a\, t^{\rm adj}_\a)^\g_{\ \b}\, \f^\b
\quad\Leftrightarrow\quad
\wt\f^\g\, t^\cR_\g = [\e^\a\, t_\a^\cR, \f^\b\, t^\cR_\b]\ ,
\ee for any (non-trivial) representation $R$. For such fields in the adjoint representation it is convenient to define
\be
\f^\cR=\f^\a t^\cR_\a \ ,
\ee
which now is an element of the (possibly complexified) Lie algebra. Then the previous relation just reads
\be
\wt \f^\cR=[\e^\cR, \f^\cR] \ .
\ee

\subsection{Gauge transformations, covariant derivative and field strength\label{gauge}}

A field in an arbitrary representation $\cR$ transforms under gauge transformations (with real parameters $\e^\a(x)$\ ) as
\be
\dd \p^l(x)=i \e^\a(x) (t^\cR_\a)^l_{\ k} \p^k(x) 
\quad \Leftrightarrow\quad
\dd\p = i \e^\cR \p \ .
\ee
For the conjugate field we have $\dd\p^\dag = -i \p^\dag \e^\cR$ if the representation is hermitian. In general, we will simply write 
$\dd\p^\dag = -i \p^\dag \e^{\Rb}$ with $\Rb=R$ for hermitian representations.

The covariant derivative of such a field $\p$ transforming in a representation $\cR$ is
\be\label{covder}
(D_\m \p)^l = \del_\m\p^l -i A_\m^\a (t^\cR_\a)^l_{\ k} \p^k 
=\del_\m\p^l -i (A_\m^\cR)^l_{\ k} \p^k 
\quad \Leftrightarrow\quad
D_\m\p = \del_\m\p - i A^\cR_\m\p \ .
\ee
If the field $\p$ is in the adjoint representation, i.e it has components $\p^\a$, we must use $A^{\rm adj}_\m$ and according to (\ref{adj}) we can then write  $D_\m \p^{\cR'} = \del_\m \p^{\cR'}-i[A_\m^{\cR'}, \p^{\cR'}]$ with $\p^{\cR'}=\p^\a t^{\cR'}_\a$, for any $\cR'$. In particular, for the gauge transformation parameters $\e$ one has ($\e^\cR=\e^\a t^\cR_\a$)
\be
D_\m \e^\cR=\del_\m \e^\cR - i[A_\m^\cR,\e^\cR] \ .
\ee
The covariant derivative (\ref{covder}) of $\p$ transforms just as $\p$ itself, provided the gauge field $A_\m$ transforms at the same time as
\be\label{Avar}
\dd A_\m^\a=\del_\m\e^\a +C^\a_{\ \b\g}A_\m^\b \e^\g 
=\del_\m\e^\a - i A^\b_\m  (t^{\rm adj}_\b)^\a_{\ \g} \e^\g \ .
\ee
Using (\ref{adj}) this can be rewritten consisely as
\be\label{Avar2}
\dd A_\m^\cR=\del_\m \e^\cR - i [A_\m^\cR,\e^\cR] = D_\m \e^\cR
\quad \Leftrightarrow\quad
\dd A_\m^\a = (D_\m \e)^\a \ .
\ee
It is then clear that if $\cL_{\rm matter}[\pb,\p,\del_\m\p,\del_\m\pb]$ is invariant under $\dd\p=i\e^\cR\p$ and $\dd\pb=-i\pb\e^\Rb$ for constant $\e$ then $\cL_{\rm matter}[\pb,\p,D_\m\p,D_\m\pb]$ is invariant under the same transformations with local $\e(x)$ if also $A_\m(x)$ transforms as in (\ref{Avar}), resp (\ref{Avar2}).

The gauge field strength is defined as the commutator of two covariant derivatives:
\be\label{covdercom}
[D_\m,D_\n]\p=-i F^\a_{\m\n} t^\cR_\a \p \equiv -i F^\cR_{\m\n} \p \ .
\ee
This is guaranteed to transform like $\p$ and hence
\be
\dd F_{\m\n}^\cR= i [\e^\cR,F^\cR_{\m\n}]
\quad \Leftrightarrow\quad
\dd F^\a_{\m\n} = C^\a_{\ \b\g} F^\b_{\m\n} \e^\g \ ,
\ee
which is just as the transformation of $A_\m$ but without the inhomogeneous term $\sim\del_\m\e$\ : $F_{\m\n}$ transforms in the adjoint representation. Computing $[D_\m,D_\n]$ yields
\be
F^\cR_{\m\n}=\del_\m A^\cR_\n - \del_\n A^\cR_\m -i[A^\cR_\m, A^\cR_\n]
\quad \Leftrightarrow\quad
F^\a_{\m\n}
=\del_\m A^\a_\n - \del_\n A^\a_\m +C^\a_{\ \b\g} A^\b_\m A^\g_\n 
\ \ .
\ee
From its definition it is easy to show that the field strength satisfies the Bianchi identity
\be\label{Bianchi}
D_{[\m} F_{\n\r]}=0 \ ,
\ee
where the brackets indicate anti-symmetrisation  of the indices (always normalized such that for an antisymmetric tensor $f_{[a_1\ldots a_p]}=f_{a_1\ldots a_p}$). Here and in the following we often suppress the labels ``$\cR$" or $\a$ unless there explicit writing is likely to avoid confusion.

\subsection{Action and field equations\label{actandfields}}

For any Lie algebra that is a direct sum of commuting compact simple and $U(1)$ subalgebras, $G=\oplus_i G_i$, there exists a real symmetric positive $g_{\a\b}$ such that $g_{\a\dd}C^{\dd}_{\ \b\g} +g_{\b\dd}C^{\dd}_{\ \a\g}=0$ and, equivalently, there exists a basis of generators for which the $C^\a_{\ \b\g}$ are totally antisymmetric in all three indices (one then often writes $C_{\a\b\g}$). In this latter basis the matrix $g$ commutes with the generators in the adjoint representation and, by Schur's lemma, the matrix $g$ must then be block-diagonal and in each block corresponding to each $G_i$ be proportional to the identity. We will call these constants of proportionality ${1\over g_i^2}$. It follows that $g_{\a\b} F^\a_{\m\n} F^{\b\m\n}$ is gauge invariant and equals $\sum_i {1\over g_i^2} \sum_{\a_i=1}^{\dim G_i} F^{\a_i}_{\m\n} F^{\a_i\,\m\n}$. By rescaling $A_\m^\a$ and $F_{\m\n}^\a$ for each $G_i$ by $g_i$ one can absorb the factors ${1\over g_i^2}$. However, the ``coupling" constants $g_i$ would appear explicitly in the expressions of the covariant derivatives, gauge field strengths etc, always accompanying the structure constants or commutators. This can be avoided if one in turn redefines the Lie algebra generators and the structure constants to include the $g_i$. Then all previous formulae remain valid, except that we have to remember that the $t_\a$ and $C^\a_{\ \b\g}$ implicitly contain the coupling constants, one $g_i$ for each simple or $U(1)$ factor $G_i$. In particular, for any simple factor $G_i$ one then has
\be\label{tanorm}
\trR t_\a t_\b=g_i^2\, C_R^{(i)}\, \dd_{\a\b} \ ,
\ee
where e.g. for $G_i=SU(N)$ we have $C_{\rm adj}=N$ and $C_N=C_{\overline N}={1\over 2}$.
The unique Lorentz and gauge invariant Lagrangian quadratic in the field strengths then is $\cL_{\rm gauge}[F_{\m\n}]=-{1\over 4} F_{\m\n}^\a F^{\a\m\n}$. Of course, there is one more possibility, $\theta_{\a\b} \e^{\m\n\r\s} F^\a_{\m\n} F^\b_{\r\s}$ but this is a total derivative. Hence the Lagrangian for matter and gauge fields is
\be\label{gmlag}
\cL=-{1\over 4} F_{\m\n}^\a F^{\a\m\n}
+\cL_{\rm matter}[\pb,\p,D_\m\p,D_\m\pb] \ .
\ee
In the present discussion of anomalies we actually do not need the precise form of the gauge field Lagrangian, only the matter part will be important. Hence we do not have to discuss the issue of gauge fixing or  what the precise form of the gauge field propagator is. 
One could even add higher-order terms like $F^\a_{\m\n}F^{\b\,\n\r}F^\g_{\r\s}F^{\dd\,\s\m} \tr t_{(\a}t_\b t_\g t_{\dd)}$ as appear e.g.~in the string theory effective action, without affecting the discussion of the anomalies. 

Finally, recall that the (classical) matter current is defined as 
\be\label{matcurr}
J_{\rm matter}^{\a\,\m}={\del\cL_{\rm matter}\over \del A_\m^\a}\ ,
\ee
and that the Euler-Lagrange equations for the gauge fields as following from the Lagrangian (\ref{gmlag}) yield
\ba\label{Euler}
\del_\m {\del\cL\over \del(\del_\m A_\n^\a)} 
= {\del\cL\over \del A_\n^\a}\quad &,&\quad
{\del\cL\over \del(\del_\m A_\n^\a)}=-F^{\a\,\m\n} \quad\ , \quad\
{\del\cL\over \del A_\n^\a}=C^\g_{\ \a\b} A_\m^\b F^{\g\,\m\n} + J_{\rm matter}^{\a\,\n} \ ,
\nonumber\\
\nonumber\\
\quad &\Rightarrow&\quad
\big(D_\m F^{\m\n}\big)^\a=-J_{\rm matter}^{\a\,\n} \ .
\ea
It then follows that the matter current is covariantly conserved if the field equations are satisfied\footnote{
We have $(D_\n J^\n_{\rm matter})^\a=-(D_\n D_\m F^{\m\n})^\a={1\over 2} ([D_\m,D_\n] F^{\m\n})^\a=-{i\over 2} F_{\m\n}^\b (t^{\rm adj}_\b)^\a_{\ \g} F^{\g\,\m\n}={1\over 2} C^\a_{\ \b\g} F^\b_{\m\n} F^{\g\,\m\n}=0$ by antisymmetry of the $C^\a_{\ \b\g}$.
}:
\be\label{currentcons}
D_\n J^\n_{\rm matter}=0 \ .
\ee
This expresses the gauge invariance because it translates the fact that the gauge field enters the gauge kinetic part of the Lagrangian only through the gauge covariant combination $F_{\m\n}$.
One of the equivalent manifestations of the anomaly is that at the quantum level $D_\n \langle J^\n_{\rm matter}\rangle \ne 0$.

\subsection{Further conventions\label{furtherconv}}

Our Minkowski space signature is $(-+++)$. The Dirac matrices then satisfy $(i\g^0)^2=(\g^j)^2=1,\ j=1,2,3$ as well as $(i\g^0)^\dag=i\g^0,\ (\g^j)^\dag=\g^j$ and, of course, $\g_0=-\g^0$. We define $\pb=\p^\dag i \g^0$ and
\be\label{g5fourdim}
\g_5=i\g_0\g_1\g_2\g_3 = -i \g^0\g^1\g^2\g^3 \quad \Rightarrow \quad
\g_5^2=1 \ , \quad \g_5^\dag=\g_5 \ ,\quad \{\g_5,\g^\m\}=0\ .
\ee
As usual, we denote $\dsl=\g^\m\del_\m$\,, $\ \Asl=\g^\m A_\m$ and $\Dsl=\g^\m D_\m\,$, etc.
The  completely antisymmetric $\e$-tensor in Minkowski space is defined as
\be\label{4depstensor}
\e^{0123}=+1 \quad \ , \quad \e_{0123}=-1 \ ,
\ee
so that we have for the Dirac trace
\be\label{epstrace}
\trD \g_5 \g^\m \g^\n \g^\r \g^\s = 4 i\, \e^{\m\n\r\s} \ .
\ee
Finally recall that when evaluating four-dimensional momentum integrals in Minkowski space the Wick rotation results in a factor of $i$ according to
\be\label{Wick}
\int \d^4 p\ f(p_\m p^\m)= i \int \d^4 p_{\rm E}\ f(p_{\rm E}^2) \ ,
\ee
where $p_0=i\, p_{\rm E}^0$. The continuation to Euclidean signature will be discussed in greater detail in subsection \ref{Euclidean}.

\section{Transformation of the fermion measure and the example of the abelian anomaly\label{fmeasure}}
\setcounter{equation}{0}

We begin by studying the probably simplest example of anomaly: the so-called abelian anomaly. This is an anomaly of chiral transformations (i.e. involving $\g_5$) of massless Dirac fermions. A nice way to understand the origin of this anomaly is as a non-invariance of the fermion measure in the functional integral under such transformations. On the other hand, a symmetry corresponds to a conserved current, and a violation of a symmetry to a current non-conservation.  In turn, such a non-conservation translates a non-invariance of some appropriately defined quantum effective action. In this section, we will study these issues and how they are related. Before doing so, let us only mention that, historically, the abelian anomaly was one of the first places an anomaly showed up, and that it played an important role in explaining the observed decay rate of a neutral pion into two photons.

\subsection{Why the matter measure matters\label{mattermeasure}}

Consider a massless complex spin ${1\over 2}$ fermion in some representation $\cR$ of the gauge group. We suppose it has a standard interaction with the gauge field, i.e. it couples to the gauge field via the covariant derivative without any occurrence of the chirality matrix  $\g_5$. The matter Lagrangian then is 
\be\label{matterlagr}
\cL_{\rm matter}[\p,\pb, D_\m\p,D_\m\pb]=-\pb \Dsl\, \p \equiv -\pb (\dsl-i \Asl^\cR) \p \ .
\ee

When using the functional integral to compute vaccum expectation values of time-ordered products of operators $\widehat\cO_i$ that involve the matter fields, like e.g.~products of (matter) currents, one may proceed in two steps: first compute the functional integral over the matter fields alone:
\be\label{matterint}
\int \cD \p \cD\pb\ \cO_1(x_1) \ldots \cO_N(x_N)\ 
e^{i \int \cL_{\rm matter}[\p,\pb, D_\m\p,D_\m\pb]} \ ,
\ee
and then do the remaining functional integral over the gauge fields. It is only in the second step that one has to deal with all the complications of gauge-fixing and ghosts, while the appearance of any anomalies is entirely related to the evaluation of (\ref{matterint}). An arbitrary $S$-matrix element, resp. Feynman diagram can be reconstructed from (matter) current correlators and gauge field (as well as ghost) propagators. Hence it is enough to evaluate (\ref{matterint}) for the case where the operators $\widehat\cO_i$ are the quantum operators corresponding to the currents $J_{\rm matter}^{\a\,\m}={\del\cL_{\rm matter}\over\del A_\m^\a}$ . Such vacuum expectation values of time-ordered products of currents can then be obtained by taking functional derivatives\footnote{
If the matter currents $J^{\a\m}$ do not only involve the matter fields but also the gauge fields, as would be the case for scalar matter, taking multiple functional derivatives would also give rise to contact terms, rather than just products of currents. Since we are interested in fermionic matter with Lagrangians like (\ref{matterlagr}) this complication does not occur here.
} 
with respect to the $A_\m^\a$ of the  ``effective" action $\wt W[A]$ defined as
\be\label{Wtilde}
\begin{array}{|c|}
\hline\\
e^{i \wt W[A]}= \int \cD \p \cD\pb\ 
e^{i \int \cL_{\rm matter}[\p,\pb, D_\m\p,D_\m\pb]} \ .\\
\\
\hline
\end{array}
\ee
We will discuss this in more detail below.

Let us now study the effect of a local transformation 
\be
\p(x)\to \p'(x)=U(x) \p(x)
\quad , \quad
\pb(x) \to \pb'(x) =\pb(x) \Ub(x) 
\quad , \quad
\Ub(x)=i\g^0 U^\dag(x) i\g^0\ ,
\ee 
where $U(x)$ is a unitary matrix acting on the indices of the representation of the gauge group and on the indices of the Clifford algebra. Of course, the matter action is {\it not} invariant under such a transformation, unless $U$ is constant or the gauge field $A_\m$ transforms appropriately. Also, the $\p$ and $\pb$ could carry  additional representations of a global symmetry group (usually called a flavor symmetry) and then, if $U$ is an element of   such a global symmetry group, it must be constant to leave the action invariant. At present, however, we are only interested in the transformation of the fermion measures $\cD\p$ and $\cD\pb$. Since these are measures for anticommuting fields they transform with the inverse Jacobian:
\be
\cD\p\to \cD\p'= (\Det \cU)^{-1}\, \cD\p 
\quad , \quad \cD\pb\to \cD\pb'= (\Det \overline\cU)^{-1}\, \cD\pb \ ,
\ee
where the operators $\cU$ and $\cUb$ are given by
\be
\bra{x} \cU \ket{y} = U(x)\, \dd^{(4)}(x-y) \quad , \quad
\bra{x} \cUb \ket{y} = \Ub(x)\, \dd^{(4)}(x-y) \ .
\ee
We will distinguish the two cases of non-chiral and chiral transformations.

\subsection{Unitary non-chiral transformation\label{unitnonchir}}

Let $U$ be a unitary non-chiral transformation (not involving $\g_5$) of the form
\be
U(x)=e^{i \e^\a(x) t_\a} \quad , \quad {\rm with}\quad t_\a^\dag=t_\a \quad {\rm and}\quad [\g^\m,t_\a]=0 \ .
\ee
The matrices $\e^\a t_\a$ could be also replaced by any generator of the global flavor symmetry group or some combination of both.  The important fact is that
\be
\Ub(x)=i\g^0 e^{-i \e^\a(x) t_\a} i\g^0=e^{-i \e^\a(x) t_\a} (i\g^0)^2=e^{-i \e^\a(x) t_\a} =U^{-1}(x) \ ,
\ee
so that
\be
\cUb=\cU^{-1} 
\quad\Rightarrow\quad
(\Det\cU)^{-1} \ (\Det\cUb)^{-1} =1 \ ,
\ee
and the fermion measure is invariant. In particular, the fermion measure is invariant under gauge transformations. One can then derive the Slavnov-Taylor identities in the usual way without having to worry about an anomalous transformation of the measures. We conclude that for matter fields that couple non-chirally to the gauge fields there are no anomalies.

The reason why we insisted on non-chiral couplings in the matter Lagrangian (\ref{matterlagr}) is the following: of course, the functional determinants $\Det\cU$ and $\Det\cUb$ should be computed using some appropriate regularization. Such a regularization corresponds to regulating the full fermion propagator in the presence of the gauge field. This must be done in a gauge invariant way if we are not to spoil gauge invariance from the beginning. As we will see below, this is problematic if the interactions include some chirality matrix or chirality projector.

\subsection{Unitary chiral transformation\label{unitchir}}

Now consider the case where
$U$ is a unitary {\it chiral} transformation, i.e. involving $\g_5$ (defined in (\ref{g5fourdim})), of the form
\be\label{chiraltra}
U(x)=e^{i \e^\a(x) t_\a \g_5} \quad , \quad {\rm with}\quad t_\a^\dag=t_\a \quad {\rm and}\quad [\g^\m,t_\a]=0 \ .
\ee
Note that since $\g_5^\dag= \g_5$ the transformation (\ref{chiraltra})  is indeed unitary, but since $\g_5$ anticommutes with $\g^0$ we now have
\be
\Ub(x)=i\g^0 e^{-i \e^\a(x) t_\a \g_5} i\g^0=e^{+i \e^\a(x) t_\a \g_5} (i\g^0)^2=e^{+i \e^\a(x) t_\a\g_5} =U(x) \ ,
\ee
so that now
\be
\cUb=\cU 
\quad\Rightarrow\quad
(\Det\cU)^{-1} \ (\Det\cUb)^{-1} =(\Det\cU)^{-2} \ ,
\ee
which does not necessarily equal unity and which we need to compute. As usual for an ultra-local integral kernel, we have $\bra{x}\cU^2\ket{y}= \int \d^4 z \bra{x}\cU\ket{z} \bra{z}\cU \ket{y}=\int \d^4 z\, U(x) \dd^{(4)}(x-z) U(z) \dd^{(4)}(z-y)=U^2(x) \dd^{(4)}(x-y)$, and similarly for all powers of $\cU$, so that $\bra{x} f(\cU)\ket{y}= f(U(x))\ \langle x\ket{y}$ and
\be
\Tr\log\cU = \int \d^4 x \bra{x} \tr\log(\cU) \ket{x}
=\int \d^4 x\, \dd^{(4)}(x-x) \tr \log(U(x))
=\int \d^4 x\, \dd^{(4)}(0) i\e^\a(x) \tr  t_\a \g_5 \ ,
\ee
where $\Tr$ is a functional and matrix trace, while $\tr$ is only a matrix trace (with respect to the $\g$ matrices and the gauge and possibly flavor representation matrices). It follows that
\be
(\Det \cU)^{-2}=e^{-2\Tr \log\cU}= e^{i \int\d^4 x\, \e^\a \cA_\a(x)}
\quad {\rm with}\quad
\cA_\a(x)=-2  \dd^{(4)}(0) \tr  t_\a \g_5 \ ,
\ee
where $\cA_\a(x)$ is called the anomaly function or simply the anomaly.

Clearly, the above expression for the anomaly is ill-defined and needs regularization. As it stands, it is the product of an infinite $\dd^{(4)}(0)$ and a vanishing $\tr \g_5 t_\a$. The former actually is
\be
\dd^{(4)}(0)=\langle x\ket{x}=\int\d^4 p\, \langle x\ket{p}\bra{p} x\rangle=\int{\d^4 p\over (2\pi)^4} e^{ip(x-y)}\Big\vert_{x=y}\ ,
\ee 
and thus is a UV divergence. 
A regularization is achieved by cutting off the large momentum contributions, e.g.~ by replacing $\int\d^4 x\, \e^\a \cA_\a(x)=-2 \Tr \cT$ where $\cT= \e^\a(\hat x) \g_5 t_\a$ by 
\be\label{TLdef}
\int\d^4 x\, \e^\a \cA_\a(x)=-2 \lim_{\L\to\infty}\Tr \cT_\L
\quad , \quad {\rm where}\quad
\cT_\L=\e^\a(\hat x) \g_5 t_\a\ f\big( (i\hat\Dsl/ \L)^2\big) \ ,
\ee
with some smooth function $f(s)$ satisfying $f(0)=1$, $f(\infty)=0$, as well as $s f'(s)=0$ at $s=0$ and at $s=\infty$. One could take e.g. $f(s)=e^{-s}$ or $f(s)={1\over s+1}$. We denoted $\hat\Dsl$ the (quantum mechanical) operator such that $\bra{\chi} \hat\Dsl \ket{x}=\Dsl \langle \chi\ket{x}$. We will discuss below why one should use the gauge-covariant $\Dsl^2$ in the cutoff rather than a simple $\del_\m\del^\m$.

While for any fixed matrix element we have $\lim_{\L\to\infty} \bra{\f_n}\cT_\L\ket{\f_m} = \bra{\f_n}\cT\ket{\f_m}$, we have for the trace
\ba\label{regtracecomp}
\Tr \cT_\L &\hskip-2.mm=\hskip-2.mm& \int \d^4 x\ \tr \bra{x} \e^\a(\hat x) \g_5 t_\a f\big( (i\hat\Dsl/ \L)^2\big) \ket{x} 
=\int \d^4 x\ \e^\a(x) \int\d^4 p\, \langle x\ket{p} \tr \g_5 t_\a \bra{p} f\big( (i\hat\Dsl/ \L)^2\big) \ket{x}
\nonumber\\
&=\hskip-2.mm&\int \d^4 x\ e^\a(x) \int{\d^4 p\over (2\pi)^4} e^{ipx} \tr \g_5 t_\a\ \underbrace{f\left( -{1\over \L^2}\left[ \g^\m \left( {\del\over \del x^\m} - i A_\m^\cR(x)\right) \right]^2\right) e^{-ipx}}
\nonumber\\
&&\hskip5.6cm \e^{-ipx} f\left( -{1\over \L^2}\left[ \g^\m \left( {\del\over \del x^\m} -i p_\m - i A_\m^\cR(x)\right) \right]^2\right) 
\nonumber
\\
&=\hskip-2.mm&\int \d^4 x\ \e^\a(x) \int{\d^4 p\over (2\pi)^4} \tr \g_5 t_\a\ f\left( -{1\over \L^2}\left[  -i \psl+\Dsl\right]^2\right) 
\nonumber\\
&=\hskip-2.mm&
\int \d^4 x\ \e^\a(x) \L^4 \int{\d^4 q\over (2\pi)^4} \tr \g_5 t_\a\ f\left( -\left[  -i \qsl+{\Dsl\over \L}\right]^2\right) \ .
\ea
One then expands $f\left( -\left[  -i \qsl+\Dsl/\L\right]^2\right)= f\left( q^2 +2i q^\m D_\m/\L - \Dsl^2/\L^2\right)$ in a Taylor series around $q^2$. A non-vanishing Dirac trace with the $\g_5$ requires at least four $\g$-matrices, while a non-vanishing limit as $\L\to\infty$ requires at most four $\Dsl/\L$. This picks out the term ${1\over 2} f''(q^2) (-\Dsl^2/\L^2)^2$ in the Taylor series, so that
\be\label{TLI}
\lim_{\L\to\infty} \Tr \cT_\L
=\int \d^4 x\ \e^\a(x)  \int{\d^4 q\over (2\pi)^4}  
{1\over 2} f''(q^2)\tr \g_5 t_\a\, (-\Dsl^2)^2\ .
\ee
Somewhat loosely speaking one could say that the regularized momentum integral is $\sim\L^4$ while the regularized $\tr t_\a \g_5$ is $\sim {1\over \L^4}$, combining to give a finite result.
The integral over $q$ in (\ref{TLI}) is easily evaluated after performing the Wick rotation: 
\ba
\int{\d^4 q\over (2\pi)^4} {1\over 2} f''(q^2) 
&=&  {i\over 2(2\pi)^4} {\rm vol}(S^3) \int_0^\infty \d q\, q^3 f''(q^2)
= {i\over 2(2\pi)^4} 2\pi^2 {1\over 2} \int_0^\infty \d\xi\, \xi f''(\xi) 
\nonumber\\
&=&  {i\over 32\pi^2}\left( \xi f'(\xi)\Big\vert^\infty_0 - \int_0^\infty \d\xi\, f'(\xi) \right) 
= {i\over 32\pi^2} \ ,
\ea
where we used the properties of $f$ at $0$ and $\infty$. On the other hand, the trace over Dirac indices and the gauge group representation involves $\Dsl^2={1\over 2} \{\g^\m,\g^\n\} D_\m D_\n + {1\over 2} [\g^\m,\g^\n] D_\m D_\n =D^\m D_\m -{i\over 4} [\g^\m,\g^\n] F_{\m\n}$. Using (\ref{4depstensor}) and (\ref{epstrace}) 
we get
\be
\quad \tr \g_5 t_\a\, (-\Dsl^2)^2 
= \left({i\over 4} \right)^2 
\trD \g_5 [\g^\m,\g^\n] [\g^\r,\g^\s]\ 
\trR t_\a F_{\m\n}F_{\r\s}
=-i \e^{\m\n\r\s} \trR t_\a F_{\m\n}F_{\r\s} \ .
\ee
Putting everything together we finally obtain 
\be\label{TLF}
\lim_{\L\to\infty} \Tr\cT_\L 
= {1\over 32\pi^2}\int \d^4 x\, \e^\a\, \e^{\m\n\r\s} \trR t_\a F_{\m\n}F_{\r\s} \ ,
\ee 
and hence for
the anomaly function
\be\label{anfunc}
\cA_\a(x)= -{1\over 16\pi^2}\e^{\m\n\r\s} \trR t_\a F_{\m\n}(x)F_{\r\s}(x) \ .
\ee
Note that this anomaly depends only on the combination $F_{\m\n}=F_{\m\n}^\a t_\a$, where the $t_\a$ include an explicit factor of the gauge coupling constant (c.f.~(\ref{tanorm})), while the $F_{\m\n}^\a$ are normalized with the canonical kinetic term as in (\ref{gmlag}). Thus, for a simple group with a single gauge coupling constant $g$, we see that the anomaly is\footnote{
This holds in perturbation theory where the leading term in $F^\a_{\m\n}\sim\del_\m A_\n^\a-\del_\n A^\a_\m$ does not give any $g$-dependence. However, for a non-perturbative configuration (like an instanton) one can well have $A_\m^\a\sim {1\over g}$.
}
$\cA_\a \sim \cO(g^3)$.
 
\subsection{A few remarks\label{remarks}}

Let us first explain why the anomaly we have just computed corresponds to a one-loop effect. One can introduce a formal loop-counting parameter by rescaling the action as $S\to {1\over \l} S$. Then in computing Feynman diagrams, the propagators get an extra factor $\l$ while the vertices get an extra ${1\over \l}$. Thus every Feynman diagram comes with a factor $\l^{I-V}$, where $I$ is the number of internal lines and $V$ the number of vertices. By a well-known relation, one has $I-V=L-1$ with $L$ being the number of loops, and one sees that $\l$ is a loop counting parameter. Since the classical action comes with a ${1\over \l}$ it is the tree-level contribution ($L=0$) to the effective action, while the anomaly, like any determinant, has no factor of $\l$ and corresponds to a one-loop contribution ($L=1$). Together with the above observation that the anomaly is of order $g^3$, we can already infer that it corresponds to a 3-point one-loop diagram, i.e. a triangle diagram. This will indeed be confirmed below.

It may seem surprising that the Jacobian for the transformation of the fermion measures, under the transformation $U$ that does not involve the gauge fields, equals $e^{i\int \d^4 x\, \e^\a \cA_\a}$ with an anomaly $\cA_\a(x)$ that does depend on the gauge fields. The reason for this is that we  used a gauge invariant regulator $f(-\Dsl^2/\L^2)$ to make sense of the otherwise ill-defined determinants. Had we used  $f(-\dsl^2/\L^2)$ instead, no gauge field would have appeared and we would have found $\cA_\a=0$ which might seem more satisfactory at first sight. However, such a regulator actually breaks gauge invariance. Indeed, the way we regulate the determinant is not just a matter of once computing $\Det\cU$. The fermion measure plays a crucial role in computing e.g. the ``effective action" $\wt W[A]$ as defined in (\ref{Wtilde}) with the result $\wt W[A] \sim \log \Det \Dsl$.
Now $\wt W[A]$ should be gauge invariant (if possible) and thus $\Det\Dsl$ should be regularized in a gauge invariant way (if possible). Consistency requires that {\it all} fermion determinants are regularized in the same way. Also, as just explained, computing determinants corresponds to computing one-loop Feynman diagrams, and the regularization of the determinants corresponds to a regularization of the fermion Feynman propagator which, again, should be done in a gauge invariant way. Once we have decided a regularization for the propagator, this will provide one and the same regularization for all fermion determinants.

Still another way to see why we must use a gauge invariant regulator is the following. One could define the fermion measure $\cD\p$ as $\prod_n \d c_n$ where the $c_n$ are the coefficients in an expansion $\p(x)=\sum_n c_n \p_n(x)$ on some orthonormal basis $\{\p_n\}$, and similarly for $\cD\pb$.
To compute $\Det\cU$ we might then determine the infinite matrix $\,\cU_{nm}=\bra{\p_n}\cU \ket{\p_m}$ and compute $\det(\cU_{nm})$. Then $\det(\cU_{nm})$ must be regularized by ``cutting off" the ``high-frequency" modes, i.e. the large eigenvalues of some appropriate operator $D$. We can take the $\p_n$ to be eigenfunctions of such $D$: $D\p_n=\l_n\p_n$, and then insert a cutoff function $f(|\l_n|/\L^2)$ when computing $\Tr\log \cU=\sum_n (\log \cU)_{nn}$. Now, we do not want the change of variables $\p\to \p'=U\p$ to break gauge invariance, and hence $\Det\cU$ should be gauge invariant (if possible) and we must use a gauge invariant regulator. Hence the $\l_n$ must be the eigenvalues of a gauge invariant operator $D$ like e.g. $\Dsl=\dsl-i\Asl$. Obviously, the gauge invariant regulator introduces a gauge-field dependence into the regularization procedure and results in a gauge-field dependent Jacobian and gauge-field dependent anomaly.\footnote{
One might still wonder what would happen if one tried to use the gauge invariant regulator $f\big( -D_\m D^\m /\L^2\big)$ instead of $f\big( -\Dslf\Dslf/\L^2\big)$ to regularize $\Tr \e^\a \g_5 t_\a$~? Then the regulator contains no $\g$-matrices and the result would vanish. Does this mean that we can find a regulator that preserves gauge invariance and chiral invariance? Certainly not. In the above computation we used a plane-wave basis to evaluate the trace over the fermionic Hilbert space. This is not fully correct since we are dealing with fermions in an external gauge field $A_\m$ and the appropriate space on which one should take the trace is precisely spanned by the eigen-functions of $i\Dslf$, not those of $D_\m D^\m$. In the above computation we ``corrected" for choosing a slightly inexact basis by using the regulator $f\big( -\Dslf\Dslf/\L^2\big)$.
}

\newpage
\subsection{The abelian anomaly and current (non)-conservation\label{currnonco}}

Although we have done the computation of the anomaly function $\cA_\a(x)$ for a rather general $U(x)=e^{i\e^\a t_\a \g_5}$, a very important class of applications concerns the case where only the $\e^\a$ corresponding to an abelian subgroup are non-vanishing. Then one simply writes
\be\label{abchir} 
U(x)=e^{i \e(x)\, t\,\g_5}\quad ,\quad t^\dag=t\ , \quad
[t,t_\a]=0 \ , \quad [t,\g_5]=0 \ .
\ee
In this case (\ref{anfunc}) reduces to
\be\label{anfuncab}
\cA(x)= -{1\over 16\pi^2}\e^{\m\n\r\s} \trR t F_{\m\n}(x)F_{\r\s}(x) \ ,
\ee
and $(\Det \cU)^{-2}=e^{i \int\d^4 x\, \e(x) \cA(x)}$. This is called the {\it abelian anomaly}. Although it is associated to a chiral transformation, the name ``chiral anomaly" will be reserved for a different anomaly to be studied later-on. Note that since $[t,t_\a]=0$, the abelian anomaly (\ref{anfuncab}) is {\it gauge invariant}. Furthermore, for {\it constant} $\e$, the transformation $\p\to U\p,\ \pb\to\pb \Ub$ is a symmetry of the matter Lagrangian (\ref{matterlagr}) since $\g^\m U=U^{-1} \g^\m$ and hence
\be
\pb\Dsl\,\p \to \pb\Ub\Dsl\, U\p=\pb U \Dsl\, U\p = \pb U U^{-1}\Dsl\,\p=\pb\Dsl\,\p \ .
\ee
As for any local transformation (with parameters $\e^a(x)$) of the fields  that is a symmetry of the Lagrangian if the parameters $\e^a$ are taken to be constant, we can associate a conserved current $J_a^\m(x)$ according to\footnote{
Note that for an abelian {\it gauge} transformation of the matter fields this is compatible with the definition (\ref{matcurr}), and in particuler yields the same sign for the current.}
\be\label{currentdef}
\dd S= \int \d^4 x\, (-J_a^\m(x)) \del_\m \e^a(x) =
\int \d^4 x\, (\del_\m J_a^\m(x))  \e^a(x)  \ .
\ee
Indeed, the variation of the action must be of this form, since we know that $\dd S=0$ if $\del_\m\e^a=0$. On the other hand, if the fields satisfy the field equations we must have $\dd S=0$ for any variation of the fields and in particular for the one induced by a local $\e^a(x)$, so that in this case 
\be
\del_\m J_a^\m(x)\big\vert_{\rm field\ equations} =0 \ .
\ee
Applying this to the above abelian chiral transformation $U(x)$ and the matter action $S_{\rm mat}[\p,\pb,A]\equiv\int \cL_{\rm matter}[\p,\pb,D_\m\p,D_\m\pb]=-\int \pb\Dsl\p$ we find
\be
J_5^\m=+ i\pb\g^\m  \g_5 t\p \ ,
\ee
and classically, i.e. if the $\p$ and $\pb$ satisfy their field equations, we know that this so-called axial current is conserved.

Let us now investigate what happens in the quantum theory and how the current non-conservation is related to the anomaly. As in the usual proof of Slavnov-Taylor identities (cf.~subsection \ref{AST} below for more details) one writes the functional integral, changes integration variables from $\p$ and $\pb$  to $\p'=U\p$ and $\pb'=\pb \Ub$ and uses the transformation properties of the action and now also of the measure:
\ba\label{cnc}
\hskip-1.2cm
\int\cD\p\cD\pb e^{i S_{\rm mat}[\p,\pb,A_\m]} 
\hskip-2.mm&=&\hskip-3.mm
\int\cD\p'\cD\pb'\ e^{i S_{\rm mat}[\p',\pb',A_\m]} 
\nonumber\\
\hskip-2.mm&=&\hskip-3.mm
\int\cD\p\cD\pb\ e^{i\int\d^4 x\, \e(x)\cA(x)} \
e^{iS_{\rm mat}[\p,\pb,A_\m] +i\int\d^4 x\, \e(x)\del_\m J_5^\m(x)}
\nonumber\\
\hskip-2.mm&=&\hskip-3.mm
\int\cD\p\cD\pb\ e^{i S_{\rm mat}[\p,\pb,A_\m]}\left[ 1+ i \int\d^4 x\, \e(x)\Big( \cA(x) + \del_\m J_5^\m(x)\Big) + \cO(\e^2)\right]  ,
\ea
from which we conclude
\be\label{J5noncons}
\begin{array}{|c|}
\hline\\
-\del_\m \langle J_5^\m(x)\rangle_A = \cA(x) 
= -{1\over 16\pi^2}\e^{\m\n\r\s} \trR t F_{\m\n}(x)F_{\r\s}(x) \ ,\\
\\
\hline
\end{array}
\ee
where $\langle\ldots\rangle_A$ indicates the vacuum expectation value computed in a fixed $A_\m$ background. Hence the axial current, though conserved classically, is not conserved in the quantum theory and its non-conservation equals {\it minus} the anomaly.

\subsection{The anomalous variation of the effective action and its relation with instantons\label{relinst}}

We will now show that the abelian anomaly represents the (anomalous) variation of an effective action under {\it chiral} transformations.
The effective action to contemplate here is not the $\wt W[A]$ defined above. Clearly, $\wt W[A]$ only depends on $A$. One might then ask whether it is invariant under gauge transformations of $A$, but at present we are concerned with chiral transformations of the fermions, not with gauge transformation. Instead one has to consider a different  effective action, namely the quantum effective action for the fermions in a fixed gauge field configuration. This is defined as follows (see the next section for more details in a similar setting): introduce sources $\chi, \overline\chi$ for the  fermions by adding $\int \d^4 x\, (\overline\chi \p+ \pb\chi)$ to the matter action, define the generating functional $\wt W[\chi,\overline\chi,A]$ for fixed $A_\m$ by doing the functional integral over $\p$ and $\pb$ (much as we did when computing $\wt W[A]$) and define the quantum effective action $\wt\G[\p_0,\pb_0;A_\m]$ by Legendre transforming with respect to $\chi$ and $\overline\chi$ for fixed $A_\m$. In the absence of any anomaly (e.g.~for the non-chiral transformations) and for constant $\e$ (so that $S_{\rm mat}$ is invariant and we really have a symmetry) this $\wt\G[\p_0,\pb_0;A_\m]$ obeys the Slavnov-Taylor identities corresponding to this symmetry, which for a {\it linear} symmetry simply are $\dd_\e\, \wt\G[\p_0,\pb_0;A_\m]=0$. As already mentioned, the proof of the Slavnov-Taylor identities uses the invariance of the fermion measures. In the presence of an anomaly (e.g.~for the chiral transformations), the fermion measures are not invariant but generate an extra term $e^{i\int \e\,\cA}$ and one instead gets, still for constant $\e\,$:
\be\label{axialeffvar}
\begin{array}{|c|}
\hline\\
\dd_\e\, \wt\G[\p_0,\pb_0;A_\m]
=\e \int\d^4 x\,\cA(x)
=- {\e\over 16\pi^2}  \int \d^4 x\, \e^{\m\n\r\s}\trR t F_{\m\n} F_{\r\s}\ .\\
\\
\hline
\end{array}
\ee 
Here $\dd_\e$ is defined to act on the $\p_0$ and $\pb_0$ in the same way the chiral transformations $\dd_\e$ acted on the $\p$ and $\pb$ before.

For $t={\bf 1}$ the right hand side of (\ref{axialeffvar}) is related to the instanton number of the gauge field. Indeed, as we will discuss in more detail in subsection \ref{charclasses}, $\int \d^4 x\, \e^{\m\n\r\s}\trR  F_{\m\n} F_{\r\s}$ does not change under smooth variations of the gauge field, since the integrand (locally) is a total derivative and the integral is only sensitive to globally non-trivial configurations. One finds for a simple gauge group\footnote{Whether the configuration with $\n=+1$ should  be called an instanton or an anti-instanton depends on the detailed conventions used in the Euclidean continuation. These (anti) instantons are non-perturbative configurations for which the ``free part" $\del_\m A_\n^\a-\del_\n A_\m^\a$ and the ``interacting part" $C^\a_{\ \b\g} A_\m^\b A_\n^\g$ in $F_{\m\n}^\a$ are of the same order in the coupling constant $g$. Since $C^\a_{\ \b\g}\sim g$, we see that $A_\m^\a\sim {1\over g}$ and hence also $F_{\m\n}^\a\sim{1\over g}$.
}
\be
\int \d^4 x\, \e^{\m\n\r\s}  F^\a_{\m\n} F^\a_{\r\s}
={64\pi^2\over g^2} \n \ , \quad \n\in{\bf Z}
\quad \Rightarrow\quad
\int \d^4 x\, \e^{\m\n\r\s}\trR  F_{\m\n} F_{\r\s}
=64\pi^2\, C_\cR\, \n \ ,
\ee
where we used (\ref{tanorm}). The integer $\n$ is called the instanton number. It then follows that 
\be
\dd_\e\, \wt\G[\p_0,\pb_0;A_\m] 
=\e \int\d^4 x\,\cA(x)
= -4\, C_\cR\, \n\, \e \ ,
\ee 
with $C_\cR$ being e.g. ${1\over 2}$ for the $N$ or $\overline N$ representations of $SU(N)$ (i.e.~for quarks). More generally, $C_\cR$ is integer or half-integer. Incidentally, this shows that (for $t={\bf 1}$) 
$\int\d^4 x\,\cA(x)$ is an even integer.

\subsection{Relation of the abelian anomaly with the index of the Dirac operator\label{relinde}}

It is useful to compute the abelian anomaly again but now directly in Euclidean signature. The result will exhibit an interesting relation with the index of the Euclidean Dirac operator $i\Dsl_{\rm E}$. 

In Euclidean signature, all $\gamma^\m$ are hermitian and $i\Dsl_{\rm E}$ is a hermitian operator so that all its eigenvalues $\l_k$ are real. Furthermore, since $t$ commutes with $\Dsl_{\rm E}$ we can choose the eigenfunctions $\vf_k$ of $i\Dsl_{\rm E}$ to be also eigenfunctions of $t$:
\be
i\Dsl_{\rm E} \vf_k=\l_k\vf_k 
\quad , \quad 
t\vf_k=t_k\vf_k
\quad , \quad 
\langle \vf_k\ket{\vf_l} \equiv \int\d^4 x_E\, \vf_k^*(x)\vf_l(x) = \dd_{kl} \ .
\ee
With appropriate boundary conditions, the $\vf_k$ form a complete basis and 
\be
{\bf 1}=\sum_k \ket{\vf_k}\bra{\vf_k}
\quad ,\quad
\Tr A=\sum_k \bra{\vf_k} A \ket{\vf_k} \ .
\ee
Now $\g_5\Dsl_{\rm E}=-\Dsl_{\rm E} \g_5$ and $[t,\g_5]=0$ imply that $\g_5\vf_k$ are still eigenfunctions of $\Dsl_{\rm E}$ and $t$ but with eigenvalues $-\l_k$ and $t_k$:
\be
i\Dsl_{\rm E}(\g_5\vf_k)=-\g_5 i\Dsl_{\rm E}\vf_k=-\g_5\l_k\vf_k=(-\l_k)(\g_5\vf_k) 
\quad ,\quad
t(\g_5\vf_k)=\g_5 t\vf_k=t_k(\g_5 \vf_k) \ .
\ee
Hence, $\vf_k$ and $\g_5\vf_k$ have the same $t$-eigenvalue but opposite $i\Dsl_{\rm E}$-eigenvalues. It follows that for $\l_k\ne 0$, $\vf_k$ and $\g_5\vf_k$ are orthogonal. In particular then, $\vf_k$ cannot be an eigenstate of $\g_5$, but we can construct $\vf_{k,\pm}={1\over 2} (1\pm\g_5) \vf_k$ which are both non-vanishing and which are eigenstates of $\g_5$. Also, although no longer eigenfunctions of $i\Dsl_{\rm E}$, they are both still eigenfunctions of $(i\Dsl_{\rm E})^2$:
\be
\l_k\ne 0 \ : \quad \vf_{k,\pm}={1\over 2} (1\pm\g_5) \vf_k \ ,
\quad
\g_5 \vf_{k,\pm}=\pm \vf_{k,\pm} \ , \quad
-\Dsl_{\rm E}^2 \vf_{k,\pm}=\l_k^2 \vf_{k,\pm} \ , \quad
t \vf_{k,\pm}=t_k \vf_{k,\pm} \ .
\ee
Hence, for $\l_k\ne 0$, the eigenfunctions of $-\Dsl_{\rm E}^2$ and $t$ come in pairs of opposite chirality.
On the other hand, if $\l_k=0$ then $\vf_k$ and $\g_5\vf_k$ have the same $i\Dsl_{\rm E}$-eigenvalue (namely 0) and we can diagonalize $\g_5$ in this $i\Dsl_{\rm E}$-eigenspace. After having done so, the $\l=0$ eigenspace contains a certain number, say $n_+$, eigenfunctions $\vf_u$ that have positive $\g_5$ eigenvalue (positive chirality) and a certain number, say $n_-$, eigenfunctions $\vf_v$ that have negative $\g_5$ eigenvalue (negative chirality):
\be 
\l_k= 0 \ : \quad \g_5\vf_u=\vf_u \ ,\  u=1,\ldots n_+ \ ,
\quad
\g_5\vf_v=-\vf_v \ , \ v=1,\ldots n_- \ ,
\quad
i\Dsl_{\rm E}\vf_u=i\Dsl_{\rm E}\vf_v=0 \ .
\ee
Note that, for $\l_k=0$, the eigenfunction $\vf_u$ and $\vf_v$  do not necessarily come in pairs of opposite chirality. It follows that when computing the regularized trace of $\g_5 t$, the contributions of all $\vf_k$ with $\l_k\ne 0$ cancel and only the zero-modes of $i\Dsl_{\rm E}$ can give a non-vanishing contribution:
\ba\label{euclTL}
\Tr \g_5 t f\left( - {\Dsl_{\rm E}^2\over \L^2}\right) 
&=& \sum_k \bra{\vf_k} \g_5 t f\left( - {\Dsl_{\rm E}^2\over \L^2}\right) \ket{\vf_k}
=\sum_k f\left({\l_k^2\over L^2}\right) t_k \bra{\vf_k} \g_5  \ket{\vf_k} 
\nonumber\\
&=& \sum_{u=1}^{n_+} f(0)\, t_u \langle\vf_u\ket{\vf_u} 
- \sum_{v=1}^{n_-} f(0)\, t_v \langle\vf_v\ket{\vf_v} 
=\sum_{u=1}^{n_+}  t_u - \sum_{v=1}^{n_-}  t_v \ ,
\ea
where we used $f(0)=1$. Here the role of the regulator $f$ is to ensure that in the infinite sum over the non-zero modes the states of opposite chirality correctly cancel. In the end, of course, we can take $\L\to\infty$ without changing the result. 

Let us now specialize to $t={\bf 1}$ so that $\sum_{u=1}^{n_+}  t_u - \sum_{v=1}^{n_-}  t_v =n_+-n_-$ is just the difference between the number of positive and negative chirality zero-modes of $i\Dsl_{\rm E}$. This is called the {\it index} of the Dirac operator, and we have shown that
\be\label{find}
{\rm index}(i\Dsl_{\rm E})\equiv n_+-n_- = \lim_{\L\to\infty} \Tr \g_5 
f\left( - {\Dsl_{\rm E}^2\over \L^2}\right)  \ .
\ee
So  we have seen that this trace over the full Hilbert space is only sensitive to the zero-modes and equals the index of $i\Dsl_{\rm E}$. Provided the Euclidean $\g_5$ is defined such that positive Euclidean chirality also corresponds to positive Minkowski chirality,   the right hand side of (\ref{find}) equals $\lim_{\L\to\infty} \Tr {\cal T}_\L$ (but without the $\e(x)$) and comparing with (\ref{TLdef}) we see that
\be\label{abanind}
\int \d^4 x\, \cA(x)= - 2\, {\rm index}(i\Dsl_{\rm E}) \ ,
\ee
where the integral of the anomaly function on the left hand side is done in Minkowskian space-time, as before.

On the other hand, our computation (\ref{regtracecomp}) allows us to express the right hand side of (\ref{find}) in terms of the gauge field strength exactly as before. There are only two differences because at present we are in the Euclidean: the $\Tr$ now contains a $\d^4 p_E$ rather than a $\d^4 p\simeq i \d^4 p_E$ and the trace over four Euclidean $\g$-matrices and the Euclidean $\g_5$ now gives $-4 \e_{\rm E}$ rather than $4i \e$. The $-i$ and $i$ compensate each other and we simply get\footnote{
In doing the Euclidean continuations involving $\e_{\m\n\r\s}$ and $\g_5$ one has to very carefully keep track of all signs. This will be discussed in some detail in section \ref{Euclidean}.
}
(cf.~eq.~(\ref{TLF}))
\be\label{ind4euc}
{\rm index}(i\Dsl_{\rm E})
={1\over 32\pi^2} \int \d^4 x_E\, \e^E_{\m\n\r\s} \trR F_E^{\m\n} F_E^{\r\s} \ .
\ee
This relation is an example of the famous Atiyah-Singer index theorem. It again shows that the r.h.s. must be an integer and invariant under smooth deformations of $A_\m$.
On the other hand (cf sect. \ref{Euclidean}), the Minkowskian continuation of the r.h.s. simply yields the same expression without the sub- or superscripts ${\rm E}$, so that finally we have again
\be
{\rm index}(i\Dsl_{\rm E})
=  {1\over 32\pi^2} \int \d^4 x\, \e_{\m\n\r\s} \trR F^{\m\n} F^{\r\s} 
=-{1\over 2}
\int \d^4 x\, \cA(x) \ .
\ee
All this matches nicely with what we found above when we related the abelian anomaly for constant $\e$ to the instanton number and found that $\int \d^4 x\, \cA(x)$ must be an even integer.

We cannot resist from deriving the index theorem in arbitrary even dimensions $d=2r$ since this is a straightforward generalization of the previous computation which we now do directly in Euclidean signature. The index is still given by (\ref{find}) with $\g_5$ replaced by the
Euclidean chirality matrix in $2r$ dimensions, $\g_{\rm E}=i^r \g_{\rm E}^1 \ldots \g_{\rm E}^{2r}$ (cf. (\ref{d30}) below). To slightly simplify the computation we will explicitly use $f(s)=e^{-s}$.
The obvious generalization of (\ref{regtracecomp}) then is
\ba
{\rm index}(i\Dsl_{{\rm E},2r})
&=&\lim_{\L\to\infty} \Tr \g_{\rm E} \,
f\left( - {\Dsl_{\rm E}^2\over \L^2}\right)
= \lim_{\L\to\infty} \int \d x_{\rm E}^{2r}\ 
\L^{2r} \int {\d^{2r} q_{\rm E}\over (2\pi)^{2r}} \tr \g_{\rm E}
\exp\left( -q_{\rm E}^2-2i {q_{\rm E}^\m D_\m^{\rm E}\over\L} +{\Dsl_{\rm E}^2\over\L^2}\right)
\nonumber\\
&=&\int \d x_{\rm E}^{2r}\ \int {\d^{2r} q_{\rm E}\over (2\pi)^{2r}} 
\ e^{-q_{\rm E}^2}\  \tr \g_{\rm E}\, {(\Dsl_{\rm E}^2)^r\over r!}
 \ .
\ea
Using $\Dsl_{\rm E}^2=D_\m^{\rm E} D^\m_{\rm E}-{i\over 2} \g_{\rm E}^\m \g_{\rm E}^\n\, F^{\rm E}_{\m\n}$ and $\trD \g_{\rm E} \g_{\rm E}^{\m_1} \ldots \g_{\rm E}^{\m_{2r}}= i^r 2^r (-)^r \e_{\rm E}^{\m_1 \ldots \m_{2r}} $ with $\e_{\rm E}^{1\ldots 2r}=+1$. we get
\be\label{dsl2r}
\tr \g_{\rm E} (\Dsl_{\rm E}^2)^r= \Big(-{i\over 2}\Big)^r \trD \g_{\rm E} \g_{\rm E}^{\m_1} \ldots \g_{\rm E}^{\m_{2r}} \ \trR F^{\rm E}_{\m_1\m_2} \ldots F^{\rm E}_{\m_{2r-1}\m_{2r}}
= (-)^r \e_{\rm E}^{\m_1 \ldots \m_{2r}} \ \trR F^{\rm E}_{\m_1\m_2} \ldots F^{\rm E}_{\m_{2r-1}\m_{2r}} \ .
\ee
On the other hand,
$\int {\d^{2r} q_{\rm E}\over (2\pi)^{2r}}\ e^{-q_{\rm E}^2}={1\over (4\pi)^r}$, so that
we obtain for the index of the Dirac operator in $2r$ Euclidean dimensions:
\be\label{index2reuc}
\begin{array}{|c|}
\hline\\
\quad
{\rm index}(i\Dsl_{{\rm E},2r})= {(-)^r\over r! (4\pi)^r}\int \d x_{\rm E}^{2r}\ 
\e_{\rm E}^{\m_1 \ldots \m_{2r}} \ \trR F^{\rm E}_{\m_1\m_2} \ldots F^{\rm E}_{\m_{2r-1}\m_{2r}} \ ,\quad
\\
\\
{\rm with} \ [D_\m^{\rm E},D_\n^{\rm E}]=-i F_{\m\n}^{\rm E} \quad {\rm and} \quad 
\g_{\rm E}=i^r \g_{\rm E}^1 \ldots \g_{\rm E}^{2r} \ .
\\
\\
\hline
\end{array}
\ee
In four dimensions, i.e. for $r=2$, we get back eq.~(\ref{ind4euc}).
Also, in the appendix,  we give an explicit example of an abelian gauge field in 2 dimensions ($r=1$) with $\int \d^2 x_{\rm E}\, \e^{\m\n}_{\rm E} F_{\m\n}^{\rm E}=-4\pi m$, $m\in {\bf Z}$ and explicitly show that with this gauge field the index equals $m$, again in agreement with (\ref{index2reuc}).
Finally, we note that the mathematical literature rather uses antihermitean field strengths $\cF_{\m\n}=-i F_{\m\n}$ (cf eq.~(\ref{iredef}) below) so that the index theorem becomes\footnote{
In most references the prefactor is $i^r$ rather $(-i)^r$ corresponding either to a chirality matrix defined as $(-i)^r \g_{\rm E}^1\ldots \g_{\rm E}^{2r}$ or to a field strength $\cF_{\m\n}$ (actually the Lie algebra generators) defined with the opposite sign convention.
}
\be\label{index3}
{\rm index}(i\Dsl_{{\rm E},2r})= {(-i)^r\over r! (4\pi)^r}\int \d x_{\rm E}^{2r}\ 
\e_{\rm E}^{\m_1 \ldots \m_{2r}} \ \trR \cF^{\rm E}_{\m_1\m_2} \ldots \cF^{\rm E}_{\m_{2r-1}\m_{2r}} \ .
\ee

\vskip1.cm
\newpage
\section{Anomalies, current non-conservation and non-invariance of the effective action\label{noneffa}}
\setcounter{equation}{0}

While the abelian anomaly is concerned with a local or global $U(1)$ symmetry that commutes with the (non-abelian) gauge symmetry, in the following, we will be mostly concerned with {\it anomalies of the (non-abelian) gauge symmetry} itself. The essential question then is whether the effective action $\wt W[A]$ obtained after integrating out the matter fields, as defined in (\ref{Wtilde}), is invariant under gauge transformations or not. In this section, we establish some general results that will be useful in later explicit computations of the anomalies. In particular, we will precisely relate the anomaly to the non-invariance of the effective action $\wt W[A]$, as well as to the non-conservation of the quantum current, and carefully work out the anomalous Ward identities with particular emphasis on  getting the signs and $i$'s correctly.

\subsection{Anomalous Slavnov-Taylor identities and non-invariance of the effective action\label{AST}}

\subsubsection{Anomalous Slavnov-Taylor identities : the general case\label{ASTgen}}

Let us first derive the anomalous Slavnov-Taylor identities for some general local symmetry\footnote{
Note that although we use a constant parameter $\e$ the symmetries are in general local ones since $F^r$ depends explicitly on $x$.
} 
acting on some set of fields $\f^r$ as
\be\label{gensym}
\f^r(x)\to {\f'}^r=\f^r(x) +\dd \f^r(x)
\quad ,\quad \dd\f^r(x)=\e\, F^r(x,\f(x)) \ ,
\ee
and under which some classical action is assumed to be invariant:
\be
S[\f^r+\e F^r]=S[\f^r] \ .
\ee
However, we suppose that the integation measure is not invariant but rather transforms as\footnote{
In the present general setting, we use the symbol $\ca$ for the anomalous transformation of the measure, reserving the symbol $\cA$ for the abelian anomaly (under chiral transformations of the fermions).
}
\be\label{meastrans}
\prod_r \cD\f^r \to \prod_r \cD(\f^r +\e F^r) 
= \prod_r \cD\f^r \ e^{i\e \int\d^4 x\ \ca(x)} \ .
\ee

Then the generating functional of connected diagrams $W[J]$ is given by
\ba
e^{i W[J]}
\hskip-2.mm&=&\hskip-2.mm 
\int  \prod_r \cD\f^r \ \exp\left[iS[\f^r] + i \int J_r(x) \f^r(x)\right]
= \int  \prod_r \cD{\f'}^r \ \exp\left[iS[{\f'}^r] + i \int J_r(x) {\f'}^r(x)\right]
\nonumber\\
\hskip-2.mm&=&\hskip-2.mm
\int  \prod_r \cD\f^r \ \exp\left[i \e \int \ca(x) + iS[\f^r] + i \int J_r(x) \f^r(x) + i \e \int J_r(x) F^r(x,\f)\right]
\nonumber\\
\hskip-2.mm&=&\hskip-2.mm
\int  \prod_r \cD\f^r \ \exp\left[iS[\f^r] + i \int J_r(x) \f^r(x)\right]
\left\{ 1 + i\e\ \int \Big( \ca(x) + J_r(x) F^r(x,\f)\Big)\right\},
\ea
where we first changed integration variables from $\f^r$ to ${\f'}^r$ and then used the invariance of the action and the anomalous transformation of the measure. It follows that
\be\label{ST1}
\int\d^4 x \, \Big( \ca(x) + J_r(x) \langle F^r(x,\Phi)\rangle_J\Big)
=0 \ ,
\ee
where $\langle \cO\rangle_J$ is the expectation value of the operator
$\cO$ in the presence of the source $J$, i.e. computed with an action $S+\int J_r \f^r$ and we denoted $\Phi^r$ the quantum operator corresponding to $\f^r$.  Note that the identity (\ref{ST1}) is true for all currents $J_r(x)$. In particular, one can take successive functional derivatives with respect to $J_{s_1}(y_1)$, \ldots $J_{s_n}(y_n)$ and then set $J=0$ to obtain relations between expressions involving various expectation values $\langle \Phi^{s_1}(y_1) \ldots \Phi^{s_n}(y_n) F^r(x,\Phi)\rangle_{J=0}$ and the anomaly function $\ca(x)$. 

We want to rewrite the identity (\ref{ST1}) in terms of the quantum effective action $\G$ defined from $W$ by a Legendre transformation. To do this Legendre transformation one first defines
\be\label{vfdef}
\vf^r(x)={\dd W[J]\over \dd J_r(x)}\equiv \langle \Phi^r(x) \rangle_J \ .
\ee
This can be solved (generically) to give $J_r$ as a function of $\vf^r$: We let $J_{\vf,r}$ be the current which is such that $\langle \Phi^r(x) \rangle_J$ equals a prescribed value $\vf^r(x)$. Then the quantum effective action $\G[\vf]$ is defined as
\be \G[\vf]=W[J_{\vf}]-\int\d^4 x\, \vf^r(x)\, J_{\vf,r}(x) \ .
\ee
Introducing a loop-counting parameter $\l$ by replacing $S\to {1\over \l}S$, as discussed above, it is not difficult to see that $\G[\vf]$ equals $S[\vf]$ at tree-level, but it also contains contributions from all loops, hence the name quantum effective action. Along the same lines one can also see that an arbitrary Green's function can be computed (in perturbation theory) by using $\G$ as an action and only computing tree diagrams with all propagators and vertices taken from $\G$. This shows that $\G$ is the generating functional of one-particle irreducible diagrams, i.e. of (inverse) full propagators and vertex functions. For these reasons the $\vf^r$ are sometimes called background or classical fields. It follows from the definition of $\G$ that $ J_{\vf,r}(x)=-
{\dd \G[\vf]\over \dd \vf^r(x)}$ and, of course, $\langle \Phi^r \rangle_{J_{\vf,r}}=\vf^r$. However, we do {\it not} have $\langle F(x,\Phi) \rangle_{J_{\vf,r}}=F(x,\vf)$, unless $F$ depends {\it linearly} on $\Phi$. Thus we can rewrite the (anomalous) Slavnov-Taylor identity (\ref{ST1}), by choosing $J$ to equal $J_\vf$, as
\be\label{ST2}
\int\d^4 x \, \Big( \ca(x) -  \langle F^r(x,\Phi)\rangle_{J_\vf} 
{\dd \G[\vf]\over \dd \vf^r(x)}\Big) =0 \ .
\ee
For a {\it linear} symmetry, i.e. with $F^r(x,\Phi)$ depending linearly on the $\Phi^s$ this simplifies as
\be\label{ST3}
\int\d^4 x \, \Big( \ca(x) -   F^r(x,\vf) 
{\dd \G[\vf]\over \dd \vf^r(x)}\Big) =0 
\qquad \text{for a linear symmetry}\ .
\ee
But $\e F^r(x,\vf)$ is just $\dd \vf^r$, so that we can rewrite this as
\be\label{ST4}
\dd_e \G[\vf] \equiv 
\int\d^4 x\, \dd\vf^r(x) {\dd \G[\vf]\over \dd \vf^r(x)}
= \int \d^4 x\ \e\, \ca(x) 
\qquad \text{for a linear symmetry}\ .
\ee 
This states that, in the absence of anomalies, the quantum effective action is invariant under the same linear symmetries as the classical action and, when anomalies are present, the variation of the quantum effective action equals the anomaly.
Of course, this remains true for a linear symmetry depending on multiple infinitesimal parameters $\e^a(x)$ in which case (\ref{ST4}) reads
\be\label{ST5}
\dd_e \G[\vf] \equiv 
\int\d^4 x\, \dd\vf^r(x) {\dd \G[\vf]\over \dd \vf^r(x)}
= \int \d^4 x\ \e^a(x)\, \ca_a(x) 
\qquad \text{for linear symmetries}\ .
\ee 

\subsubsection{Anomalous Slavnov-Taylor identities for gauge theories\label{ASTgaugeth}}

It is in the latter form (\ref{ST5}) that the anomalous Slavnov-Taylor identities are most useful in the non-abelian gauge theories, where the role of the $\f^r$ now is played by the gauge and matter fields $A_\m,\p,\pb$. However, in gauge theories the gauge invariance of the classical action $S[A,\p,\pb]$, although a linear symmetry, does not translate into the corresponding symmetry for the quantum effective action $\G[A_0,\p_0,\pb_0]$ (where $A_0,\p_0,\pb_0$ play the role of the $\vf^r$), even in the absence of anomalies. This is due to the fact that it is not the classical action which appears in the functional integral but the gauge-fixed action and the addition of the gauge fixing terms of course breaks the gauge symmetry. There are several avenues one can nevertheless pursue to exploit the gauge symmetry of the classical action:
\begin{itemize}
\item
The gauge-fixed action, although no longer gauge invariant, is BRST invariant. However, BRST symmetry is a non-linear symmetry. It follows that the (anomalous) Slavnov-Taylor identities for the BRST symmetry will hold in the form (\ref{ST2})  but this is somewhat less convenient to deal with than the simpler form (\ref{ST5}) for linear symmetries. Of course, one can still derive an infinite set of (anomalous) Ward identities between Green's functions.
\item
One may use a specific gauge fixing, called background field gauge. 
This gauge fixing breaks the gauge invariance with respect to $A_\m$ in the functional integral, as necessary e.g. to have a well-defined gauge field propagator, but this is done in such a way that $\G[A_0,\p_0,\pb_0]$ is gauge invariant with respect to gauge transformations of the so-called ``background fields" $A_0,\p_0,\pb_0$, if the functional integral measures are invariant. One can then derive (anomalous) Slavnov-Taylor identities for this quantum effective action computed using background field gauge fixing.
\item
For the present purpose of studying anomalies there is an alternative way to proceed. As discussed in the previous section, anomalies arise from the non-invariance of the fermionic matter functional integral measures, and it is enough to consider the somewhat intermediate notion of effective action $\wt W[A]$ defined in (\ref{Wtilde}) where one only does the functional integral over the matter fields. This avoids the complication of the gauge fixing procedure (and subsequent non-invariance of the gauge fixed action). It is clear from its definition that $\wt W[A]$ computes the connected vacuum to vacuum amplitude for the fermions in the presence of an ``external" gauge field $A_\m$. Alternatively, still for fixed ``external" $A_\m$ we may introduce sources $\chi,\overline\chi$ for the fermions only, compute the generating functional $W[\chi,\overline\chi; A]$ and do the Legendre transformation with respect to $\chi$ and $\overline\chi$ to get $\wt \G[\p_0,\pb_0;A]$. This is the quantum effective action for the fermions in an external gauge field $A_\m$. Clearly, setting $\p_0=\pb_0=0$ gives again the connected vacuum to vacuum amplitude for the fermions in the external field $A_\m$ so that $\wt W[A]=\wt \G[0,0;A]$. 
In the following, we will concentrate on this $\wt W[A]$.
\end{itemize}

Once we have obtained $\wt W[A]$ we must still carry out the functional integral over the gauge fields, now using as ``classical" action 
\be
\wt S[A]
=-{1\over 4}\int\d^4 x\, F_{\m\n}^\a F^{\a\m\n} + \wt W[A]\ .
\ee 
If $\wt W[A]$ is gauge invariant then so is $\wt S[A]$ and everything proceeds as usual: one goes through the standard Faddeev-Popov procedure of adding a gauge-fixing term and the ghosts (or alternatively uses the slightly more general BRST quantization of adding some BRST exact term $s_{\rm BRST} \Psi$ where $\Psi$ is some local functional of ghost number $-1$). In the end one can then prove, as usual, that the theory is renormalizable and unitary and that amplitudes between physical states do not depend on the gauge fixing. However, if $\wt W[A]$ is not gauge invariant, all this breaks down and, typically, the theory would not be renormalizable and, even worse, would contain physical states of negative norm and unitarity would be violated.

Let us now show that the gauge variation of $\wt W[A]$ is given by the anomaly and, hence, in the absence of anomalies, $\wt W[A]$ is indeed gauge invariant. Let $A'_\m=A_\m+\dd A_\m$ with $\dd A_\m=D_\m\e$ (or more explicitly, $\dd A_\m^\a=\del_\m \e^\a +C^\a_{\ \b\g} A_\m^\b \e^\g$). Then doing the by now familiar manipulations of first changing names of the integration variables from $\p,\pb$ to $\p',\pb'$ and then letting the primed fields be the gauge transformed ones and using the invariance of the classical action as well as the possible non-invariance of the fermion measures, we get
\ba
e^{i\wt W[A']}&=& \int\cD\p\cD\pb\ e^{ i S_{\rm mat}[\p,\pb,A']}
= \int\cD\p'\cD\pb'\ e^{ i S_{\rm mat}[\p',\pb',A']}
\nonumber\\
&=&\int\cD\p\cD\pb\ e^{ i \int\d^4 x\, \e^\a(x) \ca_\a(x) +
i S_{\rm mat}[\p,\pb,A]}
= e^{i \int\d^4 x\, \e^\a \ca_\a(x)} e^{i\wt W[A]}\ ,
\ea
and we conclude
\be\label{Wtildeanom}
\dd_\e \wt W[A] \equiv\wt W[A+\dd A]-\wt W[A]= \int\d^4 x\, \e^\a(x) \ca_\a(x)\ .
\ee
The left hand side is the gauge variation of $\wt W[A]$, so that the integrated anomaly equals the gauge variation of the effective action $\wt W[A]$.

\subsection{Current non-conservation and the anomaly\label{cnca}}

For {\it any} functional $\cW[A]$ of the gauge fields only, its gauge variation is given by
\ba
\dd_\e \cW[A]&\equiv&\cW[A+\dd A]-\cW[A]
=\int \d^4 x\, \dd A_\m^\a(x) {\dd\cW\over \dd A_\m^\a(x)}
=\int \d^4 x\, (D_\m \e(x))^\a {\dd\cW\over \dd A_\m^\a(x)}
\nonumber\\
&=&-\int \d^4 x\, \e(x)^\a \left( D_\m{\dd\cW\over \dd A_\m(x)}\right)_\a \ ,
\ea
where in the last step we used the fact that one can do a partial integration with the covariant derivative as if it were just an ordinary derivative. Indeed, we have\footnote{
To simplify the notation, we will assume that the $C^\a_{\ \b\g}$ are completely antisymmetric and we will then often denote them as $C_{\a\b\g}$. As discussed in sect.~\ref{actandfields} this is the case if the
gauge group is a product of commuting compact simple and $U(1)$ factors, which indeed is the case for most gauge groups one does encounter. Also, if the group is non-compact (as e.g. $SO(3,1)$) one can often study the compact version (as $SO(4)$) and ``analytically continue" the result in the end.
}
\ba
\int \d^4 x\, (D_\m \e)^\a \cG^\m_\a
&=&\int \d^4 x\, \Big(\del_\m \e^\a+ C_{\a \b\g}A_\m^\b \e^\g \Big) \cG^\m_\a
\nonumber\\
&=&\int \d^4 x\,\e^\a\Big(-\del_\m\cG_\a^\m-C_{\a\b\g} A_\m^\b \cG_\g^\m\Big)
=\ -\int \d^4 x\,\e^\a (D_\m \cG^\m)_\a \ .
\ea
Thus, equation (\ref{Wtildeanom}) can be equivalently written as
\be\label{Wtildeanom2}
\begin{array}{|c|}
\hline\\
\dd_\e \wt W[A]= \int \d^4 x\, (D_\m \e(x))^\a {\dd\wt W\over \dd A_\m^\a(x)}
=\int\d^4 x\, \e^\a(x) \ca_\a(x)
\quad\Leftrightarrow\quad
D_\m {\dd\wt W\over \dd A_\m^\a(x)} = - \ca_\a(x) \ .\\
\\
\hline
\end{array}
\ee 
The left equality states again that the integrated anomaly equals the variation of the effective action $\wt W[A]$, while the equality on the right states that  the covariant divergence of the expectation value of the non-abelian matter current equals minus the anomaly. Indeed, with the current defined as in
(\ref{matcurr}) we have
\be\label{curdef2}
{\dd\wt W\over \dd A_\m^\a(x)}= \langle J^\m_\a(x) \rangle_A \ ,
\ee
where the subscript $A$ is to remind us that the expectation value is computed with a fixed ``external" $A_\m$-field. Thus we arrive at
\be\label{currentnoncons}
\begin{array}{|c|}
\hline\\
\big(D_\m \langle J^{\m}(x) \rangle_A \big)_\a = -\ \ca_\a(x) \ .\\
\\
\hline
\end{array}
\ee
Recall that for the matter Lagrangian (\ref{matterlagr}) we have
\be
J^\m_\a= i \pb \g^\m t^\cR_\a \p \ .
\ee
These results are similar to but different from those obtained for the abelian anomaly under the chiral transformations of the fermions where we had (cf eq.~(\ref{J5noncons}))
\be\label{J5noncons2}
-\del_\m \langle  J_5^\m(x)\rangle_A = \cA(x) \ ,
\ee
while (\ref{axialeffvar}) related the axial anomaly to the variation of the 1PI effective action $\wt \G[\p_0,\pb_0,A]$ at fixed gauge field $A$.

{\it Remark :} Let us make a comment about a somewhat different definition of anomaly one can find in the earlier literature. With our definition (\ref{curdef2}) of the current, the definitions of the anomaly as the current non-conservation (\ref{currentnoncons}) or as the non-invariance of the effective action
(\ref{Wtildeanom2}) are perfectly equivalent. We will only talk about anomalies defined in this way. They are referred to as consistent anomalies in the literature (because they satisfy the Wess-Zumino consistency condition to be discussed later-on). In the earlier literature though, anomalies were mainly seen as a non-conservation of the currents. It then makes sense to change the definition of a current by adding a local expression in the gauge fields: $J^\m_\a \to J^\m_\a+X^\m_\a$. Unless $X^\m={\dd F\over \dd A_\m^\a}$ with some local functional $F$, the new current does not satisfy the definition
(\ref{curdef2}) with some modified effective action. In terms of the triangle diagrams to be discussed soon, the consistent definition automatically implements Bose symmetry between all vertices, while this is no longer the case for the modified currents. For a suitable choice of $X^\m_\a$ one may e.g. achieve that $D_\m J^\m$ is a gauge covariant expression (which is {\it not} the case for our definition of current and consistent anomaly). In this case one talks about the covariant (form of the) anomaly. Again, in these lectures we will always talk about the consistent, not the covariant, anomaly.
 
\subsection{Anomalous Ward identities\label{AWT}}

The relations derived above for the variation of the effective action or the covariant divergence of the current in an external field $A_\n$ are relations for generating functionals. Taking various derivatives with respect to the external field $A_\n$ generates an (infinite) set of identities between  Green's functions known as (anomalous) Ward identities.

Let us first recall the various signs and factors of $i$ that appear in relation with the generating functionals. Since $J^\m_\a={\dd S_{\rm mat}\over \dd A_\m^\a}$ does not involve the gauge field $A$ it follow from the definition of $\wt W[A]$ that
\be\label{Wtderiv}
{\dd \over \dd A_{\m_1}^{\a_1}(x_1)} \ldots {\dd \over \dd A_{\m_n}^{\a_n}(x_n)}\wt W \Big\vert_{A=0}= i^{n-1}\ \langle T\big( J^{\m_1}_{\a_1}(x_1) \ldots J^{\m_n}_{\a_n}(x_n) \big)\rangle_C \ ,
\ee
where the subscript $C$ indicates to take only the connected part of the current correlator. Now consider the $n$-point vertex function  $\G^{\m_1\ldots\m_n}_{\a_1\ldots\a_n}(x_1,\ldots,x_n)$ which appears as the coefficient of $A_{\m_1}^{\a_1}(x_1)\ldots$ $A_{\m_n}^{\a_n}(x_n)$ in the generating functional of 1PI diagrams $\G[A]$. More precisely, it is $i\G[A]$ which generates the 1PI diagrams. The $n^{\rm th}$ order diagram is the connected time-ordered vacuum expectation value of $n$ factors of $i \cL_{\rm int}= i J^\m_\a A_\m^\a$. Hence, we have
\be\label{gamman}
\G^{\m_1\ldots\m_n}_{\a_1\ldots\a_n}(x_1,\ldots,x_n)
=i^{n-1}\ \langle T\big( J^{\m_1}_{\a_1}(x_1) \ldots J^{\m_n}_{\a_n}(x_n) \big)\rangle_C \ ,
\ee
which equals (\ref{Wtderiv}), and thus
\be
\G^{\m_1\ldots\m_n}_{\a_1\ldots\a_n}(x_1,\ldots,x_n)
={\dd \over \dd A_{\m_1}^{\a_1}(x_1)} \ldots {\dd \over \dd A_{\m_n}^{\a_n}(x_n)}\wt W \Big\vert_{A=0} \ .
\ee

\subsubsection{The case of the abelian anomaly\label{caseabelan}}

\begin{figure}[h]
\centering
\includegraphics[width=0.25\textwidth]{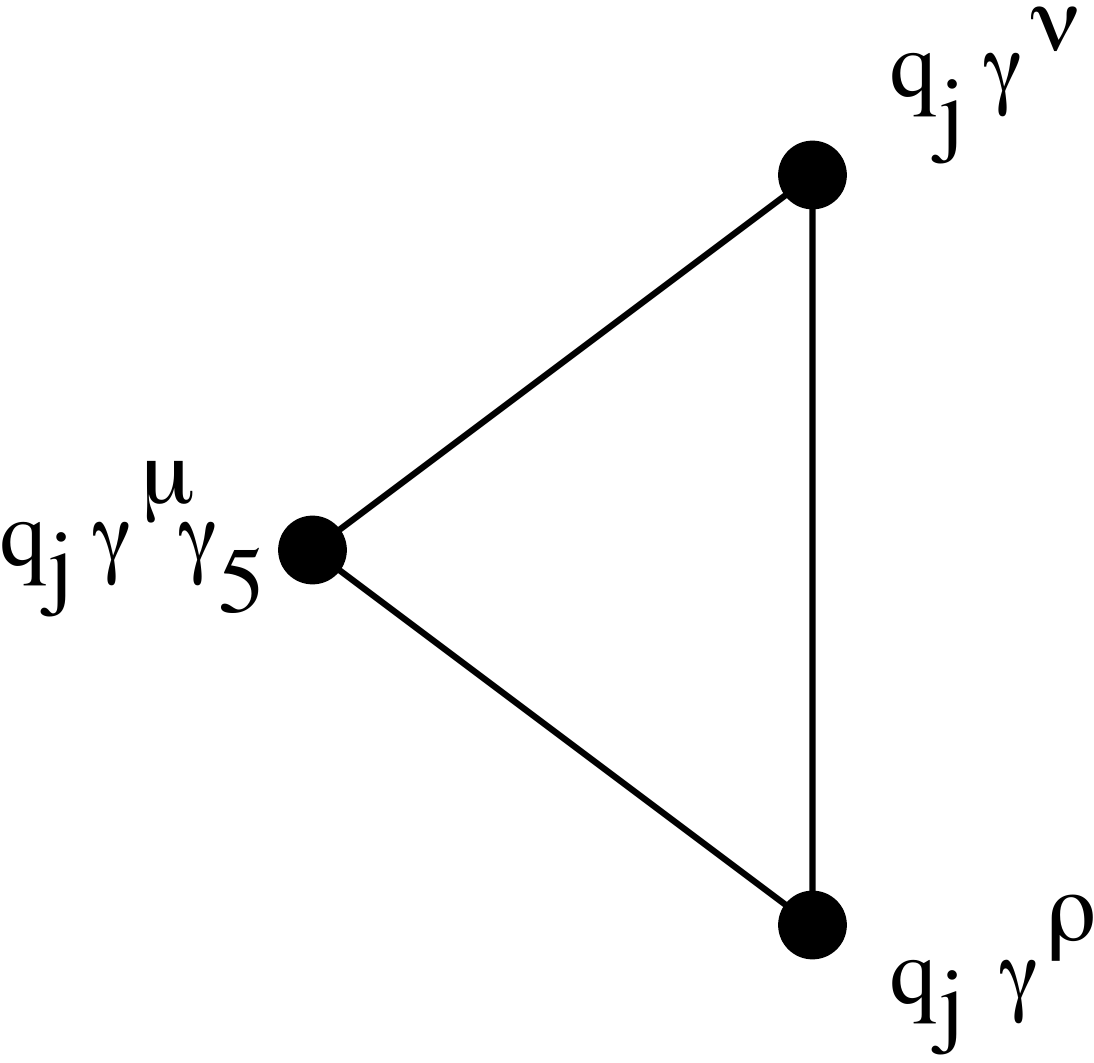}\\
\caption[]{The triangle diagram for a fermion of charge $q_j$ contributing to the abelian anomaly} \label{triangle5}
\end{figure}

As a first example, consider the abelian anomaly of eq. (\ref{J5noncons}) or (\ref{J5noncons2}) in a theory with only
a single $U(1)$ gauge field. Thus the generators $t_\a$ are replaced by a single $t$ whose eigenvalues are the $U(1)$ charges of the fields, conventionally called $q_j$. In particular, $\trR t\, t_\a t_\b \to \sum_j q_j^3$.  Then we have for the abelian anomaly given in
(\ref{anfuncab})
\ba
{\dd\over \dd A_\n(y)} {\dd\over \dd A_\r(z)} \cA(x)
&=& -{1\over 16\pi^2}\Big(\sum_j q_j^3\Big)\ {\dd\over \dd A_\n(y)} {\dd\over \dd A_\r(z)} \e^{\m\l\s\kappa} \big(2 \del_\m A_\l(x)\big)\big( 2 \del_\s A_\kappa(x)\big)
\nonumber\\
&=&-{1\over 2\pi^2}\Big(\sum_j q_j^3\Big)\ \e^{\m\n\s\r}\left({\del\over\del x^\m} \dd^{(4)}(y-x) \right){\del\over\del x^\s} \dd^{(4)}(z-x)
\nonumber\\
&=&+{1\over 2\pi^2}\Big(\sum_j q_j^3\Big)\ \e^{\n\r\l\s}\left({\del\over\del y^\l} \dd^{(4)}(y-x) \right){\del\over\del z^\s} \dd^{(4)}(z-x)\ .
\ea
By eq. (\ref{J5noncons2}) this should equal
\ba
-{\dd\over \dd A_\n(y)} {\dd\over \dd A_\r(z)} {\del\over\del x^\m}
\langle  J_5^\m(x)\rangle_A \Big\vert_{A=0}
&=&- {\del\over\del x^\m} {\dd\over \dd A_\n(y)} {\dd\over \dd A_\r(z)}\langle  J_5^\m(x)\rangle_A \Big\vert_{A=0}
\nonumber\\
&=&+{\del\over\del x^\m} \langle\  T\Big( J_5^\m(x) J^\n(y) J^\r(z)\Big)\rangle_C \ ,
\ea
where an additional minus sign arose in the last equality because according to (\ref{Wtderiv})  each ${\dd\over \dd A}$ gives an insertion of $i J$. The three-current correlator is easily computed in perturbation theory as a one-loop triangle diagram with the three currents at the vertices, see Fig.~\ref{triangle5}. In complete analogy with (\ref{gamman}), the exact relation is
\be\label{threecurrG}
\langle\  T\Big( J_5^\m(x) J^\n(y) J^\r(z)\Big)\rangle_C = - \G_5^{\m\n\r}(x,y,z) \ ,
\ee
with $i\G_5^{\m\n\r}$  computed from the triangle diagram with one vertex $i\times (i \g^\m \g_5 q_j)$ and two other vertices\footnote{
If the fermion field describes an electron, one usually writes $q=-e$ with $e>0$ and the vertices then are $+e\g^\m\g_5$, as well as $+e\g^\n$ and $+e\g^\r$, which is indeed the usual convention.
} 
$i\times (i \g^\n q_j)$ and $i\times (i \g^\r q_j)$. This will be done in section \ref{AVVFeynman}.
\vskip5.mm

\noindent
It follows that
\be
- \ {\del\over\del x^\m} \G_5^{\m\n\r}(x,y,z)
=  {1\over 2\pi^2}\Big(\sum_j q_j^3\Big)\ \e^{\n\r\l\s}\left({\del\over\del y^\l} \dd^{(4)}(y-x) \right){\del\over\del z^\s} \dd^{(4)}(z-x)\ .
\ee
Upon Fourier transforming, taking all momenta as incoming (this corresponds to $+ikx$ etc in the exponent),
\be\label{Fourier}
\int \d^4 x\,\d^4 y\,\d^4 z\, e^{+ikx+ipy+iqz}\ \G_5^{\m\n\r}(x,y,z)= (2\pi)^4 \dd^{(4)}(k+p+q)\ \G_5^{\m\n\r}(-p-q,p,q)\ ,
\ee 
we get
\be
-i (p+q)_\m \G_5^{\m\n\r}(-p-q,p,q) 
= - {1\over 2\pi^2}\Big(\sum_j q_j^3\Big)\ \e^{\n\r\l\s} p_\l q_\s \ ,
\ee
a relation we will check again below (including the precise factor and sign) by  explicitly computing the triangle Feynman diagram. More generally, for the abelian anomaly for fermions transforming in an arbitrary representation $\cR$ of a non-abelian gauge group one would have obtained
\be\label{divgamma5t}
-i (p+q)_\m \G_{5\, \b\g}^{\m\n\r}(-p-q,p,q) 
= - {1\over 2\pi^2} \big( \trR t\, t_{(\b} t_{\g)}\big)\ \e^{\n\r\l\s} p_\l q_\s \ .
\ee

One can similarly derive (anomalous) Ward identities for $n$-point vertex functions $\G_5^{\m\n_2\ldots \n_n}$ with one insertion of $\g_5$ at one of the vertices. If only $U(1)$ gauge fields are present, the abelian anomaly is bilinear in the gauge fields $A_\m$. It follows that when taking three or more functional derivatives of eq.~(\ref{J5noncons2}), the anomaly does no longer contribute. Hence, in this case, box, pentagon or higher one-loop diagrams are not affected by the abelian anomaly:
\be\label{boxpentabelian}
i\, p^{(1)}_\m\ \G_5^{\m\, \n_2 \ldots \n_n}\Big(p^{(1)},p^{(2)}, \ldots p^{(n)}\Big) = 0 
\ , \quad p^{(1)}=-\textstyle{\sum_{r=2}^n\ }  p^{(r)}
\ ,\quad n\ge 4\ \text{and only $U(1)$ gauge fields}\ .
\ee
On the other hand, if non-abelian gauge fields couple to the matter fields in the loop, i.e. if $\trR t\, t_\b t_\g \ne 0$ for generators $t_\b,\, t_\g$ of a non-abelian gauge group, then the anomaly contains terms that are bilinear, trilinear and quartic in these gauge fields and the corresponding box and pentagon diagrams could also display the abelian anomaly. For the pentagon diagram e.g.~it is straightforward to obtain 
\be\label{pentWa}
i p^{(1)}_\m\ \G^{\m\, \n_1\n_2\n_3\n_4}_{5\, \a_1\a_2\a_3\a_4} (p^{(i)})= -
{3\over 2\pi^2}  \e^{\n_1\n_2\n_3\n_4} \big( \trR t\, t_{(\b} t_{\g)}\big)
C^\b_{\ [\a_1\a_2} C^\g_{\ \a_3\a_4]} \ ,
\ee
with $p^{(1)}=-\sum_{r=1}^4 p^{(r)}$. However, for the abelian anomaly we are interested in the case $[t,t_\b]=[t,t_\g]=0$ (only then the chiral transformation is a symmetry of the classical matter Lagrangian) when the relevant Lie algebra is a sum of the $U(1)$ associated with $t$ and some non-abelian $G$ with generators $t_\b,\ t_\g$. Then the trace is
\be\label{traceU1GG}
\trR t\, t_{(\b} t_{\g)} = \sum_j q_j\ {\rm tr}_{R_j}\, t^{R_j}_\b\, t^{R_j}_\g = \sum_j q_j\, C_{R_j}\, \dd_{\b\g} \ ,
\ee
so that (\ref{pentWa}) becomes
\be\label{pentWa2}
i p^{(1)}_\m\ \G^{\m\, \n_1\n_2\n_3\n_4}_{5\, \a_1\a_2\a_3\a_4} (p^{(i)})= -
{3\over 2\pi^2}  \e^{\n_1\n_2\n_3\n_4} \sum_j q_j C_{R_j}
C^\b_{\ [\a_1\a_2} C^\b_{\ \a_3\a_4]} = 0 \ ,
\ee
because $C^\b_{\ [\a_1\a_2} C^\b_{\ \a_3\a_4]}=0$ due to the Jacobi identity. Hence the pentagon diagram is not anomalous.

\subsubsection{The case of anomalies under a non-abelian gauge symmetry\label{anWardsect}}

Now we turn to anomalies under the (non-abelian) gauge symmetry. The anomalous Slavnov-Taylor identities implied the relations (\ref{Wtildeanom2}) and the covariant current non-conservation (\ref{currentnoncons}). Let us similarly extract an anomalous Ward identity for the 3-point vertex function 
\be\label{Gmnr}
\G^{\m\n\r}_{\a\b\g}(x,y,z) 
= - \langle T\big( J^\m_\a(x) J^\n_\b(y) J^\r_\g(z)\big)\rangle 
= +{\dd\over \dd A_\m^\a(x)} {\dd\over \dd A_\n^\b(y)} 
 {\dd\over \dd A_\r^\g(z)} \wt W[A] \Big\vert_{A=0} \ ,
\ee
that one can check by a Feynman diagram computation. This time though, the functional derivative with respect to $A(y)$ does not commute with the covariant derivative $D_\m$ with respect to $x$, since the latter now contains the gauge field. We begin with the right equation (\ref{Wtildeanom2}), which we write as
\be\label{corderan}
-\ca_\a[A,x]= {\del\over \del x^\m} {\dd \wt W[A]\over \dd A_\m^\a(x)} + C_{\a\dd\e} A_\m^\dd(x) {\dd \wt W[A]\over \dd A_\m^\e(x)} \ ,
\ee
and take ${\dd \over \dd A_\n^\b(y)}$ to get
\be 
-{\dd\ca_\a[A,x]\over \dd A_\n^\b(y)}
= {\del\over \del x^\m} 
{\dd^2 \wt W[A]\over \dd A_\m^\a(x)\dd A_\n^\b(y)}
+ C_{\a\dd\e} A_\m^\dd(x) 
{\dd^2 \wt W[A]\over \dd A_\m^\e(x)\dd A_\n^\b(y)}
+ C_{\a\b\e} \dd^{(4)}(x-y){\dd \wt W[A]\over \dd A_\n^\e(x)} \ .
\ee
Taking one more derivative with respect to ${\dd \over \dd A_\r^\g(z)}$ yields
\ba
\hskip-1.cm
-{\dd^2\ca_\a[A,x]\over \dd A_\n^\b(y) \dd A_\r^\g(z)}
&=& {\del\over \del x^\m} 
{\dd^3 \wt W[A]\over \dd A_\m^\a(x)\dd A_\n^\b(y)\dd A_\r^\g(z)}
+ C_{\a\dd\e} A_\m^\dd(x) 
{\dd^3 \wt W[A]\over \dd A_\m^\e(x)\dd A_\n^\b(y) \dd A_\r^\g(z)}
\nonumber\\
&&\hskip-3.mm+\ 
C_{\a\g\e} \dd^{(4)}(x-z) 
{\dd^2 \wt W[A]\over \dd A_\r^\e(x)\dd A_\n^\b(y)}
+ C_{\a\b\e} \dd^{(4)}(x-y)
{\dd^2 \wt W[A]\over \dd A_\n^\e(x) \dd A_\r^\g(z)} \ .
\ea
Take then $A=0$ and recall that ($\trR t_\e t_\b=g_i^2 C_R^{(i)} \dd_{\e\b}$)
\be\label{matvacpol}
{\dd^2 \wt W[A]\over \dd A_\r^\e(x)\dd A_\n^\b(y)}\Bigg\vert_{A=0}
= i \langle T \big( J^\r_\e(x) J^\n_\b(y) \big)\rangle
=  \G^{\r\n}_{\e\b}(x,y) \equiv  \Pi^{\r\n}_{\e\b}(x,y) =  \dd_{\e\b} \Pi^{\r\n}_{(i)}(x,y) \ ,
\ee
which corresponds to the matter contribution to the vacuum polarization.\footnote{Of course, we are including matter loops,  but no gauge field or ghost loops.} The subscript $(i)$ on $\Pi^{\r\n}$ is to remind us that we are including the coupling constant and the normalization constant $C_R^{(i)}$ in $\Pi^{\r\n}_{(i)}$ and for a gauge group with several simple or $U(1)$ factors $G_i$ these constants differ from one factor to the other. We will also denote 
\be  
\ca_{\a,\b\g}^{\ \ \n\r}(x;y,z)=
{\dd^2\ca_\a[A,x]\over \dd A_\n^\b(y) \dd A_\r^\g(z)}\Bigg\vert_{A=0}\ ,
\ee 
and finally get the anomalous Ward identity:
\be\label{nonabWard}
-\ca_{\a,\b\g}^{\ \ \n\r}(x;y,z)=
{\del\over \del x^\m}  \G^{\m\n\r}_{\a\b\g}(x,y,z) 
+ \, C_{\a\b\g} \left[ \dd^{(4)}(x-y) \Pi_{(i)}^{\n\r}(y,z)
-  \dd^{(4)}(x-z) \Pi_{(i)}^{\n\r}(y,z)\right] \ .
\ee
In the absence of an anomaly, this is very similar to the well-known Ward identity of QED which relates the divergence of the vertex function $\del_\m\G^\m$ to the inverse of the full fermion propagator.
Again, (\ref{nonabWard}) can be checked by the computation of a one-loop, three-point amplitude, i.e. of a triangle diagram, as will be done in the next section.

Of course, taking even more derivatives with respect to $A$ before setting $A=0$ also gives similarly (anomalous) Ward identities for the 4- and 5-point functions $\G^{\m\n\r\s}_{\a\b\g\dd}$ and $\G^{\m\n\r\s\l}_{\a\b\g\dd\e}$, as well as all higher $n$-point functions. We will show in sect.~\ref{secWZBRST} that (the {\it consistent} anomaly) $\ca_\a$ only contains terms quadratic and cubic in the gauge field $A$. This implies that taking four derivatives of $\ca_\a$ with respect to $A$ yields zero, so that the Ward identities for all $n$-point functions with $n\ge 5$ are actually non-anomalous. However, due to the appearance of the {\it covariant }derivative in (\ref{corderan}), these Ward identities relate the $n$-point functions to the $(n-1)$-point functions and, e.g. for $n=5$, one has a non-anomalous Ward identity that nevertheless relates  the 5-point functions to the anomalous 4-point functions.

Taking again the Fourier transform of (\ref{nonabWard}), using analogous conventions as is (\ref{Fourier}) we get
\be\label{nonabWard2}
-i(p+q)_\m \G^{\m\n\r}_{\a\b\g}(-p-q,p,q)
= \ca_{\a,\b\g}^{\ \ \n\r}(-p-q;p,q) 
- C_{\a\b\g}\left[ \Pi_{(i)}^{\n\r}(p,-p)-\Pi_{(i)}^{\n\r}(-q,q)\right] \ .
\ee
(Of course, the two-point functions are even functions of their arguments, $\Pi_{(i)}^{\n\r}(p,-p)=\Pi_{(i)}^{\n\r}(-p,p)$.) The two terms on the right-hand-side are often referred to as the anomalous and non-anomalous contribution to the Ward identity. We will find later-on that $\ca_{\a,\b\g}^{\n\r}\sim \trR t_{(\a} t_\b t_{\g)}$ so that the anomalous contribution is the part which is completely symmetric in the Lie algebra indices, while the non-anomalous part is completely antisymmetric in these indices.\footnote{
The non-anomalous part of (\ref{nonabWard}) is also easy to obtain using current algebra arguments as follows: taking ${\del/ \del x^\m}$ of $\langle T\big( J^\m_\a(x) J^\n_\b(y) J^\r_\g(z)\big)\rangle$  gives a contribution $\langle T\big( \del_\m J^\m_\a(x) J^\n_\b(y) J^\r_\g(z)\big)\rangle$ which vanishes to lowest order in $A$, but also contributions of the type $\sim\dd(x^0-y^0) \langle T\big([J^0_\a(x), J^\n_\b(y)] J^\r_\g(z)\big)\rangle \sim \dd^{(4)}(x-y) C_{\a\b\e}\langle T\big( J^\n_\e(y) J^\r_\g(z)\big)\rangle $ which arise from $\del/\del x^0$ acting on the $\theta(x^0-y^0)$ of the time-ordering.
}

After all these general considerations it is maybe useful to recall that we have {\it not} yet determined the anomaly $\ca^\a(x)$ giving the possible non-invariance of $\wt W[A]$ under gauge transformations. What we have done is to set up a precise dictionary allowing us in the following to reconstruct the full anomaly from a single number we will extract from a one-loop triangle diagram computation, and also much later from more sophisticated considerations about the index of an appropriate Dirac operator. For now {\it suppose} the anomaly has the form
\be\label{anform32}
\ca_\a(x) = 4\, c\ \e^{\m\n\r\s} \trR t_\a t_\b t_\g\ (\del_\m A_\n^\b(x)) (\del _\r A_\s^\g(x)) + \cO(A^3) 
= c\ \e^{\m\n\r\s} \trR t_\a F_{\m\n}^{\rm lin} F_{\r\s}^{\rm lin} + \cO(A^3) \ ,
\ee
where $F_{\m\n}^{\rm lin}= \del_\m A_\n-\del_\n A_\m$ is the linearized part of the non-abelian field strength. Then, similarly to the computation we did above for the abelian anomaly, we now get
\be
\ca_{\a,\b\g}^{\ \ \n\r}(-p-q,p,q)=8\, c\ \e^{\n\r\l\s} p_\l q_\s\, 
D^\cR_{\a\b\g} \ ,
\ee
with the $D$-symbol being the symmetrized trace\footnote{
The symmetrized trace appears since in (\ref{anform32}) $\e^{\m\n\r\s}(\del_\m A_\n^\b(x)) (\del _\r A_\s^\g(x))$ is symmetric in $\b\leftrightarrow\g$. Hence, we could have replaced $\trR t_\a t_\b t_\g$ by $\trR t_\a t_{(\b} t_{\g)}$ which, because of the cyclicity of the trace, is actually symmetric in all three indices.
}
in the representation $\cR$ of three generators
\be
D^\cR_{\a\b\g}=\trR t_{(\a}t_\b t_{\g)} \ .
\ee
We conclude
\ba\label{Wardquickref}
\begin{array}{|ccc|}
\hline
& &\nonumber\\
\hskip5.mm\ca_\a(x)\ \equiv \
-(D_\m \langle J^\m(x)\rangle_A)_\a &=& \ \ c\ \e^{\m\n\r\s} \trR t_\a F_{\m\n}^{\rm lin} F_{\r\s}^{\rm lin} + \cO(A^3) \hskip0.7cm
\nonumber\\
\quad&\Leftrightarrow &\quad
\nonumber\\
\hskip0.7cm
-i(p+q)_\m \G^{\m\n\r}_{\a\b\g}(-p-q,p,q)\Big\vert_{\e-{\rm piece}}\hskip3.mm
&=&\hskip-2.0cm 8\, c\ \e^{\n\r\l\s} p_\l q_\s\, 
D^\cR_{\a\b\g} \ .\\
& &\\
\hline
\end{array}
\\
\ea

\vskip5.mm
\section{Anomalies from triangle Feynman diagrams\label{tria}}
\setcounter{equation}{0}

In the previous section, we have established precise relations, in the form of anomalous Ward identities, between (functional derivatives of) the anomaly and certain proper vertex (one-particle irreducible) functions. The relevant proper vertex functions in four dimensions are triangle diagrams. In this section we will very explicitly evaluate such triangle Feynman diagrams. We first do the computation for the abelian anomaly and confirm our results of section \ref{fmeasure}. Then we compute the triangle diagram for chiral fermions coupled to  non-abelian gauge fields, thus establishing the non-abelian gauge anomaly.

We will do the computation in Pauli-Villars regularization so that one can very explicitly see how and where the anomalies arise. We will provide many computational details. The reader who is less interested in these details may safely skip most of the calculations and directly go to the results (\ref{abeliantriangleresult}) and (\ref{nuroconserved}) for the abelian anomaly, and (\ref{GammaLdiv2}) as well as (\ref{chiralanom1}) for the non-abelian gauge anomaly.

\subsection{The abelian anomaly from the triangle Feynman diagram: AVV\label{AVVFeynman}}

We will now compute the anomalous triangle diagram with one axial current and two vector currents (AVV) and probe it for the conservation of the axial current. In the above language, we will do a Feynman diagram computation of $\G_{5\b\g}^{\ \m\n\r}$. In the next subsection, we will be interested in a very similar computation. In order to be able to easily transpose the present computation, we will replace $j_5^\m=i\pb\g^\m\g_5 t \p$ by the more general
\be
j_{5\a}^\m=i\pb\g^\m\g_5 t_\a\p \ ,
\ee
and instead compute $\G_{5\a\b\g}^{\ \m\n\r}$. We may then replace the non-abelian generator $t_\a$ by the abelian generator $t$ in the end.

\begin{figure}[h]
\centering
\includegraphics[width=0.6\textwidth]{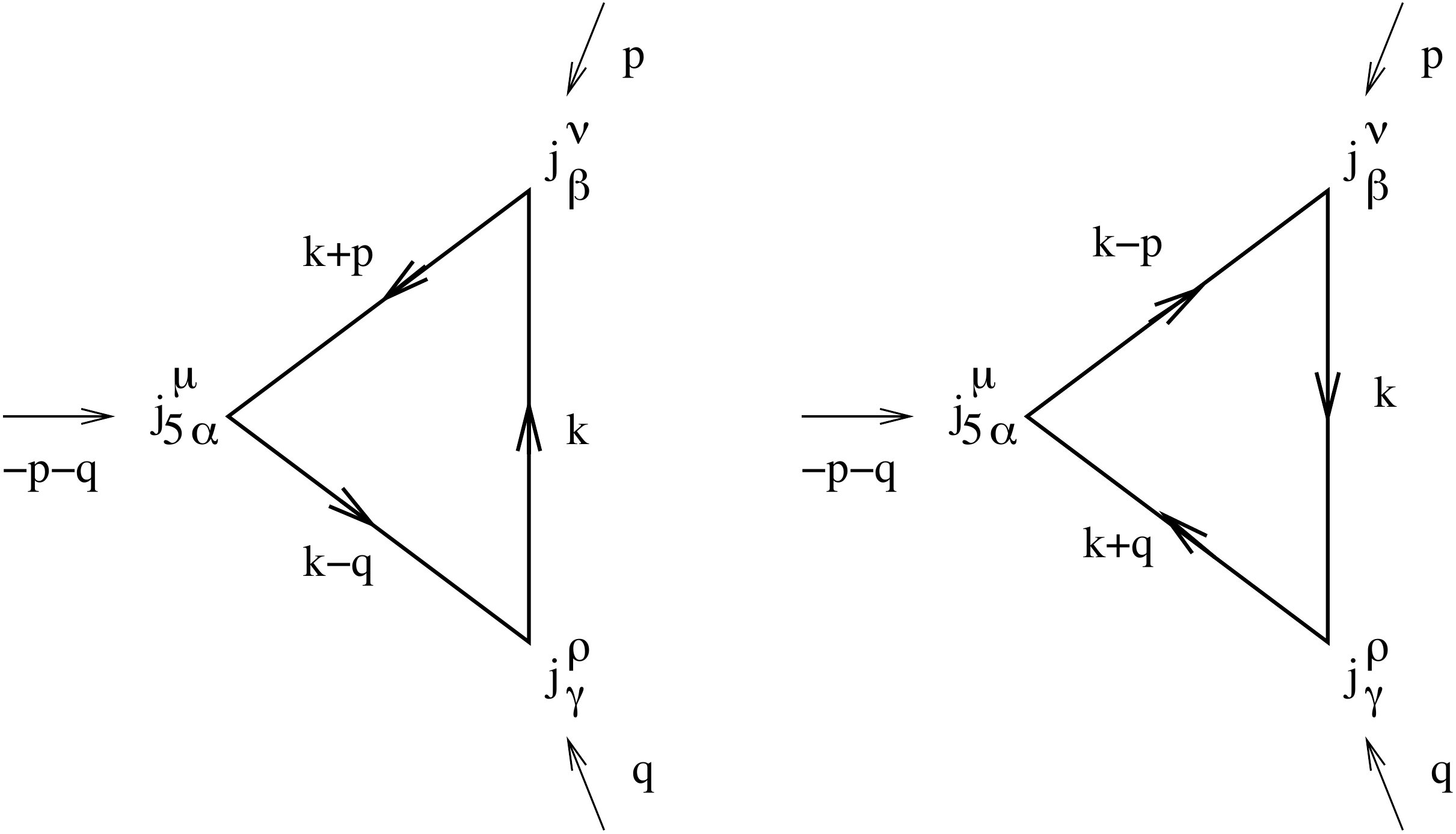}\\
\caption[]{The two triangle diagrams  contributing to the abelian anomaly} \label{triangle}
\end{figure}

As shown in Fig.~\ref{triangle}, there are two diagrams contributing to $i\G_{5\b\g}^{\ \m\n\r}$, corresponding to the two ways to contract the fermion fields contained in the currents. As explained above, the vertices contribute  $-\g^\m\g_5 t_\a$ and $-\g^\n t_\b$, resp. $-\g^\r t_\g$, while the fermion loop contributes an extra minus sign. A propagator for a fermion of momentum $k$ and mass $m$ is ${-i\over i\ksl+m-i\e} = (-i) {-i\ksl +m\over k^2+m^2-i\e}$.
We get
\ba\label{G51}
&&\hskip-1.6cm i \G_{5\a\b\g}^{\ \, \m\n\r}(-p-q,p,q)
\nonumber\\
&&\hskip-1.5cm= - \int{\d^4 k\over (2\pi)^4}\, \trD \Big\{ 
(-\g^\m\g_5)(-i) {-i(\ksl+\psl)\over (k+p)^2 -i\e}
(-\g^\n) (-i){-i\ksl\over k^2 -i\e}
(-\g^\r) (-i){-i(\ksl-\qsl)\over (k-q)^2 -i\e}\Big\} 
\trR t_\a t_\b t_\g
\nonumber\\
&&\hskip1.7cm + ( p\leftrightarrow q,\ \n\leftrightarrow\r,\ \b\leftrightarrow\g )  \ ,
\ea
or simplifying a bit (note that we get an extra minus sign from anticommuting the $\g_5$ to the left)
\ba\label{G52}
&&\hskip-0.7cm \G_{5\a\b\g}^{\ \, \m\n\r}(-p-q,p,q)
\nonumber\\
&&\hskip-0.7cm= -i\hskip-1.mm \int\hskip-1.5mm {\d^4 k\over (2\pi)^4}\, \trD \Big\{ \g_5\g^\m {(\ksl+\psl)\over (k+p)^2 -i\e}\g^\n {\ksl\over k^2 -i\e}\g^\r {(\ksl-\qsl)\over (k-q)^2 -i\e}\Big\} \trR t_\a t_\b t_\g
+ ( p\leftrightarrow q,\ \n\leftrightarrow\r,\ \b\leftrightarrow\g ) 
\ .
\nonumber\\
\ea
Of course, as it stands, the integral is divergent. Since dimensional
regularization is tricky in the presence of a $\g_5$, we will use the safer Pauli-Villars regularization. It consists in adding for each fermion another one with a large mass $M$ and opposite statistics (thus missing the minus sign accompanying the fermion loop). This amounts to subtracting from each integrand the same expression but with massive fermion propagators. If the integral (\ref{G52}) were convergent, the contribution of the regulator fields would vanish as $M\to\infty$, as it should for a sensible regularization. Instead, the divergence of (\ref{G52}) translates into an $M$-dependence of the regularized integral. At present, our regularized $\G_{5\a\b\g}^{\ \m\n\r}$  is
\be\label{gamma5reg}
\left[\G_{5\a\b\g}^{\ \m\n\r}(-p-q,p,q)\right]_{\rm reg}
= -i \hskip-1.mm\int\hskip-1.mm{\d^4 k\over(2\pi)^4}\, \Big( I^{\m\n\r}_0(k,p,q) - I^{\m\n\r}_M(k,p,q)\Big)\ \trR t_\a t_\b t_\g
 + ( p\leftrightarrow q,\ \n\leftrightarrow\r,\ \b\leftrightarrow\g ) \ ,
\ee
with
\be
I^{\m\n\r}_M(k,p,q)
=\trD \Big\{ \g_5\g^\m {\ksl+\psl+iM\over (k+p)^2 +M^2-i\e}
\g^\n {\ksl+iM\over k^2 +M^2-i\e}
\g^\r {\ksl-\qsl+iM\over (k-q)^2 +M^2-i\e}\Big\} \ .
\ee
Let us insists that, for any finite $M$, (\ref{gamma5reg}) is a finite well-defined integral, and all the usual manipulations, like shifting the integration variable, are allowed. Clearly, unlike (\ref{matterlagr}), the Lagrangian for these massive regulator fields lacks the chiral symmetry (\ref{abchir}), and this is why one finds the anomaly in the end.

We are only interested in
$(p+q)_\m \left[\G_{5\a\b\g}^{\ \m\n\r}(-p-q,p,q)\right]_{\rm reg}$ and, hence, we only need to compute $(p+q)_\m I^{\m\n\r}_M(k,p,q)$. Using
\be
\g_5(\psl+\qsl)=\g_5 (\psl+\ksl-iM) + (\ksl-\qsl-iM)\g_5 +2iM\g_5\ ,
\ee
we get
\ba\label{divIM}
(p+q)_\m I^{\m\n\r}_M(k,p,q)&=& {\trD \g_5 \g^\n  (\ksl+iM)
\g^\r (\ksl-\qsl+iM)\over [ k^2 +M^2-i\e][(k-q)^2 +M^2-i\e]}
+ {\trD \g_5 (\ksl+\psl+iM)
\g^\n (\ksl+iM) \g^\r \over [(k+p)^2 +M^2-i\e] [k^2 +M^2-i\e]}
\nonumber\\
&+&2i\,M\ {\trD  \g_5 (\ksl+\psl+iM)
\g^\n (\ksl+iM)\g^\r (\ksl-\qsl+iM)
\over  [(k+p)^2 +M^2-i\e] [k^2 +M^2-i\e] [(k-q)^2 +M^2-i\e]} \ .
\ea
Now, in the Dirac trace of the first term, only $\trD \g_5\g^\n\ksl\g^\r(\ksl-\qsl)=4i\e^{\n\l\r\s}k_\l (-q_\s)$ contributes. Then, when doing the (convergent) integral $\int\d^4 k$ (with the $M=0$ term subtracted), the $k_\l$ necessarily gets replaced by a four-vector proportional to $q_\l$, which is the only available one in this expression.\footnote{
Indeed, we have
\ba
&&\int\d^4 k \left\{ {k_\l 
\over [ k^2 +M^2-i\e][(k-q)^2 +M^2-i\e]}
- {k_\l 
\over [ k^2 -i\e][(k-q)^2 -i\e]} \right\}
\nonumber\\
&&
=  \int_0^1\d x 
\int\d^4 k \left\{ {k_\l\over [(k-x q)^2 + M^2 +x(1-x) q^2-i\e]^2}
- {k_\l\over [(k-x q)^2  +x(1-x) q^2-i\e]^2}\right\}\ ,
\ea
which is convergent.
Shifting the integration variable from $k$ to $k'=k-x q$, the integral is seen to be $\sim q_\l$.
}
When contracted with $\e^{\n\l\r\s} q_\s$ this vanishes. Similarly, the second term in (\ref{divIM}) only involves $k$ and $p$ and, after integration over $\d^4 k$, yields a contribution $\sim\e^{\l\n\s\r}p_\l p_\s=0$. Hence, only the third term in (\ref{divIM}) will contribute and, since
$\trD  \g_5 (\ksl+\psl+iM)
\g^\n (\ksl+iM)\g^\r (\ksl-\qsl+iM)=4M\, \e^{\n\r\l\s} p_\l q_\s$, we have
\be\label{divIM2}
(p+q)_\m I^{\m\n\r}_M(k,p,q)
\simeq
{8i\, M^2\ \e^{\n\r\l\s} p_\l q_\s
\over  [(k+p)^2 +M^2-i\e] [k^2 +M^2-i\e] [(k-q)^2 +M^2-i\e]}  \ ,
\ee
where $\simeq$ means equality up to the terms that vanish after integration.
Note that there is no corresponding term in $(p+q)_\m I_0^{\m\n\r}$. In this sense, the whole anomalous contribution comes from the regulator term $I_M^{\m\n\r}$. However, the vanishing of the first two terms in (\ref{divIM})  after integration is only guaranteed if we correctly consider the combination $I_0-I_M$. Using (\ref{divIM2}) and (\ref{gamma5reg}) we get
\be
-i (p+q)_\m \left[\G_{5\a\b\g}^{\ \m\n\r}(-p-q,p,q)\right]_{\rm reg}
=  8i M^2\, \e^{\n\r\l\s} p_\l q_\s\, I(p,q,M)\, \trR t_\a t_\b t_\g
 + ( p\leftrightarrow q,\ \n\leftrightarrow\r,\ \b\leftrightarrow\g ) \ ,
\ee
with
\be
I(p,q,M)= \int{\d^4 k\over (2\pi)^4} 
{1\over [(k+p)^2 +M^2-i\e] [k^2 +M^2-i\e] [(k-q)^2 +M^2-i\e]}  \ .
\ee
The regulator $M$ should be taken to $\infty$ in the end, so we only need the asymptotics of this integral for large $M$ which is easily obtained by letting $k=M\, l$:
\be\label{asympt}
I(p,q,M)\sim {1\over M^2} 
\int{\d^4 l\over (2\pi)^4} {1\over [l^2 +1-i\e]^3} 
={i\over 32\pi^2 M^2} \ ,
\ee
where the $i$ comes from the Wick rotation. Hence, we get a finite limite for $M^2 I(p,q,M)$ as we remove the regulator ($M\to\infty$), and
\ba\label{abeliantriangleresult}
-i (p+q)_\m \G_{5\a\b\g}^{\ \m\n\r}(-p-q,p,q)
&=&-{1\over 4\pi^2}\,\e^{\n\r\l\s} p_\l q_\s\,
\trR t_\a t_\b t_\g
+ ( p\leftrightarrow q,\ \n\leftrightarrow\r,\ \b\leftrightarrow\g )
\nonumber\\
&=&-{1\over 2\pi^2}\,\e^{\n\r\l\s} p_\l q_\s\,
\trR t_\a t_{(\b} t_{\g)} \ .
\ea
Upon setting $t_\a=t$, this exactly reproduces the result (\ref{divgamma5t}).

As discussed above, this result implies that the axial current $j_5^\m$ is not conserved in the quantum theory. As long as this only corresponds to an anomalous global chiral symmetry as in (\ref{abchir}), it does not lead to any inconsistency. It simply states that a certain global symmetry is broken by a quantum effect.  On the other hand, we might have probed for conservation of the two other (vector) currents $j_\b^\n$ and $j_\g^\r$. In a non-abelian gauge theory these currents do couple to the gauge fields and, as extensively discussed in the previous section, their non-conservation would signal a breakdown of the gauge invariance. Let us now check that this does not happen. Because of the symmetry under exchange of $p,\n,\b$ with $q,\r,\g$ it is enough to compute $p_\n \left[\G_{5\a\b\g}^{\ \m\n\r}(-p-q,p,q)\right]_{\rm reg}$. Now $p_\n I_M^{\m\n\r}(k,p,q)$ involves
\ba
\hskip-1.cm\big(\ksl+\psl+iM\big) \psl \big(\ksl+iM)
&=&\big(\ksl+\psl+iM\big)
\Big[ \big(\psl+\ksl-iM\big) -\big(\ksl-iM\big) \Big]\big(\ksl+iM\big)
\nonumber\\
&=&\big[ (k+p)^2+M^2\big]\big(\ksl+iM\big)
\ - \ \big(\psl+\ksl+iM\big)\big[ k^2+M^2\big] \ ,
\ea
which now leads to
\be\label{divIM3}
p_\n I^{\m\n\r}_M(k,p,q)= {\trD \g_5 \g^\m  (\ksl+iM)
\g^\r (\ksl-\qsl+iM)\over [ k^2 +M^2-i\e][(k-q)^2 +M^2-i\e]}
- {\trD \g_5 \g^\m (\ksl+\psl+iM)
 \g^\r (\ksl-\qsl+iM)\over [(k+p)^2 +M^2-i\e] [(k-q)^2 +M^2-i\e]} \ .
\ee
As compared to (\ref{divIM}) the third term is now absent. This is directly related to the absence of $\g_5$ in the current we are probing for conservation. As before, these two terms in (\ref{divIM3}) vanish upon integration over $\d^4 k$, provided we always consider the convergent combination $p_\n I^{\m\n\r}_M(k,p,q)-p_\n I^{\m\n\r}_0(k,p,q)$.\footnote{
Indeed, the first term in (\ref{divIM3}) is $\sim \e^{\m\l\r\s}k_\l q_\s$ and vanishes after the integration which replaces $k_\l$ by some  four-vector proportional to $q_\l$. In the second term, shifting the integration variable to $k'=k+p$, we see that it equals the first term with $p+q$ replacing $q$. Hence, it also vanishes after integration. 
}
We conclude that
\be\label{nuroconserved}
p_\n\, \G_{5\a\b\g}^{\ \m\n\r}(-p-q,p,q)
=q_\r\, \G_{5\a\b\g}^{\ \m\n\r}(-p-q,p,q)=0 \ .
\ee

\vskip8.mm
\subsection{Triangle diagram with chiral fermions only\label{trichir}}

\subsubsection{Chiral fermions: preliminaries\label{chirprelim}}

We begin by recalling some simple facts about chiral fermions.
Introduce the chirality projectors $P_L$ and $P_R$ as
\be
P_L={1+\g_5\over 2} \quad , \quad P_R={1-\g_5\over 2} \ .
\ee
Since $\g_5^\dag=\g_5$, these projectors are hermitian. They satisfy
$P_L \g^\m=\g^\m P_R$ and $P_R\g^\m=\g^\m P_L$, as well as $P_L P_R=P_R P_L=0$, so that in particular $P_L \g^\m P_L=0$, etc.
For any fermion field $\p$ we let
\be
\p_L=P_L\,\p \quad , \quad \p_R=P_R\, \p \quad ,\quad \p=\p_L+\p_R \ . 
\ee
The projected fields $\p_L$ and $\p_R$ are of course eigenstates of the chirality matrix $\g_5$:
\ba 
\g_5 \p_L=\p_L 
\quad &\Leftrightarrow&\quad P_L\p_L=\p_L,\ \quad P_R\p_L=0
\nonumber\\
\g_5\p_R=-\p_R
\quad &\Leftrightarrow&\quad P_R\p_R=\p_R,\ \quad P_L\p_R=0 \ .
\ea
$\p_L$ has positive chirality and is called left-handed, while $\p_R$ has negative chirality and is called right-handed. 
A fermion is called chiral if it either only has a left-handed part $\p_L$ or only a right-handed part $\p_R$.
Since the projectors are hermitian and 
$\pb=\p^\dag i\g^0$, we also have
\be
\overline{\p_L} P_L=0 \ , \quad \overline{\p_L} P_R=\overline{\p_L}
\quad , \quad 
\overline{\p_R} P_R=0 \ , \quad \overline{\p_R} P_L=\overline{\p_R}
\ .
\ee
It follows that
\be
\overline{\p_L} \p_L=\overline{\p_L} \left(P_L \p_L\right)
=\left(\overline{\p_L}P_L\right) \p_L = 0 \quad , \quad
\overline{\p_R} \p_R=\overline{\p_R} \left(P_R \p_R\right)
=\left(\overline{\p_R}P_R\right) \p_R = 0 \quad .
\ee
Thus a Dirac mass term only couples the left-handed to the right-handed part of a fermion,
\be\label{Diracmass}
\pb \p = \overline{\p_L} \p_R + \overline{\p_R} \p_L \ ,
\ee
and a chiral fermion (with either $\p_L=0$ or $\p_R=0$) cannot have a (Dirac) mass term. On the other hand,
\be
P_R \Dsl \p_L=\Dsl P_L \p_L= \Dsl \p_L
\quad , \quad
P_L \Dsl \p_R=\Dsl P_R \p_R= \Dsl \p_R \ ,
\ee
so that $\Dsl\p_L$ is right-handed and $\Dsl\p_R$ is left-handed, and the standard kinetic terms $\overline{\p_L}\Dsl\p_L$ or $\overline{\p_R}\Dsl\p_R$ are non-vanishing.

There is actually a possibility to write a mass term for a chiral fermion. As we will discuss in more detail in sect.~\ref{leftrightanti}, the charge conjugate field $\p^c=i\g^0 \cC \p^*$ (see eqs.~(\ref{chargeconj}) and (\ref{oppchir})) has the opposite chirality, i.e. if $\p$ is left-handed, then $\p^c$ is right-handed, and vice versa. Thus we can write a non-vanishing mass term of the form (\ref{Diracmass}) as
\be\label{Majmass}
m\big(\overline{\p_L} \p_L^c + \overline{\p_L^c}\p_L\big) \ .
\ee
Since $\p_L^c$ is right-handed, we could equivalently have written
$\overline{\p_R^c}\p_R  + \overline{\p_R} \p_R^c$. Note that the mass term (\ref{Majmass}) violates fermion number conservation\footnote{Although fermion number non-conservation is experimentally very much constrained, it is of course not inconsistent from a theoretical point of view.
}
which is related to the global $U(1)$ symmetry $\p\to e^{i\a}\p$.
Adopting a basis of $\g^\m$-matrices such that $\g_5=\begin{pmatrix} {\bf 1}&0\\ 0&-{\bf 1}\\ \end{pmatrix}$, one can write $\p_L=
\begin{pmatrix} \chi\\0\\ \end{pmatrix}$ with a 2-component Weyl spinor $\chi_\a$, $\a=1,2$. It is then not difficult to see that (\ref{Majmass}) can be rewritten as the standard mass term for 2-component Weyl spinors, namely\footnote{
One can take 
$\g^0=\begin{pmatrix} 0&{\bf 1}\\ -{\bf 1}&0\\ \end{pmatrix}$ and
$\g^j=\begin{pmatrix} 0&\s_j\\ \s_j&0\\ \end{pmatrix}$, so that
$\cC=-i\g^1\g^3$ after fixing an arbitrary phase. Then $\p_L^c=
\begin{pmatrix} 0\\ i\s_2\chi^*\\ \end{pmatrix}$ and $\overline{\p_L}=\p_L^\dag i\g^0 = i(0,{\chi^*}^T)$ so that $\overline{\p_L} \p_L^c= i {\chi^*}^T (i\s_2)\chi^*$ and similarly $(\overline{\p_L} \p_L^c)^*=
\overline{\p_L^c}\p_L=i \chi^T (i\s_2)\chi$.
}
\be\label{Weylmassterm}
i m \sum_{\a,\b}\big( \chi_\a \e_{\a\b}\chi_\b +  \chi^*_\b \e_{\b\a}\chi_\a^* \big) \ ,
\ee
where $\e_{\a\b}=(i\s_2)_{\a\b}$. 
More generally, one could have several fermion fields $\p_L^r$, resp. $\chi^r$, $r=1,\ldots n$ with a  mass term involving a (symmetric) mass matrix $m_{rs}$ as
\be\label{massmatrix}
\sum_{r,s} m_{rs}\big(\,\overline{\p_L^r}\, \p_L^{s,c} + \overline{\p_L^{r,c}}\,\p_L^s\,\big) 
=
i\sum_{r,s,\a,\b} m_{rs}\big(\chi_\a^r \e_{\a\b}\chi_\b^s +  (\chi_\b^r)^* \e_{\b\a}(\chi_\a^s)^* \big)
\ee
However, we will see in sect.~\ref{whichgroups} that depending on the representation $\cR$ of the gauge group (or any global symmetry group) carried by the $\p_L$, such a mass term may or may not be allowed by the gauge symmetry (or global symmetry) and, whenever it is allowed, the representation $\cR$ does not lead to any anomaly. For this reason, we may just as well continue to consider massless fermions only.
Further issues about generating masses for chiral fermions arise in spontaneously broken gauge theories like the standard model. These will be briefly discussed in sect.~\ref{anomstandard}.

\subsubsection{Matter Lagrangian for chiral fermions and ill-defined determinants\label{chidd}} 

For a massless chiral fermion one can write a gauge-invariant kinetic term as
\be\label{chiralmatlagr}
\cL_{\rm matter}^L = - \overline{\p_L} \Dsl \p_L \ ,
\ee
(and similarly for a right-handed $\p_R$).
In order to continue to use the usual Feynman rules it is useful to rewrite this in terms of a non-chiral fermion field $\p$ by inserting the chirality projector:
\be\label{chiralmatlagr2}
\cL_{\rm matter}^L = - \pb \Dsl P_L\p \ .
\ee
This shows that the propagators now are simply $(-i)\times P_L\, {-i\ksl\over k^2-i\e}$ and the vertices are $i \times (i \g^\m t_\a) P_L$. Suppose that one computes a fermion loop diagram like the above triangle diagrams. The corresponding expression is 
\ba
&&\sim \trD (i \g^\m t_\a) P_L   P_L\, {-i(\ksl+\psl)\over (k+p)^2-i\e} (i \g^\n t_\b) P_L   P_L\, {-i\ksl\over k^2-i\e} \ldots 
\nonumber\\
&&=\trD (i \g^\m t_\a) P_L  \, {-i(\ksl+\psl)\over (k+p)^2-i\e} (i \g^\n t_\b) P_L   \, {-i\ksl\over k^2-i\e} \ldots \ .
\ea
Exactly the same expression would be obtained from the same vertices
$i \times (i \g^\m t_\a) P_L$ but with the propagators being $(-i)\times\, {-i\ksl\over k^2-i\e}$ i.e. without the chirality projector. These Feynman rules would be obtained from a matter Lagrangian
\be\label{chiralmatlagr3}
\cL_{\rm matter}^{L'} = - \pb \dsl \p - \pb (-i\Asl) P_L \p
= - \pb \dsl P_R \p - \pb \Dsl P_L \p\ ,
\ee
containing a left- and a right-handed (i.e.~a non-chiral) propagating fermion, but with only the left-handed part coupling to the gauge field.

We now have two different ways to see why chiral fermions need not yield a gauge invariant effective action $\wt W[A]$. This is particularly obvious for the Lagrangian
(\ref{chiralmatlagr3}) which is manifestly not gauge invariant since only the left-handed part of the fermion field couples to the gauge field, while the right-handed part doesn't. At tree-level, with only external left-handed matter fields or gauge fields this does not manifest itself, but once we compute loops the non-gauge invariance of the right-handed part will show up as a non-invariant determinant.  Alternatively, we may consider the fully gauge invariant Lagrangian (\ref{chiralmatlagr}). There is no obvious way to define the functional integral
\be\label{chiralFI}
e^{i\wt W[A]} = \int \cD\p_L \cD \overline{\p_L} e^{-i \int  \overline{\p_L} D\hskip-2.0mm/ \p_L} \sim \Det( \Dsl P_L) \ ,
\ee
since $\Dsl P_L$ is an operator that maps left-handed fermions to right-handed fermions, which live in a different part of the Hilbert space. For such an operator there is no obvious way to define a determinant. This problem is particularly clear in the Euclidean, where we denote the operator as $(\Dsl P_L)_E$. One may instead try to consider the operator $(\Dsl P_L)_E^\dag (\Dsl P_L)_E$ which maps left-handed fermions to left-handed ones and which does have a well-defined (and gauge invariant) determinant. If $(\Dsl P_L)_E$ had a well-defined determinant we would have $\Det\big[(\Dsl P_L)_E^\dag (\Dsl P_L)_E\big]=\left\vert\Det (\Dsl P_L)_E\right\vert^2$. We can nevertheless use this to define the modulus of the ill-defined determinant and hence of the ill-defined (Euclidean) functional integral (\ref{chiralFI}).
However, this does not fix the phase which, of course, depends on the ``external" gauge field $A_\m$: there is no guarantee that one can define the phase in a satisfactory gauge invariant way. It is precisely if this cannot be done that one has an anomaly under the gauge symmetry. This argument shows that the anomaly resides in the phase of the determinant, i.e.~in the imaginary part of the Euclidean $\wt W_E[A]$. In  the second part of these lectures, in sect.~\ref{indexth}, we will make this argument more precise and exploit it to actually compute the anomaly. There we also show that when continuing from the Minkowskian to the Euclidean, the only terms in an Euclidean action that are imaginary are so-called topological terms, i.e.~terms involving $\e^{\m\n\r\s}$. Thus the present argument shows that the anomaly can only concern these $\e^{\m\n\r\s}$-terms. This is indeed what we have already observed.

\subsubsection{Feynman diagram computation of the triangle for chiral fermions : the anomalous part\label{Feynanpart}}

\begin{figure}[h]
\centering
\includegraphics[width=0.6\textwidth]{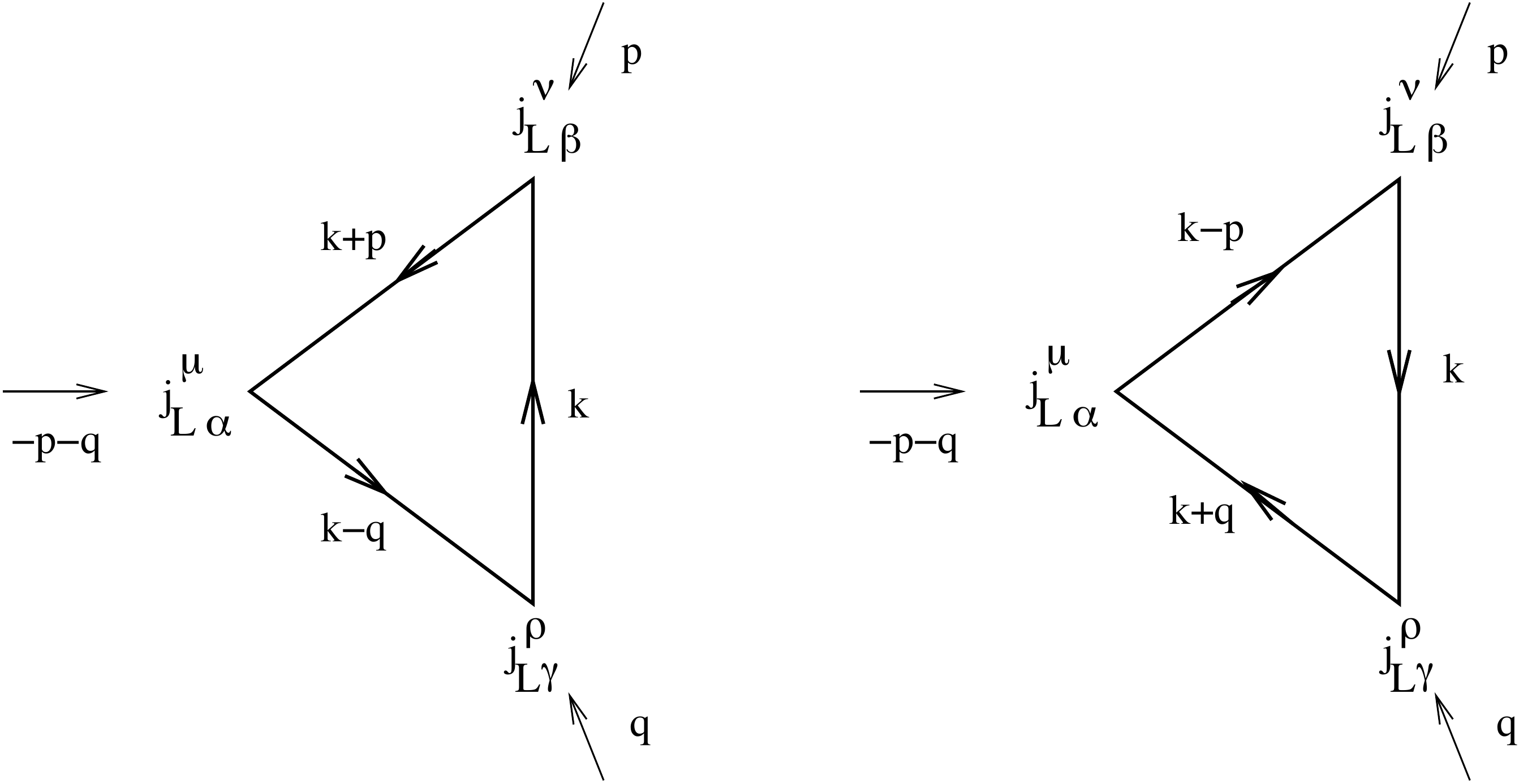}\\
\caption[]{The two triangle diagrams  contributing to the non-abelian chiral anomaly} \label{trianglechir}
\end{figure}

We will now compute the three-point vertex function $\G^{\ \,\m\n\r}_{L,\a\b\g}$ for the above matter Lagrangian (\ref{chiralmatlagr}) for a chiral left-handed fermion. As explained, we may use ordinary propagators but insert a chirality projector $P_L$ at each vertex. This amount to computing the expectation value of three left-handed currents, $\G^{\ \,\m\n\r}_{L,\a\b\g}=-\langle T\big( J_{L\a}^\m J_{L\b}^\n J_{L\g}^\r\big)\rangle$ where the currents are $J_{L\a}^\m=i\pb \g^\m t_\a P_L\p$. The corresponding two diagrams are shown in Fig.~\ref{trianglechir}.

Exactly as in the previous subsection for $\G_{5\b\g}^{\ \m\n\r}(-p-q,p,q)$, cf.~eqs. (\ref{G51}) and (\ref{G52}) (except that in (\ref{G52}) there was an extra minus sign from anticommuting the $\g_5$ to the left), we now get
\ba\label{Gmnr}
\G_{L,\a\b\g}^{\ \,\m\n\r}(-p-q,p,q)\hskip-1.mm 
&=& \hskip-1.mm i \int\hskip-1.mm 
{\d^4 k\over (2\pi)^4}\, \trD \Big\{\g^\m P_L{(\ksl+\psl)\over (k+p)^2 -i\e}\g^\n P_L {\ksl\over k^2 -i\e}\g^\r P_L{(\ksl-\qsl)\over (k-q)^2 -i\e}\Big\}\, \trR t_\a t_\b t_\g
\nonumber\\
&&\hskip2.6cm+\ ( p\leftrightarrow q,\ \n\leftrightarrow\r,\ \b\leftrightarrow\g ) 
\ .
\ea
Again, we use the Pauli-Villars method to regularize this integral. 

As is well-known, Pauli-Villars regularization preserves gauge invariance, so how then can we find an anomaly? On the other hand, for chiral fermions one cannot write a (Dirac) mass term, so how then can we use Pauli-Villars? The answers to both questions are related, of course. As discussed above, it is perfectly equivalent, at least at the level of computing the Feynman diagrams, to consider the fermions as non-chiral, i.e. with ordinary propagators, but with chiral interactions as described by the matter Lagrangian (\ref{chiralmatlagr3}). This allows us to add a mass term for the fermions, as required by the Pauli-Villars regularization. As also discussed above, this matter Lagrangian (\ref{chiralmatlagr3}) is {\it not} gauge invariant, and this is why we will find an anomaly in the end.

Since $\G_{L,\a\b\g}^{\ \,\m\n\r}(-p-q,p,q)$  has  a degree of divergence equal to one it is enough to add one regulator field of opposite statistics. However, at intermediate steps of the computation, we will encounter integrals of degree of divergence equal to two and, in order to also efficiently regularize these integrals, we add one more pair of regulator fields (one bosonic and one fermionic). Using the notation $\eta_0=\eta_2=1,\ \eta_1=\eta_3=-1$, as well as $M_0=0$, we then have
\be\label{gammareg}
\left[\G_{L,\a\b\g}^{\ \,\m\n\r}(-p-q,p,q)\right]_{\rm reg}
= i \int{\d^4 k\over(2\pi)^4}\,  \left(\sum_{s=0}^3 \eta_s 
J^{\m\n\r}_{M_s}(k,p,q) \right) \trR t_\a t_\b t_\g
\ + \ ( p\leftrightarrow q,\ \n\leftrightarrow\r,\ \b\leftrightarrow\g ) \ ,
\ee
with
\be
J^{\m\n\r}_M(k,p,q)
=\trD \Big\{ \g^\m P_L {\ksl+\psl+iM\over (k+p)^2 +M^2-i\e}
\g^\n P_L {\ksl+iM\over k^2 +M^2-i\e}
\g^\r P_L {\ksl-\qsl+iM\over (k-q)^2 +M^2-i\e}\Big\} \ .
\ee
Cancellation of the leading (quadratic) divergences just requires
$\sum_{s=0}^3 \eta_s=0$, while cancellation of the subleading  (logarithmic) divergences will require to choose the regulator masses such that $\sum_{s=1}^3 \eta_s M_s^2=0$, i.e.~$M_2^2=M_1^2+M_3^2$.

We will first show that $J^{\m\n\r}_M(k,p,q)$ equals the same expression with all $iM$'s in the {\it numerator} deleted. Indeed, pick out any piece in the Dirac trace involving such an $iM$. Due to the cyclicity of the trace it always appears as (with $\l$ and $\s$ being any of the $\m,\n,\r$)
\be
\ldots \g^\l P_L iM \g^\s P_L \ldots = \ldots \g^\l  iM P_L P_R \g^\s \ldots = 0 \ .
\ee
Thus
\ba
J^{\m\n\r}_M(k,p,q)
&=&{\trD  \g^\m P_L \big(\ksl+\psl\big)\g^\n P_L \ksl
\g^\r P_L \big(\ksl-\qsl\big)
\over [(k+p)^2 +M^2-i\e] [k^2 +M^2-i\e]  [(k-q)^2 +M^2-i\e] }
\nonumber\\
&=&{\trD  \big(\ksl-\qsl\big) \g^\m  \big(\ksl+\psl\big)\g^\n  \ksl
\g^\r P_L 
\over [(k+p)^2 +M^2-i\e] [k^2 +M^2-i\e]  [(k-q)^2 +M^2-i\e] }
\ .
\ea
Again, we are only interested in $(p+q)_\m J^{\m\n\r}_M(k,p,q)$. Thus we need 
\ba
\hskip-0.0cm\trD  \big(\ksl-\qsl\big) \big(\psl+\qsl) \big(\ksl+\psl\big)\g^\n  \ksl\g^\r P_L
&=&\trD  \big(\ksl-\qsl\big) \big(\psl+\ksl-\ksl+\qsl) \big(\ksl+\psl\big)\g^\n  \ksl\g^\r P_L
\nonumber\\
&&\hskip-3.4cm
=(k+p)^2 \trD  \big(\ksl-\qsl\big) \g^\n  \ksl\g^\r P_L
-(k-q)^2\trD \big(\ksl+\psl\big)\g^\n  \ksl\g^\r P_L
\nonumber\\
&&\hskip-3.4cm
= \big[(k+p)^2 +M^2\big]\trD  \big(\ksl-\qsl\big) \g^\n  \ksl\g^\r P_L
-\big[(k-q)^2+M^2]\trD \big(\ksl+\psl\big)\g^\n  \ksl\g^\r P_L
\nonumber\\
&&\hskip-3.0cm
+M^2 \trD \big(\psl+\qsl\big) \g^\n \ksl\g^\r P_L \ ,
\ea
yielding
\ba
(p+q)_\m J^{\m\n\r}_M(k,p,q)
&=&{\trD  \big(\ksl-\qsl\big) \g^\n  \ksl\g^\r P_L \over
[k^2 +M^2-i\e]  [(k-q)^2 +M^2-i\e] }
- {\trD \big(\ksl+\psl\big)\g^\n  \ksl\g^\r P_L \over
[(k+p)^2 +M^2-i\e] [k^2 +M^2-i\e] }
\nonumber\\
&+& M^2 {\trD \big(\psl+\qsl\big) \g^\n \ksl\g^\r P_L \over
[(k+p)^2 +M^2-i\e] [k^2 +M^2-i\e]  [(k-q)^2 +M^2-i\e] } \ .
\ea
Hence,
\ba\label{GammaLdiv}
-i(p+q)_\m\left[\G_{L,\a\b\g}^{\ \,\m\n\r}(-p-q,p,q)\right]_{\rm reg}
&=& \Big( I^{\n\r}(-q) - I^{\n\r}(p) 
+ \sum_{s=1}^3 \eta_s M_s^2 J_{M_s}^{\n\r}(p,q) \Big) 
\trR t_\a t_\b t_\g
\nonumber\\
&&+\ ( p\leftrightarrow q,\ \n\leftrightarrow\r,\ \b\leftrightarrow\g ) 
\ ,
\ea
where $I^{\n\r}(p)$ and $J_{M_s}^{\n\r}(p,q)$ are {\it convergent} integrals, allowing us in particular to shift the integration variables:
\be\label{Inr}
I^{\n\r}(p)= \int{\d^4 k\over(2\pi)^4}\, 
\sum_{s=0}^3 \eta_s \
{\trD \big(\ksl+\psl\big)\g^\n  \ksl\g^\r P_L \over
[(k+p)^2 +M_s^2-i\e] [k^2 +M_s^2-i\e] } \ ,
\ee
and
\ba\label{JMs}
J_{M_s}^{\n\r}(p,q)&=&\int{\d^4 k\over(2\pi)^4}\,
{\trD \big(\psl+\qsl\big) \g^\n \ksl\g^\r P_L \over
[(k+p)^2 +M_s^2-i\e] [k^2 +M_s^2-i\e]  [(k-q)^2 +M_s^2-i\e] }
\nonumber\\
&=&2\int_0^1\d x \int_0^{1-x} \d y \int{\d^4 k\over(2\pi)^4}\,
{\trD \big(\psl+\qsl\big) \g^\n \ksl\g^\r P_L \over
[(k+xp-yq)^2 +M_s^2+ r(p,q,x,y)-i\e]^3}
\nonumber\\
&=&2\int_0^1\d x \int_0^{1-x} \hskip-1.mm\d y\ 
\trD \big(\psl+\qsl\big) \g^\n \big( y\qsl-x\psl\big)\g^\r P_L 
\int{\d^4 k'\over(2\pi)^4}\,
{1 \over [{k'}^2 +M_s^2+ r(p,q,x,y)-i\e]^3} \ ,
\nonumber\\
\ea
where $r(p,q,x,y)= x(1-x)p^2+y(1-y)q^2 +2xy\,pq$.
We only need the large $M_s$ limits of $M_s^2 J_{M_i}^{\n\r}(p,q)$. They are all the same, (cf~(\ref{asympt}))
\ba\label{JMlimit}
\lim_{M_s\to\infty} M_s^2 J_{M_s}^{\n\r}(p,q) &=& 2\int_0^1\d x \int_0^{1-x} \d y\, 
\trD \big(\psl+\qsl\big) \g^\n \big( y\qsl-x\psl\big)\g^\r P_L 
\times {i\over 32\pi^2}
\nonumber\\
&=&{i\over 16\pi^2} {1\over 6}\,\trD \big(\psl+\qsl\big) \g^\n \big( \qsl-\psl\big)\g^\r P_L 
\nonumber\\
&=&{i\over 24\pi^2} \left[ q^\n q^\r-p^\n p^\r + {p^2-q^2\over 2} \eta^{\n\r} + i \e^{\l\n\s\r} p_\l q_\s \right]\ .
\ea

We have mentioned above in subsection \ref{chidd} (and explicitly observed for the abelian anomaly) that the anomaly is given by the terms involving $\e^{\n\r\l\s}$. We will see soon that the $I^{\n\r}(p)$ and $I^{\n\r}(-q)$ do not give rise to any such terms. Hence, inserting the $\e$-terms of (\ref{JMlimit}) into (\ref{GammaLdiv}), 
we can already state the main result of this section
\ba\label{GammaLdiv2}
-i(p+q)_\m\G_{L,\a\b\g}^{\ \,\m\n\r}(-p-q,p,q)\Big\vert_{\e -{\rm terms}}
\hskip-2.mm&=&\hskip-2.mm \sum_{s=1}^3 \eta_s\, {i\over 24\pi^2}\,  i \e^{\l\n\s\r} p_\l q_\s\,
\trR t_\a t_\b t_\g
+\ ( p\leftrightarrow q,\ \n\leftrightarrow\r,\ \b\leftrightarrow\g ) 
\nonumber\\
&=& \hskip-2.mm-{1\over 12 \pi^2}\ \e^{\n\r\l\s} p_\l q_\s\, D^\cR_{\a\b\g}
\ ,
\ea
where we used $\sum_{s=1}^3\eta_s=-\eta_0=-1$ and $D^\cR_{\a\b\g}= \trR t_\a t_{(\b}t_{\g)}$. Using our general result (\ref{Wardquickref}), we can equivalently write this as
\ba\label{chiralanom1}
\ca^L_\a(x) &=& -{1\over 96\pi^2} \e^{\m\n\r\s} \trR t_\a F_{\m\n}^{\rm lin} F_{\r\s}^{\rm lin} + \cO(A^3) 
\nonumber\\
&=&-{1\over 24\pi^2} \e^{\m\n\r\s} \trR t_\a \del_\m A_\n \del_\r A_\s + \cO(A^3) \ ,
\ea
where we added a superscript $L$ on $\ca_\a(x)$ to remind us that this anomaly is computed for left-handed fermions. We have not determined the $\cO(A^3)$ contributions to the anomaly. As discussed for the case of the abelian anomaly, they appear as anomalous contributions to the Ward identities for four-point or higher-point vertex functions. Of course, one could obtain them along similar lines, but we will not do so here. We will show later that the so-called Wess-Zumino consistency condition completely fixes these higher-order contributions in terms of  (\ref{chiralanom1}). For now, let us only say that, contrary to what the first line of (\ref{chiralanom1}) might suggest, we will find that the complete anomaly is {\it not}  $\sim \e^{\m\n\r\s}\trR t_\a F_{\m\n} F_{\r\s}$ but instead is given by 
\be\label{fullnaanom}
\ca^L_\a(x)=-{1\over 24\pi^2} \e^{\m\n\r\s} \trR t_\a\, \del_\m \Big(A_\n \del_\r A_\s -{i\over 4} A_\n [ A_\r , A_\s]\Big)\ .
\ee
Note that the absence of terms quartic in $A$  means that the Ward identities relating the 5-point and 4-point functions are not anomalous.\footnote{ Recall that  for the abelian anomaly there was a  quartic term in $A$ but the group theoretical factors combined in such a way that the corresponding anomalous contribution to the pentagon diagram was proportional to the Jacobi identity and thus vanished.} Note also that $\e^{\m\n\r\s} [A_\n,[A_\r,A_\s]]=0$ by the Jacobi identity, so that the cubic term can be rewritten as 
$\e^{\m\n\r\s} A_\n[A_\r,A_\s]= {1\over 2}\e^{\m\n\r\s} \big( A_\n[A_\r,A_\s]+ [A_\r,A_\s] A_\n\big)$, showing that this term also only involves the symmetrized trace of the three generators. Hence
\be\label{fullnaanom2}
\ca^L_\a(x)=-{1\over 24\pi^2} \e^{\m\n\r\s}  \del_\m \Big(A_\n^\b \del_\r A_\s^\g -{i\over 4} A_\n^\b [ A_\r , A_\s]^\g\Big) D^\cR_{\a\b\g}\ .
\ee

Obviously, we could just as well have done the same computation for right-handed fermions with $P_R=\half(1-\g_5)$ replacing $P_L=\half(1+\g_5)$ everywhere. As a result, the final sign in front of the $\e^{\m\n\r\s}$ would have been opposite and hence
\be\label{anLanR}
\ca_\a^R(x)=-\ca_\a^L(x) \ .
\ee

\subsubsection{The remaining terms from the triangle for chiral fermions and the anomalous Ward identity\label{remainingtria}}

We now finish our computation and explicitly evaluate $I^{\n\r}(p)$ as given in (\ref{Inr}). The reader may skip this subsection since it is not needed to get the anomaly.  We nevertheless want to show how the contributions from $I^{\n\r}(p)$ combine with those from the previous subsection to give the full anomalous Ward identity derived in sect. \ref{anWardsect}.

Introducing a Feynman parameter $x$ in eq.~(\ref{Inr}) and then shifting the integration variable from $k$ to $k-x p$ we get
\ba
I^{\n\r}(p)
&=&\int_0^1\d x \int{\d^4 k\over(2\pi)^4}\, 
\sum_{s=0}^3 \eta_s \
{\trD \big(\ksl+\psl\big)\g^\n  \ksl\g^\r P_L \over
[(k+xp)^2 +M_s^2+x(1-x)p^2-i\e]^2 }
\nonumber\\
&=&\int_0^1\d x \int{\d^4 k\over(2\pi)^4}\, 
\sum_{s=0}^3 \eta_s \
{\trD \big(\ksl+(1-x)\psl\big)\g^\n  \big(\ksl-x\psl\big)\g^\r P_L \over
[{k}^2 +M_s^2+x(1-x)p^2-i\e]^2 }
\ .
\ea
The Dirac trace appearing in the numerator equals
$\trD \big(\ksl+(1-x)\psl\big)\g^\n  \big(\ksl-x\psl\big)\g^\r P_L
=4 k^\n k^\r -2k^2\eta^{\n\r} -4x(1-x)p^\n p^\r +2x(1-x) p^2\eta^{\n\r} \ +$ terms linear in $k$. In particular, the contribution of the $\g_5$ which is $\sim\e^{\l\n\s\r} p_\l k_\s$ is linear in $k$. Upon performing the $k$-integral, all terms linear in $k$ vanish, and we can also replace  $4 k^\n k^\r \simeq k^2 \eta^{\n\r}$. As promised, $I^{\n\r}$ does not contribute an $\e^{\n\r\l\s}$-term. We get
\be
I^{\n\r}(p) = \int_0^1 \d x \Big[ -\eta^{\n\r} I_2(R) - 2x(1-x) \big( 2p^\n p^\r -p^2\eta^{\n\r}\big) I_0(R) \Big] \ ,
\ee
where $R$ is shorthand for $x(1-x) p^2$ and
\ba
I_0(R)&=&\int{\d^4 k\over(2\pi)^4}\, \sum_{s=0}^3 \eta_s 
{1\over [{k}^2 +M_s^2+R -i\e]^2 } 
= -{i\over 16\pi^2}\sum_{s=0}^3 \eta_s \log\left( M_s^2 +R\right) \ ,
\nonumber\\
I_2(R)&=&\int{\d^4 k\over(2\pi)^4}\, \sum_{s=0}^3 \eta_s 
{k^2\over [{k}^2 +M_s^2+R -i\e]^2 }
={i\over 8\pi^2} \sum_{s=0}^3 \eta_s \left( M_s^2+R\right)\log\left( M_s^2 +R\right) \ .
\nonumber\\
\ea
where we used $\sum_{s=0}^3 \eta_s =0$ and 
$\sum_{s=0}^3 \eta_s M_s^2=\sum_{s=1}^3 \eta_s M_s^2=0$ to cancel the quadratically and logarithmically divergent pieces. 
Recall also that $\eta_0=1$ and $M_0=0$ and that the other $M_s$ should be taken to infinity in the end, so that we can drop any terms that vanish in this limit. Thus
\ba
\sum_{s=0}^3 \eta_s \log\left( M_s^2 +R\right)  
&\sim& \log R +\sum_{s=1}^3 \eta_s  \log M_s^2 \ ,
\nonumber\\
\hskip-1.cm\sum_{s=0}^3 \eta_s \left( M_s^2 +R\right) \log\left( M_s^2 +R\right) 
&\sim& R\log R - R + R \sum_{s=1}^3 \eta_s  \log M_s^2+\sum_{s=1}^3 \eta_s  M_s^2 \log M_s^2 \ ,
\ea
so that
\ba
I^{\n\r}(p)&=& {i\over 4\pi^2}\int_0^1\d x \Bigg[ x(1-x)
(p^\n p^\r- p^2\eta^{\n\r}) \left( \log \Big(x(1-x) p^2\Big)+ \sum_{s=1}^3 \eta_s \log M_s^2 \right) 
\nonumber\\
&&\hskip2.6cm+ {1\over 2}\, \eta^{\n\r} 
\left(x(1-x) p^2  - \sum_{s=1}^3 \eta_s M_s^2 \log M_s^2 \right)
\Bigg]
\nonumber\\
&=&
-{i\over 24\pi^2}\Bigg[
(p^2\eta^{\n\r}-p^\n p^\r)  \left( \log p^2-{5\over 3}+ \sum_{s=1}^3 \eta_s \log M_s^2 \right) 
- {p^2\over 2} \eta^{\n\r} + 3 \eta^{\n\r}\sum_{s=1}^3 \eta_s M_s^2 \log M_s^2 \Bigg]\ .
\nonumber\\
\ea
Combining with (\ref{JMlimit}) (recalling again that $\sum_{s=1}^3\eta_s=-1$) we get
\be\label{IIJ}
I^{\n\r}(-q) - I^{\n\r}(p) 
+ \sum_{s=1}^3 \eta_s M_s^2 J_{M_s}^{\n\r}(p,q)\big\vert_{{\rm no}\ \e-{\rm terms}} 
=i\,\Pi^{\n\r}_{M}(p)-i\,\Pi^{\n\r}_{M}(q) 
\ ,
\ee
with
\be\label{PinrM}
\Pi^{\n\r}_{M}(p)=
(p^2\eta^{\n\r}-p^\n p^\r)\ {1\over 24\pi^2} 
\left( \log p^2-{8\over 3}+ \sum_{s=1}^3 \eta_s \log M_s^2 \right) \ .
\ee
Note that each $\Pi^{\n\r}_{M}(p)$ and $\Pi^{\n\r}_{M}(q)$ is transverse. Note also that the ``leading term" $\sum_{s=1}^3 \eta_s M_s^2 \log M_s^2 $ has cancelled in the difference  (\ref{IIJ}). The remaining term $\sum_{s=1}^3 \eta_s \log M_s^2$ in $\Pi^{\n\r}_{M}(p)$ is related to the usual logarithmic behavior of the vacuum polarization and is eventually cancelled by an appropriate counterterm of the gauge-field Lagrangian.

Of course, (\ref{IIJ}) is antisymmetric under exchange of $\n$ with $\r$ and $p$ with $q$, so that when adding in (\ref{GammaLdiv}) the term with $p,\n,\b\leftrightarrow q,\r,\g$ one generates the combination 
\be
\trR (t_\a t_\b t_\g - t_\a t_\g t_\b) = i C_{\dd\b\g} \trR t_\a t_\dd = i C_{\dd\b\g}\ g_i^2 C_R^{(i)} \dd_{\a\dd} = i\, g_i^2  C_R^{(i)}\, C_{\a\b\g} \ ,
\ee 
where the subscript/superscript $i$ indicates the simple or $U(1)$ factor $G_i$ to which the indices $\b,\g$ correspond.
Thus definig
\be
\Pi^{\n\r}_{M,(i)}(p)= g_i^2 C_R^{(i)} \Pi_M^{\n\r}(p) \ ,
\ee
and putting everything together we finally get
\be\label{GammaLdiv3}
-i(p+q)_\m\left[\G_{L,\a\b\g}^{\ \,\m\n\r}(-p-q,p,q)\right]_{\rm reg}
= -{1\over 12 \pi^2}\ \e^{\n\r\l\s} p_\l q_\s\, D^\cR_{\a\b\g}
- C_{\a\b\g} \left[ \Pi_{M,(i)}^{\n\r}(p) - \Pi_{M,(i)}^{\n\r}(q)\right] \ ,
\ee
which is exactly of the form of the anomalous Ward identity (\ref{nonabWard2}), provided we can indeed identify $ \Pi_{M,(i)}^{\n\r}(p)$ with the corresponding matter contribution to the vacuum polarization. However, this is easy to check. 

Indeed, we have seen that the $\g_5$ in the chirality projector $P_L$ does not contribute to the above computation of $\Pi^{\n\r}_M(p)$, hence the only effect of $P_L$ is the factor $\half$. We can compare with the well-known vacuum polarization in QED due to an electron: $\Pi_{QED}^{\n\r}(p)=(p^2\eta^{\n\r}-p^\n p^\r) \pi(p)$ with $\pi(p)= {e^2\over 2\pi^2} \int_0^1 \d x\, x(1-x) \log[ m_e^2 + x(1-x) p^2] + \ldots$, where the ellipses refer to $p$-independent terms that depend on the renormalization conditions. For $m_e=0$ this gives $\Pi_{QED}^{\n\r}(p)=(p^2\eta^{\n\r}-p^\n p^\r) {e^2\over 12\pi^2} (\log p^2 + {\rm const})$. Taking into account the $\half$ from $P_L$, as well as $\tr t t=e^2$ for QED with only electrons, i.e. $g_i^2 C_R^{(i)} \to e^2$, we see that (\ref{PinrM}), and correspondingly $\Pi^{\n\r}_{M,(i)}(p)$,  has precisely the correct normalization (and sign!) to be identified with the matter contribution to the vacuum polarization.

In conclusion, we see that $\G_{L,\a\b\g}^{\ \,\m\n\r}$, as computed from the triangle Feynman diagram, satisfies the anomalous Ward identity (\ref{nonabWard2}) and that the anomalous terms are those involving the $\e$-tensor. Since $\e^{\n\r\l\s} p_\l q_\s$ is symmetric under exchange of $(\n,\, p)$ with $(\r,\, q)$, overall Bose symmetry (the currents, resp. the gauge fields are bosonic) requires that the remaining factor must also be symmetric, and hence it must occur in the form of the symmetrized trace $D^\cR_{\a\b\g}$, as it indeed does.

\section{Locality and finiteness of the anomaly\label{lfa}}
\setcounter{equation}{0}

So far we have computed the anomaly under global chiral transformations of the fermions, i.e. the abelian anomaly, as well as the anomaly under (non-abelian) gauge transformations for chiral fermions. In both cases we have found that the anomaly $\cA(x)$, resp. $\ca_\a(x)$, or rather $\int \e(x) \cA(x)$, resp. $\int \e^\a(x) \ca_\a(x)$, which is the variation of the effective action, is a {\it local} functional\footnote{
A  functional $F[\f]$ of a field $\f$ is called {\it local} if $F[\f]=\int \d^d x f(x)$ with $f(x)$ depending only on $\f(x)$ and {\it finitely} many derivatives of $\f(x)$. Equivalently, after Fourier transforming to momentum space, the finitely many derivatives translate into a {\it polynomial} in the momenta.
}
of the gauge fields with {\it finite} coefficients, i.e. coefficients that are regulator independent. At first sight, neither of these properties is obvious: the one-loop effective action for massless fermions is a complicated non-local functional with various  divergent, i.e. regulator dependent coefficients. (Of course, finite, i.e. regulator independent expressions are obtained  after adding the appropriate counterterms.) Indeed, we have seen in the above computation that the {\it non}-anomalous part of the Ward identity involving the vacuum polarization tensors $\Pi^{\n\r}_M$ is both non-local (the $\log p^2$ terms) as well as regulator dependent (the $\sum \eta_s \log M_s^2$ terms). On the other hand, the anomalous part of the Ward identity is indeed local, i.e. polynomial in the momenta, with an $M_s$-independent coefficient. Let us now show that any anomaly must always be finite and local.

\subsection{Locality of the anomaly\label{localano}}

First recall that {\it if} the (matter-loop) Feynman diagrams can be regularized in a manifestly gauge-invariant way, the gauge invariance is manifest on the regularized 1PI vertex functions and they  can be renormalized by adding gauge-invariant counterterms. As a result, the renormalized vertex functions respect the  gauge invariance, and there are no anomalies.

If it is not possible to regularize while maintaining manifest gauge invariance (or any other invariance one is considering), then anomalies may arise. As already mentioned, one cannot regularize the propagator for a chiral fermion using the Pauli-Villars method, since a chiral fermion cannot have a mass. (Instead we used the trick to consider a non-chiral fermion but with only its left-handed fermion interacting with the gauge field - which again breaks the gauge invariance). Neither can one use dimensional regularization as for non-chiral theories, since there is no  definition of $\g_5$ in $4-\e$ dimensions that satisfies all the usual properties. 

\vskip2.mm
\begin{figure}[h]
\centering
\includegraphics[width=0.7\textwidth]{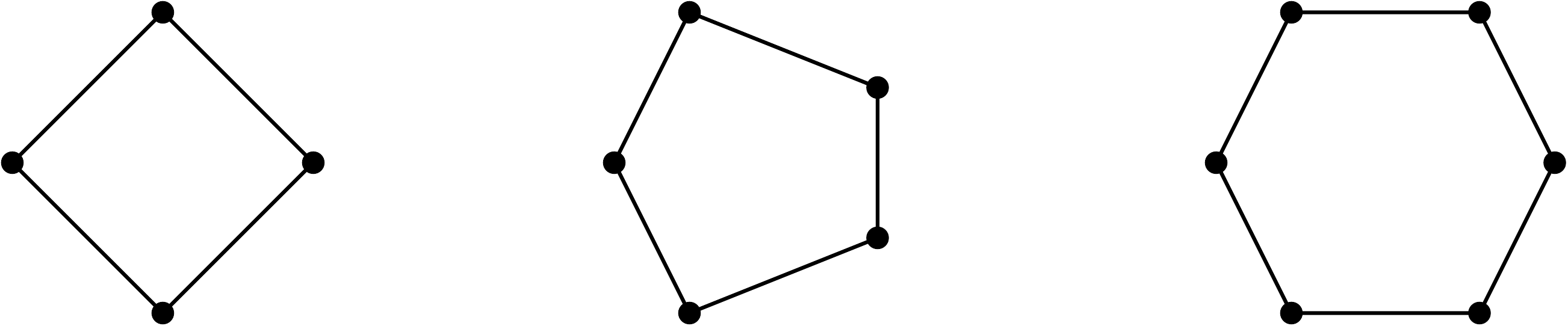}\\
\caption[]{In four dimensions, considering only fermion loops, the box diagram on the left is divergent, while the pentagon and hexagon diagrams in the middle and on the right are convergent.} \label{penthex}
\end{figure}

Clearly, a convergent amplitude on the other hand needs not to be regularized and, hence, can be computed in a manifestly gauge-invariant way. Thus it cannot be anomalous. Consider  the $n$-point one-loop vertex functions. They have $n$ fermion propagators that each behave as ${1\over k}$ for large loop momentum $k$, and $n$ vertices that do not involve any momentum.\footnote{
When considering fermions coupled to gravity, things are more complicated. The fermion-fermion-graviton interaction is $\sim e \pb \g^\m \o_\m^{ab}\g_{ab}\p$ and since the relation between $\o_\m^{ab}$ and the vielbein $e^c_\n$ involves one space-time derivative, the vertex corresponding to this fermion-fermion-graviton interaction contains one factor of momentum. This momentum, however, is the ``external" graviton momentum and we conclude that its presence does not change the power counting argument for the fermion loops.
}
Hence these diagrams have superficial degree of divergence $4-n$, and are superficially convergent for $n\ge 5$. But superficially convergent one-loop diagrams of course are convergent. Hence the pentagon, hexagon and any higher-point diagrams are convergent in four dimensions. More generally, we will see that chiral anomalies only occur in even dimensions $d=2r$ and then any $n$-point one-loop vertex function with $n\ge 2r+1$ is convergent. 
In a non-abelian gauge theory, gauge invariance relates a derivative of the $n$-point functions to $(n-1)$-point functions, as we have seen above when deriving the (anomalous) Ward identities. Thus in four dimensions, all identities for $n\ge 6$ only involve convergent one-loop diagrams and must be non-anomalous Ward identities. The identity for $n=5$ is slightly more subtle since it relates the derivative of the convergent 5-point function and the divergent 4-point functions. However, we already noted that this also is a non-anomalous Ward identity, since $\ca_\a$ does not contain any terms quartic in the gauge field $A$.

Consider now a divergent and possibly anomalous one-loop diagram (amplitude) of (superficial) degree of divergence $D$. 
Differentiating once with respect to an external momentum yields a sum of terms where in each term  one of the internal propagators is replaced according to 
\be
{\del\over\del p_\m}\left(-{1\over \ksl+\psl+\ldots}\right)
={\del\over\del p_\m}\left(-{ \ksl+\psl+\ldots\over (k+p+\ldots)^2}\right)
={(\ksl+\psl+\ldots)\g^\m(\ksl+\psl+\ldots)\over [(k+p+\ldots)^2]^2} \ ,
\ee 
and hence decreases the degree of divergence by one unit. By taking $D+1$ derivatives with respect to the external momenta then gives an expression of degree of divergence $D-(D+1)=-1$, i.e. a convergent integral. As just explained such a convergent integral does not need regularization and cannot be anomalous. Hence, taking enough derivatives of the  vertex functions gives non-anomalous expressions. Said differently, taking enough derivatives with respect to the external momenta of the anomalous part of the vertex functions yields zero. Let's be a bit more precise. Suppose the most divergent anomalous vertex function (in 4 dimensions this is the 3-point function) has degree of divergence $D$. Then  the anomalous part can be extracted form $p^{(j)}_{\m_j} \G^{\m_1\ldots\m_n}$. Due to the presence of the $p_\m^{(j)}$ we need to take one more derivative with respect to the external momenta to get zero:
\ba\label{D+2der}
&&\underbrace{ {\del\over \del p^{i_1}_{\n_1}} \ldots {\del\over \del p^{i_{D+2}}_{\n_{D+2}}}} \left( p^{(j)}_{\m_j} \G^{\m_1\ldots\m_n}\Big\vert_{\rm anom}\right) \ = 0 \quad \Leftrightarrow\quad
\underbrace{ {\del\over \del p^{i_1}_{\n_1}} \ldots {\del\over \del p^{i_{D+2}}_{\n_{D+2}}}}
\Big\vert_{{\rm fixed}\ A(p^{(i)})} \ca_\a \big(p^{(i)}, A(p^{(i)})\big)
\ =\  0 \ ,
\nonumber\\
&&\hskip4.mm D+2 \ {\rm times}\hskip52.mm D+2 \ {\rm times}
\ea
where $\ca_\a \big(p^{(i)}, A(p^{(i)})\big)$ is the Fourier transform of the anomaly $\ca_\a(x,A(x))$.
For the example of the triangle diagram  studied in detail above, i.e.~for $\G^{\quad\ \, \m\n\r}_{{\rm loop},\a\b\g}$ we have $D=1$ and $\ca_\a \big(p^{(i)}, A(p^{(i)})\big)\sim \e^{\m\n\r\s} p_\m A_\n(p) q_\r A_\s(q)$. We see indeed that taking $D+2=3$ derivatives with respect to the external momenta (at fixed $A(p)$ and $A(q)$) annihilates the anomaly. It follows in general from (\ref{D+2der}) that {\it the anomaly $\ca_\a\big(p^{(i)}, A(p^{(i)})\big)$ must be a \underline{polynomial} of degree $D+1$ in the external momenta.} We conclude~:
\be
\begin{array}{|c|}
\hline\\
\text{\it The anomaly 
is a \underline{local} functional of the gauge fields.}
\\
\\
\hline
\end{array}
\nonumber
\ee

\vskip8.mm
\subsection{Relevant and irrelevant anomalies\label{relanom}}

The anomaly, being a loop-effect, is of higher order in the coupling constant: for example, for a simple gauge group with a single coupling constant, the one-loop 3-point vertex function 
$\G^{\quad\ \, \m\n\r}_{{\rm loop},\a\b\g}$ and thus also the anomaly are of order $g^3$. (In our conventions each Lie algebra generator $t_\a$ includes a coupling constant $g$ and hence $D^\cR_{\a\b\g}\sim g^3$). As compared with the tree-level 3-point vertex function which is $\sim g$, this is higher order by a factor $g^2$. Now in any quantum field theory we are allowed to add to the classical action terms of higher-order in $g$, because classically they are ``invisible". Such terms are often generically called counterterms. Of course, this is exactly what one does in the renormalization program to cancel any divergences arising in the loops. However, we only allow to add {\it local} counterterms to the action.\footnote{By an argument very similar to the one above, one can show that the divergent terms arising from loop integrals are always local and hence can indeed be cancelled by local counterterms.} Suppose now one adds to the classical action $S_{\rm cl} = \int \d^4 x \left(-{1\over 4}F_{\m\n}^\a F^{\a\m\n} +\cL_{\rm matter}[A,\p,\pb]\right)$ a local counterterm $\D \G$ which is a 3-gauge field coupling:
\be\label{newaction}
S_{\rm cl}
\ \to \
S_{\rm cl}+ {1\over 6} \int \d^4 p\, \d^4 q\ \D \G^{\m\n\r}_{\a\b\g}(-p-q,p,q) A_\m^\a(-p-q) A_\n^\b(p) A_\r^\g(q)
\ ,
\ee
where  $\D \G\sim g^3$ must be a {\it polynomial} in $p$ and $q$. At order $g^3$, such a term has the effect of modifying the 3-point vertex function 
according to
\be
\G^{\m\n\r}_{\a\b\g}  \ \to \ \left[\G^{\m\n\r}_{\a\b\g}\right]_{\rm new} = \G^{\m\n\r}_{\a\b\g} + 
\D \G^{\m\n\r}_{\a\b\g} \ .
\ee
The question then is whether this $\left[\G^{\m\n\r}_{\a\b\g}\right]_{\rm new}$ is still anomalous or whether one can find a $\D\G$ such that $(p_\m+q_\m)\left[\G^{\m\n\r}_{\a\b\g}\right]_{\rm new}\Big\vert_{\e-{\rm piece}}=0$. If one can find such a counterterm $\D\G$, one can just use the new classical action according to 
(\ref{newaction}) and then there is no anomaly any more (at least to order $g^3$).  An anomaly that can be cancelled by the addition of a local counterterm to the classical action is called {\it irrelevant}, while an anomaly that cannot be cancelled by such an addition is called a {\it relevant} anomaly.

It is easy to see that  the above triangle anomaly for chiral fermions is a relevant anomaly.\footnote{Obviously, the same reasoning holds for the abelian anomaly.} Indeed, in  order to cancel it by the addition of a local counterterm one would need to satisfy
\be\label{countertermcond}
 -i(p_\m+q_\m) \D \G^{\m\n\r}_{\a\b\g}(-p-q,p,q)\Big\vert_{\e-{\rm piece}}  
-{1\over 12 \pi^2} \e^{\n\r\l\s} p_\l q_\s D^\cR_{\a\b\g} = 0 \ .
\ee
Since $\D \G^{\m\n\r}_{\a\b\g}$ must be polynomial in $p$ and $q$, this equation then shows that it must be linear in $p$ and $q$:
\be
\D\G^{\m\n\r}_{\a\b\g}(k,p,q)\Big\vert_{\e-{\rm piece}}  = \e^{\m\n\r\s} (a k_\s + b p_\s + c q_\s) D^\cR_{\a\b\g}\ .
\ee
Obviously, there are choices of $a,b$ and $c$ that could satisfy (\ref{countertermcond}), since it only requires $c-b={i\over 12\pi^2}$. However, as is clear from (\ref{newaction}),$\D\G^{\m\n\r}_{\a\b\g}(k,p,q)$ must be completely symmetric under permutations of $(k,\m,\a)$, $(p,\n,\b)$ and $(q,\r,\g)$. This is often referred to as the Bose symmetry of the $n$-point functions. Since $D^\cR_{\a\b\g}$ already is completely symmetric and $\e^{\m\n\r\s}$ ia completely antisymmetric, one finds $b=-a$, $c=-a$ and $c=-b$, i.e. $a=b=c=0$, and we conclude that there is no completely Bose symmetric local $\D \G^{\m\n\r}_{\a\b\g}(k,p,q)$ that can satisfy (\ref{countertermcond}). Thus, there is no way to cancel the triangle anomaly under non-abelian gauge transformation for chiral fermions by adding a local counterterm, and the anomaly is indeed relevant.

As discussed at length when we derived the anomalous Ward identities from the anomalous Slavnov-Taylor identities, the non-vanishing of 
$(p_\m+q_\m) \D \G^{\m\n\r}_{\a\b\g}(-p-q,p,q)\Big\vert_{\e-{\rm piece}}$ translates the current non-conservation which in turn is (minus) the anomaly which equals the variation of the effective action. We can then restate the definition of relevant anomalies as follows:
\be\label{relevantan}
\begin{array}{|c|}
\hline\\
\text{The anomaly is relevant} 
\quad\Leftrightarrow\quad
\int \e^\a(x) \ca_\a(x) \ \ne\  \dd_\e\, F \quad \text{with $F$ a {\it local} functional .}\\
\\
\hline
\end{array}
\ee
Obviously, if such a local $F$ exists, we can add the counterterm $-F$ to $S_{\rm cl}$ and get $\dd_\e\, [\wt W[A]]_{\rm new}= \int \e^\a(x) \ca_\a(x) - \dd_\e\, F =0$, i.e. the anomaly is cancelled by the addition of this local counterterm, i.e. is irrelevant. Otherwise, it is clearly relevant.

\subsection{Finiteness of the anomaly\label{finitano}}

Let us now explain why the anomaly necessarily had to be finite and also show that this is true more generally. We already know that, since $\G^{\m\n\r}_{\a\b\g}$ has a degree of divergence 1, the anomalous part of
$(p_\m+q_\m)\G^{\m\n\r}_{\a\b\g}(-p-q,p,q)$ must be a polynomial of
degree 2 at most in $p$ and $q$:
\be\label{genanomform}
(p_\m+q_\m)\G^{\m\n\r}_{\a\b\g}(-p-q,p,q)\Big\vert_{\rm anom\ part}
= E^{\n\r\l\s}_{\a\b\g} p_\l q_\s + 
F^{\n\r\l\s}_{\a\b\g} p_\l p_\s + 
G^{\n\r\l\s}_{\a\b\g} q_\l q_\s 
+H^{\n\r\l}_{\a\b\g} p_\l +K^{\n\r\l}_{\a\b\g} q_\l
+L^{\n\r}_{\a\b\g} \ ,
\ee
where the $E,F,G,H,K$ and $L$ must be constant Lorentz tensors, independent of the momenta. But there are no such 3-index Lorentz tensors, so that $H=K=0$. Furthermore, the anomaly is due to the presence of $\g_5$ (otherwise we could use gauge invariant dimensional regularization and there would be no anomaly) and, hence, the anomalous terms must be proportional to the $\e$-tensor. Hence, although the tensor structures $\sim Fpp, Gqq, L$ are present in the non-anomalous parts, they cannot appear in the anomalous part. Thus only $E^{\n\r\l\s}_{\a\b\g} p_\l q_\s$ can be present on the right hand side of (\ref{genanomform}). While $E^{\n\r\l\s}_{\a\b\g}$ cannot depend on the momenta, it could, a priori depend on the Pauli-Villars regulator mass\footnote{
Although we introduced 3 regulator fields to conveniently compute the non-anomalous parts, one regulator mass would have been sufficient.
} 
$M$ (or equivalently on a UV cutoff $\L$). Now, $\G^{\m\n\r}_{\a\b\g}(-p-q,p,q)$ has scaling dimension 1 (there are 3 fermion propagators each of scaling dimension $-1$ and one loop integration of scaling dimension $+4$, but the same result is obtained from considering $\int \d^4 p\, \d^4 q\, \G^{\m\n\r}_{\a\b\g}(-p-q,p,q) A_\m^\a(-p-q) A_\n^\b(p) A_\r^\g(q)$ in general), and hence $(p_\m+q_\m)\G^{\m\n\r}_{\a\b\g}(-p-q,p,q)$ has scaling dimension 2. This shows that $E^{\n\r\l\s}_{\a\b\g}$ has scaling dimension 0, and since it does not depend on the momenta, it cannot depend on the regulator mass $M$ (or UV cutoff $\L$) either. This shows that the anomalous part is necessarily finite.

This argument generalizes to arbitrary even dimensions $d=2r$. In any even dimension (and only in even dimensions) one can define a chirality matrix $\g_{2r+1}$ that anticommutes with all matrices $\g_\m$, $\m=0, \ldots 2r-1$. One can then again have massless fermions that are chiral, i.e. either left-handed or right-handed. For the same reasons as in four dimensions, there is no gauge invariant regularization for chiral fermions and anomalies may occur. On the other hand, for non-chiral fermions one may use dimensional regularization which is manifestly gauge invariant. Hence the anomalies can again be traced to the presence of the chirality matrix 
$\g_{2r+1}$ in the computations of the vertex functions.  In $2r$ dimensions the Dirac traces involving $\g_{2r+1}$ will lead to an $\e^{\m_1 \ldots \m_{2r}}$, and again the anomalous part of $\big(\sum_{j=1}^k p^{(j)}\big)_\m \G_{\a \b_1\ldots \b_k}^{\m\n_1\ldots\n_k}(-\sum_{j=1}^k p^{(j)}, p^{(1)},\ldots p^{(k)})$ must involve this 
$\e^{\m_1 \ldots \m_{2r}}$ tensor.  Since this $\e$-tensor is completely antisymmetric in all its $2r$ indices, to get a non-vanishing expression we need at least $r$ indices $\n_j$ and $r$ different momenta. Hence the minimal value for $k$ is $r$. It follows that in $2r$ dimensions the anomaly first manifests itself in the $(r+1)$-point function $\G_{\a \b_1\ldots \b_r}^{\m\n_1\ldots\n_r}$ and its anomalous part is
\be\label{dim2rGammarp1}
\Big({\textstyle\sum_{j=1}^r} p^{(j)}_\m\Big) \G_{\a \b_1\ldots \b_r}^{\m\n_1\ldots\n_r}\Big(-{\textstyle\sum_{j=1}^r} p^{(j)}, p^{(1)},\ldots p^{(r)}\Big)\Big\vert_{\rm anomalous}
= C\ \e^{\n_1\ldots\n_r \s_1 \ldots \s_r} p^{(1)}_{\s_1}\ldots p^{(r)}_{\s_r}\ D^\cR_{\a \b_1\ldots\b_r} \ , 
\ee
where the $D^\cR_{\a \b_1\ldots\b_r}$ must be the trace of a completely symmetrized product of the generators $t_\a t_{\b_1} \ldots t_{\b_r}$ and, a priori, $C$ could depend on any Lorentz scalar.
Thus in 6 dimensions the anomaly may first show up in a square diagram, in 8 dimensions in a pentagon diagram, in 10 dimensions in a hexagon diagram, etc. Now, in $2r$ dimensions an $r+1$ point one-loop diagram with only fermion propagators has a degree of divergence $D=2r - (r+1)=r-1$, and by the above argument, taking $D+2=r+1$ derivatives with respect to the external momenta of the left-hand side of (\ref{dim2rGammarp1}) must give a vanishing result. Hence the l.h.s. of (\ref{dim2rGammarp1}) must be a polynomial of at most degree $r$, and we see that the $C$ on the right-hand side must be a constant not depending on the momenta. Since the scaling dimension of the l.h.s. of (\ref{dim2rGammarp1}) is $1+(2r)-(r+1)=r$, it follows that $C$ has scaling dimension 0 and, again, since it does not depend on any momentum it cannot depend on the regulator mass, resp. UV cutoff $\L$ either. Hence $C$ is a finite numerical constant.  Similarly, the $(r+l)$-point functions with $l=2,\ldots r$ also have anomalous parts that are finite.\footnote{
Indeed, an $(r+l)$-point function has scaling dimension and degree of divergence equal to $D=2r-(r+l)=r-l$, showing that $\sum p^{(j)}_\m \G^{\m\n_1\ldots \n_{r+l-1}}\vert_{\rm anomalous}$ has scaling dimension $r-l+1$ and is a polynomial of order $r-l+1$. This is necessarily a sum of terms of the form $C_{j_1,\ldots j_{r-l+1}}\e^{\n_1\ldots\n_{r+l-1}\s_1\ldots \s_{r-l+1}} p^{(j_1)}_{\s_1}\ldots p^{(j_{r-l+1})}_{\s_{r-l+1}}$ times the trace of the generators. The coefficients $C_{j_1,\ldots j_{r-l+1}}$ then again have scaling dimension 0 and are independent of the momenta and thus also of the regulator masses or UV cutoff. 
}
We conclude:
\be\label{finiteanom}
\begin{array}{|c|}
\hline\\
\text{In any dimension $d=2r$, the anomaly is finite. It first shows up in the divergence of the $(r+1)$-} \\ 
\\
\text{point vertex function whose anomalous part is given by (\ref{dim2rGammarp1}) with a finite numerical constant $C$.} 
\\
\\
\hline
\end{array}
\nonumber
\ee

\section{Relevant fermion representations and cancellation of ano\-malies\label{frecan}}
\setcounter{equation}{0}

If a global symmetry is anomalous it only implies that  classical selection rules are not obeyed in the quantum theory and classically forbidden processes may actually occur. An example of this type of  situation is the abelian anomaly which breaks the symmetry under the global chiral transformation of massless fermions. 

On the other hand, the occurrence of an anomaly for a local (gauge) symmetry makes the gauge theory inconsistent. Indeed, once one has done the functional integral over the fermions, the starting point for quantizing the gauge fields is the ``effective classical" action $-{1\over 4}\int\d^4 x\, F_{\m\n}^\a F^{\a\m\n}  + \wt W[A]$. Gauge invariance of this effective action is necessary for unitarity, as can be seen e.g. by using the gauge invariance for going to the manifestly unitary axial gauge, or else by performing the Faddeev-Popov quantization and having the ghosts  cancel the non-physical polarizations. In the presence of an anomaly, $\wt W[A]$, and hence the ``effective classical" action is no longer gauge invariant and unitarity is violated.

We have seen that such anomalies of a non-abelian gauge symmetry generically are present if the matter includes chiral fermions. Chiral fermions seem to be very common in nature, and certainly are a main ingredient of the standard model. The consistency of the latter requires that the contributions to the anomaly of the different chiral fermions present in the model cancel each other.

In this section, we will mostly concentrate on the chiral anomaly under (non-abelian) gauge transformations and discuss which representations $\cR$ of which gauge groups have non-vanishing $D^\cR_{\a\b\g}$ symbols and how left- and right-handed particles and antiparticles contribute to the anomaly. Then we  consider explicitly one generation of the standard model with gauge group $SU(3)\times SU(2)\times U(1)$ and show that all anomalies cancel.

\subsection{Left-handed particles versus right-handed anti-particles\label{leftrightanti}}

First, recall that chiral fermions cannot have a Dirac mass term. We will discuss mass terms of the form (\ref{Majmass}), resp. (\ref{Weylmassterm}) at the end of the next subsection and show that whenever a field carries a representation such that such a mass term is allowed, this field cannot contribute to the anomaly. Hence, only massless fermions contribute to the anomaly.\footnote{
Similarly, for the abelian anomaly, the chiral transformations are symmetries only for massless fermions.
}

Let us stress that the different fermion fields contribute additively to the anomaly. Indeed, the anomaly was determined from the fermion one-loop triangle diagram, and it is clear that the different fermion contributions add up. We have already seen that right-handed fermions contribute with an opposite sign, but otherwise the contribution is universal (there is no fermion mass that could make a difference), except for the group theoretical factor 
\be
D^\cR_{\a\b\g}=\trR t_\a t_{(\b} t_{\g)}=\trR t_{(\a} t_{\b} t_{\g)}\equiv {\rm str}_R\, t_\a t_\b t_\g\ .
\ee 
Each left-handed fermion $\p_i^L$ in some representation $\cR_i^L$ contributes a $D^{\cR^L_i}_{\a\b\g}$ and each right-handed fermion in a representation $\cR^R_j$ contributes a $-D^{\cR^R_j}_{\a\b\g}$  . Adding up these individual contributions, we get for the total anomaly (cf.~(\ref{fullnaanom2}) and (\ref{anLanR}))
\ba\label{totalanomLR}
\ca_\a &=& \sum_i \ca_\a^L \big\vert_{\p_i^L} 
+\sum_j \ca_\a^R \big\vert_{\p_i^R} 
\nonumber\\&=&
-{1\over 24\pi^2} \e^{\m\n\r\s}  \del_\m \Big(A_\n^\b \del_\r A_\s^\g -{i\over 4} A_\n^\b [ A_\r , A_\s]^\g\Big) \Big(\sum_i D^{\cR_i^L}_{\a\b\g} - \sum_j D^{\cR_j^R}_{\a\b\g} \Big) \ .
\ea
Of course, we could just group together all left-handed fermions into one large (reducible) representation $\cR^L=\oplus_i \cR_i^L$ and all right-handed fermions into another representation $\cR^R=\oplus_j \cR_j^R$, so that 
\be
\sum_i D^{\cR_i^L}_{\a\b\g}
=\sum_i\, {\rm str}_{\cR_i^L}\, t^{\cR_i^L}_\a t^{\cR_i^L}_\b t^{\cR_i^L}_\g
={\rm str}_{\cR^L}\,  t^{\cR^L}_\a t^{\cR^L}_\b t^{\cR^L}_\g
=D^{\cR^L}_{\a\b\g}\ ,
\ee
and similarly for the right-handed representations. Only chiral fermions contribute to the anomaly, but formally we could also include non-chiral fermions in the sums in (\ref{totalanomLR}), since a non-chiral fermion is equivalent to a left-handed plus a right-handed fermion, both in the {\it same} representation: $\cR^L=\cR^R$ so that their contributions cancel in (\ref{totalanomLR})

When summing over all fermion species one clearly should {\it not} include particles and antiparticle separately since both are described by the same fermion field $\p$ or {\it equivalently} the charge conjugate $\p^c$. If the particle is left-handed, its antiparticle is right-handed and vice versa. But one might as well have considered the right-handed antiparticle as the particle. The question then arises whether  the contribution to the anomaly is that of a left-handed particle or a right-handed (anti)particle. As we will now show, it does not matter.

Suppose $\p$ describes a left-handed particle in some representation $\cR_L$ with generators $t^L_\a \equiv t^{\cR_L}_\a$. Its antiparticle is then described by the charge conjugate field $\p^c=i\g^0 \cC \p^*$, where  $\cC$ is the charge conjugation matrix. It satisfies
\be\label{chargeconj}
\cC (\g^\m)^T= - \g^\m\cC \quad , \quad
\cC \g_5^T= \g_5 \cC \ .
\ee
and hence since $\g_5^\dag=\g_5$ also $\cC \g_5^*=\g_5 \cC$. (Note that to show the last relation (\ref{chargeconj}) one needs to reorder the $\g^0,\g^1,\g^2,\g^3$ contained in $\g_5$ resulting in a sign $(-)^{3+2+1}=+1$. This is specific to 4 mod 4 dimensions. In 2 mod 4 dimensions instead one would have gotten an extra minus sign.) It follows that 
\be\label{oppchir}
\g_5 \p^c= \g_5 i \g^0 \cC \p^*=-i \g^0 \g_5 \cC \p^*=-i \g^0 \cC (\g_5 \p)^*= \mp \p^c \quad {\rm if} \ \g_5\p=\pm \p \ ,
\ee
so that $\p^c$ correctly describes a right-handed antiparticle if $\p$ decribes a left-handed particle, and vice versa. (This is true in 4 mod 4 dimensions, while in 2 mod 4 dimensions the above-mentioned extra minus sign implies that particles and antiparticles have the same chirality.) 

Now if $\p$ transforms in the representation $\cR_L$, this means $\dd\p=i\e^\a t^{\cR_L}_\a \p$. Then $\dd \p^*=-i \e^\a (t^{\cR_L}_\a)^* \p^*$  $= -i \e^\a (t^{\cR_L}_\a)^T \p^*$ where we used $t^*=t^T$ since our generators are hermitean. It follows that 
\be
\dd \p^c=i\g^0 \cC (\dd\p)^*= - i \e^\a  (t^{\cR_L}_\a)^T (i \g^0 \cC \p^*) = - i \e^\a  (t^{\cR_L}_\a)^T \p^c \ .
\ee
But, by definition, the right-handed $\p^c$  transforms in a representation $\cR_R$ with generators $t^{\cR_R}_\a$ according to 
\be
\dd \p^c = i \e^\a t^{\cR_R}_\a \p^c \ .
\ee
Comparing both equations, we identify
\be\label{chconjugrepres}
t^{\cR_R}_\a =- (t^{\cR_L}_\a)^T \ .
\ee

Then, if we consider the contribution to the anomaly of the fermion field associated to the left-handed particle it contains $D^{\cR_L}_{\a\b\g} = {\rm str}\, t^{\cR_L}_\a t^{\cR_L}_{\b} t^{\cR_L}_{\g}$. If instead we consider the contribution to the anomaly of the fermion field associated with the right-handed antiparticle it contains an extra minus sign due to the opposite chirality and the $D^{\cR_R}_{\a\b\g}$ instead of the $D^{\cR_L}_{\a\b\g}$:
\ba\label{Drelation4}
- D^{\cR_R}_{\a\b\g} = - {\rm str}\, t^{\cR_R}_{\a} t^{\cR_R}_{\b} t^{\cR_R}_{\g}
&=&- {\rm str}\, (-t^{\cR_L}_{\a})^T (-t^{\cR_L}_{\b})^T (-t^{\cR_L}_{\g})^T
\nonumber\\
&=& {\rm str} \big(t^{\cR_L}_{\a} t^{\cR_L}_{\b} t^{\cR_L}_{\g}\big)^T 
= {\rm str}\, t^{\cR_L}_{\a} t^{\cR_L}_{\b} t^{\cR_L}_{\g}
= D^{\cR_L}_{\a\b\g} \ ,
\ea
which is exactly the same as for the left-handed particle.
We conclude that, in four dimensions, it does not matter whether we use the field $\p$ of a left-handed particle or the field $\p^c$ of the corresponding right-handed antiparticle: we get the same contribution to the anomaly. In particular, for right-handed particles we may instead consider the left-handed antiparticles, so that we may treat all fermions as left-handed.

The previous argument holds in 4 mod 4 dimensions. Indeed, if $d=2r$ and $r$ is even, the relevant trace is $D^\cR_{\a_1 \ldots \a_{r+1}}={\rm str}_\cR\, t_{\a_1} \ldots t_{\a_{r+1}}$. Then just as in (\ref{Drelation4}) one finds
\be
-D^{\cR^c}_{\a_1\ldots \a_{r+1}}
\equiv
-D^{\cR_R}_{\a_1\ldots \a_{r+1}} = D^{\cR_L}_{\a_1\ldots \a_{r+1}}\ , 
\quad \text{in 4 mod 4 dimensions} \ ,
\ee
since we get one minus sign due to the opposite chirality and $r+1$ minus signs from $t^{\cR_R}_{a_j}=-(t^{\cR_L}_{\a_j})^T$.

In 2 mod 4 dimensions however, we have seen that particles and antiparticles have the same chirality. If the particle is in a representation $\cR$ with generators $t^\cR_\a$, it is still true that the antiparticle which is described by the charge conjugate field is in a representation $\cR^c$ with generators $t^{\cR^c}_\a = - (t^\cR_\a)^T$. On the other hand, in $d=2r$ dimensions with $r$ odd, the $D^\cR_{\a_1\ldots \a_{r+1}}$ symbol involves a symmetrized trace of an even number of generators, so that we now get 
\be\label{2mod4Drel}
D^{\cR^c}_{\a_1\ldots \a_{r+1}}= D^\cR_{\a_1\ldots \a_{r+1}} \ ,
\quad \text{in 2 mod 4 dimensions} \ .
\ee
Since particles and antiparticles have the same chirality, this shows again that it does not matter which one one uses to compute the contribution to the anomaly. In particular, in 2 mod 8 dimensions one can have Majorana-Weyl spinors, i.e. fermions that are chiral and obey $\p^c=\p$ so that they are their own antiparticles, consistent with
(\ref{2mod4Drel}). In this case however, in order not to over-count the contribution to the anomaly as if the particle were distinct from its antiparticle, one has to include a factor ${1\over 2}$ in the coefficient of the anomaly.

\subsection{Which gauge groups and which representations lead to anomalies?\label{whichgroups}}

In this subsection (except for a remark at the end) we will be specifically dealing with four dimensions again. The question of whether or not there is an anomaly then boils down to the question whether $D^\cR_{\a\b\g}$ is non-vanishing. Thus we must study for which gauge groups and which representations $\cR$ we have $D^\cR_{\a\b\g}\ne 0$.

It is useful to introduce a few definitions. 
\begin{itemize}
\item
Two representations $\cR_1$ and $\cR_2$ are {\it equivalent} if there exists a fixed matrix $S$ such that $t_\a^{\cR_1} = S t_\a^{\cR_2} S^{-1}$ for all $\a$. 
\item
For any representation $\cR$, the {\it complex conjugate representation} $\oR$ is the one with generators $t_\a^\oR$ such that $i t_\a^\oR = (i t^\cR_\a)^*$, so that the corresponding representations of the group are indeed the complex conjugate ones: $e^{i \e^\a t_\a^\oR}= \Big( e^{i \e^\a t_\a^\cR} \Big)^*$. Since our generators are hermitean we find
\be
t_\a^\oR = - \big( t_\a^\cR\big)^* = - \big( t_\a^\cR\big)^T \ .
\ee
\item
A representation $\cR$ that is equivalent to its complex conjugate representation $\oR$ satisfies
\be\label{Recc}
\big( t_\a^\cR\big)^T = - S t_\a^\cR S^{-1} \ .
\ee
Such a representation is called {\it real} if by some (fixed) similarity transformation the $t_\a^\cR$ can be made imaginary and antisymmetric (in which case it satisfies (\ref{Recc}) with $S=1$), and is called {\it pseudoreal} if not. In any case for a real or pseudoreal representation $\cR$ we have (\ref{Recc}). 
\end{itemize}

\noindent
For a real or pseudoreal representation $\cR$ we have 
\be
D^\cR_{\a\b\g}={\rm str}\, t^\cR_\a t^\cR_\b t^\cR_\g
= {\rm str}\, (t^\cR_\a)^T (t^\cR_\b)^T (t^\cR_\g)^T
= - {\rm str}\, S t_\a^\cR S^{-1} S t^\cR_\b S^{-1} S t^\cR_\g S^{-1} 
= - {\rm str}\, t^\cR_\a t^\cR_\b t^\cR_\g = - D^\cR_{\a\b\g} \ .
\ee
Hence\footnote{The relevant quantity for the anomalies in $2r$ dimensions is $D^\cR_{\a_1\ldots \a_{r+1}}$. For even $r$, i.e. in 4 mod 4 dimensions, the same argument implies the vanishing of this $D$-symbol for real or pseudoreal representations.  However, for odd $r$, i.e. in 2 mod 4 dimensions, this argument does {\it not} imply the vanishing of the $D$-symbol.
}
\be
D^\cR_{\a\b\g}=0 \quad \text{for a real or pseudoreal representation}\ \cR \ .
\ee
Obviously then:
\be
\begin{array}{|c|} 
\hline\\
\text{If a group}\ G\ \text{has only real or pseudoreal representations, its}\ D^\cR_{\a\b\g}\ \text{all vanish and there} \\
\text{cannot be any anomalies (in four dimensions) for a gauge theory with such a gauge group}\ G.\\ 
\\
\hline
\end{array}
\nonumber
\ee
This simplifies things a lot since:
\begin{itemize}
\item
All $SO(2n+1),\ n\ge 1$ (including $SU(2)\simeq SO(3)$), $SO(4n),\ n\ge 2$, $USp(2n),\ n\ge 3$, as well as the exceptional groups $G_2,\ F_4,\ E_7$ and $E_8$ only have real or pseudoreal representations and hence have $D_{\a\b\g}=0$. The same is true for any direct product of these groups.
\item
The groups $SO(4n+2),\ n\ge 1$ and $E_6$ also have $D_{\a\b\g}=0$ for {\it all} their representations, even though they do admit representations that are neither real nor pseudoreal.
\item
Only $SU(n),\ n\ge 3$ and $U(1)$ or product groups involving these factors have representations with $D^\cR_{\a\b\g}\ne 0$.
\end{itemize}

\noindent
This means that if the gauge group is $G_1\times G_2 \times \ldots \times G_k$, with each $G_i$ being a simple or $U(1)$ factor, then at least one of the $G_i$ must be $SU(n),\ n\ge 3$ or $U(1)$ in order that there are representations of the product group with non-vanishing $D^\cR_{\a\b\g}$. 

\begin{figure}[h]
\centering
\includegraphics[width=0.25\textwidth]{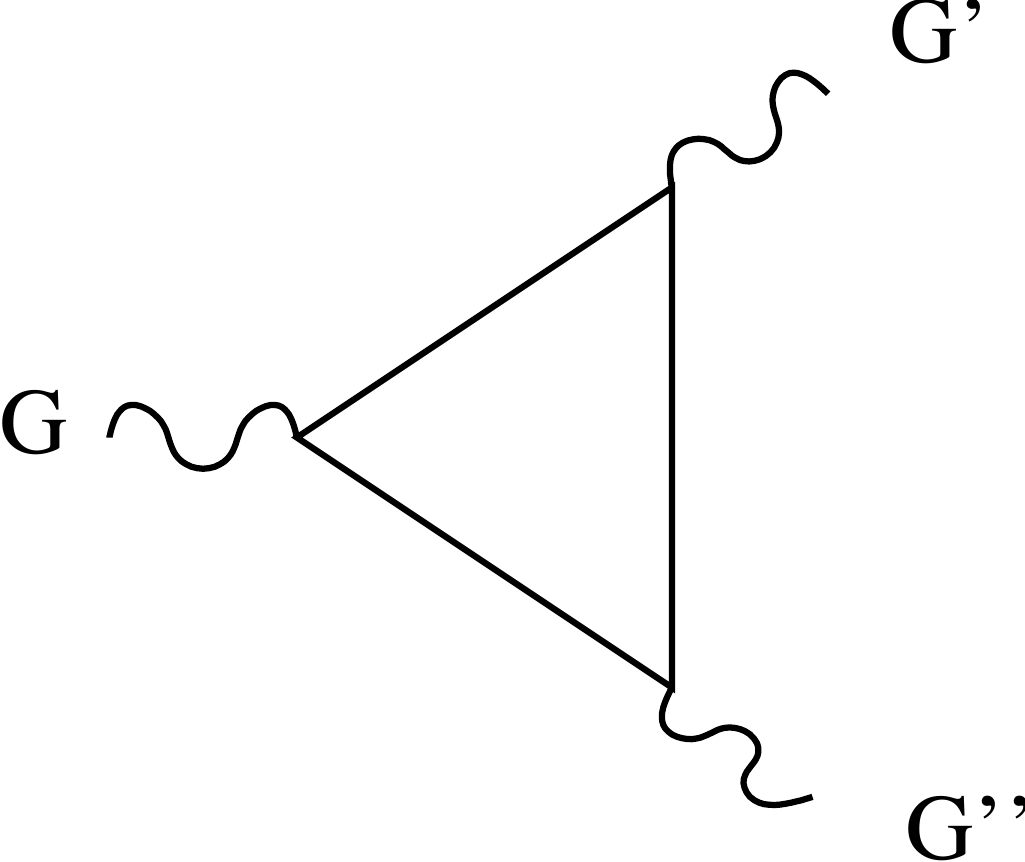}\\
\caption[]{The triangle diagram coupling to gauge bosons of the gauge groups $G$, $G'$ and $G''$.} \label{producttriangle}
\end{figure}

In general, $t_\a$ can be a generator of a simple or $U(1)$ factor called $G$, $t_\b$ of a factor called $G'$ and $t_\g$ of a factor called $G''$. The anomalies corresponding to these different possibilities are referred to as $G-G'-G''$ anomalies. They are probed by computing the triangle diagram that would couple to one $G$ gauge boson, one $G'$ gauge boson and one $G''$ gauge boson, see Fig. \ref{producttriangle}.

\vskip5.mm
\noindent
If we denote by $G_s$ a simple Lie algebra (so that for any of its generators $t_\a$ one has $\tr t_\a=0$), then we have the following possibilities
\begin{itemize}
\item
$U(1)-U(1)-U(1)$: Here $D_{\a\b\g}\to \tr t\, t\, t = \sum_i q_i^3$.
\item 
$U(1)-G_s-G_s$: Any representation $\cR$ of the product group $U(1)\times G_s$ decomposes into a sum $\cR=\oplus_j (q_j,\cR_j)$ where, of course, all states within each (irreducible) representation $\cR_j$ of $G_s$ have the same $U(1)$ charge $q_j$. Thus we have (cf.~(\ref{traceU1GG}))
\be
D^\cR_{\a\b\g}\to \trR t\, t_\b t_\g = \sum_j q_j\, {\rm tr}_{R_j} t_\b t_\g=g^2\, \sum_j q_j\, C_{R_j}\dd_{\b\g}\ ,
\ee
where $g$ is the gauge coupling constant for the gauge group $G_s$. 
It follows that the corresponding anomaly can occur for {\it any} simple $G_s$ if the corresponding fermions in the representation $\cR_j$ have a non-vanishing $U(1)$-charge $q_j$. 
\item
$U(1)-G_s-G'_s$ or $G_s-G_s-G_s'$ for $G_s\ne G'_s$: Here the trace factorizes into a $\tr t_\b$ and a $\tr t_\g$ which both vanish, hence $D_{\a\b\g}=0$.
\item
$G_s-G_s-G_s$: Here $D_{\a\b\g}\ne 0$ only for $G_s=SU(n),\ n\ge 3$.
\end{itemize}

\begin{figure}[h]
\centering
\includegraphics[width=0.25\textwidth]{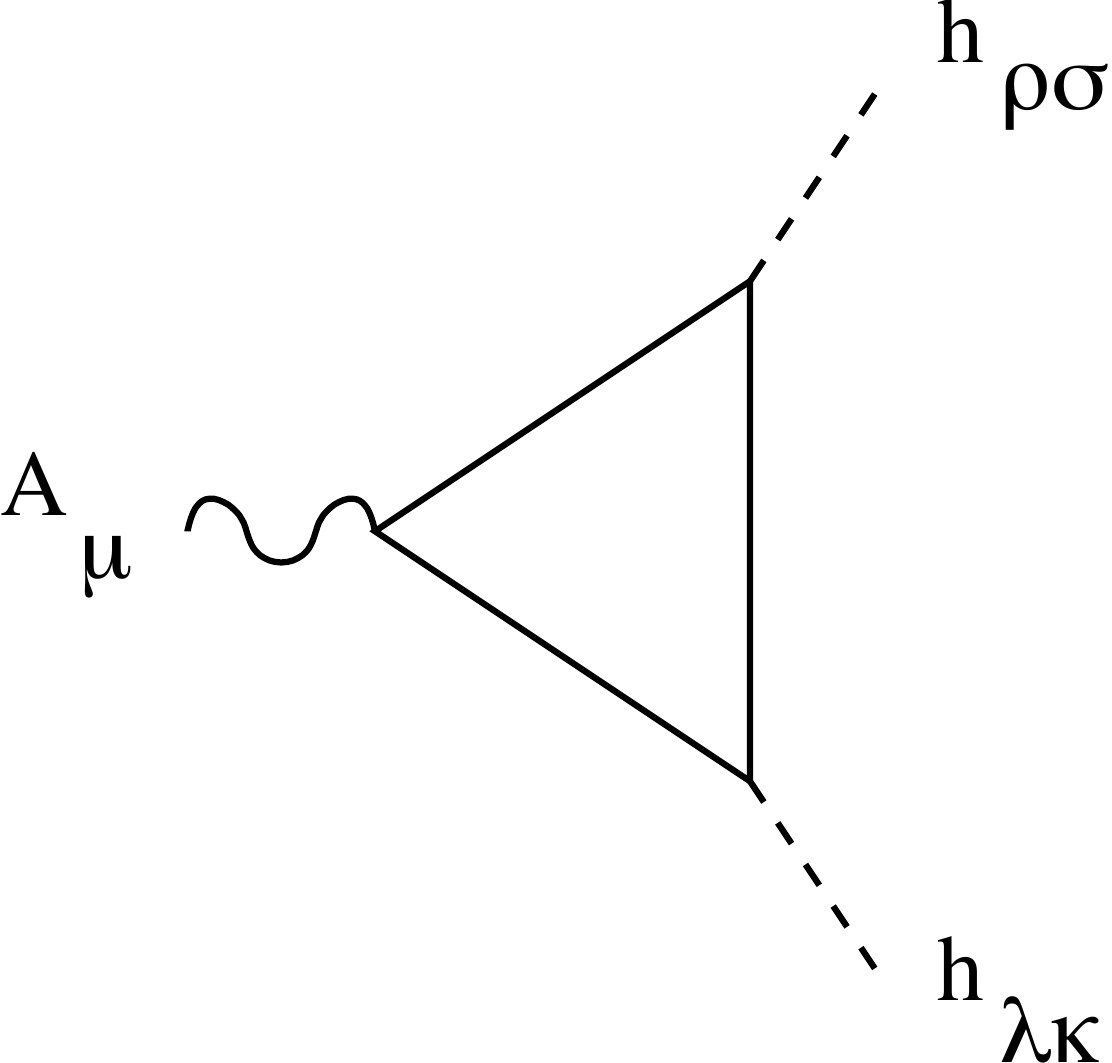}\\
\caption[]{The triangle diagram responsible for the mixed $U(1)$-gravitational anomaly.} \label{gravanom}
\end{figure}
\vskip5.mm

Note that for the purpose of studying {\it gravitational anomalies} one may probe for anomalies under local Lorentz transformations, which can be considered as $SO(3,1)$, resp. $SO(4)$ gauge transformations. This will be explained in more detail in section \ref{gravmixed}. At present, we already see from our discussion above that there are no purely gravitational anomalies (i.e. no $SO(4)-SO(4)-SO(4)$ anomalies) in 4 dimensions. However, in the presence of a $U(1)$ factor in the gauge group there are $U(1)-SO(4)-SO(4)$ anomalies, also called mixed $U(1)$-gravitational anomalies. They correspond to a triangle diagram with fermions  coupling to one $U(1)$ gauge field and to two gravitons, as shown in Fig.~\ref{gravanom}, i.e.~to the time-ordered expectation value of the $U(1)$ current $j^\m$ and two energy-momentum tensors $T_{\r\s}$ and $T_{\l\k}$. Since all particles couple universally to gravity, the relevant $D$-symbol, $\tr t\, t^{SO(4)}_A t^{SO(4)}_B\sim \dd_{AB} \tr t$ is simply the sum of all $U(1)$ charges with the appropriate multiplicities. Just as for the gauge anomalies, this anomaly must cancel in order to consistently couple gravity to matter charged under the $U(1)$. Of course, as noted above, in $d=2$ mod 4 dimensions the relevant $D$-symbol with only $SO(d)$ generators does not have to vanish and there can also be pure gravitational anomalies.

Finally, let us show that {\it only massless particles can contribute to the anomaly}. First of all, a massive particle with a standard Dirac mass term $m\pb \p$ necessarily is non-chiral, i.e. $\p=\p_L+\p_R$. (Indeed, as we have seen in sect.~\ref{chirprelim},  in $m\pb \p$ only the terms $m\overline{\p_L}\p_R + m \overline{\p_R}\p_L$ are non-vanishing and one cannot write a Dirac mass term with only a chiral field.) Such a non-chiral field cannot contribute to the anomaly.  However, we have also seen in sect.~\ref{chirprelim} that, for a single chiral field $\p_L$, one can nevertheless write a ``Majorana" mass term of the form $\overline{\p_L} \p_L^c + \overline{\p_L^c}\p_L$ (cf.~(\ref{Majmass}) or equivalently (\ref{Weylmassterm})). The important point is that such a mass term is not always compatible with the gauge invariance (or any
global symmetry one wants to impose). Indeed, as discussed in the previous subsection, under a transformation $\dd \p_L=i\e^\a t_\a^{\cR_L}$ we have
$\dd \p_L^c=-i\e^\a (t_\a^{\cR_L})^T \p_L^c$ since our generators are hermitian so that $(t_\a^{\cR_L})^T = (t_\a^{\cR_L})^*$. Also recall that $\dd\overline{\p_L}= -i\e^\a \overline{\p_L} t_\a^{\cR_L}$. It follows that
\be
\dd \big( \overline{\p_L} \p_L^c + \overline{\p_L^c}\p_L \big)
= -i \e^\a  \overline{\p_L} \Big( t_\a^{\cR_L} + (t_\a^{\cR_L})^T  \Big) \p_L^c
+ i \e^\a \overline{\p_L^c} \Big( (t_\a^{\cR_L})^T + t_\a^{\cR_L} \Big) \p_L  \ .
\ee
There can be no cancellations between the first terms and the second terms on the r.h.s. because the first only involves the components of $\p_L^*$ and the second only the components of $\p_L$. Thus both terms must vanish separately and we conclude that
\be\label{tttransp}
(t_\a^{\cR_L})^T + t_\a^{\cR_L}=0\ ,
\ee
i.e. the representation must be real. But we have seen above that real representations have vanishing $D_{\a\b\g}$ and do not contribute to the anomaly. It is not difficult to extend this argument to the more general mass terms involving several chiral fields and a symmetric mass matrix $m_{rs}$ as in (\ref{massmatrix}). Then if $\cR_r$ is the representation carried by $\p_L^r$ one finds that whenever the matrix element $m_{rs}$ is non-vanishing, the representation $\overline{\cR_r}$ is equivalent to the representation $\cR_s$ (i.e. $-(t_\a^{\cR_r})^T =S t_\a^{\cR_s} S^{-1}$) so that, again, the anomaly either vanishes (if $r=s$) or cancels between the two representations (if $r\ne s$).

There is one more point to be discussed before we can conclude that only massless fermions can contribute to the anomaly: even though we have shown that the relevant $D$-symbols for massive fermions always vanish, we still must show that for a chiral particle with a ``Majorana" mass, the anomaly again is given by some expression times the relevant $D$-symbol. Looking back out our Feynman diagram triangle computation, it is clear that even if we use appropriate massive propagators for the fermions (we actually did for the Pauli-Villars regulator fields), the 3 vertices still each contribute a generator $t^\cR$ yielding a $\trR t_\a t_\b t_\g$ for one of the two Feynman diagrams and a $\trR t_\a t_\g t_\b$ for the other. Any anomalous part must be accompanied by an $\e^{\n\r\l\s}$ contracted with the only available momenta $p_\l$ and $q_\s$. Bose symmetry then requires that the second diagram equals the first one with $(p,\n,\b)$ and $(q,\r,\g)$ exchanged, resulting indeed in a factor $\trR t_\a t_{(\b}t_{\g)}=D^\cR_{\a\b\g}$.

These arguments straightforwardly generalize to $d=2r=$ 4 mod 4 dimensions, since in these dimensions $\p_L^c$ is again right-handed and we can write the same ``Majorana" mass term (\ref{Majmass}), find again that the representation $\cR_L$ must be real and hence that $D^{\cR_L}_{\a_1\ldots \a_{r+1}}=0$. On the other hand, for $d=2r=$ 2 mod 4 dimensions, $\p_L^c$ is still left-handed and $\overline{\p_L} \p_L^c$ vanishes, just as does $\overline{\p_L} \p_L$, so that chiral fermions can have neither Dirac nor Majorana masses is 2 mod 4 dimensions. 
Thus we conclude in general:
\be
\begin{array}{|c|} 
\hline\\
\text{Only massless particles can contribute to the anomaly.}\\ 
\\
\hline
\end{array}
\nonumber
\ee
This  is very fortunate since it allows us to study the question of anomalies in theories describing the known elementary particles without the need to know which  heavy particles might be discovered at some very high energy.

\subsection{Anomaly cancellation in the standard model\label{anomstandard}}

Here the gauge group is $SU(3)\times SU(2)\times U(1)$ and from the above discussion, a priori, we can have anomalies for\hfill\break
$\bullet \quad SU(3)\times SU(3)\times SU(3)$,\hfill\break 
$\bullet \quad SU(3)\times SU(3)\times U(1)$, \hfill\break
$\bullet \quad SU(2)\times SU(2)\times U(1)$, \hfill\break
$\bullet \quad U(1)\times U(1)\times U(1)$, \hfill\break
$\bullet \quad$mixed $U(1)$-gravitational.\hfill\break
We must know which representations appear. Since each generation of quarks and leptons repeat the same representations it is enough to look at the first generation. There are the left-handed neutrino $\n_e$ and left-handed electron which form a doublet of $SU(2)$, denoted $\begin{pmatrix} \n_e \\ e \\ \end{pmatrix}_L$, as well as the right-handed electron $e_R$ which is an $SU(2)$ singlet. According to our previous discussion, we can equivalently consider the left-handed positron $(e_R)^c\simeq (e^c)_L$. Similalry, there is the left-handed quark doublet $\begin{pmatrix} u \\ d \\ \end{pmatrix}_L$ of $SU(2)$ and the right-handed $SU(2)$ singlets $u_R$ and $d_R$ which we describe as left-handed singlets $(u_R)^c$ and $(d_R)^c$. Of course, the quarks are in a {\bf 3} of $SU(3)$ and hence $(u_R)^c$ and $(u_R)^c$ are in a $\overline{\bf 3}$. The table summarizes all the left-handed particles/antiparticles.

\begin{table}[htdp]
\centering
\begin{tabular}{|c|c|c|c|}
\hline
& & &\\
 & $SU(3)$ repres. & $SU(2)$ repres. & $U(1)$ hypercharge \\
& & &\\
\hline
& & &\\
$\begin{pmatrix} \n_e \\ e \\ \end{pmatrix}_L$ &{\bf 1} &{\bf 2}  &${1\over 2}$ \\
& & &\\
\hline
& & &\\
$(e_R)^c$&{\bf 1}  &{\bf 1}  & $-1$\\
& & &\\
\hline
& & &\\
$\begin{pmatrix} u \\ d \\ \end{pmatrix}_L$ &{\bf 3} &{\bf 2}  &$-{1\over 6}$ \\
& & &\\
\hline
& & &\\
$(u_R)^c$&$\overline{\bf 3}$ &{\bf 1}  & ${2\over 3}$\\
& & &\\
\hline
& & &\\
$(d_R)^c$&$\overline{\bf 3}$ &{\bf 1}  & $-{1\over 3}$\\
& & &\\
\hline
\end{tabular}
\vskip3.mm
\caption{All the left-handed particles/antiparticles of one generation in the standard model}\vskip3.mm
\label{default}
\end{table}

\subsubsection{Unbroken phase}

Of course, most of these particles are actually massive due to their interactions with the scalar (Higgs) field that acquires a vacuum expectation value and is responsible for the electro-weak symmetry breaking. We will discuss this issue below. For the time being, we assume that the couplings of the fermions to the scalar field are taken to vanish\footnote{
In the standard model these couplings all are independent parameters which one must adjust to fit the experimentally observed fermion masses. From the theoretical point of view, it is perfectly consistent to set these couplings to zero.
} 
so that the fermions indeed are massless.\footnote{
More physically, at energies well above the electro-weak symmetry breaking scale $\sim g v$ (where $g$ is the SU(2) coupling constant and $v$ the scalar field expectation value) the theory is in the ``unbroken phase" and the fermions can indeed be considered as massless.
}

Let us now work out the contributions to the $D_{\a\b\g}$ for the different anomalies. We will denote the gauge coupling constants of $SU(3)$, $SU(2)$ and $U(1)$ by $g_s$, $g$ and $g'$ respectively. We get:
\begin{itemize}
\item
$SU(3)\times SU(3)\times SU(3)$ :
The total (reducible) $SU(3)$ representations that occurs is (in the order of the table) $R=({\bf 1}+{\bf 1}) + {\bf 1} + ( {\bf 3}+{\bf 3}) + \overline{\bf 3}+\overline{\bf 3}$ which is real, since $\overline{\bf 3}$ is the complex conjugate of ${\bf 3}$. Hence $D^\cR_{\a\b\g}=0$, and there is no $SU(3)\times SU(3)\times SU(3)$-anomaly in the standard model. Of course, the generators for the trivial representation ${\bf 1}$ simply vanish, and only the quarks (and antiquarks) contribute. The relevant representation then is $R=( {\bf 3}+{\bf 3}) + \overline{\bf 3}+\overline{\bf 3}$ and we reach the same conclusion.

\item
$SU(3)\times SU(3)\times U(1)$ : As just mentioned, only the quarks and antiquarks contribute. If we call $t$ the $U(1)$ hypercharge generator and $t_\a$ the $SU(3)$ generators we have 
\ba
\tr t^\cR_\a t^\cR_\b t^\cR &=& 2\times \tr t^3_\a t^3_\b \times \big(-{1\over 6}\, g'\big) +  \tr t^{\overline{3}}_a t^{\overline{3}}_b \times \big( {2\over 3}\, g'\big) +  \tr t^{\overline{3}}_a t^{\overline{3}}_b \times \big( -{1\over 3}\, g'\big)
\nonumber\\
&=& g_s\, g'\, C_3\, \dd_{\a\b} \left( 2\times \big(-{1\over 6}\big)+ {2\over 3}-{1\over 3}\right) = 0  \ ,
\ea
where we used $C_{\overline{3}}=C_3$.

\item
$SU(2)\times SU(2)\times U(1)$ :
Here only the $SU(2)$ doublets can contribute, and calling $t_\a$ now the $SU(2)$ generators, we get
\be
\hskip3.mm
\tr t^\cR_\a t^\cR_\b t^\cR 
=\tr t^2_\a t^2_\b \times \big({1\over 2}\, g'\big)
+ 3\times \tr t^2_\a t^2_\b \times \big(-{1\over 6}\, g'\big)
= g\, g'\, C_2 \, \dd_{\a\b} \left(  {1\over 2}+ 3 \times \big(-{1\over 6}\big)\right) = 0 \ .
\ee

\item
$U(1)\times U(1)\times U(1)$ :
\be
\hskip6.mm 
\tr t^\cR t^\cR t^\cR = 2\times \big({1\over 2}\, g'\big)^3 + \big( -g'\big)^3 + 3\times 2\times \big(-{1\over 6}\, g'\big)^3 + 3\times \big({2\over 3}\, g' \big)^3 + 3\times \big(-{1\over 3}\, g'\big)^3 = 0 \ .
\ee

\item
mixed $U(1)$-gravitational anomalies :
We have seen that they are proportional to the sum of all $U(1)$ charges:
\be
\hskip15.mm\tr t^\cR=2\times {1\over 2}\, g' + (-g') + 3\times 2\times \big(-{1\over 6}\, g'\big)+ 3\times \big({2\over 3}\, g' \big) + 3\times \big(-{1\over 3}\, g'\big) = 0 \ .
\ee
\end{itemize}

\noindent
Thus all possible anomalies cancel for every generation of the standard model. If in one generation a quark (or any other particle) were missing, one would get non-vanishing anomalies (not for $SU(3)\times SU(3)\times SU(3)$, but for the three other combinations).

\subsubsection{Broken phase}
Let us now discuss the issue that the fermions are actually not massless but get masses due to their couplings to the scalar field. This scalar field $\f=\begin{pmatrix} \f^+\\ \f^0\\ \end{pmatrix}$ is an $SU(2)$ doublet and has a potential such that it develops a vacuum expectation value, which one can take as $\langle\f^+\rangle=0,\ \langle \f^0\rangle=v$, spontaneously breaking the $SU(2)\times U(1)$ symmetry to a single $U(1)$ which corresponds to the electromagnetic gauge symmetry. Its generator $t_{\rm em}$ is a combination of the original $t_3$ and the hypercharge $t$, namely 
\be\label{elhyper}
t_{\rm em}= {e\over g}\, t_3 - {e\over g'}\, t\ ,
\ee 
where $-e<0$ is the electric charge of an electron. Specifically, the interaction of the neutrino and electron fields with the scalar is
\be\label{Lenfi}
\cL_{e\n\f}=-g_e \overline{\begin{pmatrix} \n_e\\ e\\ \end{pmatrix}_L}
\begin{pmatrix} \f^+\\ \f^0\\ \end{pmatrix} e_R + h.c.
\ee
There are similar terms for the interactions of the quarks with the scalar field but with different and independent coupling constants. We may well assume that we keep these coupling constants zero for the time being and only switch on the interaction $\cL_{e\n\f}$ with a non-vanishing $g_e$. Inserting the vacuum expectation value for $\f$ this gives
\be\label{Lenfiv}
\cL_{e\n\f}\big\vert_{\f=\langle\f\rangle}=-(g_e v)\ \overline{e_L}\,
e_R + h.c. \ ,
\ee
which is just a standard Dirac mass term for a non-chiral electron, the (chiral) neutrino remaining massless. Of course, such a term is not compatible with the original $SU(3)\times SU(2)\times U(1)$ symmetry and with the representations as given in the above table. However, from a low-energy perspective, the symmetry is broken to the ``low-energy gauge group" $SU(3)\times U(1)_{\rm em}$ and the mass term (\ref{Lenfiv}) clearly is compatible with this symmetry. Let us then check that there are no anomalies with respect to this $SU(3)\times U(1)_{\rm em}$. The electron being massive, it  does not contribute to the anomaly any more. (Indeed, $e_L$ and $(e_R)^c$ are $SU(3)$ singlets and have opposite electric charge, so that their contributions to $\tr t_\a t_\b t_{\rm em}$ or $\tr t_{\rm em} t_{\rm em} t_{\rm em}$ always cancel.) The left-handed neutrino is an $SU(3)$ singlet and has zero electric charge so it does not contribute to any anomaly either. Finally, for the quarks, $u_L$ and $u_R$ the electric charges are obtained using the relation (\ref{elhyper}) and the hypercharge values of $t/g'$ given in the table. One finds that $u_L$ and $u_R$ have $t_{\rm em} = {2\over 3}\, e$ and $d_L$ and $d_R$ have 
$t_{\rm em} = -{1\over 3}\, e$. Of course, $(u_R)^c$ has $t_{\rm em} = -{2\over 3}\, e$ and $(d_R)^c$ has $t_{\rm em} = {1\over 3}\, e$. Then one finds for their contributions to the different anomalies:
\begin{itemize}
\item 
$SU(3)\times SU(3)\times SU(3)$ : The representation is $( {\bf 3}+{\bf 3}) + \overline{\bf 3}+\overline{\bf 3}$ which is real, and there is no anomaly.
\item
$SU(3)\times SU(3)\times U(1)$ : We have 
$\tr t^\cR_\a t^\cR_\b t^\cR_{\rm em}
=\tr t^3_\a t^3_\b \times \big({2\over 3}  - {1\over 3} \big)e +  \tr t^{\overline{3}}_a t^{\overline{3}}_b \times \big( -{2\over 3}+ {1\over 3} \big) e= e\,  g_s\, C_3\, \dd_{\a\b} \big({2\over 3}  - {1\over 3} -{2\over 3} + {1\over 3}\big) =0$, where we used again $C_{\overline{3}}=C_3$.
\item
$U(1)\times U(1)\times U(1)$ :
Here simply
$\tr t^\cR_{\rm em} t^\cR_{\rm em} t^\cR_{\rm em} = e^3\ \Big(
\big({2\over 3}\big)^3+ \big(-{1\over 3}\big)^3+\big(-{2\over 3}\big)^3+\big({1\over 3}\big)^3 \Big)=0$.
\end{itemize}
Thus, again, all anomalies cancel. Actually, this should have been obvious since $u_L$ and $(u_R)^c$ together form a real representation of $SU(3)\times U(1)_{\rm em}$, and similarly for $d_L$ and $(d_R)^c$. This also shows that one can add different (Dirac) mass terms for the $u$ and $d$ quarks, as are indeed generated by the coupling to the scalar vacuum expectation value.

\newpage
{\huge\centerline{\bf Part II :}
\vskip3.mm
\centerline{\bf Gauge and gravitational anomalies}
\vskip3.mm
\centerline{\bf in arbitrary dimensions}}
\vskip6.mm

\noindent
Having completed our rather detailed study of anomalies in four-dimensional (non-abelian) gauge theories, we now turn to various more formal developments. On the one hand, we will develop the tools to characterize and compute relevant gauge anomalies in arbitrary (even) dimensions and, on the other hand, we will extend this formalism to also include gravitational anomalies in a generally covariant theory. Such gravitational anomalies can be viewed either as anomalies of the diffeomorphisms or as anomalies of local Lorentz transformations.
This will allow us in the end to study some prominent examples of cancellation of gauge and gravitational anomalies in ten dimensions.

\section{Some formal developments: differential forms and characteristic classes in arbitrary even dimensions\label{formaldev}}
\setcounter{equation}{0}

We have seen that the chiral anomaly for a {\it left-handed}, i.e. positive chirality fermion in four dimensions is given by (cf eq.~(\ref{chiralanom1}))
\be\label{chiralanom3}
\ca_\a(x) = -D_\m \langle j^\m_\a \rangle
= -{1\over 24\pi^2} \e^{\m\n\r\s} \trR t_\a \del_\m A_\n \del_\r A_\s + \cO(A^3) \ ,
\ee
or equivalently by
\be\label{chiralanom4}
\dd_\e \G[A] = \int \d^4 x\, \e^\a(x) \ca_\a(x) 
= -{1\over 24\pi^2} \int\d^4 x\, \e^{\m\n\r\s} \trR \e\, \del_\m A_\n \del_\r A_\s + \cO(A^3) \ ,
\ee
with $\e=\e^\a t^\cR_\a$. From now on, we will use the more common symbol $\G[A]$ for the effective action $\wt W[A]$. As noted in section \ref{trichir}, the $\cO(A^3)$-terms could have been determined from a square diagram computation, but we will get them below from the powerful consistency conditions.

The appearance of $\e^{\m\n\r\s}$ in (\ref{chiralanom3}) and (\ref{chiralanom4}) is characteristic of differential forms. It will indeed prove very useful to reformulate these expressions in terms of differential forms. So far we have been working in flat space-time, but the use of differential forms will allow us to extend most results to curved space-time in a straightforward way. Indeed, differential forms are naturally defined on a curved manifold without the need of using the metric explicitly.  For completeness, and also to fix our normalizations,  we will  briefly review some basic notions about differential forms. 
The reader familiar with these notions can safely skip the first subsection \ref{diffforms}.

\subsection{Differential forms in arbitrary dimensions, exterior derivative and de Rham cohomology\label{diffforms}}

Consider a $d$-dimensional space-time which may be a curved manifold.
One introduces the symbols $\d x^\m$ and a wedge product $\d x^\m \wedge \d x^\n = - \d x^\n \wedge \d x^\m$. One then has e.g.
\be 
\d x^\m \wedge \d x^\n\wedge \d x^\r
=-\ \d x^\m \wedge \d x^\r\wedge\d x^\n
=+\  \d x^\r \wedge \d x^\m \wedge \d x^\n \ .
\ee
Such a wedge product is clearly completely antisymmetric in all indices.
A general $p$-form $\xi^{(p)}$ is a sum 
\be
\xi^{(p)}= {1\over p!}\, \xi_{\m_1 \ldots \m_p} \,
\d x^{\m_1}\wedge\ldots \wedge \d x^{\m_p} \ ,
\ee
where the coefficients are completely antisymmetric tensors of degree $p$. Obviously, due to the antisymmetry, in $d$ dimensions the maximal degree of a form is $p=d$.
The wedge product of a $p$-form $\xi^{(p)}$ with a $q$-form $\zeta^{(q)}$ is then defined in an obvious way:
\ba
\xi^{(p)}\wedge \zeta^{(q)}
&=&{1\over p!}\, {1\over q!}\, 
\xi_{\m_1 \ldots \m_p} \,\zeta_{\n_1\ldots \n_q}\,
\d x^{\m_1}\wedge\ldots \wedge \d x^{\m_p}\wedge\d x^{\n_1}\wedge\ldots\wedge\d x^{\n_q}
\nonumber\\
&=&{1\over p!q!}\, \xi_{[\m_1 \ldots\m_p}\zeta_{\n_1\ldots\n_q]}
\d x^{\m_1} \wedge\ldots \wedge\d x^{\n_q} \ ,
\ea
and yields a $p+q$ form. 
It follows from the above properties that this wedge product is anticommutative if $p$ and $q$ are both odd and commutative otherwise (provided the coefficients $\xi_{\m_1 \ldots \m_p}$ and $\zeta_{\n_1\ldots \n_q}$ are $c$-numbers.) Note that one does not always explicitly write the symbol $\wedge$ since the product of two differential forms is always meant to be the wedge product unless otherwise stated.

It is important to note that under coordinate transformation, the coefficients $\xi_{\m_1\ldots \m_p}$ of a $p$-form transform as a covariant (antisymmetric) tensor, while the $\d x^{\m_1}\wedge\ldots\wedge \d x^{\m_p}$ obviously transform as a contravariant (antisymmetric) tensor. It follows that the $p$-form $\xi^{(p)}$ transforms as a scalar. This is one of the reasons why it is very convenient to deal with differential forms.

The \underline{exterior derivative $\d=\d x^\m \del_\m$} acts on a $p$-form as
\ba
\d \xi^{(p)} &=& \d x^\m \del_\m \Big({1\over p!}\, \xi_{\m_1 \ldots \m_p} \, \d x^{\m_1}\wedge\ldots \wedge \d x^{\m_p} \Big)
= {1\over p!}\, \del_{[\m}\xi_{\m_1\ldots\m_p]}\, \d x^\m \wedge \d x^{\m_1}\wedge\ldots \wedge \d x^{\m_p} 
\nonumber\\
&=& {1\over (p+1)!} \underbrace{
\left( \del_\m \xi_{\m_1\ldots \m_p} - \del_{\m_1} \xi_{\m \m_2\ldots \m_p} + \ldots \right)}
 \d x^\m \wedge \d x^{\m_1}\wedge\ldots \wedge \d x^{\m_p} \equiv \zeta^{(p+1)} \ .
\nonumber\\
&&\hskip3.cm p+1\ {\rm terms}
\ea
Hence if $\zeta^{(p+1)} =\d \xi^{(p)}$, then the coefficients of $\zeta^{(p+1)} ={1\over (p+1)!}\, \zeta_{\m_1\ldots\m_{p+1}} \d x^{m_1}\wedge\ldots \wedge \d x^{\m_{p+1}}$ are given by $\zeta_{\m_1\ldots\m_{p+1}} = \underbrace{
\left( \del_\m \xi_{\m_1\ldots \m_p} - \del_{\m_1} \xi_{\m \m_2\ldots \m_p} + \ldots \right)}_{p+1 {\rm terms}}$. A most important property of the exterior derivative is its nilpotency, i.e. $\d^2=0$. Indeed,
\be
\d \d \xi^{(p)} = \d \left( {1\over p!}\, \del_{[\m}\xi_{\n_1\ldots\n_p]}\, \d x^\m \wedge \d x^{\n_1}\wedge\ldots \wedge \d x^{\n_p} \right)
={1\over p!}\, \del_{[\r} \del_\m \xi_{\n_1\ldots \n_p]}\d x^\r \wedge\d x^\m \wedge \d x^{\n_1}\wedge\ldots \wedge \d x^{\n_p} 
=0 \ ,
\ee
since $\del_\r\del_\m - \del_\m\del_\r =0$.

\vskip2.mm
A $p$-form $\xi^{(p)}$ is called {\it closed} if $\d \xi^{(p)}=0$. It is called {\it exact} if there exists a (globally well-defined) $(p-1)$-form $\o^{(p-1)}$ such that $\xi^{(p)}=\d \o^{(p-1)}$. In many cases such a $\o^{(p-1)}$ may exist only locally but be not globally well-defined. Since $\d^2=0$ it follows that every exact form is also closed. The converse, however, is not true in general.  

{\it de Rham cohomology :} one is interested in determining the closed $p$-forms modulo exact ones: 
The so-called de Rham cohomology group $H^{(p)}$ is the set of all closed $p$-forms $\xi^{(p)}$ modulo exact ones, i.e. the set of all $\xi^{(p)}$ such that $\d\xi^{(p)}=0$, subject to the equivalence relation 
\be
\wt\xi^{(p)}\simeq \xi^{(p)}\quad {\rm if}\quad \wt\xi^{(p)}= \xi^{(p)}+ \d \o^{(p-1)}\quad \text{for some $(p-1)$-form}\ \o^{(p-1)}\ .
\ee
Obviously, $H^{(p)}$ is a vector space and its dimension is called the $p^{\rm th}$ Betti number $b_p$.  Actually, the $H^{(p)}$ depend crucially on the {\it topology} of the manifold one is considering and the $b_p$ are called the Betti numbers of the manifold.

Let us look at an example which should be familiar from classical electrodynamics. Let space (or space-time) be ${\bf R}^4$. The exterior derivative of a scalar $\f$ is $\d \f=\del_\m\f\, \d x^\m$ which is just a gradient. If $A=A_\m \d x^\m$ is a one-form, then 
\be
\d A={1\over 2} (\del_\m A_\n - \del_\n A_\m) \d x^\m \d x^\n \equiv F
\quad {\rm with}\quad F_{\m\n}=\del_\m A_\n - \del_\n A_\m \ .
\ee
If $A=\d \f$ is exact (``pure gauge") then $\d^2=0$ implies that $A$ is closed, i.e. $F=\d A=0$. Conversely, if $F=\d A=0$, $A$ is closed and we then usually conclude that $A$ is pure gauge: $A=\d \f$ for some $\f$, i.e. $A$ is exact. This conclusion indeed holds in ${\bf R}^4$ which is topologically trivial so that every closed $p$-form is exact (for $p=1,2,3,4$). 

\begin{figure}[h]
\centering
\includegraphics[width=0.25\textwidth]{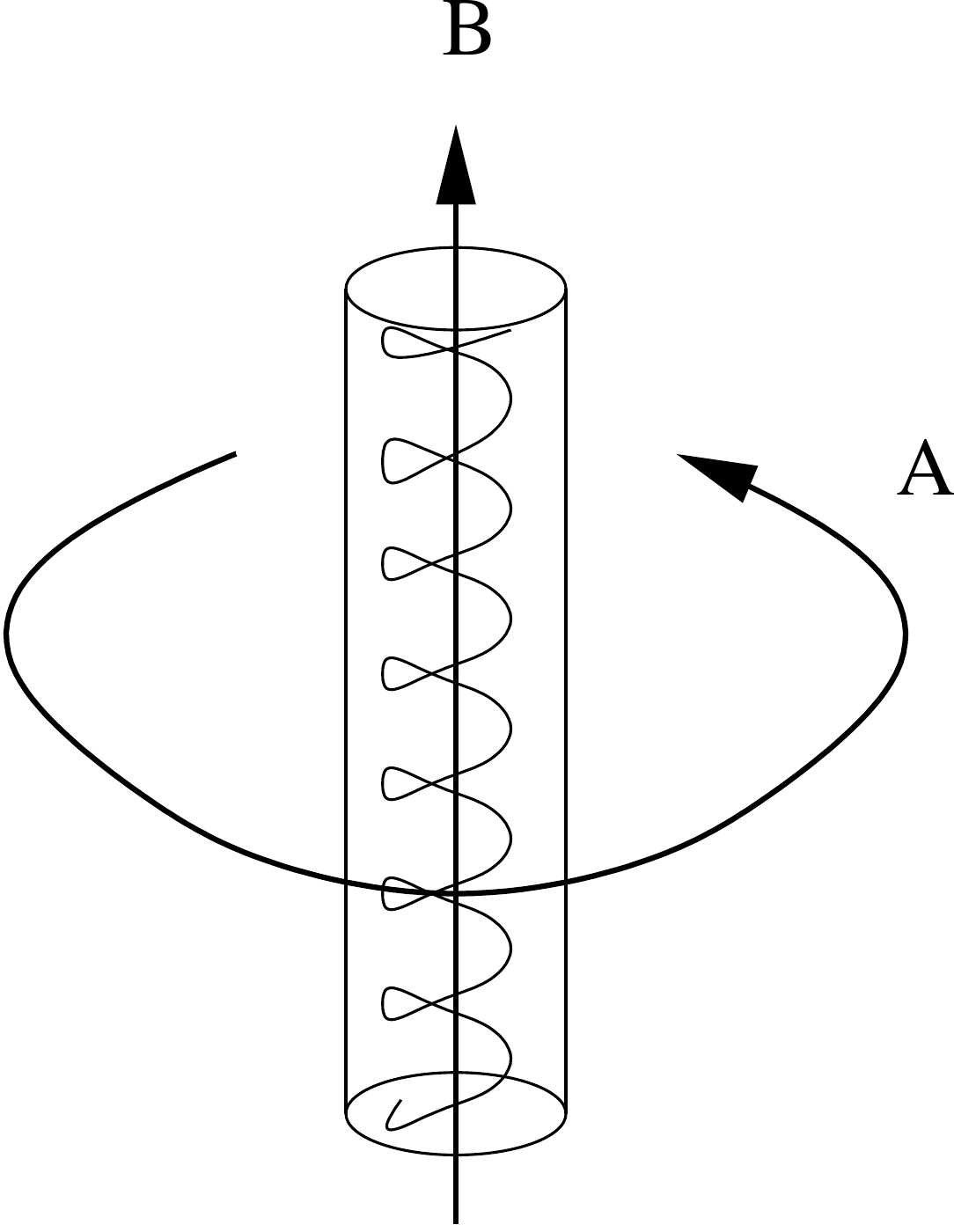}\\
\caption[]{An infinite solenoid creates a magnetic field confined to the interior of the solenoid.} \label{solenoid}
\end{figure}
\vskip5.mm

Consider now another example from classical electrodynamics which involves a topologically non-trivial space. We look at a static situation (so that we can neglect time and effectively have a 3-dimensional space only). The components of the magnetic field then are $B^x=F_{23},\ B^y=F_{31}$\break 
and $B^z=F_{12}$. Take as our manifold the space outside an infinitely long solenoid:\break 
$\cM={\bf R}^3\backslash \{ \text{a cylinder around the $z$-axis}\}$. Physically, a current in the solenoid produces a magnetic field confined to the {\it interior} of the solenoid so that everywhere in $\cM$ the magnetic field vanishes: $F=\d A=0$ and $A$ is closed. However, we cannot conclude that this implies that $A$ is exact, i.e. $A=\d\f$ with a well-defined (i.e. single-valued) scalar field $\f$. Indeed, we know by Stoke's theorem that for any closed curve $\cC$ the integral $\oint_{\cC} A$ equals the flux of the magnetic field through the surface spanned by $\cC$. If we take $\cC$ to surround once the solenoid, this flux is non-vanishing. On the other hand if we had $A=\d \f$ then we would get, again by Stokes theorem,  $\oint_{\cC} A = \oint_{\cC} \d \f = 0$, in contradiction with the above. It is easy to see what happens. If the modulus of the magnetic field produced by the solenoid is called ${\cal B}$, in cylindrical coordinates $(z,\r,\varphi)$ the gauge field  is $A={{\cal B}\over 2\pi} \d\varphi$. This is well-defined everywhere in $\cM$. Of course, $\d\varphi$ is ill-defined on the $z$-axis, but this is not part of our manifold $\cM$. One sees that we correctly have $F=\d A=0$ everywhere in $\cM$. However, $A$ is not exact: although formally $A=\d \left({{\cal B}\over 2\pi}\varphi\right)$ the expression in the brackets is not a well-defined 0-form since $\varphi$ is  not single-valued on $\cM$.

{\it Integration of differential forms :}
Note that on a $d$-dimensional manifold $\d x^{\m_1}\wedge \ldots \d x^{\m_d}$ is completely antisymmetric in all $d$ indices and hence proportional to the``flat" $\e$-tensor $\wh\e$ normalized as $\wh\e^{01\ldots (d-1)}=+1$. Hence we have
\be\label{volform}
\d x^{\m_1}\wedge \ldots \d x^{\m_d}= \wh \e^{\,\m_1\ldots \m_d}\ \d x^0 \wedge \ldots \d x^{d-1} \equiv \wh \e^{\,\m_1\ldots \m_d}\ \d^d x \ .
\ee
On a curved manifold, the true $\e$-tensor and the pseudo-tensor $\wh\e$ are related by
\be
\e_{\m_1\ldots\m_d} = \sqrt{-g}\ \wh\e_{\m_1\ldots\m_d}
\quad , \quad 
\e^{\m_1\ldots\m_d} = {1\over \sqrt{-g}}\ \wh\e^{\ \m_1\ldots\m_d}
\ee
where, of course, $g$ stands for $\det g_{\m\n}$. Then (\ref{volform}) is rewritten as
\be\label{volform2}
\d x^{\m_1}\wedge \ldots \d x^{\m_d}=\e^{\,\m_1\ldots \m_d}\  
\sqrt{-g}\ \d^d x\ .
\ee 
It follows that for any $d$-form one has
\be\label{rewrite}
\xi^{(d)}={1\over d!}\, \xi_{\m_1\ldots \m_d} \d x^{\m_1} \ldots \d x^{\m_d}= 
{1\over d!}\, \xi_{\m_1\ldots \m_d} \e^{\,\m_1\ldots \m_d}\ 
\sqrt{-g}\ \d^d x \ .
\ee
Since $\xi_{\m_1\ldots \m_d} \e^{\,\m_1\ldots \m_d}$ is a scalar and 
$\sqrt{-g}\ \d^d x$ the volume element, it is clear that we can directly integrate any $d$ form over the $d$-dimensional manifold or over any $d$-dimensional submanifold. (Note  that the original definition of the differential form $\xi^{(d)}$ does not involve the metric, only our rewriting (\ref{rewrite}) does.) Similarly, any $p$-form with $p\le d$ can be directly integrated over any $p$-dimensional submanifold $S_{(p)}$:
\be
\int_{S_{(p)}} \xi^{(p)} 
\ee
is a well-defined scalar, i.e. invariant under changes of the coordinates used. {\it Stokes's theorem} then can be written as
\be
\int_{S_{(p)}} \d \zeta^{(p-1)} = \int_{\del S_{(p)}} \zeta^{(p-1)} \ ,
\ee
where $\del S_{(p)}$ is the $(p-1)$-dimensional manifold which is the boundary of the manifold $S_{(p)}$.
Consider again the example of electrodynamics with $F=\d A$ in 4 dimensions. Then $F\wedge F$ is a 4-form and we have
\be\label{FwedgeF}
F\wedge F= {1\over 4} F_{\m\n} F_{\r\s} \d x^\m \d x^\n \d x^\r \d x^\s = {1\over 4 } \e^{\m\n\r\s}F_{\m\n} F_{\r\s}\ \sqrt{-g}\ \d^4 x \ ,
\ee
an expression familiar from our abelian anomaly computations. (Of course, no $\sqrt{-g}$ appeared there since we worked in flat space-time.)

{\it Hodge dual :}
For any $p$-form $\xi^{(p)}$ one defines the Hodge dual $*\xi^{(p)}$, which is a $(d-p)$-form, as
\ba\label{Hodge}
*\xi^{(p)}&=&{1\over p!}\, \xi_{\m_1\ldots \m_p}  
*(\d x^{\m_1}\ldots \d x^{\m_p}) \ , 
\nonumber\\
*(\d x^{\m_1}\ldots \d x^{\m_p})
&=&{1\over (d-p)!}\, \e^{\m_1\ldots \m_p}_{\ \ \ \ \ \ \ \n_{p+1}\ldots \n_d}\ \d x^{\n_{p+1}}\ldots  \d x^{\n_d} 
\nonumber\\
&=&{1\over (d-p)!}\, g^{\m_1 \n_1} \ldots g^{\m_p \n_p} 
\e_{\n_1\ldots \n_p \n_{p+1}\ldots \n_d}\ \d x^{\n_{p+1}}\ldots  \d x^{\n_d} \ .
\ea
Clearly, to define the Hodge dual one needs the metric. Note that $*(*\xi^{(p)})=-(-)^{p(d-p)}\xi^{(p)}$. Since the Hodge dual is a $(d-p)$-form, $\xi^{(p)} \wedge * \xi^{(p)}$ is a $d$-form and, as we have seen above,  must be proportional to the volume form $\sqrt{-g}\, \d^d x$. Indeed, using (\ref{rewrite}) and (\ref{Hodge}) it is straightforward to find
\be\label{xiwedgehodgexi}
\xi^{(p)}\wedge *\xi^{(p)}= {1\over p!} \xi^{\m_1\ldots \m_p} \xi_{\m_1\ldots \m_p} \sqrt{-g}\,  \d^d x \ ,
\ee
where, of course, $\xi^{\m_1\ldots \m_p}= g^{\m_1 \n_1}\ldots g^{\m_p \n_p} \xi_{\n_1\ldots \n_p}$. This can be used to rewrite the kinetic term in an action involving antisymmetric tensors using the corresponding differential forms, e.g.
\be\label{kineticforms}
-{1\over 4} \int \d^d x\ \sqrt{-g}\,  \ F^{\m\n} F_{\m\n}
=-{1\over 2} \int F\wedge *F \ .
\ee

\subsection{Non-abelian gauge fields as differential forms and the gauge bundle\label{gaugebundle}}

In a (non-abelian) gauge theory one defines the gauge connection one-form and field strength two-form as
\be
A=A_\m \d x^\m = A_\m^\a  t_\a \d x^\m 
\quad , \quad
F={1\over 2} F_{\m\n} \d x^\m \d x^\n = {1\over 2} F_{\m\n}^\a t_\a\d x^\m \d x^\n  \ .
\ee
These one-forms are matrix-valued. It follows that such one-forms no longer simply anticommute with each other. Instead one has e.g.
\be
A^2 \equiv A\wedge A = A_\m A_\n \d x^\m \d x^\n 
= {1\over 2} (A_\m A_\n - A_\n A_\m) \d x^\m \d x^\n 
={1\over 2} [A_\m , A_\n ] \d x^\m \d x^\n \ .
\ee
This allows us to write the non-abelian field strength as
\be
F={1\over 2} F_{\m\n} \d x^\m \d x^\n
= {1\over 2} \left( \del_\m A_\n-\del_\n A_\m - i [A_\m, A_\n]\right) \d x^\m \d x^\n = \d A -i A^2 \ .
\ee
Gauge transformations are now written as
\be
\dd A=\d \e -i [A,\e] 
\quad , \quad
\dd \p = i\e \p \ ,
\ee
where it is understood, as usual, that $\e=\e^\a t_\a$  and the $t_a$ are in the appropriate  representation. One also defines a covariant exterior derivative as
\be
{\rm D}=\d x^\m D_\m = \d - i A \ .
\ee

It is convenient to absorb the $i$'s appearing all over the place by a redefinition of the generators, gauge connection and field strength as follows. We let $T_\a =-i t_\a$. The $T_\a$ now are antihermitian 
generators, $T^\dag_\a = - T_\a$ satisfying $[T_\a,T_\b]=C^\g_{\ \a\b} T_\g$ (with the same real structure constants as before). If we also let $\sa_\m^\a=-i A_\m^\a$ and define (as before) $\sa_\m=\sa_\m^\a t_\a$ we have $\sa_\m=-i A_\m^\a t_\a= A_\m^\a T_\a$ so that $\sa\equiv \sa_\m \d x^\m=A_\m^\a T_\a \d x^\m=-i A$. Furthermore, we let $\cF=-i F$ so that $\cF=\d\sa+\sa^2$, without the $i$, indeed. Similarly one redefines the infinitesimal parameters of the gauge transformations as $v\equiv v^\a t_\a=-i \e^\a t_\a= \e^\a T_\a =-i \e$ so that the gauge transformations now read $\dd\p=-v\p$ and $\dd\sa=\d v + [\sa,v]$.  Finally, the covariant derivative becomes ${\rm D}=\d+\sa$. Let us summarize
\be\label{iredef}
\begin{array}{|c|}
\hline\\
\quad
\sa=-i A \quad , \quad \cF=-i F \quad , \quad \cF=\d\sa+\sa^2
\quad , \quad {\rm D}=\d+\sa
\quad\\
\\
v=-i\e \quad , \quad \dd\p=-v\p \quad , \quad \dd\sa=\d v + [\sa,v]\ .
\\
\\
\hline
\end{array}
\ee
As before (cf. the corresponding discussion in section \ref{Notation}), if the $\sa$ in the covariant derivative acts on a $p$-form field in the adjoint representation like e.g. $\cF$, this action can be rewritten as a commutator (if $p$ is even) or anticommutator (if $p$ is odd). Thus the Bianchi identity for $\cF$ can be simply written as
\be\label{Bianchi}
{\rm D}\cF=\d\cF+\sa\cF-\cF\sa = 0 \ .
\ee

Let us now show how to rewrite the anomaly obtained above using the language of differential forms. Just as we obtained (\ref{FwedgeF}), we have
\be
\d A \d A = \del_\m A_\n \del_\r A_\s \d x^\m \d x^\n \d x^\r \d x^\s = \del_\m A_\n \del_\r A_\s \e^{\m\n\r\s} \sqrt{-g}\, \d^4 x \ .
\ee
Again, in flat space-time the $\sqrt{-g}$ is not needed, and this is why it did not appear when we computed the anomaly. We then rewrite the anomaly for a left-handed (positive chirality) fermion (\ref{chiralanom4}) as
\ba\label{anomdiffform}
\dd_\e \G[A] = \int \d^4 x \sqrt{-g}\  \e^\a \ca_\a
&=& -{1\over 24\pi^2} \int \d^4 x \sqrt{-g}\   \e^\a \e^{\m\n\r\s} \trR t_\a \del_\m A_\n \del_\r A_\s + \cO(A^3)
\nonumber\\
&=& -{1\over 24\pi^2} \int \trR \e\, \d A \d A + \cO(A^3)
= {i\over 24\pi^2} \int \trR v\, \d \sa \d \sa + \cO(\sa^3)
\ .
\nonumber\\
\ea
\vskip-3.mm
\noindent
This is clearly a convenient and compact notation.

\begin{figure}[h]
\centering
\includegraphics[width=0.70\textwidth]{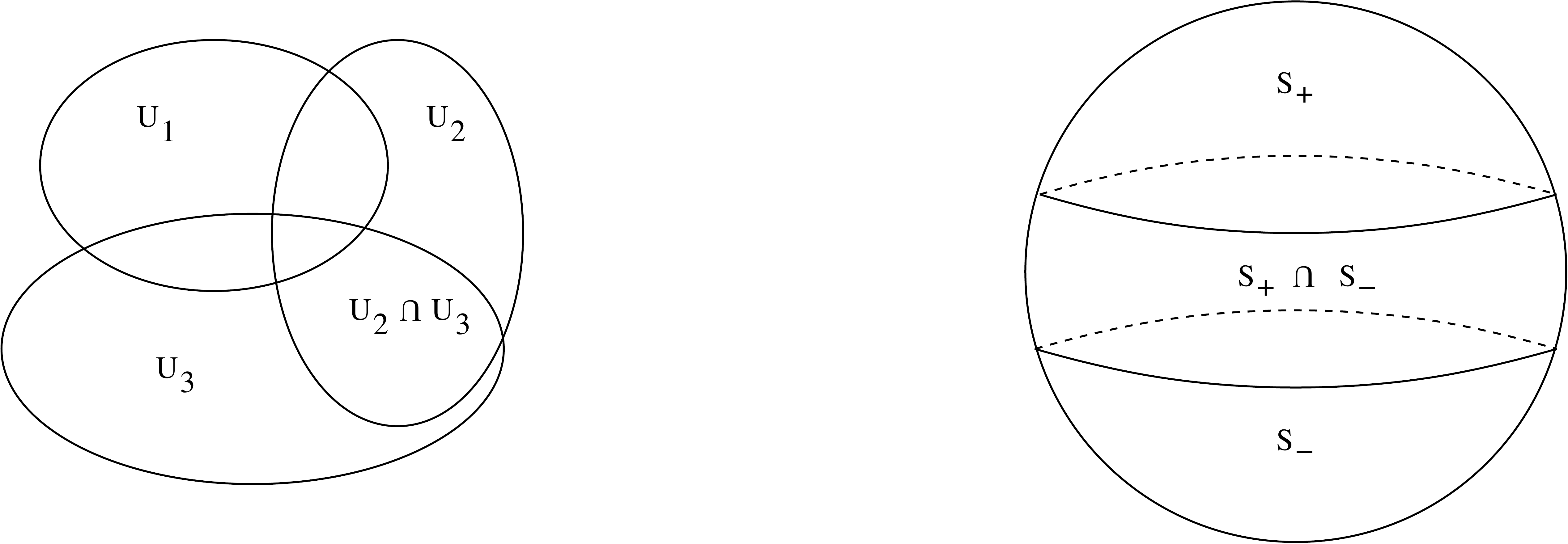}\\
\caption[]{Coordinate patches on some manifold (left) and the two standard patches $S_+$ and $S_-$ on the two-sphere $S^2$ that overlap on the equator ribbon (right).} \label{patches}
\end{figure}

Whenever the space-time manifold is topologically non-trivial one must cover it by a (finite) number of coordinate patches $U_i$, each diffeomorphic to an open subset of ${\bf R}^d$ (see Fig.~\ref{patches}). Then to define a vector or  tensor field, one defines it on each patch separately, together with appropriate transition rules on the overlaps $U_i\cap U_j$. As noted above, a $p$-form should transform as a scalar, i.e. it has trivial transition functions. Consider now an {\it abelian} gauge theory. The field strength is a 2-form $\cF_{(i)}$ defined on each patch $U_i$ and we must have $\cF_{(i)}=\cF_{(j)}$ on the overlaps $U_i\cap U_j$. However, this does not imply that the gauge connections should also be related in this simple way. Indeed, we can only require that $\sa_{(i)}$ be related to $\sa_{(j)}$ on the overlap by a gauge transformation. More generally, in a {\it non-abelian} gauge theory we allow the $\sa_{(i)}$ and $\sa_{(i)}$ to be related by a (finite) gauge transformation
\be\label{transfct}
\sa_{(i)}= g_{ij}^{-1} (\sa_{(j)} + \d ) g_{ij}
\quad \Rightarrow\quad
\cF_{(i)}= g_{ij}^{-1} \cF_{(j)} g_{ij} \qquad {\rm on}\ U_i\cap U_j \   .
\ee
These transition rules define the gauge bundle. The gauge group-valued $g_{ij}$ are called the {\it transition functions}. They encode the topological information contained in the gauge bundle.

\vskip5.mm
\begin{figure}[h]
\centering
\includegraphics[width=0.25\textwidth]{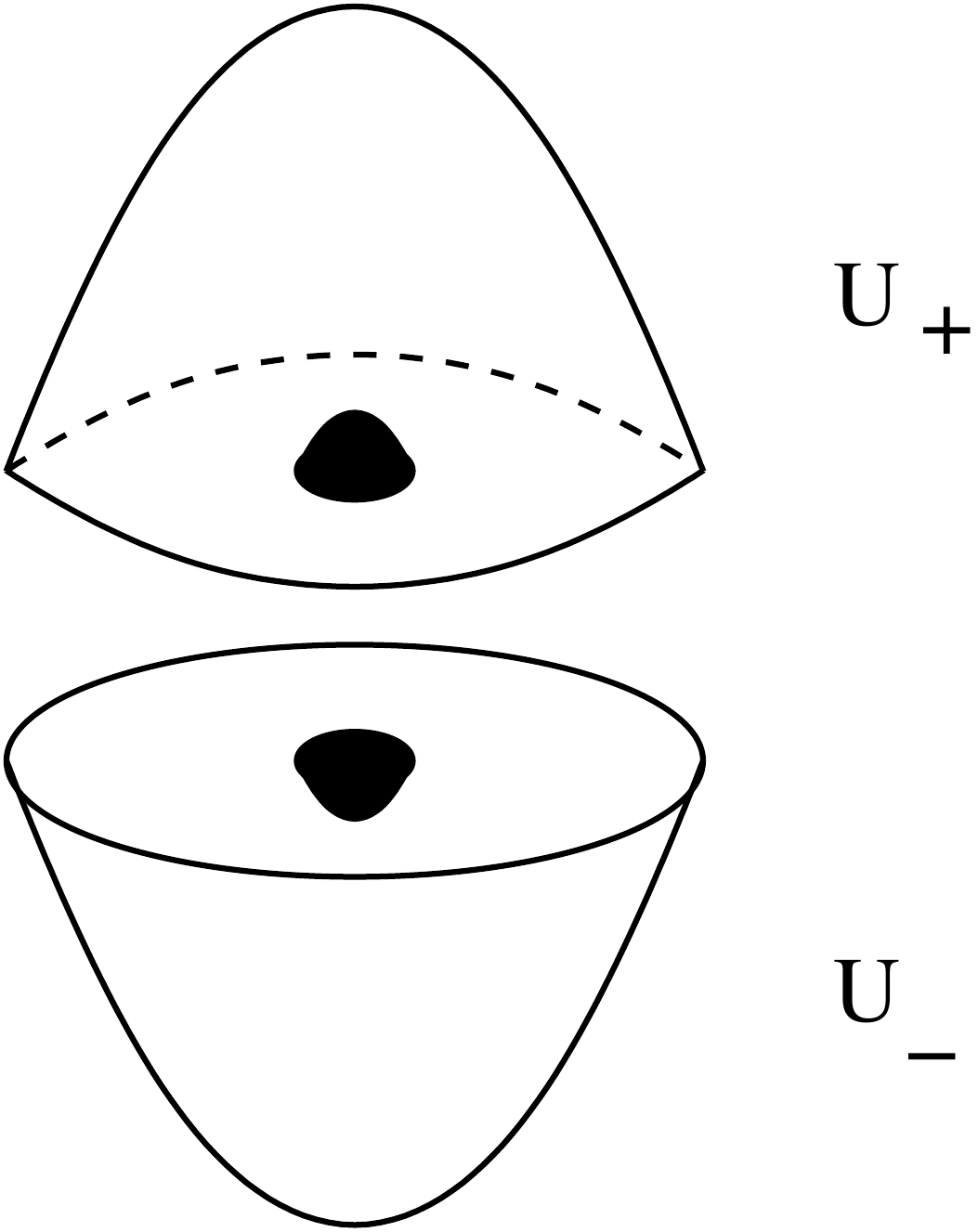}\\
\caption[]{A sketch of the two patches $U_+$ and $U_-$ on ${\bf R}^3$ with a little ball around the origin deleted.} \label{monop}
\end{figure}

As an example, consider again a $U(1)$ gauge theory and restrict to static configurations so that the problem becomes 3-dimensional. Let the manifold be $\cM={\bf R}^3 \backslash\, \{r\le r_0\}$, i.e. ordinary space with a (little) ball of radius $r_0$ around the origin excised.
One introduces two coordinate patches $U_\pm$ as sketched in Fig.~\ref{monop}, corresponding to the upper half and lower half space:
\ba\label{UpUm}
U_+&=&\{ (r,\theta,\vf),\ r\ge r_0,\ 0\le \theta\le {\pi\over 2} + \dd\}
\nonumber\\
U_-&=&\{ (r,\theta,\vf),\ r\ge r_0,\ {\pi\over 2} - \dd\le \theta\le\pi \} \ .
\ea
Their overlap is
\be
U_+\cap U_- = \{ (r,\theta,\vf),\ r\ge r_0,\ {\pi\over 2} - \dd\le \theta\le {\pi\over 2} + \dd \} \ ,
\ee
which is not simply connected. On this manifold $\cM$ one can construct the following $U(1)$ gauge bundle, called the {\it monopole bundle} for reasons that will be clear soon. On $U_\pm$ one defines\footnote{
For $U(1)$ theories it is more convenient to continue to use the real gauge fields $A$ and $F$ rather than the imaginary $\sa$ and $\cF$.
}
\ba
{\rm on}\ U_+ \ : \quad A_+&=&\g (1-\cos\theta)\d\vf
\quad\ \ \Rightarrow\quad F_+=\g \sin\theta \d\theta\d \vf \ ,
\nonumber\\
{\rm on}\ U_- \ : \quad A_-&=&\g (-1-\cos\theta)\d\vf
\quad \Rightarrow\quad F_-=\g \sin\theta \d\theta\d \vf \ .
\ea
Note that $\d\vf$ is well-defined everywhere except on the $z$-axis, i.e. $\theta=0$ or $\theta=\pi$. Now, $U_+$ contains the half-line $\theta=0$, but $1-\cos\theta$ vanishes there. Similarly, $U_-$ contains the half-line $\theta=\pi$, but here $-1-\cos\theta$ vanishes. This also shows why we need to use two diferent $A_+$ and $A_-$, since extending e.g. $A_+$ to all of $\cM$ would result in a string-like singularity along the half-line $\theta=\pi$. (This is precisely the Dirac string singularity which appears in the older treatments of the magnetic monopole configuration.) Thus we have two perfectly well-defined, non-singular gauge connections $A_\pm$ yielding the same field strength $F=F_+=F_-$ in the overlap $U_+\cap U_-$. However, me must still make sure that on the overlap $A_+$ and $A_-$ are related by a gauge transformation. For $U(1)$ this requires
\be
 i\, g_{+-}^{-1}\, \d g_{+-}=A_+-A_-= 2 \g \d\vf \quad {\rm on}\ U_+\cap U_-\ ,
\ee
with solution
\be
g_{+-}=\exp\left(-2i\g \vf\right) \quad {\rm on}\ U_+\cap U_-\ .
\ee
This is a well-defined, single-valued function on  $U_+\cap U_-$ only if 
\be
\g={k\over 2} \quad , \quad k\in {\bf Z} \ .
\ee
Of course, this is due to the first homotopy group of $U_+\cap U_-$ being ${\bf Z}$.
To see the physical meaning of this solution one computes the flux of the magnetic field through any two-sphere $S^2$ in $\cM$ centered at $r=0$. Using our differential forms, this flux is given by $\int_{S^2} F$. Comparing Figures \ref{patches} and \ref{monop} we see that one can separate  $S^2$ into an upper a half-sphere $S_+$ contained in $U_+$ and a lower half sphere $S_-$ contained in $U_-$. (With respect to Fig. \ref{patches} we now take the overlap of $S_+$ and $S_-$ to be just the equator circle.) Then we compute
\be
\int_{S^2} F\equiv \int_{S_+} F_+ + \int_{S_-} F_- = {k\over 2} \int_{S^2} \sin\theta \d\theta\d\vf = {k\over 2}\, 4\pi =2\pi k \ .
\ee
By definition, this flux equals the magnetic charge contained inside the sphere. Since $F_\pm=\d A_\pm$ everywhere in $U_\pm$ this magnetic charge cannot be located in $\cM$ and, of course, it is interpreted as a magnetic monopole at the origin $r=0$. The remarkable fact is that from purely topological reasoning one finds that this magnetic charge is quantized since $k\in{\bf Z}$. It is useful to compute $\int_{S^2} F$ again in a different way, using Stoke's theorem and the fact that the boundaries of $S_+$ and $S_-$ both are the same circle, but with an opposite orientation. We have
\ba
\int_{S^2} F
&=& \int_{S_+} F_+ \ \ + \int_{S_-} F_- 
= \int_{S_+} \d A_+ \ \ + \int_{S_-} \d A_-
= \int_{\del S_+} A_+ \ \ + \int_{\del S_-}  A_-
\nonumber\\
&=&\int_{S^1} (A_+ - A_-) = \int_{S^1} k \d\vf = 2\pi k \ .
\ea
This way of proceeding has the advantage to show that the value of this integral only depends on $A_+-A_-$ on the overlap, which is entirely given in terms of the transition function of the gauge bundle. In particular, if the transition function is trivial, so that $A_+=A_-\equiv A$ can be globally defined, we have $F=\d A$ globally. Obviously then $F$ is exact and $\int_{S^2} F=0$.

Another interesting well-known example is related to {\it instantons}
in (Euclidean) ${\bf R}^4$: Of course, ${\bf R}^4$ is not compact, and in order to get configurations with finite action one requires that far away from the origin  the gauge connection $A$ asymptotes to a pure gauge. This can be reformulated by using two patches, $U_<$ containing all points within a certain large radius, and $U_>$ containing all points outside this radius. In $U_<$ we then keep $A_<=A$ as our gauge connection. However in $U_>$ we use a gauge transformed $A_>$ which is such that it vanishes everywhere at infinity. This is possible precisely because the original $A$ was pure gauge at infinity. Since $A_>$ vanishes everywhere at infinity, we can effectively replace the non-compact $U_>$ by a compact $\wt U_>$. As a result, we now have a compact manifold, which topologically is $S^4$. The price to pay is to have two different gauge-connections $A_>$ and $A_<$ that are related by a gauge transformation on the overlap. The overlap is topologically an $S^3$ and these gauge transformations thus are maps from $S^3$ into the gauge group. Such maps are classified by $\pi^3(G)$ which equals ${\bf Z}$ for any simple $G$. This shows that instantons (or more precisely the instanton bundles) are classified by an integer.

\subsection{Characteristic classes, Chern-Simons forms and descent equations\label{charclasses}}

\subsubsection{Characteristic classes\label{charcl}}

We have just seen some simple examples of characteristic classes. More generally we have:\hfill\break 
$\bullet$ A characteristic class $P$ is a local form on the compact manifold\footnote{
We will assume that the manifold (or the relevant submanifold) is compact, or else that the behaviour of the fields ``at infinity" is such that we can effectively treat the manifold as compact, as we did in our above discussion of instantons.
} 
$\cM$ that is constructed from the curvature or field strength $\cF$ and is such that its integral over the manifold (or a submanifold) is sensitive to non-trivial topology, i.e. to non-trivial transition functions only. 

\begin{figure}[h]
\centering
\includegraphics[width=0.25\textwidth]{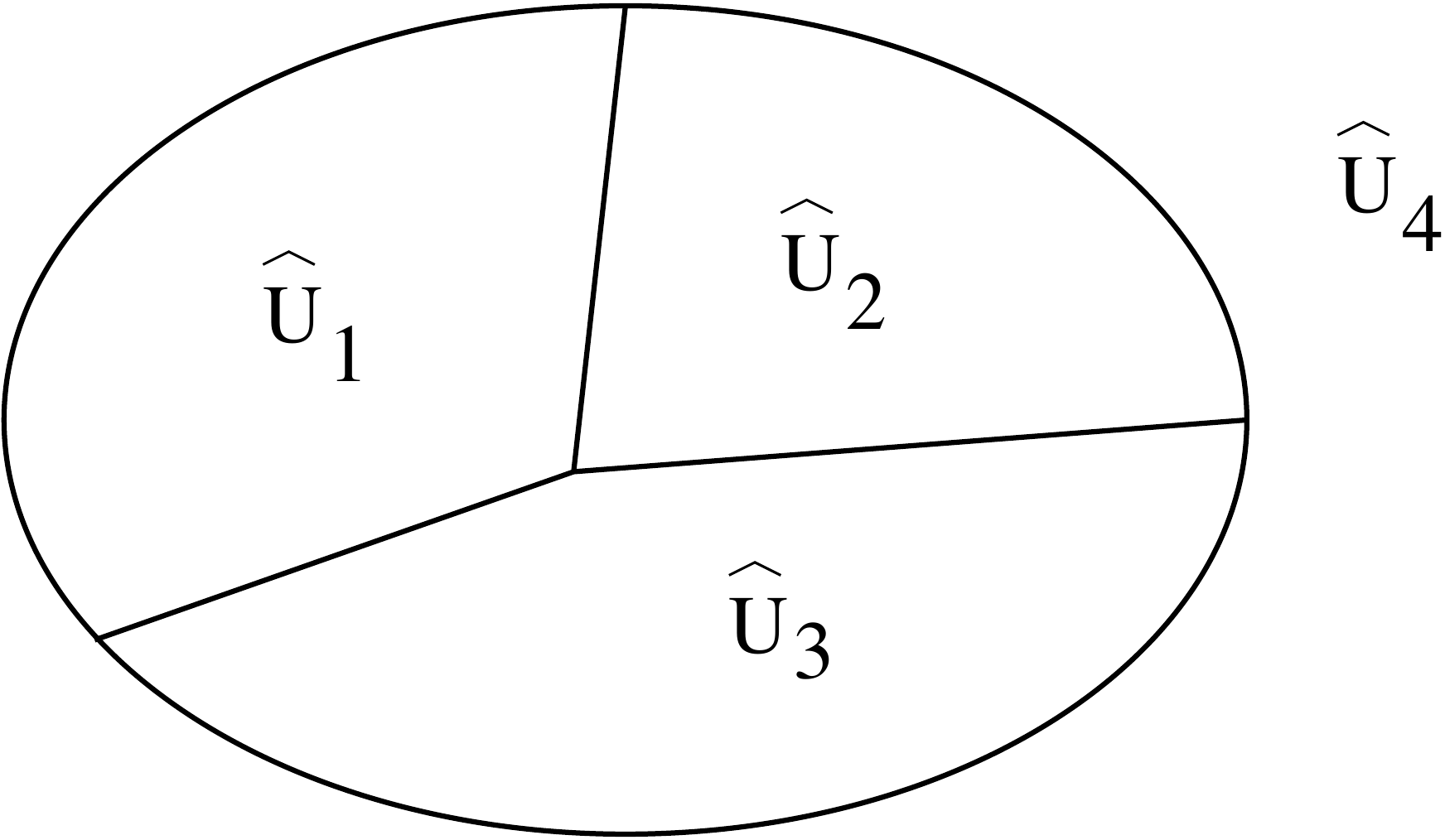}\\
\caption[]{The overlaps of the $d$-dimensional $\wh U_i$ are $(d-1)$-dimensional.} \label{patchpartition}
\end{figure}

\vskip2.mm
\noindent
The latter property is due to the fact that $P$ is closed but not exact. As discussed above, a closed form is locally exact, but need not be globally exact. Indeed, on every topologically trivial patch $U_i$, $\d P_{(i)}=0$ implies the existence of a $Q_{(i)}$ such that $P_{(i)}=\d Q_{(i)}$. Each $Q_{(i)}$ is well defined on its $U_i$, but there is no guarantee that the different $Q_{(i)}$ can be patched together to yield a globally defined $Q$. We have seen this very explicitly for the monopole bundle with $P=F$ and $Q_\pm=A_\pm$. Just as in this example, consider computing the integral $\int_\cM P$. We may reduce the $U_i$ to $\wh U_i\subset U_i$ such that $\cM=\cup_i \wh U_i$ and the $(d-1)$-dimensional boundary of $\wh U_i$ is just the sum of those  intersections $\wh U_i \cap \wh U_j$ that are non-empty, cf. Fig.~\ref{patchpartition}:
\be
\del \wh U_i = \sum_{j\ne i} \wh U_i \cap \wh U_j \ ,
\ee

\vskip5.mm

\noindent
It follows that the integral of $P$ over the manifold $\cM$ is given by
\be\label{StokesQi}
\int_\cM P = \sum_i \int_{\wh U_i} P_{(i)} =
\sum_i \int_{\wh U_i} \d Q_{(i)}
= \sum_i \int_{\del\wh U_i} Q_{(i)}
= \sum_{i<j} (\pm) \int_{\wh U_i \cap \wh U_j} (Q_{(i)}-Q_{(j)}) \ ,
\ee
where the last equality arises because each overlap $\wh U_i \cap \wh U_j$ arises twice in the sum, as the boundary of $\wh U_i$ with $\wh U_j$, and as the boundary of $\wh U_j$ with $\wh U_i$. The $\pm$ depend on the precise conventions adopted for the orientations of the
$\wh U_i \cap \wh U_j$. In any case, eq.~(\ref{StokesQi}) shows that the integral of the closed form $P$ over the manifold $\cM$ only depends on the transition functions between the $Q_{(i)}$ and the $Q_{(j)}$ on the overlaps $\wh U_i \cap \wh U_j$.

We are interested in the case where $P$ is a (gauge) invariant polynomial  of $\cF$, i.e. $P(g^{-1} \cF g)=P(\cF)$ for any $g\in G$. In practice, we will consider
\be
P_m(\cF)=\tr \cF^m \equiv \tr \underbrace{\cF \wedge \ldots \wedge \cF}_{m\ {\rm times}}\ .
\ee
In fact, {\it any} invariant polynomial can be constructed from sums of products of these $P_m$'s. Then
\begin{itemize}
\item
$P_m$ is closed.
\item
Integrals of $P_m$ are topologically invariant, i.e. they are invariant under deformations of the $\sa$ that preserve the transition functions and they depend only on the latter. 
\end{itemize} 
Let us first show that $P_m$ is closed. Recall the Bianchi identity (\ref{Bianchi}) which states $\d\cF=\cF\sa-\sa\cF$. Furthermore, cyclicity of the trace implies $\tr\xi^{(p)}\zeta^{(q)} = (-)^{pq} \tr  \zeta^{(q)}\xi^{(p)}$ for any matrix-valued $p$ and $q$ forms $\xi^{(p)}$ and $\zeta^{(q)}$. It follows that
\be
\d P_m= \d \tr \cF^m = m \tr (\d\cF) \cF^{m-1} 
=m \tr (\cF\sa-\sa\cF)\cF^{m-1}=0 \ .
\ee
To show that the integral of $P_m$ is invariant under deformations of the $\sa$ that preserve the transition functions, consider two gauge connections $\sa_1$ and $\sa_0$ with the {\it same} transition functions. Let $\cF_1$ and $\cF_0$ be the corresponding field strengths. We want to show that 
\be\label{PF1PF2}
P_m(\cF_1)-P_m(\cF_0)= \d R\ ,
\ee 
with some globally defined $(2m-1)$-form $R$. Since the manifold is assumed to be compact it then follows from Stokes's theorem that
\be\label{Pminv}
\int_\cM P_m(\cF_1) -\int_\cM P_m(\cF_2) =\int_\cM \d R = 0 \ .
\ee 
To show (\ref{PF1PF2}) we let 
\be
\sa_t=\sa_0+t(\sa_1-\sa_0) \quad , \quad \cF_t=\d\sa_t+\sa_t^2
\quad , \quad t\in [0,1] \ .
\ee
Then
\ba
{\del\over\del t} \cF_t&=& \d (\sa_1-\sa_0) + (\sa_1-\sa_0)\sa_0+\sa_0(\sa_1-\sa_0) + 2 t (\sa_1-\sa_0)^2
\nonumber\\
&=& \d (\sa_1-\sa_0) + (\sa_1-\sa_0)\sa_t + \sa_t (\sa_1-\sa_0)
\nonumber\\
&=& {\rm D}_t (\sa_1-\sa_0) \ ,
\ea
where ${\rm D}_t$ stands for the covariant exterior derivative that involves the gauge connection $\sa_t$. Note in particular the Bianchi identity for $\cF_t$ is $D_t\cF_t=0$. It follows that
\be
{\del\over\del t} P_m(\cF_t) = m\ \tr {\del\cF_t\over\del t} \cF_t^{m-1}
= m\ \tr \left( {\rm D}_t (\sa_1-\sa_0)\right) \cF_t^{m-1} 
= m\ {\rm D}_t \tr (\sa_1-\sa_0) \cF_t^{m-1} \ .
\ee
Under a gauge transformation, both $\sa_1$ and $\sa_0$ transform inhomogeneously, but their difference transforms covariantly, just as does the field strength $\cF_t$. Then $ \tr (\sa_1-\sa_0) \cF_t^{m-1}$ is invariant under gauge transformations, and the covariant derivative of an invariant quantity is just the ordinary derivative. Thus we arrive at
\be\label{Pmderiv}
{\del\over\del t} P_m(\cF_t) = m\ \d \tr (\sa_1-\sa_0) \cF_t^{m-1} \ .
\ee
Now the important point is that $\sa_1$ and $\sa_0$ have the same transition functions on overlapping patches, cf.~(\ref{transfct}), so that the inhomogeneous terms drop out and $\sa_1^{(i)}-\sa_0^{(i)} = g_{ij}^{-1} (\sa_1^{(j)}-\sa_0^{(j)}) g_{ij}$, just as $\cF_t^{(i)} = g_{ij}^{-1} \cF_t^{(j)} g_{ij}$. It follows that $\tr (\sa_1-\sa_0)\cF_t^{m-1}$ has transition functions equal to 1 and is globally defined. Integrating (\ref{Pmderiv}) with respect to $t$ from 0 to 1 we get (\ref{PF1PF2}) with a globally defined\footnote{
We write $\int d t(\ldots)$ rather than $\int \d t (\ldots)$ to emphasize that $d t$ is not a 1-form.}
\be
R=m \int_0^1 d t \tr (\sa_1-\sa_0)\cF_t^{m-1}\ ,
\ee
and we conclude that (\ref{Pminv}) indeed  holds, i.e.
\be\label{PF1equalsPF2}
\int_{\cM} P_m(\cF_1)=\int_{\cM} P_m(\cF_0) \ ,
\ee 
and the $P_m(\cF)$ are characteristic classes.
 
\subsubsection{Chern-Simons forms\label{CSf}}
 
Since the $P_m(\cF)$ are closed $2m$-forms they must be {\it locally} exact, i.e.
\be\label{Pmlocalexact}
P_m(\cF)= \d Q_{2m-1}(\sa_{(i)},\cF_{(i)}) \qquad \text{{\it locally}\ on each}\ U_i \ .
\ee
The previous computation precisely shows how to obtain suitable  
$(2m-1)$-forms $Q_{2m-1}(\sa_{(i)},\cF_{(i)})$.  Indeed, we have shown that
\be
P_m(\cF_1)=P_m(\cF_0) 
+ \d \Big( m \int_0^1 d t \tr (\sa_1-\sa_0)\cF_t^{m-1}\Big)
\ee
for every $\sa_1$ with the same transition functions as $\sa_0$. Locally, within each patch $U_i$, there is no issue of transition functions and we can simply take $\sa_0=0$ and $\sa_1=\sa$ (which is meant to be the $\sa$ on the patch $U_i$, i.e. $\sa_{(i)}$) so that $\cF_0=0$ and $\cF_1=\cF$, as well as $\sa_t=t\sa$ and $\cF_t=t \d\sa + t^2 \sa^2$. We then get  (\ref{Pmlocalexact}) with 
\ba\label{CSforms}
Q_{2m-1}(\sa, \cF)&=& m \int_0^1 d t\, \tr \sa\,\cF_t^{m-1}
= m \int_0^1 d t\, \tr \sa\,\big( t \d \sa + t^2\sa^2\big)^{m-1}
\nonumber\\
&=& m \int_0^1 d t\, t^{m-1} \tr  \sa\,\Big( \cF +(t-1) \sa^2\Big)^{m-1}\ .
\ea
$Q_{2m-1}(\sa, \cF)$ is called the Chern-Simon $(2m-1)$-form. Of course, the defining relations (\ref{Pmlocalexact}) determine the $Q_{2m-1}$ only up to adding some exact form. It is nevertheless customary to call Chern-Simons forms the expressions given by (\ref{CSforms}). Let us explicitly compute the Chern-Simons 3- and 5-forms and check that their exterior derivatives indeed yield the characteristic classes $P_2$ and $P_3$. We have
\ba
Q_3&=&2\int_0^1 d t \, \tr \sa\, \big( t\d\sa + t^2\sa^2\big)
= \tr \big(\sa\d\sa+{2\over 3}\sa^3\big)= \tr\big( \sa\cF-{1\over 3} \sa^3\big)
\\
Q_5&=&3\int_0^1 d t \, \tr \sa\, \big( t\d\sa + t^2\sa^2\big)^2
= \tr \big(\sa\d\sa\d\sa+{3\over 2}\sa^3\d\sa+{3\over 5} \sa^5\big)
\nonumber\\
&&\hskip5.cm
=\tr\big( \sa\cF^2-{1\over 2} \sa^3\cF+{1\over 10} \sa^5\big)\ .
\ea
Let us first check that $\d Q_3=P_2$. First note that 
\be
\tr \sa^{2k}=0 \ ,
\ee
since, using the anticommutation of odd forms, as well as the cyclicity of the trace, $\tr \sa \sa^{2k-1}=-\tr \sa^{2k-1}\sa$.
Similarly, $\tr\sa\d\sa\sa=-\tr\d\sa\sa^2$, etc. It follows that
\be
\d Q_3=\tr\Big( \d\sa\d\sa+{2\over 3} \d\sa\sa^2 
-{2\over 3}\sa\d\sa\sa+{2\over 3}\sa^2\d\sa\Big)
=\tr\Big( \d\sa\d\sa+2\d\sa\sa^2\Big)=\tr \cF^2\ .
\ee
Next, we check that $\d Q_5=P_3$ (here we need to use $\tr\sa\d\sa\sa\d\sa=0$):
\ba
\d Q_5
&=& \tr\Big( \d\sa\d\sa\d\sa +{3\over 2} \sa^2\d\sa\d\sa-{3\over 2}\sa\d\sa\sa\d\sa+{3\over 2} \d\sa\sa^2\d\sa +5\times {3\over 5}\sa^4\d\sa\Big)
\nonumber\\
&=&\tr\Big( (\d\sa)^3 + 3\sa^2(\d\sa)^2 + 3 \sa^4\d\sa\Big) = \tr\cF^3 \ .
\ea
Finally, note from (\ref{CSforms}) that our normalization of the Chern-Simons forms always is such that
\be\label{CSnorm}
Q_{2m-1}(\sa,\cF)=\tr \sa \cF^{m-1} +\ldots = \tr \sa (\d\sa)^{m-1} +\ldots \ ,
\ee
where $+\ldots$ stands for terms with less factors of $\cF$ or less derivatives.
 
\subsubsection{Descent equations\label{desceq}}

Contrary to the characteristic classes $P_m$, the Chern-Simons forms $Q_{2m-1}$ are not gauge invariant. However, one can use the invariance of the $P_m$ to characterize the gauge variation $\dd Q_{2m-1}$ of the Chern-Simons form $Q_{2m-1}$ as follows:
\be\label{deltaQclosed}
\d\, \dd\, Q_{2m-1} = \dd\, \d\, Q_{2m-1} = \dd\, P_m(\cF)=0 \ ,
\ee
since  taking the gauge variation ($\dd$) obviously commutes with taking the exterior derivative ($\d$). We see that (on each patch $U_i$ where it is well-defined) the gauge variation of $Q_{2m-1}$ is a closed $(2m-1)$-form. Hence, it is locally exact, and on each patch $U_i$ we have
\be\label{descinf}
\dd\, Q_{2m-1}(\sa_{(i)},\cF_{(i)})= \d\, Q_{2m-2}^1(v,\sa_{(i)},\cF_{(i)}) \qquad \text{on each} \ U_i \ .
\ee
Note that this applies to {\it infinitesimal} gauge transformations with parameter $v$. For {\it finite} gauge transformations with some $g(x)\in G$ one has a more complicated relation. Indeed, if we let (on $U_i$) $\sa^g_{(i)} = g^{-1} (\sa_{(i)} + \d ) g$, one can show that
\be\label{descfin}
Q_{2m-1}(\sa^g_{(i)},\cF^g_{(i)}) - Q_{2m-1}(\sa_{(i)},\cF_{(i)})
=Q_{2m-1}(g^{-1}\d g,0)+ \d (\ldots) \ ,
\ee
where e.g. $\int_{S^{2m-1}} Q_{2m-1}(g^{-1}\d g,0) \sim \int_{S^{2m-1}}  \tr \big( g^{-1}\d g\big)^{2m-1}$ computes the number of times $g: S^{2m-1}\to G$ maps the basic sphere $S^{2m-1}$ to a (topological) sphere in $G$, i.e. it computes the homotopy class of $g$ in $\Pi_{2m-1}(G)$. In particular, for every compact, connected simple Lie group $G$ one has $\Pi_3(G)={\bf Z}$, and also $\Pi_{2m-1}(SU(n))={\bf Z}$ for all $n\ge m\ge 2$. For an infinitesimal transformation, the homotopy class of $g$ is trivial and $Q_{2m-1}(g^{-1}\d g,0)$ is exact, so that (\ref{descfin}) is compatible with (\ref{descinf}).

Let us again work out the examples of $m=2$ and $m=3$. Recall that $\dd\sa=\d v+[\sa,v]$ and $\dd\cF=[\cF,v]$. It follows that $\dd \cF^k=[\cF^k,v]$ and
\be
\dd  \sa^l 
= \d v \sa^{l-1} + \sa \d v \sa^{l-2} + \ldots \sa^{l-1}\d v
+[\sa^l,v] \ ,
\ee
so that
\be\label{AlFk}
\dd \tr \sa^l \cF^k=\tr \Big(\d v \sa^{l-1} + \sa \d v \sa^{l-2} + \ldots \sa^{l-1}\d v\Big)\cF^k \ ,
\ee
since $\tr [\sa^l\cF^k,v]=0$. Explicitly, we find
\be
\dd Q_3= \dd \tr \Big(\sa\cF-{1\over 3} \sa^3\Big) 
= \tr \big( \d v \cF- \d v \sa^2\big) = \tr \d v \d \sa
\ ,
\ee
(note that the terms $\sim\tr\d v \sa^2$ have cancelled !) and we conclude that $\dd Q_3= \d Q_2^1$ with
\be\label{Q21}
Q_2^1=\tr v\, \d\sa \ ,
\ee
modulo some exact form. We could e.g. add a term  $\sim\d \tr v\sa$ to $Q_2^1$, resulting in the presence of a term involving $\d v$. It is customary to fix this ambiguity in such a way that $Q_{2m-1}^1$ only involves $v$ and not $\d v$.  Similarly, for $Q_4^1$ we get
\be
\dd Q_5=\dd \tr\Big( \sa\cF^2-{1\over 2} \sa^3\cF+{1\over 10} \sa^5\Big)
=\tr\Big(\d v \cF^2 -{1\over 2} \d v\,(\sa^2\cF+\sa\cF\sa+\cF\sa^2) +{1\over 2} \d v \sa^4 \Big).
\ee
Again, the terms $\sim \tr \d v \sa^4$ cancel as they indeed must in order that we can write $\dd Q_5$ as $\d(\ldots)$. The remaining terms are
\be
\dd Q_5=\tr \d v \Big( \d\sa\d\sa +{1\over 2} \big( \d\sa \sa^2 - \sa\d\sa \sa + \sa^2 \d \sa \big) \Big)
=\tr \d v \Big( \d\sa\d\sa +{1\over 2} \d \sa^3 \Big) \ ,
\ee
and we conclude
\be\label{Q41}
Q_4^1
=\tr v\, \d \big( \sa \d\sa +{1\over 2} \sa^3\big) \ ,
\ee
(again up to an exact term).

As is clear from (\ref{AlFk}), when computing $\dd Q_{2m-1}$ all commutator terms cancel and $\dd Q_{2m-1}$ is linear in $\d v$, so that we may write
\be\label{CSvarbis}
\dd Q_{2m-1}(\sa,\cF)= Q_{2m-1}(\sa+\d v,\cF)-Q_{2m-1}(\sa,\cF) \ .
\ee
Hence $\d Q_{2m-2}^1$ must be linear in $\d v$ (and not contain $v$ without derivative) and thus $Q_{2m-2}^1$ must be of the general form $Q_{2m-2}^1=\tr v \d \big(\ldots\big)$ (up to exact terms), which we observed indeed for $Q_2^1$ and $Q_4^1$.
One can actually prove the general formula
\be
Q_{2m-2}^1(v,\sa,\cF)= m (m-1) \int_0^1 \d t (1-t) \tr v\, \d \big( \sa \cF_t^{m-2}\big) \ .
\ee
Clearly, for $m=2$ and $m=3$ this reproduces eqs (\ref{Q21}) and (\ref{Q41}). Note also that the normalization is always such that
\be\label{Q1norm}
Q_{2m-2}^1(v,\sa,\cF)=\tr v\, (\d\sa)^{m-1}+\ldots = \tr v\, \cF^{m-1} +\ldots \ ,
\ee
where $+\ldots$ stands for terms with less factors of $\cF$ or with less derivatives. We will sometimes use instead
\be\label{Q1AQ1sa}
Q_{2m-2}^1(\e,A,F)= m (m-1) \int_0^1 \d t (1-t) \tr \e\, \d \big( A F_t^{m-2}\big) = i^m Q_{2m-2}^1(v,\sa,\cF)\ ,
\ee
which is related to $\tr F^m=i^m \tr \cF^m=i^m P_m(\cF)$ by descent. Of course, its normalization is such that $Q_{2m-2}^1(\e,A,F)=\tr \e\, (\d A)^{m-1}+\ldots = \tr \e\, F^{m-1}+\ldots$.

We will derive one more relation for $Q_{2m-2}^1(v,\sa,\cF)$ that will be useful later-on. As we have just seen, we can write $Q_{2m-2}^1(v,\sa,\cF)=\tr v\, \d q_{2m-2}(\sa,\cF)$ with some $q_{2m-2}(\sa,\cF)$. Then eq.~(\ref{CSvarbis}) yields $Q_{2m-1}(\sa+\d v,\cF)-Q_{2m-1}(\sa,\cF)=\d Q_{2m-2}^1(v,\sa,\cF)=\tr \d v\, \d q_{2m-2}(\sa,\cF)$. Both sides of this equation are linear in $\d v\equiv \wh v$ and one concludes $Q_{2m-1}(\sa+\wh v,\cF)-Q_{2m-1}(\sa,\cF)
=\tr \wh v\, \d q_{2m-2}(\sa,\cF)$. But we have just seen that this is $Q_{2m-2}^1(\wh v,\sa,\cF)$. Hence
\be\label{CSvar3}
Q_{2m-1}(\sa+\wh v,\cF)-Q_{2m-1}(\sa,\cF)
= Q_{2m-2}^1(\wh v,\sa,\cF) \ .
\ee
This is an algebraic identity which holds whether $\wh v=\d v$ or not.

We already noted that the $Q_{2m-2}^1$ are only defined modulo an exact form. They are actually also defined modulo addition of the gauge variation of some $(2m-2)$-form:  recall that the $Q_{2m-1}$ were only defined up to $Q_{2m-1}\to Q_{2m-1}+ \d \a_{2m-2}$. Then
$\dd Q_{2m-1}\to \dd Q_{2m-1} + \dd \d \a_{2m-2}=\dd Q_{2m-1} + \d \dd \a_{2m-2}$, so that $Q_{2m-2}^1\to Q_{2m-2}^1 + \dd \a_{2m-2}$. Thus altogether the ambiguity in determining $Q_{2m-2}^1$ is
$Q_{2m-2}^1 \simeq Q_{2m-2}^1 + \dd \a_{2m-2} + \d \b_{2m-3}$.
Let us summarize:
\be\label{descentsummary}
\begin{array}{|c|}
\hline\\
\text{characteristic classes :}\qquad
\d \,P_m = \dd\, P_m = 0 \ , \hskip4.cm\\
\\
\text{descent equations :}\quad
P_m = \d\, Q_{2m-1} \quad , \quad \dd\, Q_{2m-1} = \d \, Q_{2m-2}^1
\qquad \text{locally on each}\ U_i\\
\\
Q_{2m-2}^1 \ \simeq\  Q_{2m-2}^1 + \dd \a_{2m-2} + \d \b_{2m-3} \ .\\
\\
\hline
\end{array}
\ee
 
We will show below that the non-abelian gauge anomaly in $d=2k$ dimensions is given by $c\ Q_{2k}^1$ (i.e. $m=k+1$) where $c$ is some appropriate normalization constant. More precisely, we will show that $\dd\, \G = c\int Q_{2k}^1$. Clearly, the freedom to add to $Q_{2k}^1$ an exact term $\d \b_{2k-1}$ does not change $\dd\,\G$. On the other hand, we have also seen that we are allowed to add to $\G$ a {\it local}, possibly non gauge invariant counterterm $c\int \a_{2k}$. This amounts to adding $\dd\a_{2k}$ to $Q_{2k}^1$. Thus the ambiguity in the definition of $Q_{2k}^1\equiv Q_{2m-2}^1$ has an exact counterpart in the definition of the non-abelian gauge anomaly. On the other hand, the characteristic class or invariant polynomial $P_m$ is defined without ambiguity: different equivalent $Q_{2m-2}^1$ correspond to the same $P_m$. This will imply that different equivalent forms of the anomaly in $2k$ dimensions can be characterized invariantly by one and the same $P_{k+1}$ which is a $(d+2)$-form in $d+2$ dimensions.

\section{Wess-Zumino consistency condition, BRST cohomology and descent equations\label{secWZBRST}}
\setcounter{equation}{0}

In this section, we will study and constrain the form of any anomaly under infinitesimal (non-abelian) gauge transformations in arbitrary even dimensions $d=2r$. (Recall that in odd dimensions there are no chiral fermions that could lead to an anomaly under infinitesimal gauge transformations.) This will lead to the Wess-Zumino consistency conditions which are most simply expressed using a BRST formalism. The solutions to these consistency conditions naturally are given in terms of the $Q_{2r}^1(v,\sa,\cF)$ studied in the previous section and which are related via the descent equations to the characteristic classes $\tr \cF^{r+1}$. In particular, this will fix all terms of higher order in $\sa$, once the coefficient of the leading term $\tr v (\d\sa)^r$ is known.

To establish these results, we will consider certain local functionals of the gauge and possibly ghost fields. In this section, in order to be able to freely integrate by parts, we will assume that either (i) the $2r$-dimensional space-time manifold $\cM$ is compact and the gauge fields are globally defined (i.e. have trivial transition functions $g_{ij}$ between different patches $U_i$ and $U_j$) or (ii) the gauge fields have non-trivial transition functions $g_{ij}$ between different patches but the parameters $\e^\a(x)$, resp. the ghost fields $\o^\a(x)$ are non-vanishing only on a single patch $U_{i_0}$ or (iii) the space-time manifold is topologically ${\bf R}^{2r}$ with globally defined gauge fields and $\e^\a(x)$, resp $\o^\a(x)$ vanish sufficiently quickly as $|x|\to\infty$.  In any case, we then have
\be\label{exactiszero}
\int_{\cM} \d \tr \o (\ldots)=0 \ ,
\ee 
with $(\ldots)$ any $(2r-1)$ form made from the gauge fields and their derivatives.

\subsection{Wess-Zumino consistency condition\label{WZconscond}} 

Recall that we defined the anomaly $\ca_\a$ as the gauge variation of the effective action (for the gauge fields, obtained after doing the functional integral over the matter fields):
\be\label{andef33}
\dd_\e \G[A] = \int \e^\a(x) \ca_\a(x)
\quad \Leftrightarrow \quad
\ca_\a(x) = - \left( D_\m {\dd\over \dd A_\m(x)}\right)_\a \G[A] \ .
\ee
The fact that the anomaly $\ca_\a$ is a certain derivative of some functional constrains its form (just like any gradient of a scalar field is constraint to have vanishing curl, i.e. $\xi=\d f$ is constraint by $\d \xi=0$ or in components $\del_\m \xi_\n - \del_\n \xi_\m=0$.) This constraint for the anomaly is known as the Wess-Zumino consistency condition. Historically, different definitions of the anomaly
were used (cf our remark at the end of sect.~\ref{cnca}). An anomaly that satisfied the Wess-Zumino consistency condition was called consistent anomaly.  With our definition (\ref{andef33}) of the anomaly, this condition is automatically satisfied.

To derive the consistency condition, define the operator
\be
\cG_\a(x)= - \left( D_\m {\dd\over \dd A_\m(x)}\right)_\a
= - {\del\over \del x^\m} {\dd\over \dd A_\m^\a(x)}
- C_{\a\dd\g} A_\m^\dd(x) {\dd\over \dd A_\m^\g(x)} \ ,
\ee
so that 
\be\label{aaGG}
\ca_\a(x)=\cG_\a(x) \G[A] \ .
\ee
Let us compute the algebra satisfied by the $\cG_\a$. First note that ${\del\over \del x^\m} {\dd\over \dd A_\m^\a(x)}$ and ${\del\over \del y^\n} {\dd\over \dd A_\n^\b(y)}$ obviously commute. One then finds
\ba\label{Galgebra}
[\cG_\a(x), \cG_\b(y)]&=&
\left[ {\del\over \del x^\m} {\dd\over \dd A_\m^\a(x)}
+ C_{\a\dd\g} A_\m^\dd(x) {\dd\over \dd A_\m^\g(x)}\ ,\ 
{\del\over \del y^\n} {\dd\over \dd A_\n^\b(y)}
- C_{\b\e\k} A_\n^\e(x) {\dd\over \dd A_\n^\k(y)}\right]
\nonumber\\
&=& C_{\b\a\k} \left({\del\over \del x^\m} \dd^{(4)}(x-y) \right)
{\dd\over \dd A_\m^\k(y)} 
+C_{\a\dd\g} A_\m^\dd(x) C_{\b\g\k} \dd^{(4)}(x-y) {\dd\over \dd A_\m^\k(y)}
\nonumber\\
&&\hskip-3.mm-
C_{\a\b\g} \left({\del\over \del y^\n} \dd^{(4)}(x-y) \right)
{\dd\over \dd A_\n^\g(x)}
-C_{\b\e\k} A_\n^\e(y) C_{\a\k\g} \dd^{(4)}(x-y) {\dd\over \dd A_\n^\g(x)}\ .
\ea
To safely evaluate the two terms involving space-time derivatives of $
\dd^{(4)}(x-y)$ we multiply them with two test functions:\footnote{Again, if necessary may assume that these test functions have support on a single patch so that we can safely integrate by parts.}
\ba
&&\hskip-1.cm
\int \d^4 x\, \d^4 y\ \vf^\a(x) \p^\b(y) 
\left(C_{\b\a\k} \left({\del\over \del x^\m} \dd^{(4)}(x-y) \right)
{\dd\over \dd A_\m^\k(y)} - C_{\a\b\g} \left({\del\over \del y^\n} \dd^{(4)}(x-y) \right) {\dd\over \dd A_\n^\g(x)}\right)
\nonumber\\
&=&C_{\a\b\g} \int \d^4 x\ 
\left( \del_\m \vf^\a(x) \p^\b(x) {\dd\over \dd A_\m^\g(x)} 
+ \vf^\a(x) \del_\n\p^\b(x) {\dd\over \dd A_\n^\g(x)}\right)
\nonumber\\
&=&-C_{\a\b\g} \int\d^4 x\ \vf^\a(x) \p^\b(x) {\del\over\del x^\m} 
{\dd\over \dd A_\m^\g(x)}
\nonumber\\
&=& \int \d^4 x\, \d^4 y\ \vf^\a(x) \p^\b(x) \left( -C_{\a\b\g} \dd^{(4)}(x-y){\del\over\del x^\m} {\dd\over \dd A_\m^\g(x)} \right)\ .
\ea
The two other terms in (\ref{Galgebra}) are easily combined using the Jacobi identity. The final result then is
\be\label{Galgebra2}
[\cG_\a(x), \cG_\b(y)]
=\dd^{(4)}(x-y) C_{\a\b\g} \left( - {\del\over \del x^\m} {\dd\over \dd A_\m^\g(x)}
- C_{\g\dd\k} A_\m^\dd(x) {\dd\over \dd A_\m^\k(x)} \right)
=\dd^{(4)}(x-y) C_{\a\b\g}  \cG_\g(x)\ ,
\ee
or, if we let $\cT_\a(x)=i\cG_\a(x)$\ :
\be
[\cT_\a(x), \cT_\b(y)]=i C_{\a\b\g} \dd^{(4)}(x-y) \cT_\g(x) \ ,
\ee
which is the local version of the gauge algebra. Now apply the identity (\ref{Galgebra2}) to $\G[A]$ and use (\ref{aaGG}) to get
\be\label{WZcond}
\cG_\a(x) \ca_\b(y) - \cG_\b(y) \ca_\a(x) = C_{\a\b\g} \dd^{(4)}(x-y) \ca_\g(x)  .
\ee
This is the Wess-Zumino consistency condition. It constrains the form of the anomaly. Of course, as already mentioned, the full anomaly defined as the variation of the effective action must satisfy this condition. In practice, however, one only computes part of the anomaly, e.g. in four dimensions the piece bilinear in the gauge fields, and then uses this condition to obtain the missing pieces. Indeed, we will show that the consistency condition is strong enough to completely determine the form of the anomaly, up to an overall coefficient which cannot (and should not) be fixed by the WZ condition which is linear in $\ca$.

\subsection{Reformulation in terms of BRST cohomology\label{reformBRST}}

The anomaly $\ca_\a(x)$ is some local functional of the gauge fields and their derivatives. To emphasize this we may write $\ca_\a(x,A)$. 
Then the gauge variation of the effective action is $\dd_\e \G=\int\d^4 x\, \e^\a(x) \ca_\a(x,A)$. We want to reformulate this as the BRST transformation of the effective action. Recall that on the gauge field $A_\m$ the BRST operator $s$ acts as $s A_\m = D_\m \o$ which is like a gauge transformation but with the ghost $\o^\a(x)$ replacing the gauge parameter $\e^\a(x)$. Similarly for any functional $F[A]$ of the gauge fields only one has
\be
s F[A] = \int\d^4 x\, (D_\m \o(x))^\a {\dd\over \dd A_\m^\a(x)} F
=  \int\d^4 x\ \o^\a(x) \left( -D_\m {\dd\over \dd A_\m(x)}\right)_\a F \equiv \int\d^4 x\ \o^\a(x) \cG_\a(x) \ F\ .
\ee
Taking $F$ to be the effective action $\G$ and comparing with the definition of the anomaly we see that
\be\label{sgamma}
s \G= \int\d^4 x\ \o^\a(x) \ca_\a(x,A)\equiv\ca[\o,A] \ .
\ee
Since the BRST operation $s$ is nilpotent, $s^2=0$, it follows from 
the previous equation that
\be
s \ca[\o,A] = s^2\G =0\ . 
\ee
We now show that the condition $s \ca[\o,A]=0$ is precisely equivalent to the Wess-Zumino consistency condition. To do so, we need the action of $s$ on the ghost field: $s \o^\a=-{1\over 2} C_{\a\b\g} \o^\b \o^\g$ and the fact that $s(\o^a \ca_\a)=(s \o^\a) \ca_\a - \o^\a (s \ca_\a)$. We then get
\ba
s \ca[\o,A]&=& \int \d^4 x\, \Big( (s\o^a(x)) \ca(x,A) - \o^\a(x) \, s \ca_\a(x,A) \Big)
\nonumber\\
&=& \int \d^4 x\, \left(-{1\over 2} C_{\a\b\g} \o^\b(x) \o^\g(x)\ca_\a(x,A)
- \o^\a(x) \int\d^4 y\, \o^\b(y)\cG_\b(y) \ca_\a(x,A)\right)
\nonumber\\
&=&\int \d^4 x\,\d^4 y\, \Big( -{1\over 2} \o^\a(x) \o^\b(y)\Big)
\Big( C_{\a\b\g} \dd^{(4)}(x-y) \ca_\g(x,A) 
\nonumber\\
&&\hskip 5.4cm
+\ \cG_\b(y)\ca_\a(x,A) 
-\cG_\a(x) \ca_\b(y,A) \Big) \ ,
\ea
where we used the anticommutation of $\o^\a(x)\o^\b(y)$ to antisymmetrize  $\cG_\b(y)\ca_\a(x,A)$ with respect to  $(\b,y)\leftrightarrow (\a,x)$.
We see that the vanishing of $s\ca[\o,A]$ for arbitrary ghost fields  is equivalent to the Wess-Zumino condition (\ref{WZcond}):
\be\label{WZBRST}
s \ca[\o,A]=0 \quad \Leftrightarrow\quad \text{Wess-Zumino condition}\ .
\ee
Thus the anomaly $\ca[\o,A]$ is a BRST-closed functional of ghost number one. Suppose that we can find some {\it local} functional $F[A]$ of ghost number zero such that $\ca[\o,A]=s F[A]$. As before, locality means that $F$ should be the integral of sums of products of $A(x)$ and its derivatives at the same point $x$. As discussed above, the anomaly is (at least) order $g^2$ with respect to the classical action, and one could add a ``counterterm" $-F$ to the classical action, $\D S=-F$, resulting at order $g^2$ in a $\D\G=-F$. Then $s(\G+\D\G)=s(\G-F)=\ca-s F=0$ so that addition of such a counterterm would eliminate the anomaly. Recall that a relevant anomaly is one that cannot be eliminated by the addition of a local counterterm. Said differently, we are always free to add such local counterterms, and an anomaly is only defined as an equivalence class with respect to the equivalence relation 
\be
\ca[\o,A] \simeq \ca[\o,A]+ s F[A] \quad \text{with local}\ F[A]\ .
\ee
Then a relevant anomaly is given by a an $\ca[\o,A]$ such that $s \ca[\o,A]=0$ and $\ca[\o,A]\ne s F$, i.e. $\ca[\o,A]\simeq\hskip-4.mm/ \ 0$ : a relevant anomaly is BRST-closed but not BRST-exact. Thus
\vskip2.mm
\be
\begin{array}{|c|}
\hline\\
\text{{\it Relevant anomalies} are given by non-trivial BRST cohomology classes at ghost number one}\\ 
\text{on the space of {\it local} functionals.}\\
\\
\hline
\end{array}
\nonumber
\ee

\newpage
\subsection{Determining the higher order terms of the anomaly in $d=4$\label{dethigherord}}

From our triangle computation in section~\ref{tria} we know that the contribution to the anomaly of a left-handed, i.e. positive chirality fermion in the representation $\cR$  is, cf.~(\ref{anomdiffform})
\be\label{anomwsa}
\ca[\o,A] = -{1\over 24\pi^2}\int \trR \o\,\d A\d A + \cO(A^3)
={i\over 24\pi^2}\int \trR w\,\d \sa\d \sa + \cO(\sa^3) \equiv \ca[w,\sa]\ ,
\ee
where we used (\ref{iredef}), as well as an analogous redefinition for the ghost field:\footnote{
The notation $\ca[w,\sa]$ is somewhat inexact. We clearly do not mean it to be the same functional as $\ca[\o,A]$ with the arguments $\o$ and $A$ replaced by $w$ and $\sa$. What we mean is obvious from (\ref{anomwsa}): $\ca[w,\sa]$ equals $\ca[\o,A]$ but we indicate that we express everthing in terms of $w$ and $\sa$. This is different from the convention adopted for $Q_{2r}^1$ where we really had $Q_{2r}^1(v,\sa,\cF)=Q_{2r}^1(-i\e,-iA,-iF)=(-i)^{r+1} Q_{2r}^1(\e,A,F)$.
}
\be
w=-i\o \ .
\ee
Since the components $\o^\a$ or $w^\a$ of the ghost field anticommute among themselves (as well as with all other fermionic fields), and since the $\d x^\m$  anticommute among themselves, too, it is natural to also make the {\it choice} that the $\d x^\m$ anticommute with the ghost fields:
\be 
\d x^\m\ w=-w\ \d x^\m \ .
\ee 
Then in particular $[A_\m,\o]\d x^\m= -A\o-\o A\equiv - \{ A,\o\}$ or $[\sa,w]\d x^\m=-\{\sa,w\}$. It follows that the BRST transformation of $\sa$ is $s \sa= s \sa_\m \d x^\m = \big(\del_\m w + [\sa_\m,w]\big) \d x^\m =-\d w - \{\sa,w\}$. Of course, this subtlety only affects the BRST transformation of forms of odd degree. In summary:
\be\label{BRSTtransf}
s\sa=-\d w-\{\sa,w\} \quad , \quad s \cF=[\cF,w]
\quad , \quad s w = - w w \ .
\ee
With these conventions it is easy to see that 
\be 
s\ \d = - \d\ s \ .
\ee
It then follows e.g. 
\ba\label{compBRST}
s(\d\sa)&=&-\d (s\sa)=-\d\big(-\d w - \{\sa,w\}\big)= \d\{\sa,w\}=[\d\sa,w]-[\sa,\d w] \ ,
\nonumber\\
s\sa^2&=&(s\sa)\sa - \sa (s\sa)= -\d w\sa+\sa\d w +[\sa^2,w] 
\nonumber\\
s\sa^3&=&(s\sa^2)\sa+\sa^2(s\sa)=-\d w\sa^2+\sa\d w\sa-\sa^2\d w - \{ \sa^3,w\}
\nonumber\\
s\sa^4&=&(s\sa^2)\sa^2+\sa^2(s\sa^2)=-\d w\sa^3+\sa\d w\sa^2-\sa^2\d w\sa + \sa^3\d w + [\sa^4,w] 
\ .
\ea

Let us now show that imposing $s \ca[\o,A]=0$ completely determines all $\cO(A^3)$-terms in the anomaly (\ref{anomwsa}) in terms of the $\tr w\d\sa\d\sa$ term. First recall from our general arguments in section \ref{lfa} that the anomaly must be a local functional of $\sa$ and obviously be linear in $w$. Moreover, the BRST operator $s$ changes the scaling dimensions uniformly by one unit, so that all terms in $s\int\tr w\d \sa\d\sa$ have the same scaling dimension, namely $5+1$. They can only be cancelled by terms of the same dimension arising from some $s\int \tr w(\ldots)$ where $(\ldots)$ then must have dimension 4. This allows terms $\sim\sa^4$ and terms with three $\sa$'s and one derivative. Also since we know that the anomaly must involve the $\e^{\m\n\r\s}$ tensor, we must indeed be able to write these terms as differential 4-forms. Thus the most general ansatz is
\be\label{anomansatz}
\ca[w,\sa]= {i\over 24\pi^2}	\int \tr w \big( \d\sa\d\sa+b_1\,\sa^2\d\sa+b_2\,\sa\d\sa\sa+b_3\, \d\sa\sa^2+c\,\sa^4\big) \ .
\ee
First note that, using the fourth equation (\ref{compBRST}), we get
\ba
s\tr w\sa^4&=&-\tr w^2\sa^4 - \tr w (s\sa^4) 
=-\tr w^2\sa^4 - \tr w [\sa^4,w] + \cO(w,\d w) \nonumber\\
&=&\tr w^2\sa^4  +  \cO(w,\d w)\ ,
\ea
where $\cO(w,\d w)$ stands for all terms involving $w$ and $\d w$.
Observe that in $s\ca[w,\sa]$ there can be no other term $\sim \tr w^2\sa^4$ since all other terms involve at least one $\d\sa$ or a $\d w$. 
We conclude that necessarily $c=0$. 

Next, a straightforward computation yields
\ba\label{AAdAterms}
s\tr w\sa^2\d\sa&=& 
\tr\big( w^2\sa^2\d\sa+\d w w\sa^3+\d w\sa w\sa^2\big)
+\tr\big( w\d w\sa\d\sa-w\sa\d w\d\sa\big)\ ,
\nonumber\\
s\tr w\sa\d\sa\sa&=&
\tr\big( w^2\sa\d\sa \sa+w\sa\d w\sa^2+\d w\sa w\sa^2\big)
+ \tr\big( w\d w\d\sa\sa-\d w w\sa\d\sa\big)\ ,
\nonumber\\
s\tr w \d\sa\sa^2&=&
\tr \big( w^2\d\sa\sa^2-w\d w\sa^3 +w\sa\d w\sa^2\big)
+\tr \big( -\d w w\d\sa\sa+w\d\sa\d w\sa\big)\ .
\ea
For each expression we separated the terms involving one derivative (and 3 $\sa$'s) from those involving two derivatives (and 2 $\sa$'s). Concentrate first on the terms with only one derivative. No other terms involving only one derivative can arise from $s \tr w\d\sa\d\sa$, so that the terms involving only one derivative in (\ref{AAdAterms}) have to cancel each other or add up to an exact form. Obviously, they cannot cancel for any choice of $b_1, b_2, b_3$, and we must try to fix the $b_i$'s to get an exact term. 
Observe that we must get exact forms separately for terms with the ordering $\tr w\sa w\sa^2$ (with one derivative somewhere) and for terms $\tr w^2 \sa^3$ (with one derivative somewhere). It is then easy to see that this requires
\be\label{bivalues}
b_1=-b_2=b_3\equiv b \ ,
\ee
so that the $\tr w\sa w\sa^2$ terms cancel and the $\tr w^2 \sa^3$ terms yield the exact form $b\,  \d \tr w^2\sa^3$. Note that the values (\ref{bivalues}) of the $b_i$ are  such that the corresponding three terms  in (\ref{anomansatz}) combine into $b \tr w \d(\sa^3)$.
Next, we look at the terms in (\ref{AAdAterms})  involving two derivatives. Multiplying them with the corresponding $b_i$ according to (\ref{bivalues}) and adding them up gives
\be\label{contr1}
b \tr \big( -w\d w\d\sa^2 - \d w w\d\sa^2 -w\sa\d w\d\sa + w\d\sa\d w\sa \big)
=b \tr \big( -2 \d w\d w \sa^2+\d w\sa\d w\sa\big) + \d \tr\big(\ldots\big) \ .
\ee
Finally, one similarly finds after a slightly lengthy but straightforward computation:
\ba\label{contr2}
s \, \tr w\d\sa\d\sa
&=&\tr\big( w^2\d\sa\d\sa -w\d w\sa\d\sa + w\sa\d w\d\sa -w\d\sa\d w\sa + w\d\sa\sa\d w \big)
\nonumber\\
&=&\tr \big(  \d w\d w\sa^2 -\d w\sa\d w\sa\big) 
+\d \tr\big(\ldots\big) \ .
\ea
Now, $\d w\sa$ is an anticommuting 2-form and we have $\tr (\d w\sa)(\d w\sa)=-\tr (\d w\sa)(\d w\sa)=0$. Obviously then, if and only if
\be
b={1\over 2}\ ,
\ee
the two contributions (\ref{contr1}) and (\ref{contr2}) add up to an exact form:
\be
s \tr w \big(\d\sa\d\sa+{1\over 2}\d \sa^3\big)
=\d \big(\ldots\big) \ ,
\ee
and we finally conclude that
\be\label{conssol}
\ca[w,\sa]={i\over 24\pi^2} \int \trR w\, \d \big( \sa\d\sa+{1\over 2} \sa^3\big) = -{1\over 24\pi^2}\int \trR \e\, \d\big( A\d A-{i\over 2} A^3\big) \ .
\ee
The important result is that the Wess-Zumino condition determines the anomaly $\ca$ completely just from knowing the piece quadratic in $\sa$ (up to irrelevant BRST exact terms, of course).

Comparing with our discussion on characteristic classes and descent equations we see that 
\be 
\ca[w,\sa]={i\over 24\pi^2} \int Q_4^1(v\equiv w,\sa) \ .
\ee
This is not a coincidence as we will now show.

\subsection{Descent equations for anomalies in $d=2r$ dimensions\label{descsec}}

The present discussion will be for a space-time $\cM$ having an arbitrary even dimension $d=2r$. Then the maximal degree of a form is $2r$. We will nevertheless need to define forms of degree $2r+2$. This can be justified as follows.

In $2r$ dimensions the gauge field is given by the one-form
$\sa=\sum_{\m=0}^{2r-1} \sa_\m(x^\n)\, \d x^\m$.
Imagine one introduces two extra parameters on which the gauge field depends, say $\t$ and $\r$. We then have a family of gauge fields $\sa_\m(x^\n,\t,\r)$. This is what is done e.g. in the classical paper by Alvarez-Gaum\'e and Ginsparg \cite{AGG} to study the variation of the phase of ${\rm Det}(\Dsl P_L)$ as the parameters $\t$ and $\r$ are varied in a certain way. We will discuss this in more detail in the next section. Here, let us only say that this construction involves an auxiliary $(2r+2)$-dimensional manifold  $\cM\times {\cal D}$ (where ${\cal D}$ is the two-dimensional disk parametrized by $\r$ and $\t$), as well as adding a piece $\sa_\r \d\r +\sa_\t\d\t$ to $\sa_\m\d x^\m$ so that $\sa$ becomes a genuine one-form in $(2r+2)$ dimensions. Similarly, the exterior derivative gets an additional piece so that it becomes $\d=\d x^\m \del_\m +\d\r \del_\r + \d\t \del_\t$. In any case, we will consider $\sa$ and $\cF$ to be the gauge field one-form and corresponding field strength two-form on a $(2r+2)$-dimensional manifold (with space-time $\cM$ being a certain $2r$-dimensional submanifold). In particular, all previous relations now hold in $(2r+2)$ dimensions, e.g. $\cF=\d\sa+\sa^2$ with the $(2r+2)$-dimensional exterior derivative $\d$.

Recall from our general arguments that the anomaly must involve the $\e^{\m_1\ldots\m_{2r}}$ and hence can be expressed as a $2r$-form. More precisely,
\be
\ca[w,\sa]=c\int_{\cM} q_{2r}^1(w,\sa) \ ,
\ee
where $q_{2r}^1(w,\sa)$ is a $2r$-form of ghost-number one, i.e. linear in $w$. The Wess-Zumino consistency condition for a relevant anomaly immediately generalizes to arbitrary dimensions and can be again written as
\be
s\, \ca[w,\sa]=0 \quad , \quad \ca[w,\sa]\ne s\big(\ldots\big) \ .
\ee
A convenient way to obtain solutions to this BRST cohomology problem is to use the descent equations, as we will now show. Recall that
\ba\label{descent3}
P_{r+1}=\tr \cF^{r+1} \quad , \quad
\d P_{r+1}=s P_{r+1} = 0
\nonumber\\
\Rightarrow \ P_{r+1}=\d Q_{2r+1} \quad , \quad s Q_{2r+1}=\d Q_{2r}^1 \quad {\rm (locally)} \  .
\ea
The {\bf claim} we want to prove is the following:\hfill\break

\noindent
$\bullet$\ $\int Q_{2r}^1(w,\sa)$ is a representative of the BRST cohomology at ghost number one and thus a solution of the Wess-Zumino condition. Moreover, the anomaly must be of the form
\be
\ca[w,\sa]=c \int Q_{2r}^1(w,\sa) + s\big(\ldots\big) \ ,
\ee
with $Q_{2r}^1$ being a solution of the descent equations (\ref{descent3}). If $c\ne 0$ the anomaly is relevant.

\noindent
{\it Proof:} 
First recall (or simply admit) that at ghost number zero the BRST cohomology consists of gauge invariant functionals of the gauge field only.\footnote{Note that a BRST invariant functional of ghost number zero can involve the gauge fields, $n\ge0$ ghost fields and the same number $n$ of antighost fields. The statement is that whenever $n>0$ such functionals are BRST exact.} Recall also that there are no gauge invariant forms of odd degree and, hence, for odd form degrees the BRST cohomology at ghost number zero is empty. Furthermore, for ghost number zero, an even form of the field strength only, like $P_{r+1}$ is a non-trivial representative of the BRST cohomology. Indeed, $P_{r+1}$ is gauge and hence BRST invariant, and it cannot be obtained as $s \a_{2r+1}^{-1}$ (since $\a_{2r+1}^{-1}$ necessarily must contain at least one antighost field $\o_\a^*$ and $s\o_\a^*\sim h_\a$ which is not present in $P_{r+1}$). 

After these preliminaries let us show that $s\int Q_{2r}^1=0$ and $Q_{2r}^1\ne s \a_{2r}$. From the descent equation $s Q_{2r+1}=\d Q_{2r}^1$ it immediately follows that
\be
0=s \big(s Q_{2r+1})=s \big(\d Q_{2r}^1 \big)= - \d \big( s Q_{2r}^1\big) \ .
\ee
Note that  the $(2r+2)$-dimensional ``space-time" may have non-trivial topology and require several patches with non-trivial transition functions for the gauge fields. However, according to our  general discussion at the beginning of this section, the $2r$-form $Q_{2r}^1$ which is linear in the ghost fields and similarly $s\, Q_{2r}^1$ which is bilinear in the ghost fields are assumed to be globally defined.
Now the topology of the $(2r+2)$-dimensional space can be chosen such that every closed $2r$-form is exact.\footnote{
Without proving this statement, let us nevertheless illustrate it. First of all, it is only a statement about $2r$-forms, not about  forms of any other degree. Nevertheless, if $\cM$ is compact, e.g. $S^{2r}$ as is often used in Euclidean signature in the connection with instantons, there always is the volume $2r$-form on $\cM$ which is trivially closed on $\cM$ and not exact. The statement is that one can add the two extra dimensions in such a way that this no longer holds. As an example, consider $r=1$ and $\cM=S^2$ with volume form ${\rm vol}_2=\sin\t\d\t\d\vf$. One can add the two extra dimensions such that one is just the radial direction with coordinate $r$ and the other is just a copy of ${\bf R}$. The resulting $2+2$ dimensional space simply is ${\bf R}^3\times {\bf R}$ which clearly dos not have any closed 2-forms that are not exact. Indeed, $\sin\t\d\t\d\vf$ now is singular at $r=0$ and  no longer is a well-defined 2-form.
} 
Then
\be
 s Q_{2r}^1=\d\a_{2r-1}^1 \quad \Rightarrow\quad
 s\int_\cM Q_{2r}^1 = \int_\cM \d\a_{2r-1}^2  =0 \ ,
\ee
where we used (\ref{exactiszero}) on $\cM$. Hence $\int_\cM Q_{2r}^1$ is BRST closed.

Let us show that $Q_{2r}^1 \ne s \a_{2r}$. Suppose on the contrary that $Q_{2r}^1 = s \a_{2r}$, with $\a_{2r}$ being necessarily a functional of the gauge fields only. This would imply $s Q_{2r+1}=\d Q_{2r}^1 =\d s \a_{2r} = -s\d \a_{2r}$, and thus $s\big( Q_{2r+1}+\d \a_{2r}\big)=0$. But as recalled above, at ghost number zero the BRST cohomology is trivial for odd forms. Moreover, there is no BRST exact odd form of ghost number zero (and not involving the antighost or the $h$-field) either.  Hence $Q_{2r+1}+\d\a_{2r}=0$, i.e. $Q_{2r+1}=-\d \a_{2r} \Rightarrow \d Q_{2r+1}=0$. But this cannot be true since $\d Q_{2r+1}=P_{r+1}\ne 0$.
We conclude that $Q_{2r}^1$ cannot be BRST exact. It remains to show that also $\int Q_{2r}^1$  cannot be BRST exact. Suppose we had $\int Q_{2r}^1=s\int \b_{2r}=\int s\b_{2r}$. This would imply $Q_{2r}^1+\d \g_{2r-1}^1=s\b_{2r}$ for some $\g_{2r-1}^1$. Now $Q_{2r}^1$ was defined by the descent equations only up to adding an exact form, so that $\widetilde Q_{2r}^1=Q_{2r}^1+\d\g_{2r-1}^1$ is just as good, and our above argument shows that $\widetilde Q_{2r}^1$ cannot be BRST exact. Hence we cannot have $\int Q_{2r}^1=s\int \b_{2r}$.
This concludes the proof.

We can rephrase this result as follows:
\be\label{anomdesc}
\begin{array}{|c|}
\hline\\
\text{In $d=2r$ dimensions the anomaly is}\\
\\
\ca[w,\sa]=c\int Q_{2r}^1(w,\sa) + s\big(\ldots\big)\ .\\
\\
\text{It corresponds via the descent equations to $c\, P_{r+1}$.}\\
\\
\text{ If and only if $c=0$ the anomaly is irrelevant, i.e $\ca[w,\sa]=s\big(\ldots\big)$.}
\\
\\
\hline
\end{array}
\ee
The ``ambiguity" of adding BRST exact terms to the anomaly corresponds exactly to the possibility of adding (non gauge invariant) counterterms to the action.

The characterization (\ref{anomdesc}) of anomalies shows that all relevant anomalies are coded in the $P_{r+1}$ together with the corresponding coefficients $c$ which depend on the chiral field content of the theory. If ${\cal S}$ is the set of all chiral fields $\p_i$, one defines the total {\it anomaly polynomial}
\be
I_{2r+2}^{\rm total} = \Big( \sum_{\p_i\in {\cal S}} c(\p_i) \Big) P_{r+1} = \sum_{\p_i\in {\cal S}}  I^{(i)}_{2r+2} \ ,
\ee
where, obviously, $I^{(i)}_{2r+2} = c(\p_i) P_{r+1}$. For the example of a left-handed (i.e. positive chirality) fermion in 4 dimensions we had $c={i\over 24 \pi^2}$ and hence 
\be\label{I6pos}
I^{\rm pos.chirality\, fermion}_6 = {i\over 24\pi^2} \tr \cF^3 \ .
\ee
We have $\tr \cF^3=\cF^\a \cF^\b \cF^\g D_{\a\b\g}$, and this is how the $D$-symbol arises. In $2r$ dimensions, we similarly have
\be
\tr \cF^{r+1} 
= \cF^{\a_1} \ldots \cF^{\a_{r+1}} D_{\a_1 \ldots \a_{r+1}} \quad , \qquad 
D_{\a_1 \ldots \a_{r+1}} =\tr t_{(\a_1} \ldots t_{\a_{r+1})} \ .
\ee

\vskip5.mm
\newpage
\section{Relation between anomalies and index theorems\label{indexth}}
\setcounter{equation}{0}

We have already seen in section \ref{fmeasure} that the abelian anomaly in four dimensions was related to the index of the four-dimensional Dirac operator $\Dsl$ which was $\sim\tr t \cF \cF$. On the other hand, we have seen in the previous section that the chiral non-abelian anomaly in four (or $2r$) dimensions is related via descent to $\tr \cF^3$ (or $\tr\cF^{r+1}$), which in turn is proportional to the index of a six-dimensional (or $(2r+2)$-dimensional) Dirac operator $\Dsl$

It is the purpose of this section to outline how and why the anomaly in $2r$ dimensions is given by the index of a Dirac operator in two more dimensions. We will mostly follow the classical work by Alvarez-Gaum\'e and Ginsparg \cite{AGG}. We will not be able to give too many details or justify all the statements. Nevertheless, we will try as much as possible to make them plausible at least.

We will use Euclidean signature throughout this section. The continuation between Euclidean and Minkowski signature is particularly subtle in the context of anomalies and anomaly cancelation due to the appearance of the $\e^{\m_1\ldots \m_{2r}}$-tensor and the necessity to carefully distinguish factors of $i$ from factors of $-i$.

\subsection{Continuation to Euclidian signature\label{Euclidean}}

Let us  first discuss the Euclidean continuation in some detail.\footnote{
The present discussion is similar to the one in \cite{BM}, but not all of our present conventions are the same as there.
}
While the functional integral in the Minkowskian contains
$e^{iS_{\rm M}}$, the Euclidean one contains $e^{-S_{\rm E}}$ where the kinetic terms in $S_{\rm E}$ must be non-negative.
This implies the choice
\be\label{d15} S_{\rm M}=i\, S_{\rm E} \qquad , \qquad
x^0=-i\, x^0_{\rm E} \quad\Leftrightarrow \quad x^0_{\rm E}=i x^0  \ . 
\ee 
Obviously then, for a lower index we have $\del_0=i\del^{\rm E}_0$.
However, for a Euclidean manifold
$\cM_{\rm E}$ it is natural to index the coordinates from 1 to $d$,
not from $0$ to $d-1$. One could, of course, simply write
$ix^0=x^0_{\rm E}\equiv x^D_{\rm E}$, as is done quite often in the literature. The problem then is for even $d=2r$ that 
$\d x^0_{\rm E}\w \d x^1\w\ldots\d x^{2r-1} 
= -\ \d x^1\w\ldots\d x^{2r-1} \w \d x^{2r}_{\rm E}$ and if 
$(x^0_E,\ldots x^{2r-1})$ was a right-handed coordinate system then 
$(x^1,\ldots x^{2r}_E)$ is a left-handed one. This problem is solved by shifting the indices of the coordinates as 
\be\label{d16} 
i\, x^0 = x^0_{\rm E} = z^1 \ , \quad x^1=z^2 \ , 
\quad \ldots \ , \quad x^{D-1} = z^D \ . 
\ee 
This is equivalent to a specific choice of orientation on the Euclidean manifold $\cM_{\rm E}$. In particular, we impose 
\be\label{d16a} 
\int \sqrt{g}\, \d z^1\w \ldots\w\d z^d 
= + \int \sqrt{g}\, \d^d z \ge 0 \ . 
\ee 
Then, of course, for any tensor we similarly shift the indices, e.g.
$C_{157}=C^{\rm E}_{268}$ and $C_{034}=i\, C^{\rm E}_{145}$, but we still have $G_{\m\n\r\s}\, G^{\m\n\r\s} = G^{\rm E}_{jklm}\, G_{\rm
E}^{jklm}$ as usual. In particular, for any  $p$-form
\be\label{d17}
\xi={1\over p!}\, \xi_{\m_1\ldots\m_p}\,
 \d x^{\m_1}\w\ldots\w\d x^{\m_p}
={1\over p!}\, \xi^{\rm E}_{j_1\ldots j_p}\,
 \d z^{j_1}\w\ldots\w\d z^{j_p} =\xi^{\rm E} \ ,
\ee
and for $p=d$ we have
\be\label{d18}
\int_{\cM_{\rm M}} \xi = \int_{\cM_{\rm E}} \xi^{\rm E}\ .
\ee
Since anomalies are given by such integrals of  $d$-forms, it is most important to know the exact sign in this relation.
Finally, note that the Minkowski relation
(\ref{volform2}) becomes
\ba\label{d19}
&&\d x^{\m_1}\wedge \ldots \d x^{\m_d}=\e^{\,\m_1\ldots \m_d}\  
\sqrt{-g}\ \d^d x\ 
\quad {\rm with} \quad
\e^{0\ldots (d-1)}= {1\over \sqrt{-g}}
\nonumber\\
\quad \Leftrightarrow\quad
&&\d z^{j_1}\w \ldots\w\d z^{j_d}=
+\, \e_{\rm E}^{j_1\ldots j_d}\, \sqrt{g_{\rm E}}\, \d^d z
\quad {\rm with} \quad
\e_{\rm E}^{1\ldots d}={1\over \sqrt{g_{\rm E}}} \ .
\ea
The Hodge dual of a $p$-form $\xi^{\rm E}$ is defined as in (\ref{Hodge})
but using $\e_{\rm E}$. It then follows that
$\st(\st\xi_{\rm E})=(-)^{p(d-p)}\, \xi_{\rm E}$ (with an additional
minus sign with respect to the Minkowski relation) and, as in
the Minkowskian, eq. (\ref{xiwedgehodgexi}), we have
\be\label{kineticform}
\xi_{\rm E}\w \st\xi_{\rm E}
= {1\over p!}\, \xi^{\rm E}_{j_1\ldots j_p}\,
\xi_{\rm E}^{j_1\ldots j_p}\, \sqrt{g_{\rm E}}\ \d^d z\ .
\ee

We are now ready to  give the Euclidean continuation of some general Minkowski action
\be\label{Minkaction}
S_{\rm M}
= \int_{\cM_{\rm M}} \d^d x \sqrt{-g}\, \Big( -{\a\over 2 p!}\, \xi^{\m_1\ldots\m_p}\xi_{\m_1\ldots\m_p} 
+ {\beta\over d!}\, \e^{\m_1\ldots \m_d}  \zeta_{\m_1\ldots \m_d}\Big)
= \int_{\cM_{\rm M}} \Big(-{\a\over 2}\,  \xi^{(p)}\w\st\xi^{(p)} 
+ \beta\, \zeta^{(d)}\Big) \ .
\ee
Using eqs. (\ref{d15}) to (\ref{kineticform}), the corresponding Euclidean action is
\be\label{Euclaction}
S_{\rm E}
= \int_{\cM_{\rm E}} \d^d z \sqrt{g_{\rm E}}\ \Big( {\a\over 2 p!}\, \xi^{j_1\ldots j_p}_{\rm E}\xi_{j_1\ldots j_p}^{\rm E}
- i \, {\beta\over d!}\, \e_{\rm E}^{\m_1\ldots \m_d}  \zeta_{\m_1\ldots \m_d}^{\rm E}\Big)
= \int_{\cM_{\rm E}} \Big({\a\over 2}\,  \xi^{(p)}_{\rm E}\w\st\xi^{(p)}_{\rm E} 
- i\, \beta\, \zeta^{(d)}_{\rm E}\Big) \ .
\ee
Note that only the so-called topological terms in the action, i.e. those terms that involve the $\e$-tensor, acquire an explicit factor $-i$ in the Euclidean continuation. Actually, the non-topological terms (those with coefficient $\a$) get two factors of $-i$, one from
$S_{\rm E}=-i S_{\rm M}$ and one from $\d z^1\equiv\d x^0_{\rm E}=-i \d x^0$, resulting in a minus sign. On the other hand, the topological terms only get one factor of $-i$ from $S_{\rm E}=-i S_{\rm M}$, lacking the second factor since (\ref{d18}) does not involve any $i$.

Recall that the anomaly is the variation of the effective action and it always involves the $\e$-tensor, i.e. it is a topological term. The previous argument shows that its Euclidean continuation must be purely imaginary. This then shows that the anomalous part  of the Euclidean effective action must reside in its imaginary part. This is in perfect agreement with a general  argument we will give below that only the imaginary part of the Euclidean effective action can be anomalous.

According to these remarks, and as we will indeed confirm below,
the anomalies of the Euclidean effective action are of the form 
\be\label{d25E}
\delta \G_{\rm E}= -
i\, \int_{M_{\rm E}^{2r}} \hat I^1_{2r} \ ,
\ee 
where $\hat I^1_{2r}$ is a real $d=2r$-form. It corresponds to a variation of the Minkowskian effective action given by
\be\label{d25} 
\delta \G_{\rm M} =
\int_{M_{\rm M}^{2n}} \hat I^1_{2n} 
\ee
where now $\hat I^1_{2n}$ is rewritten in Minkowski coordinates
according to (\ref{d17}). 

There is one more subtlety that needs to be settled when discussing the relation between the Minkowski and the Euclidean
form of the anomaly:  we have to know how the chirality matrix $\g$ (the analogue of $\g_5$) is continued from the Euclidean to the Minkowskian and vice versa. The continuation of the
$\g$-matrices is dictated by the continuation of the coordinates
we have adopted (cf (\ref{d16})):
\be\label{d30}
i\, \g^0_{\rm M}=
\g^1_{\rm E} \ , \quad \g^1_{\rm M}= \g^2_{\rm E} \ ,
\quad \ldots
\quad \g^{2n-1}_{\rm M}= \g^{2n}_{\rm E} \ .
\ee
In accordance
with ref. \cite{AGG} we define the  Euclidean
chirality matrix  $\g_{\rm E}$ in $2r$
dimensions as
\be\label{g5eucl}
\g_{\rm E} = i^r \g^1_{\rm E} \ldots  \g^{2r}_{\rm E} \ .
\ee
For the sign convention of the Minkowskian chirality matrix there are several different conventions in the literature. Here we choose
\be\label{g5mink}
\g_{\rm M}=i^{r+1} \g^0_{\rm M}\ldots \g^{2r-1}_{\rm M} \ .
\ee
Then both $\g_{\rm M}$ and $\g_{\rm E}$ are hermitian. In particular, for $r=2$, i.e. $d=4$ we recover our definition (\ref{g5fourdim}) of $\g_5$. Taking into account (\ref{d30}) we have
\be\label{d32}
\g_{\rm M}=   \g_{\rm E} \ ,
\ee
i.e. what we call positive (negative) chirality in Minkowski space is also called positive (negative) chirality in Euclidean space.\footnote{The sign convention for $\g_{\rm M}$ in (\ref{g5mink}) is opposite from the one in ref. \cite{AGG}, but has the advantage of leading to $\g_{\rm M}=\g_{\rm E}$ rather than $\g_{\rm M}=-\g_{\rm E}$. 
Since we will take \cite{AGG} as the
standard reference for computing anomalies in the Euclidean, we
certainly want to use the same convention for $\g_{\rm E}$. On the
other hand, we have somewhat more freedom to choose a sign
convention for $\g_{\rm M}$. Note that for $d=10$ the definition (\ref{d32}) yields $\g_{\rm
M}= -\g^0_{\rm M}\ldots \g^9_{\rm M}$ which is opposite from the one used in \cite{GSW}. It is also such that it agrees with the definition used in \cite{POL} for $D=4$ and 8, but is opposite for $D=2$, 6 and 10.} 
In particular, in 4 dimensions ($r=2$) we have
\be
\g_{\rm E}=-\g^1_{\rm E}\g^2_{\rm E}\g^3_{\rm E}\g^4_{\rm E}
\quad \Rightarrow \quad 
\trD \g_{\rm E} \g^j_{\rm E}\g^k_{\rm E}\g^l_{\rm E}\g^m_{\rm E}
=-4\e_{\rm E}^{jklm} \ .
\ee

We can now give the anomalous variation of the {\it Euclidean} effective action for a positive chirality fermion in four dimensions, as obtained from our result (\ref{anomdiffform}) of the one-loop computation, the consistency condition (cf. (\ref{conssol})) and eqs. (\ref{d25E}), (\ref{d25}). We get
\be\label{Euclanom4D}
\dd\G_E = + {1\over 24\pi^2} \int \tr v \d \big( \sa\d\sa+{1\over 2}\sa^3\big) = + {1\over 24\pi^2} \int  Q_4^1(v,\sa,\cF) \ .
\ee
Note that this is purely imaginary.

Finally let us make a remark on our convention (\ref{d16}) which is different from what is mostly done in the literature. In four dimensions, ref.~\cite{Weinbook} e.g. continues as $i x^0=x^4_E$ and $x^j=x^j_E,\ j=1,2,3$. Since the definition $\e_E^{1234}=+1$ is always adopted, it follows that the Minkowskian continuation of a topological term like $\int \d^4 x_{\rm E}\, \e_{\rm E}^{jklm} F^{\rm E}_{jk} F^{\rm E}_{lm}$ then differs by a sign from our convention. On the other hand, the same sign difference also appears in the continuations of the chirality matrices, so that any statement relating a chiral anomaly to a characteristic class is independent of theses conventions.

\subsection{Defining the phase of the determinant of $D\hskip-2.9mm/\hskip1.mm_+$\label{phasedefdet}}

We are again interested in doing the functional integral over the chiral fermions. To be definite, we take a positive chirality fermion
\be
\g\p=+\p \ ,
\ee
where $\g\equiv \g_{\rm E}$. In Euclidean space, $\ov\p$ no longer is related to the hermitian conjugate of $\p$ but is an independent (negative chirality) spinor.  We take its normalization such that the Euclidean action for the fermions reads
\be
S_{\rm matter}^{\rm E} = \int \d^d z \sqrt{g_{\rm E}}\ \ov\p\, i \Dsl_+ \p \ ,
\ee
where 
\be
\Dsl_+\equiv \Dsl\, {1+\g\over 2} = \g^j (\del_j - i A_j)\
{1+\g\over 2} 
= \g^j (\del_j + \sa_j)\ {1+\g\over 2} 
\ .
\ee
We will call ${\cal H}_\pm$ the Hilbert spaces of positive, resp. negative chirality spinors. Clearly, $\Dsl_+$ maps positive chirality spinors to negative chirality ones, i.e. ${\cal H}_+$ to ${\cal H}_-$. 

The Euclidean effective action then is
\be\label{eucleffac}
e^{-\G[\sa]}\equiv e^{-\wt W[\sa]} 
= \int \cD\p \cD\ov\p\ \exp\Big(-\int\d^d z \sqrt{g_{\rm E}}\ \ov\p\, i \Dsl_+ \p\Big) \ .
\ee
As already discussed in section \ref{chidd}, this functional integral is ill-defined. Indeed, it would be $\Det (i\Dsl_+)$, but determinants are only well-defined for operators $M$ for which ${\rm Im}(M)\subset {\rm Def}(M)$, while ${\rm Def}(i\Dsl_+)={\cal H}_+$ and ${\rm Im}(i\Dsl_+)={\cal H}_-$. One may consider $(i\Dsl_+)^\dag = i {1+\g\over 2} \Dsl=i\Dsl {1-\g\over 2}=i\Dsl_-$ and compute instead the well-defined determinant $\Det (i\Dsl_- i\Dsl_+)= \Det \big((i\Dsl_+)^\dag i\Dsl_+\big)$ which formally would equal $|\Det(i\Dsl_+)|^2$. We may indeed use this as a definition for the {\it absolute value} of  (\ref{eucleffac}), i.e. the real part of $\G[\sa]$. This shows that we may unambiguously define the real part of the Euclidean effective action.

Alternatively, just as in section \ref{chidd}, we may artificially add negative chirality fermions that do not couple to the gauge fields and define
\be
\wh D=\Dsl_+ + \dsl_- = \g^j (\del_j +\sa_j) {1+\g\over 2} + \g^j \del_j {1-\g\over 2} = \g^j \big( \del_j +\sa_j {1+\g\over 2}\big) \ ,
\ee
so that
\be
e^{-\G[\sa]}\equiv e^{-\wt W[\sa]} = \Det\big( i\wh D\big) \ ,
\ee
up to an irrelevant multiplicative constant.
Obviously, this Dirac operator $\wh D$ which acts on ${\cal H}={\cal H}_+\oplus{\cal H}_-$ is not gauge invariant, but does have a well defined (non gauge invariant) determinant. With respect to the decomposition of 
${\cal H}$ into ${\cal H}_+$ and ${\cal H}_-$ the Dirac operator $\wh D$ has the ``off block-diagonal" form $\begin{pmatrix} 0 & \Dsl_+ \\ \dsl_- & 0 \\ \end{pmatrix}$ and
\be
(i\wh D)^\dag  (i \wh D)  = \begin{pmatrix} i \dsl_+ i\dsl_- & 0 \\ 0 & i \Dsl_- i \Dsl_+ \\ \end{pmatrix}\ ,
\ee
so that, upon taking the determinant, we get
\ba
|\Det (i\wh D)|^2 
&=& \Det(i \dsl_+ i\dsl_- ) \ \Det \big( i \Dsl_- i \Dsl_+ \big)
= {\rm const} \times 
\Det \begin{pmatrix} 0 & i\Dsl_+ \\ i\Dsl_- & 0 \\ \end{pmatrix} 
\nonumber\\
&=& {\rm const} \times \Det\big( i\Dsl_+ + i\Dsl_-\big)
={\rm const} \times \Det \big( i \Dsl\big)
\ea
Now, (as long as the gauge field is such that there is no zero-mode) $\Det\big( i\Dsl\big)$ is real and positive and we conclude
$|\Det (i\wh D)|= \Big( \Det \big( i \Dsl\big)\Big)^{1/2}$, up to an irrelevant real, positive multiplicative constant. Thus
\be
e^{-\G[\sa]}=\Det (i\wh D)= \Big( \Det \big( i \Dsl\big)\Big)^{1/2} e^{i \Phi[\sa]} \ .
\ee
Clearly, $\Big( \Det \big( i \Dsl\big)\Big)^{1/2}$ is gauge invariant, and we see again that all non-gauge invariance is contained in the phase $\Phi[\sa]$, i.e. in the imaginary part of the Euclidean effective action. As we have seen in eq. (\ref{Euclaction}) the imaginary part  of this effective action are the topological terms involving the $\e^{j_1\ldots j_{2r}}$-tensor. Thus:\\
\be
\begin{array}{|c|}
\hline\\
\text{The real part of the Euclidean effective action always is gauge invariant.}\\
\\
\text{Only the imaginary part can be anomalous.}\\
\\
\text{The imaginary part corresponds precisely to the topological terms $\sim \e^{j_1\ldots j_{2r}}$.}
\\
\\
\hline
\end{array}
\nonumber
\ee
\vskip4.mm

Suppose that the Euclidean space $\cM$ is $S^{2r}$ or at least has the topology of $S^{2r}$. (This includes in particular the case of ${\bf R}^{2r}$ with gauge fields such that the gauge field strengths vanish sufficiently fast at infinity, as is the case e.g. for instanton configurations.) We fix some ``reference" gauge field $\sa(x)$, $x\in 
\cM$ and introduce a parameter $\t\in [0,2\pi]$. Let $g(\t,x)\in G$ be some a family of gauge group elements such that $g(0,x)=g(2\pi,x)={\bf 1}$. We may then view $g$ as being defined on $[0,2\pi]\times S^{2r}$ with all points $(0,x)$ identified, and similarly all points $(2\pi,x)$ identified. Thus we are not dealing with $[0,2\pi]\times S^{2r}$ but with $S^{2r+1}$. We now define a gauge field $\sa^\t$ as the gauge transformed of $\sa(x)$ by the element $g(\t,x)$:
\ba\label{Athetadef}
\sa^\t(x) \equiv \sa(\t,x) &=& g(\t,x)^{-1} \big( \d +\sa(x)\big) g(\t,x) \ , 
\nonumber\\
{\rm or} \quad A^\t(x) \equiv A(\t,x) &=& g(\t,x)^{-1} \big(\, i\, \d +A(x)\big) g(\t,x) \ ,
\qquad \d=\d x^j \del_j  \ .
\ea
It is important to note that $\d$ does {\it not} include a piece $\d\t\del_\t$ and that $\sa^\t$, resp. $A^\t$ still is a $2r$-dimensional gauge field. The corresponding Dirac operator $\wh D(\sa^\t)$ also still is a Dirac operator on $\cM=S^{2r}$ and not on $S^{2r+1}$. We have (still assuming that $i \Dsl(\sa)$ has no zero modes)
\be
\Det \big( i \wh D(\sa^\t)\big) = \Big(\Det\big( i \Dsl(\sa^\t)\big) \Big)^{1/2} e^{i\Phi[\sa^\t]}
=\Big(\Det\big( i \Dsl(\sa)\big) \Big)^{1/2} e^{i\Phi[\sa^\t]} \ ,
\ee
where the second identity holds since $\Det\big( i \Dsl(\sa)\big)$ is gauge invariant. We will write $\Phi[\sa^\t(x)]\equiv \Phi[\sa(x),\t]$. Note that the boundary conditions on $g(\t,x)$ imply  $\sa(2\pi,x)=\sa(0,x)=\sa(x)$, so that
\be
\Phi[\sa,2\pi]=\Phi[\sa,0]+ 2\pi m \ , \qquad m\in {\bf Z}\ ,
\ee
or
\be
\int_0^{2\pi} \d\t\ {\del \Phi[\sa,\t]\over \del\t} = 2\pi m \ .
\ee
Clearly, if the phase $\Phi[\sa]$ is gauge invariant, then $\Phi[\sa(x),\t]$ cannot depend on $\t$ and we have $m=0$. On the other hand, if $m\ne 0$ there is no way $\Phi[\sa]$ can be gauge invariant and we have an anomaly. One can very explicitly relate the anomaly to the integer $m$ as follows:
\be
i\, {\del\Phi[\sa,\t]\over\del\t} = - {\del \G[\sa^\t]\over\del\t} 
=- \int \d^{2r} z \sqrt{g_{\rm E}} \  {\del (\sa^\t)^\a_j (x)\over \del\t}\ {\dd \G[\sa^\t]\over \dd (\sa^\t)^\a_j (x)} \ .
\ee
We need to compute
\be\label{vthetadef}
{\del (\sa^\t)^\a_j \over \del\t} = \big( \del_j v +[\sa_j^\t,v]\big)^\a\equiv \big(D_j^\t v\big)^\a
\ , \quad {\rm where}\ v\equiv v^\t(x)\equiv v(\t,x)=g^{-1}(\t,x) {\del\over \del\t} g(\t,x) \ ,
\ee
so that
\be
i\, {\del\Phi[\sa,\t]\over\del\t} = -  \int \d^{2r} z \sqrt{g_{\rm E}} \  \big(D_j^\t v^\t\big)^\a(x) {\dd \G[\sa^\t]\over \dd (\sa^\t)^\a_j (x)}
=  \int \d^{2r} z \sqrt{g_{\rm E}} \ (v^\t)^\a(x) \left( D^\t_j {\dd \G[\sa^\t]\over \dd (\sa^\t)^\a_j (x)}\right)_\a \ .
\ee
The right hand side of course equals minus the anomaly under a gauge transformation $\dd \sa^\t={\rm D}^\t v^\t$ or, equivalently, $\dd A^\t={\rm D}^\t \e^\t$ with $\e^\t=i v^\t$. Finally, we get
\be\label{mvanom}
m={1\over 2\pi} \int_0^{2\pi} \d\t {\del\Phi[\sa,\t]\over\del\t}
= {1\over 2\pi i} \int_0^{2\pi} \d\t \int \d^{2r} z \sqrt{g_{\rm E}} \ (v^\t)^\a(x) \left( D^\t_j {\dd \G[\sa^\t]\over \dd (\sa^\t)^\a_j (x)}\right)_\a \ .
\ee

\vskip5.mm
\subsection{Relation with the index of a Dirac operator in $2r+2$ dimensions\label{indextheor}}

The anomaly with gauge transformation parameter $v$ and normalized as in (\ref{mvanom}) equals the integer $m$. We will now outline how, following \cite{AGG}, this integer $m$ can be related to the index of an appropriate Dirac operator in $2r+2$ dimensions.

First remark that $g(\t,x)$ is a map from $S^{2r+1}$ into $G$, and such maps are characterized by their homotopy class in $\Pi_{2r+1}(G)$. (For most simple compact groups $G$ we have $\Pi_{2r+1}(G)={\bf Z}$, while for product groups one has $\Pi_{2r+1}(G_1\times \ldots \times G_k)=\Pi_{2r+1}(G_1)\times\ldots\times\Pi_{2r+1}(G_k)$.) Consider two maps $g_1(\t,x)$ and $g_2(\t,x)$ that can be continuously deformed into each other, i.e. that correspond to the same homotopy class in $\Pi_{2r+1}(G)$. Then one can also continuously deform the corresponding $v_1(\t,x),\ A^\t_1(x)$ and $v_2(\t,x),\ A^\t_2(x)$ into each other. It follows that one can continuously deform $\Phi_1[A,\t]$ into $\Phi_2[A,\t]$ and that both configurations must correspond to the same integer $m$. Thus the integer $m$ characterizes the homotopy class of $g$ in $\Pi_{2r+1}(G)$.

On the other hand, $m$ is some sort of winding number around the circle $S^1$ parametrized by $\t$. To compute this winding number, one extends the $A^\t$ to a two-parameter family $A^{\r,\t}$ of gauge fields, where now $\r\in [0,1]$ and $(\r,\t)$ parametrize a disc, with $A^{\r,\t}(x)\vert_{\r=1}=A^\t$ living on the boundary of the disk.
Equation (\ref{mvanom}) can then be interpreted as computing the winding number around the disc. One can then show \cite{AGG} that
\begin{itemize}
\item
The winding number $m$ around the boundary of the disk equals the sum of local winding numbers around the points in the interior of the disc where $\Det\big( i \wh D(A^{\r,\t})\big)$ vanishes. (Recall that on the boundary of the disc where $A^{\r,\t}=A^\t$ is a gauge transformed of $A$, $\ i\wh D(A^{\r,\t})$ has no zero-modes.)
\item
The zeros of $\Det\big( i \wh D(A^{\r,\t})\big)$ are in one-to-one correspondence with the zero-modes of a $(2r+2)$-dimensional  Dirac operator $i\Dsl_{2r+2}$ such that the winding numbers $\pm 1$ equal the $\pm 1$ chirality of the zero-modes.
\end{itemize}
Thus one concludes\footnote{
Recall that the index of a Dirac operator ${D\hskip-2.4mm/\hskip1.mm}$ is defined as the number of positive chirality zero-modes minus the number of negative chirality zero-modes. Similarly, for the Weyl operator ${D\hskip-2.4mm/\hskip1.mm}_+={D\hskip-2.4mm/\hskip1.mm} (1+\g)/2$ the index is defined as the number of zero-modes of ${D\hskip-2.4mm/\hskip1.mm}_+$ minus the number of zero-modes of $({D\hskip-2.4mm/\hskip1.mm}_+)^\dag$.
}
\be\label{mequalsindex}
{1\over 2\pi} \int_0^{2\pi} \d\t {\del\Phi[A,\t]\over\del\t}
\equiv m = \ind\big( i \Dsl_{2r+2}\big) \ ,
\ee
or
\be\label{indanom}
\ind\big( i \Dsl_{2r+2}\big)
={1\over 2\pi i}\int_0^{2\pi} \d\t \int \d^{2r} z \sqrt{g_{\rm E}} \ (v^\t)^\a(x) \left( D^\t_j {\dd \G[\sa^\t]\over \dd (\sa^\t)_j (x)}\right)_\a \ .
\ee
To compute this index we need to know
what is the relevant Dirac operator $\Dsl_{2r+2}$ and $(2r+2)$-dimensional manifold.
 
The manifold is $S^2\times S^{2r}$ with $S^2=S^2_+\cup S^2_-$, as shown in Fig.~\ref{patches} in sect.~\ref{gaugebundle}. We identify $S^2_+$ with the disc with coordinates $(\r,\t)$, and $S^2_-$ with some other disc with coordinates $(\s,\t)$. Thus the $(2r+2)$-dimensional manifold is constructed from the two ``patches" $S^2_+\times S^{2r}$ and $S^2_-\times S^{2r}$, and the $(2r+2)$-dimensional gauge field $\sa(x,\r,\t)$ must be specified on both patches, with a transition function that is a gauge transformation. 
We take it to be
\ba
\sa(x,\r,\t)\equiv \sa_+(x,\r,\t)&=& f(\r)\, g^{-1}(x,\t) \big(\sa(x)+\d +\d\t \del_\t \big) g(x,\t) \ , 
\quad {\rm on}\ S^2_+\times S^{2r} \ ,
\nonumber\\
\sa(x,\s,\t)\equiv \sa_-(x,\s,\t)&=& \sa(x) \ , 
\hskip6.4cm {\rm on}\ S^2_-\times S^{2r} \ .
\ea
The $C^\infty$ function $f(\r)$ is chosen such that (i) $f(\r)\sim \r$ for small $\r$, i.e. close to the center of the disc, so that $f(\r)\, \d\t$ is well-defined even at $\r=0$, and (ii) $f(\r)=1$ for $\r$ close to 1, i.e. on some annulus which is the fattened boundary of the disk. This latter condition ensures that on the overlap between 
$S^2_+\times S^{2r}$ and $S^2_-\times S^{2r}$, which is an annulus times $S^{2r}$, we have $\sa_+(x,\r,\t)=g^{-1}(x,\t) \big(\sa(x)+\d +\d\t \del_\t \big) g(x,\t) = g^{-1}(x,\t) \big(\sa(x)+\d +\d\t \del_\t +\d\r \del_\r\big) g(x,\t)$ which is indeed the gauge transformed (in the $(2r+2)$-dimensional sense) of $\sa(x)=\sa_-(x,\r,\t)$. In particular, on the overlap we have $\cF_+=g^{-1}\cF_-\, g$, where $\cF_+=(\d +\d\t \del_\t +\d\r \del_\r)\sa_+ + \sa_+^2$ on all $S^2_+\times S^{2r}$, and $\cF_-=(\d +\d\t \del_\t +\d\r \del_\r)\sa_-+\sa_-^2=\d\sa+\sa^2$ 
on all $S^2_-\times S^{2r}$. 

In eq.~(\ref{index2reuc}) we have computed the index of the Dirac operator in $d=2r$-dimensional {\it flat} Euclidean space. Here we need the index of the Dirac operator constructed with the above gauge fields on the $(2r+2)$-dimensional manifold $S^2\times S^{2r}$. We know from the standard Atiyah-Singer index theorem that it is still given by the same expression (\ref{index2reuc}) or (\ref{index3}), except for the  replacement $2r\to 2r+2$, namely\footnote{
To compare with (\ref{index3}) which contains a factor ${1\over (4\pi)^{r+1}}$ rather than ${1\over (2\pi)^{r+1}}$, note that $\cF={1\over 2} \cF_{\m\n}\d x^\m \d x^\n$, and $\int\tr \cF^{r+1}={1\over 2^{r+1}}\int\tr \cF_{\m_1\m_2}\ldots \cF_{\m_{2r+1}\m_{2r+2}}\e^{\m_1\ldots\m_{2r+2}}\sqrt{g}\, \d^{2r+2}x$.
} 
\be\label{AStheorem}
\ind\big( i \Dsl_{2r+2}(\sa)\big) = {(-i)^{r+1}\over (r+1)! (2\pi)^{r+1}}
\int_{S^2\times S^{2r}} \tr \cF^{r+1} \ .
\ee
Of course, the precise prefactor $(-i)^{r+1}$ results from our normalization of the antihermitean Lie algebra generators and the definition of the Euclidean chirality matrix $\g_{\rm E}$.
The field strentgh $\cF$ is meant to be $\cF_+$ or $\cF_-$ as constructed above.

Note that on a general curved manifold $\cM$ there is also an extra factor $\wh A(\cM)$, which will play an important role in section \ref{gravmixed} when studying gravitational anomalies. Here, however, $\wh A(S^2\times S^{2r})=1$. Thus the index is given by a characteristic class as studied in sect. \ref{charclasses}.
Let us then use the descent equations on this $(2r+2)$-dimensional manifold. As emphasized in sect. \ref{charclasses}, the descent equations hold locally on each patch $S_+^2\times S^{2r}$ and $S_-^2\times S^{2r}$. Thus
\ba\label{indexCS}
\ind\big( i \Dsl_{2r+2}(\sa)\big)
&=&{(-i)^{r+1}\over (r+1)! (2\pi)^{r+1}} \Big[
\int_{S^2_+\times S^{2r}}\hskip-1.mm \d Q_{2r+1}(\sa_+,\cF_+) 
+\int_{S^2_-\times S^{2r}}\hskip-1.mm \d Q_{2r+1}(\sa_-,\cF_-) \Big]
\nonumber\\
&=& {(-i)^{r+1}\over (r+1)! (2\pi)^{r+1}}
\int_{S^1\times S^{2r}} \hskip-2.mm Q_{2r+1}(\sa_+,\cF_+)  \ ,
\ea
since the contribution from the boundary of $S_-^2\times S^{2r}$ vanishes. Indeed, since $\sa_-(x,\s,\t)=\sa(x)$ the Chern-Simons form $Q_{2r+1}(\sa_-,\cF_-)$ cannot have a $\d\t$-piece and, hence, $\int_{S^1\times S^{2r}} Q_{2r+1}(\sa_-,\cF)=0$. Note also that the orientation of $S_+^2$ and its boundary $S^1$ are such that $\int_{S_+^2} \d (\ldots) = + \int_{S^1} (\ldots)$.
On the other hand, on $S^1\times S^{2r}$ we have 
\be
\sa_+(x,\r,\t)=g^{-1}(x,\t) \big(\sa(x)+\d +\d\t \del_\t \big) g(x,\t)=\sa^\t(x) + \d\t \, v^\t(x) \ ,
\ee
with $\sa^\t$ and $v^\t(x)\equiv v(x,\t)$ defined in (\ref{Athetadef}) and (\ref{vthetadef}). Since $\sa_+$ is the gauge transformed of $\sa$ (in a $(2r+1)$-dimensional sense) we also have $\cF_+=g^{-1}(x,\t) \big( \d\sa+\sa^2\big) g(x,\t)=\d\sa^\t +\sa^\t\sa^\t\equiv\cF^\t$. It follows that 
\be
\int_{S^1\times S^{2r}} Q_{2r+1}(\sa_+,\cF_+)
=\int_{S^1\times S^{2r}} \Big[ Q_{2r+1}(\sa^\t+\d\t\, v^\t,\cF^\t)
- Q_{2r+1}(\sa^\t,\cF^\t) \Big] \ ,
\ee
where we added a second term that does not contribute to the integral since it does not have any $\d\t$-piece. We can now use (\ref{CSvar3}) to rewrite the integrand as $Q_{2r}^1(\d\t\, v^\t,\sa^\t,\cF^\t)=\d\t\, Q_{2r}^1(v^\t,\sa^\t,\cF^\t)$. Inserting this in (\ref{indexCS}) yields for the index
\be\label{indexCS2}
\ind\big( i \Dsl_{2r+2}(\sa)\big)
={(-i)^{r+1}\over (r+1)! (2\pi)^{r+1}}
\int_{S^1} \d\t \int_{S^{2r}} Q_{2r}^1(v^\t,\sa^\t,\cF^\t) \ .
\ee

Comparing equations (\ref{indanom}) and (\ref{indexCS2}) for the index yields
\be
\int_{S^1} \d\t \int_{S^{2r}} \d^{2r} z \sqrt{g_{\rm E}} \ (v^\t)^\a(x) \left( D^\t_j {\dd \G[\sa^\t]\over \dd (\sa^\t)_j (x)}\right)_\a 
= {(-i)^r\over (r+1)! (2\pi)^r}
\int_{S^1} \d\t \int_{S^{2r}} Q_{2r}^1(v^\t,\sa^\t,\cF^\t) \ .
\ee
Now the $\t$-dependence of $v^\t(x)=v(\t,x)$ is fairly arbitrary and we must have equality of the expressions even without integrating over $\t$. Then fixing some value of $\t$ and calling the corresponding values of $v^\t$, $\sa^\t$ and $\cF^\t$ simply $v$, $\sa$ and $\cF$ we finally get the (Euclidean) anomaly as the variation of the Euclidean effective action:
\be\label{anomwithcoeff1}
\dd_v \G_{\rm E}[\sa] =
- \int_{S^{2r}} \d^{2r} z \sqrt{g_{\rm E}} \ v^\a(x) \left( D_j {\dd \G_{\rm E}[\sa]\over \dd \sa_j (x)}\right)_\a 
= -{(-i)^r\over (r+1)! (2\pi)^r} \int_{S^{2r}} Q_{2r}^1(v,\sa,\cF) \ .
\ee
We have added a subscript ``E" to emphasize that we have been dealing with the Euclidean effective action throughout this section. For $r=2$, i.e. four dimensions, the r.h.s. is simply $+{1\over 24 \pi^2} \int Q_4^1(v,\sa,\cF)$, in perfect agreement with the result (\ref{Euclanom4D}) which was derived from our explicit one-loop computation!

We may rewrite the result (\ref{anomwithcoeff1}) also in terms of the BRST variation of the Euclidean effective action by replacing the gauge parameter $v$ by the ghost field:
\be\label{eucan2r}
s\, \G_{\rm E}[\sa]\equiv \ca_{\rm E}[w,\sa]=
-{(-i)^r\over (r+1)! (2\pi)^r} \int_{S^{2r}} Q_{2r}^1(w,\sa,\cF) \ .
\ee
Thus we have rederived  (the Euclidean version of) eq.~(\ref{anomdesc}) providing an explicit value for the coefficient $c$. However, before comparing with our previous results for the constant $c$ in four dimensions we have to continue back to Minkowski space.

We have argued above that the anomalous part of the Euclidean action should be purely imaginary. Let us check that this is indeed so. We have $Q_{2r}^1(v,\sa,\cF)=Q_{2r}^1(-i\e,-i A,-i F)$\break $=(-i)^{r+1} Q_{2r}^1(\e,A,F)$ with real $ Q_{2r}^1(\e,A,F)$ (cf.~(\ref{Q1AQ1sa})). Thus we can write eq.~(\ref{anomwithcoeff1}) as
\be\label{anomwithcoeff2}
\dd_\e \G_{\rm E}[A] =
- \int_{S^{2r}} \d^{2r} z \sqrt{g_{\rm E}} \ \e^\a(x) \left( D_j {\dd \G_{\rm E}[A]\over \dd A_j (x)}\right)_\a 
= {-i\, (-)^{r+1}\over (r+1)! (2\pi)^r} \int_{S^{2r}}  Q_{2r}^1(\e,A,F) \ ,
\ee
or using $\dd \G_{\rm E}=-i \int \hat I_{2r}^1$, cf.~eq.~(\ref{d25E})
\be\label{I2r1hat}
\hat I_{2r}^1 = {(-1)^{r+1}\over (r+1)! (2\pi)^r}  Q_{2r}^1(\e,A,F) \ .
\ee
Continuing to Minkowski signature as in (\ref{d25}) we get
\be\label{Minkanomformula}
\dd \G_M=\int \hat I_{2r}^1 = {(-1)^{r+1}\over (r+1)! (2\pi)^r} \int 
 Q_{2r}^1(\e,A,F) \ .
\ee
We can again check this against our one-loop computation for $r=2$, e.g.~in the form (\ref{anomdiffform}), and find perfect agreement. 
Let us emphasize that all this is for a positive chirality Dirac spinor, positive chirality being defined by our conventions for the chirality matrices. Also, in $2r=2\ {\rm mod}\ 8$ dimensions a chiral spinor can also obey a Majorana (reality) condition. Since two chiral Majorana spinors are equivalent to one chiral Dirac spinor, to get the anomaly for a single positive chirality Majorana spinor one must include an additional factor ${1\over 2}$ in (\ref{Minkanomformula}).

\newpage
\subsection{Gauge anomalies in $2r$ dimensions\label{gaugean2r}}

Let us summarize the results obtained so far for the anomalous variation of the effective action under non-abelian gauge transformations. We use the antihermitean fields $\sa, \cF$ and gauge transformation parameters $v$, related to the hermitean ones by $\sa=-iA$, $\cF=-iF$, $v=-i\e$, so that $\cF=\d\sa+\sa^2$ and $\dd\sa=\d v+[\sa,v]$, cf. eq.~(\ref{iredef}). The characteristic classes are $P_m=\tr \cF^m$ and the $Q_{2m-1}(\sa,\cF)$ and $Q_{2m-2}^1(v,\sa,\cF)$ are defined by the descent equations $P_m=\d Q_{2m-1}$ and $\dd Q_{2m-1}=\d Q_{2m-1}^1$, cf. eq.~(\ref{descentsummary}). Finally, positive or negative chirality is defined with respect to the Euclidean or Minkowskian chirality matrices $\g_{\rm E}$ or $\g_{\rm M}$ defined in 
(\ref{g5eucl}) and (\ref{g5mink}). Then
\be\label{an2rsum}
\begin{array}{|c|}
\hline\\
\quad\dd\, \G_{\rm E}=-i \int \hat I_{2r}^1
\quad \Leftrightarrow \quad
\ca[v,\sa]= \dd\, \G_{\rm M}= \int \hat I_{2r}^1 \ ,\quad
\\
\\
\hline
\end{array}
\ee
where $\hat I_{2r}^1$ is given by the descent equations
\be\label{Idescent}
\begin{array}{|c|}
\hline\\
\quad\hat I_{2r+2}=\d\, \hat I_{2r+1} \quad , \quad
\dd\, \hat I_{2r+1} = \d\, \hat I_{2r}^1 \ ,\quad
\\
\\
\hline
\end{array}
\ee
and $\hat I_{2r+2}$ is the characteristic class $P_{r+1}$ including the appropriate prefactor, cf eq.~(\ref{eucan2r}):
\be\label{I2r+2hat}
\hat I_{2r+2}={(-i)^{r+1}\over (r+1)! (2\pi)^r} P_{r+1}
={(-i)^{r+1}\over (r+1)! (2\pi)^r} \tr \cF^{r+1} \ .
\ee
Note that this is $2\pi$ times the ``index density", i.e. $\int\hat I_{2r+2}$ is $2\pi$ times the index:
\be\label{gaugeindexdensity}
\begin{array}{|c|}
\hline\\
\quad\hat I_{2r+2}= 2\pi \times \text{index density} \ .\quad
\\
\\
\hline
\end{array}
\ee
It is convenient to rewrite this in terms of the Chern character
\be\label{Chern}
{\rm ch}(\cF)= \tr \exp\left( {i\over 2\pi} \cF\right) 
= \sum_{k=0} {i^k\over k! (2\pi)^k} \tr \cF^k \ ,
\ee
which is a formal sum of forms of even degree, as
\be\label{I2r+2hatbis}
\begin{array}{|c|}
\hline\\
\quad\hat I_{2r+2}=(-)^{r+1}\, 2\pi\ [{\rm ch}(\cF)]_{2r+2}
=2\pi\ [{\rm ch}(-\cF)]_{2r+2} \ ,\quad
\\
\\
\hline
\end{array}
\ee
where the notation $[\ldots]_{2r+2}$ instructs us to pick only the form of degree $2r+2$.

\newpage
\section{Gravitational and mixed gauge-gravitational anomalies\label{gravmixed}}
\setcounter{equation}{0}

In this section, we will extend the formalism to be able to discuss also gravitational anomalies that may arise in generally covariant theories. These are anomalies of the diffeomorphism symmetry or, alternatively, of the local Lorentz symmetry. A classical reference on gravitational anomalies is \cite{AGW}.

\subsection{Some basic formalism for describing gravity\label{basicform}}

If the space-time has a non-trivial geometry described by a metric $g_{\m\n}$ we introduce local orthonormal frames (``vielbeine") $e_\m^a$ such that
\be\label{getarel}
g_{\m\n}=e_\m^a e_\n^b\, \eta_{ab} \ ,
\ee
where $\eta_{ab}$ is the flat Minkiowski-space metric ${\rm diag}(-+\ldots +)$ or in Euclidean signature simply $\dd_{ab}$. One can then consider diffeomorphisms and local Lorentz transformations separately. Accordingly one has coordinate tensors
$\S^{\m\n}_{\ \ \r}$, $g_{\m\n}$, etc and frame tensors $\S^{ab}_{\ \ c}$, $\eta_{ab}$, etc. They are related by (\ref{getarel}) and
\be
\S^{ab}_{\ \ c}= e^a_\m e^b_\n E^\r_c\, \S^{\m\n}_{\ \ \r}
\quad {\rm where}\quad
E_c^\r e^c_\s = \dd_\s^\r \ , \quad e^a_\s E^\s_b=\dd^a_b \ .
\ee

\begin{itemize}
\item
One defines the \underline{covariant derivative ${\rm D}=\d + [\o, \ ]$ for frame tensors}  as
\be
({\rm D}\S)^{ab}_{\ \ c} = \d \S^{ab}_{\ \ c} +\o^a_{\ d}\,\S^{db}_{\ \ c}
+\o^b_{\ d}\, \S^{ad}_{\ \ c} - (-)^p\, \S^{ab}_{\ \ d}\,\o^d_{\ c} \ ,
\ee
where $p$ is the form degree of $\S^{ab}_{\ c}$.
Here $\o^a_{\ b}= (\o_\m)^a_{\ b}\, \d x^\m$ is a connection 1-form analogous to $(\sa^\a t_\a)^k_{\ l}$. As a matrix, $\o$ is an element of the Lie algebra of $SO(d-1,1)$, resp. $SO(d)$.
\item
The \underline{covariant derivative $\nabla=\d+[\G, \ ]$ for coordinate tensors} is defined as
\be
(\nabla \S)^{\m \n}_{\ \ \r}= \d\S^{\m\n}_{\ \ \r} +\G^\m_{\ \s} \S^{\s\n}_{\ \ \r} + \G^\n_{\ \s} \S^{\m\s}_{\ \ \r} -(-)^p \S^{\m\n}_{\ \ \s} \G^\s_{\ \r} \ ,
\ee
where $\nabla=\d x^\m\, \nabla_\m$ and $\G^\n_{\ \r}= \d x^\m\, \G_{\m\r}^\n$.
\item
The \underline{compatibility} of the two definitions of covariant derivative, i.e. $({\rm D}\S)^{ab}_{\ \ c} =e^a_\m e^b_\n E^\r_c\,     (\nabla \S)^{\m \n}_{\ \ \r}$  is guaranteed by requiring the covariant derivative of $e^a_\m$ to vanish:
\be
\d e^a_\m +\o^a_{\ b} e^b_\m - e^a_\r \G^\r_{\ \m}=0 \ .
\ee
This also implies $\nabla g_{\m\n}=0$.
\item
The \underline{curvature 2-form} is equivalently defined as
\be
R^a_{\ b} =\d \o^a_{\ b} + \o^a_{\ c}\o^c_{\ b} 
\quad {\rm or} \quad R=\d\o+\o^2
\ee
or as
\be
R^\m_{\ \n}=\d \G^\m_{\ \n}+\G^\m_{\ \r} \G^\r_{\ \n} 
\quad {\rm with}\quad
R^\m_{\ \n}=E^\m_a e^b_\n R^a_{\ b} \ .
\ee
Note that, just as $\o$, the curvature 2-form also is an $SO(d-1,1)$-, resp. $SO(d)$-matrix.
\item
A \underline{local Lorentz transformations} acts as
$e^a_\m \to (L^{-1})^a_{\ b} e^b_\m$ or, using  matrix notation,
\be\label{locLor}
e\to L^{-1} e \quad , \quad
\o\to L^{-1} (\o+\d) L \quad , \quad
R\to L^{-1} R L \ ,
\ee
or for infinitesimal local Lorentz transformations 
$L^a_{\ b}=\dd^a_{\ b} + \vh^a_{\ b}$ we have 
\be
\dd^{\rm L}_\vh e^a_\m=-\vh^a_{\ b} e^b_\m \quad , \quad
\dd^{\rm L}_\vh \o^a_{\ b} = \d \vh^a_{\ b} + \o^a_{\ c} \vh^c_{\ b} - \vh^a_{\ c} \o^c_{\ b} \quad , \quad
\dd^{\rm L}_\vh R^a_{\ b}= R^a_{\ c} \vh^c_{\ b} - \vh^a_{\ c} R^c_{\ b} \ ,
\ee 
or again in matrix notation
\be
\dd^{\rm L}_\vh e=-\vh e \quad , \quad
\dd^{\rm L}_\vh \o=\d\vh + [\o,\vh]={\rm D}\vh \quad , \quad
\dd^{\rm L}_\vh R = [R,\vh] \ .
\ee
\end{itemize}

\noindent
All this is completely analogous to the gauge theory case where we had $\dd^g_v \sa = \d v +[\sa,v]={\rm D} v$ and $\dd^g_v \cF=[\cF,v]$. Also we had $\sa=A_\m^\a (-i t_\a) \d x^\m$ with anti-hermitian $(-it_\a)$ while here we have $\o=\o_\m \d x^\m= \o_\m^{(cd)} T_{(cd)}\d x^\m$ with $\big( T_{(cd)}\big)^a_{\ b} = {1\over 2} (\dd^a_c\dd_{db} - \dd^a_d\dd_{cb})$ a real and antisymmetric, hence also anti-hermitian matrix.\footnote{
Strictly speaking, $T_{(cd)}$ is antisymmetric only in the Euclidean. In Minkowski signature, one must replace $\dd^{ca}$ by $\eta^{ca}$ etc, which introduces some extra minus signs.}
The $T_{(cd)}$ are the generators of the defining (fundamental) representation of $SO(d)$, resp. $SO(d-1,1)$.

We can then transpose almost all formula about characteristic classes and descent equations from gauge theory  to the case of local Lorentz symmetry by the replacements
\be
\sa\to\o \quad , \quad
\cF\to R \quad , \quad 
v\to\vh \ .
\ee
In particular one has (the superscript L stands for ``local Lorentz")
\be
P_m^{\rm L}=\tr R^m \quad , \quad 
P_m^{\rm L}= \d Q_{2m-1}(\o,R)\quad , \quad
\dd^{\rm L}_\vh Q_{2m-1}=\d Q_{2m-2}^1(\vh,\o,R) \ ,
\ee
with in particular $Q_3(\o,R)= \tr\big( \o\d\o+{2\over 3} \o^3\big)$ and $Q_2^1(\vh,\o,R)= \tr \vh\d\o$. An important property of the generators $T_{(cd)}$ is their antisymmetry which implies that the trace of an odd power vanishes, hence 
\be
P^{\rm L}_{2k+1}=\tr R^{2k+1} =0 \ .
\ee

\subsection{Purely gravitational anomalies of chiral spin ${1\over 2}$ fields\label{puregravano}}

Despite all the similarities between gauge transformations and local Lorentz transformations, a chiral fermion coupled to gravity in a non-trivial geometry is not just the same thing as a fermion coupled to an $SO(d)$ gauge field on a flat background. On the one hand, the fermions do not transform in the fundamental representation of $SO(d)$ or in any of its tensor products, but in a spin representation with generators
$T^{(cd)}_{\rm spin 1/2}={1\over 4} [\g^c,\g^d]={1\over 2} \g^{cd}$ so that the relevant Dirac operator is
\be\label{Diracgrav}
i\Dsl= i \g^\m \Big( \del_\m +A_\m^\a (-it_\a) +{1\over 4} \o^{cd}_\m \g^{cd}\Big) \ .
\ee
On the other hand, one must appropriately take into account the curved geometry, which manifests itself in $\g^\m=E^\m_a \g^a$, where the $\g^a$ are the (ordinary) flat space $\g$-matrices.

In the {\it absence} of a gauge field $A_\m$ the index of this Dirac operator on a curved manifold $\cM$ of dimension $d=2m$ is given by the so-called Dirac genus $\widehat A(\cM)\ $
\be\label{indexAroof}
{\rm ind} \big( i \Dsl\big\vert_{\sa=0}\big) = \int \left[\widehat A(\cM)\right]_{2m} \ ,
\ee
where $[\ldots]_{2m}$ indicates to pick only the form of degree $2m$, and
\ba\label{Diracgenus}
\widehat A(\cM)
&=&
1+{1\over (4\pi)^2}{1\over12}\tr R^2
+{1\over(4\pi)^4}\left[{1\over360}\tr R^4
+{1\over288}(\tr R^2)^2\right]\nonumber\\
&&+{1\over(4\pi)^6}\left[{1\over5670}\tr R^6
+{1\over4320}\tr R^4\tr R^2+{1\over10368}(\tr R^2)^3\right]+\ \ldots \ .
\ea
Here the traces are to be taken in the fundamental representation of $SO(d)$, resp. $SO(d-1,1)$.  Explicitly, one has $R^a_{\ b}={1\over 2} R^a_{\ b\m\n} \d x^\m \d x^\n$ and e.g. $\tr R^2= R^a_{\ b}R^b_{\ a} = {1\over 4} R^a_{\ b\m\n} R^b_{\ a\r\s} \d x^\m \d x^\n \d x^\r \d x^\s$, etc. Note again that $\widehat A(\cM)$ only involves forms of degrees that are  multiples of 4.

The purely gravitational anomalies in $d=2r$ dimensions are again related via descent equations to the index of the Dirac operator extended to $2r+2$ dimensions by a construction similar to the one given above for gauge theories. We will not go into details here but only mention that this construction does ``not add any additional curvature contribution" and the descent is done with $[\widehat A(\cM_{2r})]_{2r+2}\,$. This is only non-vanishing if $2r+2$ is a multiple of 4, i.e. $2r+2=4(k+1)$ or $d=2r=4k+2$
Hence it is only for these dimensions that one can have purely gravitational anomalies, i.e. in $d=2,6,10,\ldots$ dimensions. In particular, in 4 dimensions there are no purely gravitational anomalies.
\be\label{gravanomdim}
\begin{array}{|c|}
\hline\\
\text{Purely gravitational anomalies only exist in $d=2r=4k+2$ dimensions.}
\\
\\
\hline
\end{array}
\ee
So let the dimension be given by (\ref{gravanomdim}). The purely gravitational anomaly for a positive chirality spin-${1\over 2}$ fermion then is given by
\be
\dd^{\rm L}\, \G_{\rm E}=-i \int \hat I_{2r}^1({\rm grav})
\quad \Leftrightarrow \quad
\ca[\vh,\o]= \dd^{\rm L}\, \G_{\rm M}= \int \hat I_{2r}^1({\rm grav}) \ ,
\ee
where $\hat I_{2r}^1({\rm grav})$ is given by the descent equations
\be\label{Idescentgrav}
\hat I_{2r+2}({\rm grav})=\d\, \hat I_{2r+1}({\rm grav}) 
\quad , \quad
\dd^{\rm L}\, \hat I_{2r+1}({\rm grav}) 
= \d\, \hat I_{2r}^1({\rm grav}) \ ,
\ee
and
\be
\hat I_{2r+2}({\rm grav})=2\pi \left[\widehat A(\cM_{2r})\right]_{2r+2} \ .
\ee
Note that this is again of the form (\ref{gaugeindexdensity}), i.e. $2\pi$ times the relevant index density. We insisted that this anomaly was for a chiral spin-${1\over 2}$ fermion since there are other fields that can have gravitational anomalies, although they do not couple to the gauge field. They will be considered in a later subsection.

Let us look at the example of 2 dimensions. Purely gravitational anomalies in 2 dimensions play an important role in string theory in relation with the conformal anomaly on the world sheet. If we let $r=1$, i.e. $k=0$ in the above equations, we get 
\be\label{grav2dim}
\hat I_4({\rm grav})= {1\over 96 \pi} \tr R^2
\quad ,\quad
\hat I_{3}({\rm grav})= {1\over 96 \pi}\tr(\o\d\o+{2\over 3}\o^3)
\quad ,\quad
\hat I_2^1({\rm grav})= {1\over 96 \pi}\tr \vh\d\o \ .
\ee
We conclude that the anomalous variation under a local Lorentz transformation with parameter $\vh$ of the  Minkowskian effective action for a positive chirality spin ${1\over 2}$ fermion in two dimensions is
\be\label{grav2dact}
\dd^{\rm L} \G_{\rm M}= {1\over 96 \pi} \int \tr \vh\d\o \ .
\ee
Just as in section \ref{noneffa} one can relate this anomaly to the non-conservation of the corresponding current which is the energy-momentum tensor. The latter actually is the current that corresponds to diffeomorphisms rather than local Lorentz transformations. However, one can show that anomalies with respect to local Lorentz transformations are equivalent to anomalies with respect to diffeomorphisms, and (\ref{grav2dact}) indeed corresponds to
\be
\del_\m \langle T^{\m\n}\rangle\big\vert_\o \ne 0 \ ,
\ee
or more explicitly in 2 dimensions
\be
{\del\over \del x^\m} \langle T^{\m\n}(x) T^{\r\s}(y) \rangle\big\vert_{\o=0} \ne 0 \ .
\ee
(Of course, this last result being for $\o=0$ even holds on flat two-dimensional Minkowski space.)
The precise form of the right hand side can be obtained from (\ref{grav2dact}) along the same lines as in section \ref{noneffa}. On the other hand it is not difficult  to compute the two-point function of the energy-momentum tensor for a two-dimensional chiral fermion explicitly and check the result. (This computation, together with an instructive discussion, can be found in \cite{AGW}.)

\subsection{Mixed gauge-gravitational anomalies\label{mixeggravabo}}

\subsubsection{Arbitrary (even) dimensions\label{arbevendim}}

In the presence of both gauge fields and gravity, the index of the Dirac operator (\ref{Diracgrav}) is simply given by (cf. eqs.(\ref{AStheorem}), (\ref{Chern}) and (\ref{indexAroof}))
\be\label{indexmixed}
{\rm ind} \big( i \Dsl(\sa,\o)_{2r+2}\big) = (-)^{r+1}\int \left[\widehat A(\cM)\ {\rm ch}(\cF) \right]_{2r+2} 
= \int \left[\widehat A(\cM)\ {\rm ch}(-\cF) \right]_{2r+2} \ ,
\ee
We noted above that the minus sign appearing here for even $r$ is required by our precise conventions (definition of the chirality matrices, generators of the Lie algebra, etc.) and is confirmed by the explicit triangle computation for $r=2$.

In complete analogy with the above, the full gauge, gravitational and mixed gauge-gravitational anomalies for a positive chirality spin-${1\over 2}$ fermion in a representation $\cR$ of the gauge group then are again given by descent with respect to the index density:
\be\label{an2rsummixed}
\begin{array}{|c|}
\hline\\
\ (\dd^{\rm gauge}+\dd^{\rm L})\, \G_{\rm E}=-i \int \hat I_{2r}^1({\rm gauge,grav})
\\
\\
\quad \Leftrightarrow \quad
\\
\\
\ca[v,\sa,\vh,\o]= (\dd^{\rm gauge}+\dd^{\rm L})\, \G_{\rm M}= \int \hat I_{2r}^1({\rm gauge,grav}) \ ,\
\\
\\
\hline
\end{array}
\ee
where $\hat I_{2r}^1({\rm gauge,grav})$ is given by the descent equations
\be\label{Idescentmixed}
\begin{array}{|ccc|}
\hline & &\\
\quad \hat I_{2r+2}({\rm gauge,grav})&=&\d\, \hat I_{2r+1}({\rm gauge,grav}) \ , \quad
\\
& &\\
\quad 
(\dd^{\rm gauge}+\dd^{\rm L})\, \hat I_{2r+1}({\rm gauge,grav}) 
&=& \d\, \hat I_{2r}^1({\rm gauge,grav}) \ ,\quad
\\
& &\\
\hline
\end{array}
\ee
and
\be\label{gaugegravIhat}
\begin{array}{|c|}
\hline\\
\quad \hat I_{2r+2}({\rm gauge,grav}) 
= 2\pi \left[\widehat A(\cM)\ {\rm ch}(-\cF) \right]_{2r+2}\ , \quad
\\
\\
\hline
\end{array}
\ee
where ${\rm ch}(-\cF) =\trR \exp(-{i\over 2\pi} \cF)$, and $\widehat A$ was given in (\ref{Diracgenus}).

\subsubsection{The example of 4 dimensions\label{4dimex}}

Although there are no purely gravitational anomalies in four dimensions, there are mixed gauge-gravitational anomalies since
\ba\label{mixed4}
&&\hskip-1.5cm2\pi \left[\widehat A(\cM)\ {\rm ch}(-\cF) \right]_{6}
\nonumber\\
&&= 2\pi \Bigg[\left(1+{1\over (4\pi)^2}{1\over12}\tr R^2\right)
\left( 1-{i\over 2\pi}\trR \cF -{1\over 2(2\pi)^2} \trR \cF^2 +{i\over 6(2\pi)^3}\trR \cF^3 \right) \Bigg]_6
\nonumber\\
&&=\ {i\over 24\pi^2}\ \trR \cF^3\ -\ {i\over 192 \pi^2}\ \trR\cF\ \tr R^2 \ .
\ea
The first term just reproduces the anomaly polynomial for pure gauge anomalies as extensively discussed in sect. \ref{descsec}, see e.g. eq.~(\ref{I6pos}). The second term represents the mixed gauge-gravitational anomaly. Since for any simple Lie algebra $\trR \cF=0$, only the $U(1)$ parts can contribute, in which case 
\be
\trR \cF=\trR F (-i t)= -i\, \sum_{i,s} q_i^{(s)}\ F^{(s)} \ ,
\ee
where the superscript $s$ labels different $U(1)$ factors (if present) and $i$ runs over the different (positive chirality) ``individual fields"\footnote{
As discussed in sect. \ref{frecan}, negative chirality particles are treated as positive chirality anti-particles so that the relevant minus sign appears through the opposite $U(1)$ charges $q_i^{(s)}$.} 
that are contained in the representation $\cR$. Hence the mixed part of (\ref{mixed4}) is
\be\label{mixed4a}
\hat I_6({\rm mixed})=2\pi \left[\widehat A(\cM)\ {\rm ch}(-\cF) \right]_{6}^{\rm mixed}
=-{1\over 192\pi^2}\, \Big(\sum_{i,s} q_i^{(s)}\, F^{(s)}\Big) \, \tr R^2 \ .
\ee

Next, we apply the descent equations to obtain  the form of the anomaly. As already noted in sect. \ref{formaldev}, there is no unique way to do the descent, and in sect. \ref{secWZBRST} we interpreted this ambiguity as the possibility to change the form of the anomaly by adding a local counterterm to the effective action (without being able to remove a relevant anomaly altogether). The mixed gauge-gravitational anomaly in 4 dimensions provides a nice example where different ways to do the descent will either lead to an effective action that is not gauge invariant or to one that is not local Lorentz invariant. Below, we will exhibit a local counterterm that allows to interpolate between both possibilities. A first possibility to do the descent is
\ba\label{gaugedesc}
\hat I_6({\rm mixed})&=& \d \left( -{1\over 192\pi^2}\, \Big(\sum_{i,s} q_i^{(s)} A^{(s)}\Big) \, \tr R^2 \right)
\ ,
\nonumber\\
 \dd \left( -{1\over 192\pi^2}\, \Big(\sum_{i,s} q_i^{(s)} A^{(s)}\Big) \, \tr R^2 \right)
&=& \d \left( -{1\over 192\pi^2}\, \Big(\sum_{i,s} q_i^{(s)} \e^{(s)}\Big) \, \tr R^2 \right) \ .
\ea
Of course, $\dd$ is meant to be $\dd^{\rm gauge} + \dd^{\rm L}$, but the expression is invariant under local Lorentz transformations, only $\dd^{\rm gauge}$ is effective. One may call this the descent in the gauge sector.
Alternatively, one may write
\ba\label{gravdesc}
\hat I_6({\rm mixed})&=& \d \left( -{1\over 192\pi^2}\, \Big(\sum_{i,s} q_i^{(s)} F^{(s)}\Big) \, \tr \big( \o\d\o +{2\over 3}\o^3\big) \right)
\ ,
\nonumber\\
 \dd \left( -{1\over 192\pi^2}\, \Big(\sum_{i,s} q_i^{(s)} F^{(s)}\Big) \, \tr \big( \o\d\o +{2\over 3}\o^3\big)  \right)
&=& \d \left( -{1\over 192\pi^2}\, \Big(\sum_{i,s} q_i^{(s)} F^{(s)}\Big) \, \tr \vh \d\o \right) \ .
\ea
This time in $\dd=\dd^{\rm gauge} + \dd^{\rm L}$,  only $\dd^{\rm L}$ is effective, and we may call this the gravitational descent.
We conclude that there are (at least) two ways to define the effective action such that either
\be
\dd \,\G_{\rm M}^{(1)}=
-{1\over 192\pi^2}\int \Big(\sum_{i,s} q_i^{(s)} \e^{(s)}\Big) \, \tr R^2 \ ,
\ee
or
\be
\dd \,\G_{\rm M}^{(2)}=
-{1\over 192\pi^2}\int \Big(\sum_{i,s} q_i^{(s)} F^{(s)}\Big) \, \tr \vh \d\o \ .
\ee
Clearly, $\G_{\rm M}^{(1)}$ is invariant under local Lorentz transformations, but is not gauge invariant, and $\G_{\rm M}^{(2)}$ is gauge invariant but not invariant under local Lorentz transformations.
We can interpolate between both by adding to the effective action a local counterterm
\be
\D \G_{\rm M}= -{1\over 196\pi^2} \int \Big(\sum_{i,s} q_i^{(s)} A^{(s)}\Big) \,  \tr \big( \o\d\o +{2\over 3}\o^3\big) \ .
\ee
Indeed, we have
\ba
\dd\, \D \G_{\rm M}&=& -{1\over 196\pi^2} \int \left[\Big(\sum_{i,s} q_i^{(s)} \d\e^{(s)}\Big) \,  \tr \big( \o\d\o +{2\over 3}\o^3\big) 
+\Big(\sum_{i,s} q_i^{(s)} \ A^{(s)}\Big) \d \tr \vh \d\o \right] 
\nonumber\\
&=& {1\over 196\pi^2} \int \left[\Big(\sum_{i,s} q_i^{(s)} \e^{(s)}\Big) \,  \tr R^2 
-\Big(\sum_{i,s} q_i^{(s)} F^{(s)}\Big)  \tr \vh \d\o \right]
\ ,
\ea
so that
\be
\dd \left(\G_{\rm M}^{(1)}+\, \D \G_{\rm M}\right)=\dd\, \G_{\rm M}^{(2)} \ .
\ee
More generally, we may add a counterterm $\l \D \G_{\rm M}$ for arbitrary real $\l$ resulting in an effective action that is neither gauge nor local Lorentz invariant. Nevertheless, the anomaly is always characterised by the unique $\hat I_6({\rm mixed})$.

Recall that the issue of anomaly cancellation can be discussed before doing the descent, simply in terms of the invariant polynomial $\hat I_6({\rm mixed})$. The condition for $\hat I_6({\rm mixed})$ to vanish is simply $\sum_{i,s} q_i^{(s)}=0$ for every $U(1)_{(s)}$. This is the same condition as for the vanishing of any $U(1)\times G\times G$ gauge anomaly, and in particular is satisfied in the standard model, as discussed in sect. \ref{frecan}.

\subsection{Other chiral fields with gravitational anomalies\label{Othchirfi}}

Finally, we give the relevant anomaly polynomials for the two other sorts of chiral fields that can give rise to anomalies. These are chiral spin-${3\over 2}$ fermions, like the gravitino, and self-dual or anti self-dual antisymmetric tensor fields in $4k+2$ dimensions. To understand why the latter give rise to anomalies it is enough to recall that in $2r=4k+2$ dimensions such tensor fields can be constructed from a pair of spin-${1\over 2}$ fields of the same chirality (while in $4k$ dimensions it would require two spinors of opposite chirality). By definition, a self-dual antisymmetric tensor field $H^{\m_1\ldots\m_r}$  satisfies
\be\label{selfdual}
H_{\rm M}^{\m_1\ldots\m_r}={1\over r!}\, \e^{\m_1\ldots\m_r}_{\hskip8.mm \n_1\ldots\n_r}\, H_{\rm M}^{\n_1\ldots\n_r} 
\quad \Leftrightarrow\quad
H_{\rm E}^{\m_1\ldots\m_r}={i\over r!}\, \e^{\m_1\ldots\m_r}_{{\rm E}\hskip6.mm \n_1\ldots\n_r}\, H_{\rm E}^{\n_1\ldots\n_r} \ ,
\ee
where the subscripts M and E refer to Minkowski and Euclidean signature, respectively. For anti self-dual fields one would have an extra minus sign. If $\chi$ and $\p$ are two positive chirality spinors in $2r=4k+2$ dimensions it is indeed easy to see\footnote{
It is enough to show in flat Minkowski space that this $H_{\rm M}$ satisfies $H_{\rm M}^{01\ldots (r-1)}={1\over r!} \e^{01\ldots (r-1)}_{\hskip12.mm \n_r\ldots \n_{2r-1}} H_{\rm M}^{\n_r\ldots \n_{2r-1}}$ $= + H_{\rm M}^{r\ldots (2r-1)}$. But (using $r=2k+1$) we have
\ba
H_{\rm M}^{01\ldots (r-1)}
&=& \pb \g^0 \ldots \g^{r-1}\chi
= \pb \g^0 \ldots \g^{r-1}\g_{\rm M}\chi
= i^{r+1}\pb \g^0 \ldots \g^{r-1} \g^0 \ldots \g^{r-1} \g^r\ldots \g^{2r-1}\chi
\nonumber\\
&=&i^{r+1} (-)^{r(r-1)/2} (-) \pb \g^r\ldots \g^{2r-1}\chi
=+\, \pb \g^r\ldots \g^{2r-1}\chi
=\, +\, H_{\rm M}^{r\ldots (2r-1)}\ .
\ea
} 
that
\be
H_{\rm M}^{\m_1\ldots\m_r}= \pb \g^{[\m_1} \ldots \g^{\m_r]}\chi
\ee
is indeed self-dual, i.e. satisfies (\ref{selfdual}).

Consider first a positive chirality
spin-${3\over2}$ field. Such a field is obtained from a positive
chirality spin-${1\over2}$ field with an extra vector index by
subtracting the spin-${1\over2}$ part. The extra vector index can be treated in analogy with an $SO(d)$ gauge symmetry and leads
to an additional factor for the index density
\be
\tr \exp\left({i\over2\pi}{1\over2}R_{ab}T^{ab}\right)=\tr
\exp\left({i\over2\pi}R \right)
\end{equation}
since the vector representation is
$(T^{ab})_{cd}=\delta^a_c\delta^b_d-\delta^a_d\delta^b_c$. Hence the index for the relevant $2m$-dimensional operator is
\begin{equation}
{\rm ind}(iD_{3\over2})=\int_{M_{2m}}\left[\hat A(M_{2m})\left(\tr
\exp\left({i\over2\pi}R \right)-1\right)\ {\rm ch}(-\cF)\right]_{2m}.
\ee
Note that for a gravitino the factor ${\rm ch}(-\cF)$ is absent since it does not couple to any gauge group.

Next, consider a self-dual rank $r$ antisymmetric tensor field $H$ in $2r=4k+2$ dimensions.
Such antisymmetric tensors fields cannot
couple to the gauge group. Since it can be constructed from a
pair of positive chirality spinors, it turns out that the index is simply
$\hat A(M_{2m})$ multiplied by $\tr
\exp\left({i\over2\pi}{1\over2}R_{ab}T^{ab}\right)$, where
$T^{ab}={1\over2}\gamma^{ab}$ as appropriate for the
spin-${1\over2}$ representation. Note that the trace over the
spinor representation gives a factor $2^n$ in $2n$ dimensions.
There is also an additional factor ${1\over2}$ from the chirality
projector of this second spinor and another factor ${1\over2}$
from a reality constraint ($H$ is real) so that
\be
{\rm ind}(iD_A)
={1\over4}\int_{M_{2m}} \left[\hat A(M_{2m})
\tr \exp\left({i\over2\pi}{1\over4}R_{ab}\g^{ab}\right)
\right]_{2m}
={1\over4}\int_{M_{2n}}[L(M)]_{2m}.
\label{indDA}
\ee
$L(M)$ is called the Hirzebruch polynomial, and the subscript on
$D_A$ stands for ``antisymmetric tensor".
(Note that, while $\hat A(M_{2m})\tr
\exp\left({i\over2\pi}{1\over4}R_{ab}\gamma^{ab}\right)$ carries
an overall factor $2^m$,
$L(M_{2m})$ has a factor $2^k$ in front of each
$2k$-form part. It is only for $k=m$ that they coincide.)

Explicitly one has
\ba\label{ArooftrR}
\hat A(M_{2m})\left(\tr {\rm e}^{{i\over2\pi}R} -1\right)
\hskip-2.mm&=&\hskip-2.mm
(2m-1)+{1\over (4\pi)^2}{2m-25\over12}\tr R^2
+{1\over(4\pi)^4}\left[{2m+239\over360}\tr R^4
+{2m-49\over288}(\tr R^2)^2\right]
\nonumber\\
&&\hskip-1.5cm+{1\over(4\pi)^6}\left[{2m-505\over5670}\tr R^6
+{2m+215\over4320}\tr R^4\tr R^2
+{2m-73\over10368}(\tr R^2)^3\right]+\ldots
\ea
and
\ba\label{Hirzebruch}
L(M_{2m})
&=&1-{1\over (2\pi)^2}{1\over6}\tr R^2
+{1\over(2\pi)^4}\left[-{7\over180}\tr R^4
+{1\over72}(\tr R^2)^2\right]
\nonumber\\
&&
+\ {1\over(2\pi)^6}\left[-{31\over2835}\tr R^6
+{7\over1080}\tr R^4\tr R^2-{1\over1296}(\tr R^2)^3\right]+\ldots
\ea
The corresponding anomaly polynomials then are
(for positive chirality, respectively  self-dual antisymmetric tensors)
\ba\label{Ihathalf}
\hat I_{2r+2}^{spin{3\over2}}&=&2\pi \left[\hat A(M_{2r})\
\left(\tr \exp\left({i\over2\pi}R\right)-1\right)\ {\rm
ch}(-\cF)\right]_{2r+2}\label{Ihat3half}\\
\hat I_{2r+2}^{A}&=&2\pi \left[\left(-{1\over2}\right){1\over4}\
L(M_{2r})\right]_{2r+2}.\label{IhatA}
\ea
The last equation contains an extra factor
$\left(-{1\over2}\right)$ with respect to the index (\ref{indDA}).
The minus sign takes into account the Bose rather than Fermi
statistics, and the $1\over2$ corrects the $2^{r+1}$ to $2^r$
which is the appropriate dimension of the spinor representation on
$M_{2r}$ while the index is computed in $2r+2$ dimensions. Note again
that in the cases of interest, the spin-${3\over2}$ gravitino is
not charged under the gauge group and the
factor of ${\rm ch}(-\cF)$ then is absent in (\ref{Ihat3half}).

\newpage
\section{Anomaly cancellation in ten-dimensional type IIB supergravity and in the (field theory limits of) type I and heterotic superstrings\label{anomcanc}}
\setcounter{equation}{0}

In this last section, we will derive some prominent examples of anomalies and their cancellations in certain ten-dimensional quantum field theories. The conditions for anomaly cancellation in these ten-dimensional theories typically constitute over-determined systems of equations. Quite amazingly, these systems nevertheless not only admit non-trivial solutions, but these solutions also are relatively simple and actually correspond (in most cases) to the low-energy limits of the known ten-dimensional superstring theories \cite{GSW,POL}. 

\subsection{The ten-dimensional anomaly polynomials\label{tendimanopol}}

As repeatedly emphasized, relevant anomalies are characterized by a non-vanishing anomaly polynomial $\hat I_{2r+2}$ which is the sum of all individual contributions $\sum_j \hat I_{2r+2}({\rm field}\ j)$.
So far we have given these individual contributions for positive chirality (resp. self-dual tensor fields). The contributions for negative chirality (anti self-dual tensor fields) have the opposite signs. Recall that in 4 mod 4 dimensions, particles and antiparticles have opposite chirality and we could describe a negative chirality particle as a positive chirality antiparticle in the appropriate charge conjugate representation of the gauge group, cf. eq.~(\ref{chconjugrepres}).  On the other hand, in 2 mod 4 dimensions, particles and antiparticles have the same chirality, and we must treat negative chirality (anti)particles as such. In 2 mod 8 dimensions, fermions can be Majorana-Weyl, being their own anti-particles.

Specializing the previous formulae to ten dimensions, i.e. $r=5$, and a gravitino which is a spin-${3\over 2}$ field without gauge interactions, we get from (\ref{ArooftrR}) to (\ref{IhatA})
\be\label{I12}
\hat I_{12}^{gravitino}={1\over 64(2\pi)^5} \left[
-{11\over126}\tr R^6
+{5\over96}\tr R^4\tr R^2
-{7\over1152}(\tr R^2)^3
\right]
\ee
and
\be
\hat I_{12}^{A}={1\over8 (2\pi)^5}\left[{31\over2835}\tr R^6
-{7\over1080}\tr R^4\tr R^2+{1\over1296}(\tr R^2)^3\right] \ ,
\ee
and for a spin-${1\over 2}$ field from (\ref{gaugegravIhat})
\ba\label{I12spinhalf}
\hat I_{12}^{spin\, 1/2}&=&
{1\over 64 (2\pi)^5} \big(\trR 1\big)
\left[{1\over5670}\tr R^6
+{1\over4320}\tr R^4\tr R^2+{1\over10368}(\tr R^2)^3\right]
\nonumber\\
&-&{1\over 32 (2\pi)^5} \big(\trR \cF^2\big)
\left[{1\over360}\tr R^4
+{1\over288}(\tr R^2)^2\right]
\nonumber\\
&+&{1\over 1152 (2\pi)^5} \big(\trR \cF^4\big)\ \tr R^2 
\ -\ {1\over 720 (2\pi)^5} \trR \cF^6 \ .
\ea
As just mentioned, in ten dimensions one can have Majorana-Weyl spinors, i.e. chiral fields that in addition obey a reality condition. All our $\hat I$ are for complex positive chirality spinors. If the spinors are Majorana positive chirality one has to include an additional factor ${1\over 2}$ in $\hat I_{12}^{spin\, 1/2}$ and in $\hat I_{12}^{gravitino}$. Of course, there is no extra factor of ${1\over 2}$ for 
$\hat I_{12}^{A}$ (we already included one in (\ref{IhatA}) to take into account the reality of the antisymmetric tensor field).

In a given theory containing various chiral or (anti) self-dual fields, one has to add all individual anomaly polynomials of these fields to get the total anomaly polynomial. The theory is free of anomalies, i.e. the anomalies cancel, if this total anomaly polynomial vanishes.

\subsection{Type IIB supergravity in ten dimensions\label{IIsugra}}

Consider a ten-dimensional theory that may involve a certain number of Majorana-Weyl gravitinos (spin ${3\over 2}$), Majorana-Weyl spin ${1\over 2}$ fields and (anti) self-dual antisymmetric tensor fields. We assume here that there is {\it no gauge group}, as is the case for IIB supergravity.\footnote{
A word on terminology: in any dimension, a theory with the minimal amount of supersymmetry is referred to as ${\cal N}=1$, with twice the minimal amount of supersymmetry as ${\cal N}=2$, etc. In ten dimensions, an ${\cal N}=1$ supergravity has one Majorana-Weyl gravitino and an ${\cal N}=2$ supergravity has two Majorana-Weyl gravitinos. In the latter case, if both gravitinos have opposite chirality, the supergravity is called IIA (it is non-chiral and trivially free of anomalies), while if the two gravitinos have the same chirality, the supergravity is called IIB.
} 
The case with gauge group is more complicated and will be treated in a later subsection. Let
$n_{3/2}$, resp. $n_{1/2}$ be the number of positive chirality Majorana-Weyl gravitinos, resp. spin ${1\over 2}$ fields,
minus the number of negative chirality ones. Similarly let $n_A$ be the number of self-dual minus the number of anti self-dual antisymmetric tensor fields. Then the total anomaly polynomial for such a theory is
\ba
\hat I_{12}^{\rm total}(n_{3/2},n_{1/2},n_A)
\hskip-2.mm&=&\hskip-2.mm{n_{3/2}\over 2}\ \hat I_{12}^{gravitino} 
+ {n_{1/2}\over 2}\ \hat I_{12}^{spin\, 1/2} 
+n_A\ \hat I_{12}^{A}
\nonumber\\
&&\hskip-1.5cm={1\over 128 (2\pi)^5}\Bigg\{ { -495 n_{3/2} + n_{1/2} +992 n_A \over 5670} \tr R^6 
+\,{ 225 n_{3/2} + n_{1/2} -448 n_A\over 4320} \tr R^4\, \tr R^2
\nonumber\\
&&\hskip1.7cm+\,{-63 n_{3/2} +n_{1/2} + 128 n_A \over 10368} (\tr R^2)^3
\Bigg\} \ .
\ea
The vanishing of this total anomaly polynomial constitutes a homogenous linear system of 3 equations in 3 variables. In general such a system has only the trivial solution $n_{3/2}=n_{1/2}=n_A=0$.
Amazingly, however, as first observed in \cite{AGW}, the 3 equations are not linearly independent and do admit  non-trivial solutions. Moreover, these solutions are very simple, namely
\be\label{anom10dsol}
n_{3/2}=2 n_A \quad , \quad n_{1/2}=-2 n_A \ .
\ee
The simplest case, $n_A=1$ corresponds to one self-dual antisymmetric tensor field, a pair of positive chirality Majorana-Weyl gravitinos and a pair of negative chirality Majorana-Weyl spin-${1\over 2}$ fields. This is precisely the (chiral) field content of ten-dimensional type IIB supergravity! Hence, not only is type IIB supergravity in ten dimensions free of gravitational anomalies, it is also the simplest chiral theory in ten dimensions in which gravitational anomaly cancellation occurs.

\newpage
\subsection{Anomaly cancellation by inflow and Green-Schwarz mechanism\label{inflow}}

In sect.~\ref{relanom} we have discussed that a relevant anomaly is one that cannot be removed by adding a local counterterm to the (effective) action. In sect.~\ref{reformBRST} we have seen that the freedom to add a local counterterm corresponds to changing the representative within the {\it same} BRST cohomology class. Finally in sect.~\ref{descsec} we have seen that the descent equations always associate the {\it same} invariant polynomial to different forms of the anomaly that differ only by addition of a local counterterm. This showed that a non-vanishing anomaly polynomial $\hat I_{2r+2}$ indicates a relevant anomaly, i.e. an anomaly that cannot be removed by a local counterterm.

In all these considerations it was always understood that the local counterterm is constructed solely from the gauge fields $A$ and $F$ (and the gravitational connection $\o$ and curvature $R$), the fermions having been integrated out. There is, however, the possibility that the theory contains one or more extra fields that do transform under gauge (or local Lorentz) transformations and that their classical action contains non gauge invariant terms, or that we add such terms as counterterms.
 
The simplest example is a so-called axion field $a$ in a 4 dimensional $U(1)$ gauge theory. The gauge anomaly in such a $U(1)$ theory simply is 
\be\label{U1anom}
\dd \G[A]=-{1\over 24\pi^2} \sum_j q_j^3\int\e\, \d A \d A 
= -{1\over 24\pi^2} \sum_j q_j^3\int\e\, F F \qquad (\text{for a $U(1)$-theory})\ .
\ee 
Suppose we add a ``counterterm" 
\be\label{axioncounter}
\D \G[a,A]={1\over 24\pi^2} \sum_j q_j^3\int a\, F F  \ ,
\ee 
and ``declare" that the axion field $a$ transforms under a gauge transformation as 
\be\label{axionshift}
a\to a+\e \ .
\ee 
Then obviously $\dd \D \G[a,A]={1\over 24\pi^2} \sum_j q_j^3\int\e\, F F$ and 
\be\label{axioncancel}
\dd \big( \G[A]+\D\G[a,A]\big) =0 \ .
\ee
Does this mean that we can always eliminate the anomaly (\ref{U1anom}) simply this way? Of course, things are not this simple. To add a term like (\ref{axioncounter}) one needs to have a good reason to introduce an additional field $a$ which should, in principle, correspond to an observable particle. Also, the naive kinetic term for such a field, $\int (-\del_\m a \del^\m a - M^2 a^2)$ is {\it not}  invariant under the transformation (\ref{axionshift}), although its variation vanishes on-shell.

One should note an important point. The anomaly is a one-loop effect. 
This manifests itself in (\ref{U1anom}) as the coefficient $\sum_j q_j^3$ which is smaller than a typical interaction term in the classical action (that would be $\sim q$) by a factor of a charge squared, or coupling constant squared. Alternatively, one could introduce $\hbar$ as a loop-counting parameter, and then an $L$-loop term would be accompanied by a factor $\hbar^{L-1}$: the anomaly has a $\hbar^0$, while a usual classical action gets multiplied by ${1\over\hbar}$. Thus an anomaly cancelling counterterm in the classical action must include an explicit factor of $\hbar$.

Another mechanism to cancel a relevant anomaly is available in certain geometric settings. Suppose our four-dimensional space-time $\cM_4$ is just the boundary of some five-dimensional space-time $\cM_5$, just as the one-dimensional circle is the boundary of the two-dimensional disc. We write $\cM_4=\del \cM_5$. Suppose that the (non-abelian) gauge fields actually live on the five-dimensional $\cM_5$, while the chiral matter fields only live on its boundary  $\cM_4$. This then leads to the usual gauge anomaly with 
\be
\dd \G[A] = -{1\over 24\pi^2} \int_{\cM_4} 
\trR \e\, \d \big( A\d A-{i\over 2} A^3\big)=
-{1\over 24\pi^2} \int_{\cM_4} Q_4^1(\e,A,F)\ .
\ee We may then add
a counterterm that {\it only depends on the gauge fields} but is defined on the five-dimensional $\cM_5$, namely
\be
\D\G[A]={1\over 24\pi^2} \int_{\cM_5} Q_5(A,F) \ ,
\ee
where $Q_5(A,F)$ is the Chern-Simons 5-form, related to $Q_4^1$ by the descent equation
\be
\dd Q_5 = \d Q_4^1 \ .
\ee
Then we have, using Stoke's theorem,
\be
\dd \D \G[A]= {1\over 24\pi^2} \int_{\cM_5} \dd Q_5(A,F) 
={1\over 24\pi^2} \int_{\cM_5} \d Q_4^1(\e,A,F)
={1\over 24\pi^2} \int_{\cM_4} Q_4^1(\e,A,F) = - \dd \G[A] \ ,
\ee
so that this five-dimensional counterterms indeed cancels the anomaly on the four-dimensional space-time $\cM_4$. This mechanism is called anomaly cancellation by {\it inflow}, as the relevant variation ``flows" from the five-dimensional bulk into the four-dimensional boundary.

In string and M-theory there are many occurrences of even-dimensional manifolds embedded in a higher-dimensional ``bulk" space. Typically, there are chiral fields living on the even-dimensional manifolds leading to gauge and/or gravitational anomalies. Complete cancellation of these anomalies often requires additional contributions generated by inflow from the bulk.  The classical example of Green-Schwarz anomaly cancellation in the type I or heterotic superstrings, on the other hand, involves a non-trivial transformation of a rank-2 antisymmetric tensor field, somewhat similar to the mechanism displayed in eqs (\ref{axioncounter}) to (\ref{axioncancel}). Let us look at this case in more detail.

\subsection{Anomaly cancellation in the (field theory limits of) type I $SO(32)$ and $E_8\times E_8$ heterotic superstrings\label{anocanhet}}

Consider now a ten-dimensional theory that is an $\cN=1$ supergravity coupled to $\cN=1$ super Yang-Mills theory with gauge group $G$.
This is the low-energy limit of type I or heterotic superstring theories \cite{GSW,POL}. The supergravity multiplet contains a positive chirality\footnote{
Of course, the overall chirality assignment is conventional, and we could just as well reverse {\it all} chiralities.
}
Majorana-Weyl  gravitino, a negative chirality spin ${1\over 2}$ fermion, as well as the graviton, a scalar (called dilaton) and a two-index antisymmetric tensor field $B_{\m\n}$ or equivalently two-form $B$. The super Yang-Mills multiplet contains the gauge fields $A_\m^\a$ and the gauginos $\chi^\a$ that are positive chirality Majorana-Weyl spin ${1\over 2}$ fields. Supersymmetry requires that the latter are in the same representation as the gauge fields $A_\m^\a$, namely in the adjoint representation.

An important point in the construction of this theory is that the $B$-field is not invariant under gauge (and local Lorentz) transformations. The consistent coupling of the supergravity and the super Yang-Mills theories requires $H=\d B-\b Q_3(\sa,\cF)$ to be gauge invariant, where $Q_3$ is the gauge Chern-Simons 3-form. (The precise value of the constant $\b$ depends on the representation over which the trace is taken to define the Chern-Simons 3-form and the normalization of the $B$ and $H$ fields. We fix the trace to be in some reference representation - for which we simply write $\tr$ - and then rescale $B$ and $H$ to set $\b=1$.) Although not visible at the tree-level, it turns out that one must also include a gravitational Chern-Simons form and
\be\label{HBrel}
H=\d B - Q_3(\sa,\cF)+\tilde\b\, Q_3(\o,R) \equiv \d B-Q_3^{\rm YM} + \tilde\b\, Q_3^{\rm L} \ ,
\ee
with $\tilde\b$ to be fixed below.
Thus $H$ will be invariant if
\be\label{Btransf}
(\dd^{\rm gauge}+\dd^{\rm L}) B
=Q_2^1(v,\sa,\cF) -\tilde\b\, Q_2^1(\vh,\o,R)
\equiv Q_2^{{\rm YM},\, 1}-\tilde\b\,  Q_2^{{\rm L},\, 1}\ .
\ee
It is then possible to construct a non-invariant counterterm, often called the Green-Schwarz term,
\be\label{GSterm}
\D\G=\int B\wedge X_8 \quad , \quad 
(\dd^{\rm gauge}+\dd^{\rm L}) X_8=0 \ ,
\ee
with some appropriate gauge and local Lorentz invariant closed 8-form $X_8$ constructed from the gauge and gravitational characteristic classes, i.e. from $\trR F^n$ and $\tr R^m$. As in sect.~\ref{desceq} we then have
\be
X_8=\d X_7 \quad ,\quad
(\dd^{\rm gauge}+\dd^{\rm L}) X_7 = \d X_6^1 \ .
\ee
Of course, this does not determine $X_7$ or $X_6^1$ uniquely. If e.g.
$X_8=\trR \cF^2 \tr R^2$ one could take $X_7=\l\, Q_3^{\rm YM} (\tr R^2)+(1-\l)\,(\trR \cF^2)\,   Q_3^{\rm L}$ for arbitrary $\l$. In any case, we have
\be\label{GScontrib}
(\dd^{\rm gauge}+\dd^{\rm L}) \D\G
= \int (Q_2^{{\rm YM},\, 1}- \tilde\b\, Q_2^{{\rm L},\, 1})\wedge X_8 
= \int (Q_2^{{\rm YM},\, 1}- \tilde\b\, Q_2^{{\rm L},\, 1})\wedge \d X_7
= - \int (\dd Q_3^{{\rm YM}}- \tilde\b\,\dd Q_3^{{\rm L}})\wedge X_7 \ .
\ee
Upon using the descent equations, this corresponds to an invariant 12-form polynomial
\be\label{deltaI}
\D \hat I_{12} = (\tr \cF^2 - \tilde\b\,\tr R^2) \wedge X_8 \ .
\ee
We conclude that 
\be
\begin{array}{|c|}
\hline
\\
\text{{\it If} the total anomaly polynomial $\hat I_{12}^{\rm total}$ does not vanish but takes a {\it factorized} form as in
(\ref{deltaI}) one}\\
\\
\text{can cancel the anomaly by adding the counterterm $\D\G$ of (\ref{GSterm}) with the appropriate coefficient.}\\
\nonumber\\
\hline
\end{array}
\ee
So let's see whether and when such a factorization occurs.

From the chiral field content we immediately get
\be
\hat I_{12}^{\rm total}= \hat I_{12}^{gravitino} - \hat I_{12}^{spin\, 1/2}\vert_{\cR=1} +\hat I_{12}^{spin\, 1/2}\vert_{\cR={\rm adj}} \ ,
\ee
with $\hat I_{12}^{gravitino}$ given in (\ref{I12}) and $\hat I_{12}^{spin\, 1/2}\vert_{\cR=1}$ given by (\ref{I12spinhalf}) with $\cF=0$ and $\trR 1 =1$, while $\hat I_{12}^{spin\, 1/2}\vert_{\cR={\rm adj}} $ is given by (\ref{I12spinhalf}) with $\cR$ being the adjoint representation. It is customary to write ${\rm tr}_{\rm adj}\equiv \Tr$ so that in particular $\Tr 1={\rm dim} G$. Thus we get
\ba\label{I12withgauge}
\hat I_{12}^{\rm total}&=&{1\over 64 (2\pi)^5} \left[ {{\rm dim}G-496\over 5670} \tr R^6 + {{\rm dim}G+224\over 4320} \tr R^4 \tr R^2 + {{\rm dim}G-64\over 10368} (\tr R^2)^3 \right]
\nonumber\\
&-&{1\over 32 (2\pi)^5} \big(\Tr \cF^2\big)
\left[{1\over360}\tr R^4
+{1\over288}(\tr R^2)^2\right]
+{1\over 1152 (2\pi)^5} \big(\Tr \cF^4\big)\ \tr R^2 
\ -\ {1\over 720 (2\pi)^5} \Tr \cF^6 \ .
\nonumber\\
\ea
Obviously, this cannot vanish, and and one has to rely on the Green-Schwarz mechanism to achieve anomaly cancellation. The latter requires (\ref{I12withgauge}) to have a factorized form like (\ref{deltaI}) so that it can be cancelled by addition of the countertem (\ref{GSterm}). Clearly, the $\tr R^6$ and the $\Tr \cF^6$ terms must be absent to have factorization. On the one hand, the $\tr R^6$ term is absent precisely if
\be\label{dimcond}
{\rm dim}G=496 \ .
\ee
On the other hand, the $\Tr \cF^6$ cannot vanish. However in certain cases it can be entirely be expressed as a combination of $\Tr \cF^4 \Tr \cF^2$ and $(\Tr \cF^2)^3$, so that one may get factorization if furthermore the coefficient have appropriate values.

\subsubsection{$SO(n)$ groups\label{SOngroups}}

Let us note without proof that for $G=SO(n)$ one has the following relations between the adjoint traces $\Tr$ and the traces in the fundamental (vector) representation, denoted $\tr$:
\be\label{SOntracerel}
\Tr \cF^2 = (n-2) \tr \cF^2 \ ,\quad
\Tr \cF^4 = (n-8) \tr \cF^4 + 3 (\tr \cF^2)^2 \ , \quad
\Tr \cF^6 = (n-32) \tr \cF^6 + 15 \tr \cF^2 \tr \cF^4 \ .
\ee
Thus we can reexpress the $\Tr \cF^6$ in terms of $\tr \cF^2 \tr \cF^4$ and $(\tr \cF^2)^3$, resp. in terms of $\Tr \cF^2 \Tr \cF^4$ and $(\Tr \cF^2)^3$ precisely if
\be
n=32 \ .
\ee
This singles out $SO(32)$ as only possiblity among the $SO(n)$-groups.
Amazingly, ${\rm dim} SO(32)=496$, so that (\ref{dimcond}) is also satisfied! As for the case studied in sect.~\ref{IIsugra}, anomaly cancellation imposes an overdetermined system of equations which nevertheless admits a solution. Let us now concentrate on $G=SO(32)$
and use the relations (\ref{SOntracerel}). 
Then (\ref{I12withgauge}) becomes
\ba\label{I12withgaugeA}
\hat I_{12}^{\rm total}&=&{1\over 64 (2\pi)^5} \left[ {1\over 6} \tr R^4 \tr R^2 + {1\over 24} (\tr R^2)^3 \right]
-{1\over 64 (2\pi)^5} \big(\tr \cF^2\big)
\left[{1\over 6}\tr R^4
+{5\over 24}(\tr R^2)^2\right]
\nonumber\\
&+&{1\over 384 (2\pi)^5} \Big[8\tr \cF^4+ (\tr \cF^2)^2\Big]\ \tr R^2 
-\ {1\over 48 (2\pi)^5} \tr \cF^4 \tr \cF^2
\nonumber\\
&=& {1\over 384 (2\pi)^5} \left[\tr R^2-\tr \cF^2\right]
\left[ \tr R^4 +{1\over 4} (\tr R^2)^2 -\tr \cF^2 \tr R^2 +8 \tr \cF^4 \right] \ ,
\ea
which is indeed of the required factorized form. Again was by no means obvious {\it a priori}. 

We see that it is then enough to choose $\tilde\b=1$ in (\ref{HBrel}) and (\ref{Btransf}), let the reference trace in (\ref{deltaI}) be the trace in the fundamental representation of $SO(32)$ and take
\be 
X_8= {1\over 384 (2\pi)^5} 
\left[ \tr R^4 +{1\over 4} (\tr R^2)^2 -\tr \cF^2 \tr R^2 +8 \tr \cF^4 \right] \ ,
\ee
to achieve
\be
\D \hat I_{12} + \hat I_{12}^{\rm total} =0 \ ,
\ee
i.e. cancel the anomaly by the Green-Schwarz mechanism.

\subsubsection{$E_8\times E_8$ and other groups\label{E8E8}}

We have already seen that among the $SO(n)$ groups, anomaly cancellation can only occur and does indeed occur for $n=32$. What about other groups ? In any case we need ${\rm dim}G=496$, but the group need not be simple, i.e. it can be a product of several simple and/or $U(1)$ factors. For $G=G_1\times G_2 \times \ldots$ one has for the traces in the adjoint representation $\Tr \cF^r=\Tr \cF_1^r + \Tr \cF_2^r + \ldots$. A simple example which has the right dimension is $G=E_8\times E_8$, since ${\rm dim} E_8=248$. Also, for each $E_8$ one has
\be
\Tr \cF^4 = {1\over 100}(\Tr\cF^2)^2 \quad , \quad
\Tr \cF^6={1\over 7200} (\Tr \cF^2)^3 \ .
\ee
Although $E_8$ does not have a ``vector" representation, it is useful to define  $\tr \cF^2$ simply by $\tr \cF^2 = {1\over 30} \Tr \cF^2$, so that the previous relations become
\be
\Tr \cF^4 = 9\,(\tr\cF^2)^2 \quad , \quad
\Tr \cF^6={75\over 20} (\tr \cF^2)^3 \ ,
\ee
and for $E_8\times E_8$
\be
\Tr \cF^4 =\Tr \cF_1^4 +\Tr \cF_2^4= 9\,(\tr\cF_1^2)^2 +9\,(\tr\cF_2^2)^2\quad , \quad
\Tr \cF^6=\Tr\cF_1^6 +\Tr\cF_2^6={75\over 20} (\tr \cF_1^2)^3+{75\over 20} (\tr \cF_2^2)^3 \ ,
\ee
Inserting this into (\ref{I12withgauge}) we get
\ba\label{I12withgaugebis}
\hat I_{12}^{\rm total}&=&{1\over 64 (2\pi)^5} \left[ {1\over 6} \tr R^4 \tr R^2 + {1\over 24} (\tr R^2)^3 \right]
-{1\over 64 (2\pi)^5} \big(\tr \cF_1^2+\tr\cF_2^2\big)
\left[{1\over 6}\tr R^4
+{5\over 24}(\tr R^2)^2\right]
\nonumber\\
&+&{1\over 128(2\pi)^5} \Big[(\tr \cF^2_1)^2+(\tr\cF_2^2)^2\Big]\ \tr R^2 
\ -\ {1\over 192 (2\pi)^5} \Big[(\tr \cF_1^2)^3+(\tr\cF_2^2)^3\Big] 
\ .
\ea
Again, this can be factorized as
\ba
\hat I_{12}^{\rm total}&=&
{1\over 384 (2\pi)^5} \Big[ \tr R^2-\tr\cF_1^2-\tr\cF_2^2\Big]
\Big[ \tr R^4 +{1\over 4} (\tr R^2)^2 - \tr R^2 (\tr\cF_1^2+\tr\cF_2^2) 
\nonumber\\
&&\hskip6.3cm
-2\tr\cF_1^2\tr\cF_2^2 + 2(\tr\cF_1^2)^2 + 2 (\tr\cF_2^2)^2 \Big] 
\ \ ,
\ea
so that one achieves anomaly cancellation by choosing the Green-Schwarz term, again with $\tilde\b=1$, but now with an 
\be\label{X8het}
X_8={1\over 384 (2\pi)^5}\Big[ \tr R^4 +{1\over 4} (\tr R^2)^2 - \tr R^2 (\tr\cF_1^2+\tr\cF_2^2) 
-2\tr\cF_1^2\tr\cF_2^2 + 2(\tr\cF_1^2)^2 + 2 (\tr\cF_2^2)^2 \Big] 
\ .
\ee

There are two more possibilities to achieve factorization of 
$\hat I_{12}^{\rm total}$ which we now briefly discuss. Clearly, we must keep ${\rm dim G}=496$.  One possibility then is to take $G=E_8\times U(1)^{248}$, the other being $G=U(1)^{496}$. Now, for any product of $U(1)$ groups, the traces are always $\trR F^k=\sum_s \sum_i (q_i^{(s)})^k$ with $q_i^{(s)}$ being the charge of the $i^{\rm th}$ particle with respect to the $s^{\rm th}$ $U(1)$. Here, however, the gauginos are in the adjoint representation which means that $q_i^{(s)}=0$. (``Photinos" just as photons carry no electric charge.) Hence $\Tr_{\text{adj of $U(1)^{248}$}} F^k=0,\ \forall k>0$. As a result,  we can use {\it all} formula previously established for $E_8\times E_8$ and simply set $\tr \cF_2^k=0,\ \forall k>0$, if the gauge group is $G=E_8\times U(1)^{248}$ or  set $\tr\cF^k_1=\tr \cF_2^k=0,\ \forall k>0$, if the gauge group is $G=U(1)^{496}$. In particular, $X_8$ is still given by (\ref{X8het}) with the appropriate replacements. One can actually scan all 496-dimensional groups and show that there are no other possibilities to achieve factorization of $\hat I_{12}^{\rm total}$.

It turns out that for type I superstrings, the gauge group $SO(32)$
is also singled out by other typically string-theoretic consistency conditions, while for the heterotic superstring, similar stringy consistency conditions single out $SO(32)$ and $E_8\times E_8$. On the other hand, the somewhat more trivial solutions to the anomaly cancellation conditions, $E_8\times U(1)^{248}$ and $U(1)^{496}$ do not seem to correspond to any consistent superstring theory.

\section{Concluding remarks\label{Conclusions}}

We hope to have conveyed the idea that anomalies play an important role in quantum field theory and their cancellation has been and still is a valuable guide for constructing coherent theories. The more formal treatment of anomalies makes fascinating contacts with several branches of modern mathematics.

\vskip1.5cm
\noindent
{\large\bf Acknowledgements}
\vskip2.mm
\noindent
We gratefully acknowledge support by the EU grants MRTN-CT-2004-005104 and MRTN-CT-2004-512194, as well as by the french ANR grant ANR(CNRS-USAR) no.05-BLAN-0079-01.

\appendixA{}
\section{Appendix : An explicit two-dimensional illustration of the index theorem\label{2dindex}}

In this appendix, we will explicitly compute the index of the Dirac operator on  two-dimensional flat Euclidean space in a $U(1)$ instanton background and show how it indeed coincides with the general prediction of the index theorem.

We use cartesian coordinates $x$ and $y$, or equivalently polar coordinates $r=\sqrt{x^2+y^2}$ and $\vf=\arctan {y\over x}$. A two-dimensional $U(1)$ instanton configuration should asymptote to a pure gauge for large $r$, i.e. $A\sim g^{-1} i \d g$ with $g(\vf)=e^{i m \vf}$, $m\in {\bf Z}$. Thus a non-singular configuration with the appropriate asymptotics is 
\be
A=-m{r^2\over r^2+1}\d\vf={m\over x^2+y^2+1}(y\d x-x\d y)
\quad\Rightarrow\quad
F=-2m{r\d r\d\vf\over (r^2+1)^2} = -2m {\d x\d y\over (x^2+y^2+1)^2} \ .
\ee
Note that $F={1\over 2} F_{ij}\,\d x^i \d x^j=F_{12}\,\d x \d y$ so that $F_{12}=-2m{1\over (x^2+y^2+1)^2}$. It is straightforward to compute the integral of $F$ over the plane:
\be
{1\over 2\pi} \int_{{\bf R}^2} F = - m \ .
\ee

Next we study the Dirac operator in this background gauge field. In two Euclidean dimensions we take $\g^1=\s_x$, $\g^2=-\s_y$ so that the chirality matrix is $\g_{\rm E}=i \g^1\g^2=-i\s_x\s_y=\s_z$, and
\be
\Dsl=\s_x (\del_x-i A_x) -\s_y (\del_y-i A_y)
= \begin{pmatrix}
0 & \del_x+i\del_y -i A_x + A_y \\
\del_x-i\del_y - i A_x-A_y & 0 \\
\end{pmatrix} \ .
\ee
Using the explicit form of the gauge field $A_x=m{y\over x^2+y^2+1}$ and $A_y=-m{x\over x^2+y^2+1}$, and switching to complex coordinates
$z=x+i y$, $\zb=x-i y$, we simply get
\be
\Dsl=\begin{pmatrix}
0 & 2\del_\zb - m{z\over z\zb+1}\\
2\del_z + m{\zb\over z\zb+1} & 0 \\
\end{pmatrix} \ .
\ee
Let us \underline{assume that $m>0$.}
A positive chirality zero-mode $\begin{pmatrix} \chi \\ 0 \\ \end{pmatrix}$ of $\Dsl$ satisfies
\be
2\del_z\chi + m {\zb\over z \zb+1} \chi=0 \ .
\ee
The solutions of this differential equation are
\be
\chi(z,\zb)={g(\zb)\over (1+z\zb)^{m/2}} \ ,
\ee
with $g$ a holomorphic function of its argument. These solutions
are square normalizable, $\int |\chi|^2 \d z\d\zb <\infty$, only if the holomorphic function $g$ is a polynomial of degree $m-2$ at most and ``marginally" square normalizable if $g$ is a polynomial of degree $m-1$. This has $m$ independent coefficients, yielding $m$ linearly independent zero-modes of positive chirality. On the other hand, a negative chirality zero-mode 
$\begin{pmatrix} 0 \\ \hat\chi \\ \end{pmatrix}$ of $\Dsl$ satisfies
\be
2\del_\zb \hat\chi - m {z\over z \zb+1} \hat\chi=0 \ ,
\ee
with solutions
\be
\hat\chi(z,\zb)=h(z) (1+z\zb)^{m/2} \ ,
\ee
again with a holomorphic $h(z)$. Obviously, for $m\ge0$, none of these zero-modes is (marginally) nomalizable.

It follows that the index of $i\Dsl$, which is the number of linearly independent, (marginally) normalizable positive chirality zero-modes minus the number of linearly independent, (marginally) normalizable negative chirality zero-modes is $m-0=m$.
This result was derived for $m> 0$. For $m<0$, the roles of positive and negative chirality  are reversed: there are no positive chirality (marginally) normalizable zero-modes, while there are $|m|=-m$ negative chirality ones, yieding again an index $0-|m|=m$. Thus:
\be
{\rm ind}(i\Dsl)=m=-{1\over 2\pi} \int F= -{1\over 4\pi} \int \e_{\rm E}^{\m\n} F_{\m\n} \ ,
\ee
in agreement with eq.~(\ref{index2reuc}) for $r=1$.

\vskip0.5cm

\end{document}